%% file: SuspensionPIV.tex
\RequirePackage{fix-cm}
\RequirePackage{amsmath}
\documentclass[twocolumn]{article}

\usepackage{graphicx}
\usepackage{latexsym}
\usepackage{tikz}
\usepackage{tikzscale}
\usepackage{pgfplots}
\pgfplotsset{compat=newest}
\usepackage{epsfig}
\usepackage[round,semicolon,authoryear]{natbib}
\usepackage{url}
\usepackage{amsmath}
\usepackage{amsfonts}
\usepackage{amssymb}
\usepackage{siunitx}
\usepackage{graphicx}
\usepackage{color}
\usetikzlibrary{arrows.meta}
\usetikzlibrary{arrows}
\usepackage[defaultlines=4,all]{nowidow}
\usepackage[misc]{ifsym}
\usepackage{selinput}
\usepackage{bm}
\usepgfplotslibrary{external}
\usepackage{changes}
\usepackage{paralist}
\usetikzlibrary{patterns}
\usepackage{hyperref}
\usepackage{caption}
\usepackage{subcaption}
\usepackage{upgreek}
\usepackage{textgreek}
\usepackage{selinput}
\SelectInputMappings{adieresis={ä}, germandbls={ß}}
\DeclareSIUnit[]
\pixel{px}

\usepackage{titlesec}
\titleformat*{\section}{\large\bfseries}
\titleformat*{\subsection}{\normalsize\bfseries}
\titleformat*{\subsubsection}{\normalsize\bfseries}
\titleformat*{\paragraph}{\normalsize\bfseries}
\titleformat*{\subparagraph}{\normalsize\bfseries}

\begin{document}
\title{On the micro PIV accuracy and reliability utilizing suspension particles of large, non-Gaussian particle image}

\author{Sebastian Blahout\(^{1}\)\footnote{\Letter\ blahout@sla.tu-darmstadt.de} , Simon R. Reinecke\(^2\), Harald Kruggel-Emden\(^2\), Jeanette Hussong\(^1\)\\\\
\(^1\) Institute for Fluid Mechanics and Aerodynamics\\
Technical University of Darmstadt\\
64287 Darmstadt, Germany\\\\
\(^2\) Chair of Mechanical Process Engineering and Solids Processing\\
Technische Universität Berlin\\
10587 Berlin, Germany
}

\date{Dated: \today}

\maketitle

\begin{abstract}
Optical investigations on the dynamics of dense suspensions are challenging due to reduced optical accessibility. Furthermore, the suspension particle image size can strongly deviate from the optimal particle image size for PIV measurements. Optical accessibility can be achieved by refractive index matching of surface labelled suspension particles. This results in particle images that are transparent in the particle image center, but fluoresce at the particle image rim, resulting in ring-shaped particle images. In the present study the influence of particle image size of such ring-shaped particle images is compared with Gaussian and plateau-shaped particle images. Particles of Gaussian image shape result from fully labelled particles with small image diameters and are commonly used in PIV measurements. Such particles are also utilized for the determination of the continuous phase velocities in the experimental part of the present study. With increasing image diameter, fully labelled particles are observed to assume plateau-shaped particle images.\\
Monte Carlo simulations show that ring-shaped particle images have a superior behavior, i.e. they assume a reduced displacement estimation error for noisy as well as for noise-free image data, compared to Gaussian and plateau-shaped particle images. This is also true for large particle image diameters when particle images are intersected at interrogation window borders.\\
The detectability is similar for all three particle image shapes as long as particles do not intersect with the interrogation window border. Interestingly, for intersected particles of large image diameter, ring-shaped particle images show a slightly improved detectability compared to particle images of Gaussian and plateau shape. Combined with a good optical accessibility, this allows to perform simultaneous cross-correlation evaluations on large ring-shaped particle images and fluid tracers with Gaussian particle images that are two orders of magnitude smaller compared to suspension particle images.\\
Consequently, ring-shaped particle images are successfully utilized to measure the suspension bulk dynamics by means of micro Particle Image Velocimetry (\textmu PIV). Measurements are performed on a suspension containing 5 Vol.-\% surface labelled, refractive index matched \SI{60}{\micro\meter} PMMA particles. Simultaneously, \textmu PIV measurements of the carrier liquid flow are performed utilizing \SI{1.19}{\micro\meter} fluorescent PS particles. Individual velocity profiles of suspension particles and the carrier liquid are determined to calculate slip velocities between the suspension phase and the carrier liquid. Measurement results reveal a parabolic shape of the velocity profiles of both phases with a slip velocity of 7.4\% at the position of maximum streamwise velocity in a \SI{533}{\micro\meter} high trapezoidal channel.\\
Overall, the present study demonstrates theoretically and experimentally that the usage of suspension particles with ring-shaped images is suitable for \textmu PIV measurements to gain detailed insights into suspension bulk dynamics.
\end{abstract}

\section{Introduction}\label{intro}
An established method for non-invasive flow measurements is the so called Particle Image Velocimetry (PIV). If optical accessibility is granted, it can be applied to any type of particle seeded flow to trace the fluid motion, if particles behave as ideal fluid tracers. If the particle size is large compared to the characteristic channel dimensions, slip and migration effects may occur and PIV can be utilized to measure the bulk dynamics of particles themselves in a carrier liquid. PIV is commonly applied to situations of high particle seeding density. It is suitable to measure transient flow fields, also in combination with other optical measurement techniques \citep{Skarman1996a,Funatani2004,Kordel2016a}. In classical PIV applications, fluid tracers should assume a small image diameter of only a few pixels \citep{Willert1996a,Westerweel1997b}. Without astigmatism effects their intensity profile will be typically Gaussian \citep{Adrian1991,Willert1991a}. The accuracy and reliability of PIV measurements for Gaussian particle images is a function of various parameters, such as the amount of particles per interrogation window, the particle image density, the particle image diameter, the signal to noise ratio, out-of-plane as well as in-plane loss-of-pairs and velocity gradients \citep{Adrian1991,Willert1996a,Westerweel1997b}.\\
A measure for the accuracy of a cross-correlation result is the displacement estimation error. It is known to minimize for particle image diameters of \mbox{\(D_{PI} = 2 - \SI{3}{\pixel}\)}, while it grows for larger diameters due to random errors \citep{Westerweel1997b}. While PIV is typically used to quantify flow fields, it is applicable to any kind of displacement field as long as the particle image shape and particle group formation do not change significantly between correlated frames.\\
A measure for the reliability of PIV measurements is the detectability, which is the ratio of the highest to the second highest correlation peak value. It can be understood as the probability, that the highest correlation peak corresponds to the real particle image displacement \citep{Adrian1991}. Detailed information regarding the detectability and the displacement estimation error are given in Sect. \ref{sec:Theory}.\\
Due to a reduced optical accessibility of suspension flows, PIV measurements of multiphase flows were often limited to dilute suspensions \citep{Koutsiaris1999b} or bubble flows of low gas volume fractions \citep{Lindken2002}. For optical investigations of suspension flows with higher volume fractions, a refractive index matching of the liquid to the solid phase can be done to increase the optical accessibility \citep{Wiederseiner2011a,Blanc2013a}. The concept of refractive index matching can be applied to every liquid-solid combination with a transparent liquid and solid phase. Solely difficulties in handling of highly flammable or toxic liquids may restrict the choice of materials \citep{Hassan2008a,Wiederseiner2011a}. Hence, this concept was already successfully applied for Particle Tracking Velocimetry (PTV) \citep{Wang2008a}, Laser Doppler Velocimetry (LDV) \citep{Haam2000b}, Laser Induced Fluorescence (LIF) \citep{Chen2005a} or Astigmatism Particle Tracking Velocimetry (APTV) measurements in multiphase flows. An advantage of the PIV measurement technique compared to other optical, non-invasive measurement techniques like, e.g., the APTV approach is that no sophisticated calibration technique is required and all suspension particles can be labelled and, therefore, contribute to the measurement result. In the present study, we apply refractive index matching of a ternary carrier fluid to suspension particles to gain optical access. This mixture is used as a basis for synthetic particle image generation (see Sect. \ref{sec:SIG}) as well as for suspension flow measurements (see Sect. \ref{sec:Experiments}.)\\
To assess the suitability of cross-correlation based measurement techniques like PIV for measurements in dense suspensions, we evaluate the displacement estimation error and the detectability of Gaussian, ring- and plateau-shaped particle images of different particle image diameters by means of Monte Carlo simulations that are based on synthetically generated particle images (see Sects. \ref{sec:PID} and \ref{sec:ResDetectability}). These show that the accuracy and reliability of particles with large image diameter strongly depends on the particle image shape. While a Gaussian intensity distribution is characteristic for small particle image diameters \citep{Willert1991a}, particles assume rather a plateau-shaped intensity distribution for large image diameters (see also Sect. \ref{sec:SIG}). To separate the size effect from the shape effect for PIV measurements, we study the behaviour of both, Gaussian and plateau-shaped particle images over the whole investigated image size regime. Ring-shaped intensity profiles may emerge for refractive index matched particles with fluorescent surface labelling. This situation may be encountered for optical investigations of dense particle laden flows but also in situations where particles naturally contain mainly carrier liquid such as, e.g., hydrogel particles \citep{Byron2013a} or cells \citep{Lima2006b}.\\
For small particle images, an in-plane loss-of-pairs refers to particles that are located only in one interrogation window of the first or second frame \citep{Keane1992}. However, when considering large particle image sizes, the contribution of intersected particle images may become relevant, too. This effect is investigated for Gaussian, ring- and plateau-shaped particle images in Sect. \ref{sec:PID-Cut}.\\
To understand the dynamics of suspended particles, the relative motion between particles and surrounding fluid must be evaluated. Relative motion may, e.g., occur when the particle diameter is in the order of magnitude of the characteristic channel dimensions and, therefore, is much larger than for tracer particles commonly used in PIV measurements. To demonstrate the suitability of utilizing ring-shaped particle images to measure such suspension dynamics, micro Particle Image Velocimetry (\textmu PIV) measurements are performed (see Sect. \ref{sec:Experiments}). As, additionally, small tracer particles are suspended to the flow, a simultaneous evaluation of the velocity profiles of both, the continuous and the particulate phase is possible and slip velocities can be calculated.

\section{Theoretical background}\label{sec:Theory}
This section deals with the mathematical description of the cross-correlation of two images. The choice of a suitable peak fit estimator is discussed for the determination of displacement vectors (Sect. \ref{sec:CC&PeakFitEstimator}). Afterwards, the concept of detectability and displacement estimation error are introduced (Sect. \ref{sec:Validation}).

\subsection{Cross-correlation and peak fit estimator}\label{sec:CC&PeakFitEstimator}
Two singly exposed images \(I_1(\textbf{X})\) and \(I_2(\textbf{X})\) recorded at times \(t_0\) and \mbox{\(t_0 + \Delta t\)} are considered, where \(\textbf{X}\) is a matrix that contains the center positions of all particle images in the corresponding image such that the cross-correlation \(R(s)\) of the two images \(I_1(\textbf{X})\) and \(I_2(\textbf{X})\) can be defined as \citep{Keane1992}:
\begin{equation}
R(\textbf{s}) = \int_{W_1} I_1(\textbf{X}) I_2(\textbf{X}+\textbf{s})\ d\textbf{X}\label{eq:CC}
\end{equation}
\mbox{}\\
Here, \(\textbf{s}\) is the separation vector, by which the second image \(I_2(\textbf{X})\) has to be shifted to match the particle positions of the first image, i.e. \mbox{\(\textbf{s} = - \Delta \textbf{X}_P\)}. According to \citet{Keane1992} the estimator for the cross-correlation can be decomposed into three characteristic parts for singly exposed double-frame images:
\begin{equation}
R(\textbf{s}) = R_C(\textbf{s}) + R_D(\textbf{s}) + R_F(\textbf{s})\label{eq:CC-decomposed}
\end{equation}
\mbox{}\\
\(R_C(\textbf{s})\) and \(R_F(\textbf{s})\) contain the convolution of the mean intensities and the fluctuating noise of \(I_1(\textbf{X})\) and \(I_2(\textbf{X})\), respectively. \(R_D(\textbf{s})\) results from the relative displacement of particle images between \(I_1(\textbf{X})\) and \(I_2(\textbf{X})\) and is hence on referred to as displacement correlation peak. As shown by \citet{Willert1991a}, \(R_D(\textbf{s})\) assumes a Gaussian peak shape for particle images of Gaussian intensity distribution. To increase the detection accuracy of the center position of the displacement correlation peak and therefore of the whole PIV measurement to sub pixel range, an appropriate peak fit estimator can be used \citep{Willert1991a}.\\
Usually, a Gaussian peak fit estimator \(\hat{\epsilon}_G\) is used to interpolate the center position of \(R_D(\textbf{s})\) \citep{Westerweel1997b}:
\begin{equation}
\hat{\epsilon}_G = \frac{\ln \left( R_{-1}\right) - \ln\left(R_{+1}\right)}{2 \left[\ln\left(R_{-1} \right) + \ln\left(R_{+1} \right) - 2 \ln\left(R_0 \right)\right]},\label{eq:GaussPeakFit}
\end{equation}
\mbox{}\\
with \mbox{\(R_0 = R_{D}(\Delta \textbf{X}_p)\)} and \mbox{\(R_{\pm1} = R_{D}(\Delta \textbf{X}_p \pm 1)\)}. A Gaussian peak fit estimator is generally used as it is more insensitive to peak-locking effects compared to other peak fit estimators.

\subsection{Detectability and displacement estimation error}\label{sec:Validation}
The quality of PIV measurements or more precisely, the accuracy with which the displacement correlation peak position is determined and the probability that the position of a displacement correlation peak is equivalent to the real particle image displacement, can be evaluated through the displacement estimation error and the detectability. Thus, they are a measure to describe the accuracy and the reliability of a cross-correlation result.\\
The detectability of a displacement correlation peak can be interpreted as the probability that a displacement correlation peak is correctly identified as valid and, therefore, corresponds to the real particle image displacement. It is defined as the ratio of the highest to the second highest correlation peak value of a correlation map \(R(\textbf{s})\) \citep{Coupland1988a,Keane1990a,adrian2011particle}:
\begin{equation}
D = \frac{R(s)_{max,1}}{R(s)_{max,2}}\label{eq:D}
\end{equation}
\mbox{}%
The displacement estimation error can be expressed as the variance of its corresponding peak fit estimator. For a Gaussian peak fit estimator it can be described in a general mathematical form as \citep{Westerweel1997b}:
\begin{equation}
\text{var}\left\lbrace\hat{\epsilon}_G \right\rbrace \approx \sum_{i=-1}^{+1} \sum_{j=-1}^{+1} \frac{\partial \epsilon_G}{\partial R_i} \frac{\partial \epsilon_G}{\partial R_j}\ \text{cov} \left\lbrace R_i, R_j \right\rbrace
\label{eq:EE-general}
\end{equation}
\mbox{}\\
In case of a non-fractional displacement, (\ref{eq:EE-general}) can be expressed as follows:
\begin{equation}
\text{var} \left\lbrace \hat{\epsilon}_G \right\rbrace \approx \epsilon_1 \cdot \epsilon_2,
\label{eq:EE}
\end{equation}
\mbox{}\\
with both terms \(\epsilon_1\) and \(\epsilon_2\) representing different properties of the displacement correlation peak.\\
The first term \(\epsilon_1\) is the squared derivative of the Gaussian peak fit estimator \(\hat{\epsilon}_G\) given in (\ref{eq:GaussPeakFit}):
\begin{align}
\epsilon_1&=\left(\frac{\partial \epsilon_G}{\partial R_{\pm 1}} \right) ^2 =\nonumber \\  
&\left(\frac{4 \cdot \left[\ln(R_{\mp 1})-\ln(R_{0})\right]}{R_{\pm 1} \left[2\cdot\ln(R_{\pm 1}) - 4 \ln(R_{0}) + 2 \cdot \ln(R_{\mp 1}) \right]^2}\right)^2 \label{eq:EE-1}
\end{align}
\mbox{}\\
If the correlation peak flattens out, i.e. \mbox{\(R_{\pm1} \rightarrow R_0\)}, the denominator of (\ref{eq:EE-1}) approaches zero sooner than the nominator and \(\epsilon_1\), and therefore also \mbox{\(\text{var} \left\lbrace \hat{\epsilon}_G\right\rbrace\)}, goes to infinity.\\
The second term \(\epsilon_2\) reads:
\begin{equation}
\epsilon_2 = \left[\text{var}\left\lbrace R_{-1}\right\rbrace + \text{var} \left\lbrace R_{+1} \right\rbrace - 2 \text{cov} \left\lbrace R_{-1}, R_{+1} \right\rbrace \right]
\end{equation}
\mbox{}\\
It is a measure of (i) the spread of the normalized correlation peak slope i.e. its width as well as (ii) the spread of the correlation peak symmetry of an ensemble. For an ensemble with only perfectly symmetric correlation peaks (that is \mbox{\(R_{+1}=R_{-1}\)}) the second term and (\ref{eq:EE}), respectively, will approach zero for a finite correlation peak width.

\section{Synthetic particle image generation}\label{sec:SIG}
\begin{figure}[ht]
    \flushright
    \begin{subfigure}{.3\linewidth}
        \subcaption{}
        \includegraphics[width=.9\textwidth]{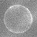}
        \label{fig:CompSI_subfig:RawRing}
    \end{subfigure}%
    \begin{subfigure}{.3\linewidth}
        \subcaption{}
        \includegraphics[width=.9\textwidth]{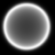}
        \label{fig:CompSI_subfig:SynthRing}
    \end{subfigure}%
    \begin{subfigure}{.3\linewidth}
        \subcaption{}
        \includegraphics[width=.9\textwidth]{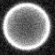}
        \label{fig:CompSI_subfig:SynthRingNoise}
    \end{subfigure}\\%
    \vspace{0.2em}%
    \begin{subfigure}{.3\linewidth}
        \subcaption{}
        \includegraphics[width=.9\textwidth]{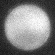}
        \label{fig:CompSI_subfig:RawPlateau}
    \end{subfigure}%
    \begin{subfigure}{.3\linewidth}
        \subcaption{}
        \includegraphics[width=.9\textwidth]{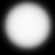}
        \label{fig:CompSI_subfig:SynthPlateau}
    \end{subfigure}%
    \begin{subfigure}{.3\linewidth}
        \subcaption{}
        \includegraphics[width=.9\textwidth]{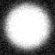}
        \label{fig:CompSI_subfig:SynthPlateauNoise}
    \end{subfigure}\\%
    \vspace{0.2em}%
    \begin{subfigure}{.3\linewidth}
        \subcaption{}
        \includegraphics[width=.9\textwidth]{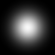}
        \label{fig:CompSI_subfig:SynthGauss}
    \end{subfigure}%
    \begin{subfigure}{.3\linewidth}
        \subcaption{}
        \includegraphics[width=.9\textwidth]{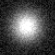}
        \label{fig:CompSI_subfig:SynthGaussNoise}
    \end{subfigure}%
    \caption{\subref{fig:CompSI_subfig:RawRing} Recorded, ring-shaped particle image before preprocessing; \subref{fig:CompSI_subfig:SynthRing} synthetically generated, ring-shaped particle image; \subref{fig:CompSI_subfig:SynthRingNoise} synthetically generated, ring-shaped particle image with 8.5\% image noise; \subref{fig:CompSI_subfig:RawPlateau} Recorded, plateau-shaped particle image; \subref{fig:CompSI_subfig:SynthPlateau} synthetically generated, plateau-shaped particle image; \subref{fig:CompSI_subfig:SynthPlateauNoise} synthetically generated, plateau-shaped particle image with 8.5\% image noise; \subref{fig:CompSI_subfig:SynthGauss} synthetically generated, Gaussian particle image; \subref{fig:CompSI_subfig:SynthGaussNoise} synthetically generated, Gaussian particle image with 8.5\% image noise.}
    \label{fig:CompSI}
\end{figure}

\begin{figure}[ht]
\input{Figures/SIG/RadIntFunc.tikz}
\caption{Radial, fitted intensity distributions of Gaussian, ring- and plateau-shaped particle images.}
\label{fig:RadIntFuncs}
\end{figure}
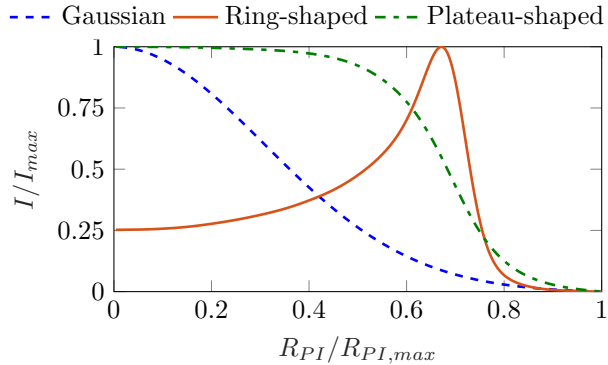

\begin{figure}[ht]
\input{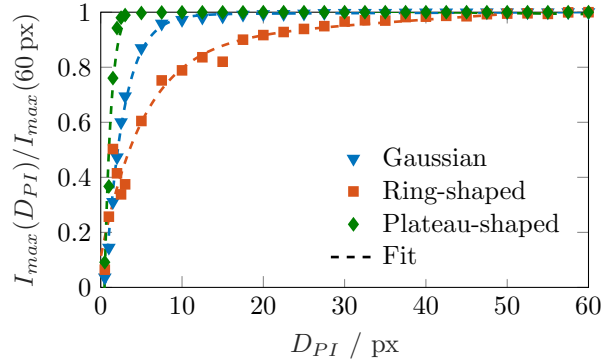}
\caption{Maximum intensity values as a function of the particle image diameter for Gaussian, ring- and plateau-shaped particle images.}
\label{fig:MaxIntVal}
\end{figure}

\begin{figure}[ht]
    \begin{subfigure}[b]{.475\linewidth}
        \centering%
        \subcaption{}%
        \input{Figures/SIG/Final_Halo_PID=60.tikz}%
        \label{fig:SynthImages_subfig:PID60}%
    \end{subfigure}\hspace{.5em}%
    \begin{subfigure}[b]{.475\linewidth}
        \centering%
        \subcaption{}%
        \input{Figures/SIG/Cut50-3_Halo_PID=60.tikz}%
        \label{fig:SynthImages_subfig:PID60-Cut}%
    \end{subfigure}\hfill%
    \begin{subfigure}[t]{1\columnwidth}%
        \centering%
        \input{Figures/SIG/Colorbar_BlackWhite.tikz}%
    \end{subfigure}
    \caption{Synthetic sample images of \subref{fig:SynthImages_subfig:PID60} five ring-shaped particle images and \subref{fig:SynthImages_subfig:PID60-Cut} five ring-shaped particle images including three intersected particle images (\mbox{\(K_{5} = 3\)}).}
    \label{fig:SynthImages}
\end{figure}
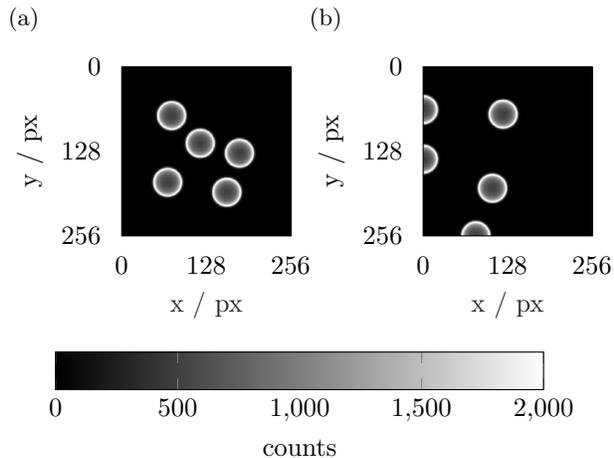

In the present study we investigate the influence of the particle image diameter on the cross-correlation result for different particle image shapes. For this, synthetic particle images are generated with Gaussian, ring- and plateau-shaped intensity profiles. While Gaussian intensity profiles are based on an analytical function, ring- and plateau-shaped profiles are taken from fit functions of experimental image data. For this, self-labelled and commercially labelled PMMA particles with a nominal particle diameter of \mbox{\(D_p = \SI{60}{\micro\metre}\)} are suspended in a carrier liquid and recorded under laser illumination. Examples of both, ring- and plateau-shaped particle images, are shown in Fig. \ref{fig:CompSI_subfig:RawRing} and \subref{fig:CompSI_subfig:RawPlateau}, respectively.
\mbox{}\\
Recorded particle images assume a diameter of \mbox{\(D_{PI} \approx \SI{70}{\pixel}\)}. Obviously, self-labelled PMMA particles assume a ring-shaped particle image, while commercially labelled particles show a plateau-shaped intensity profile. To achieve ring-shaped particle images, PMMA particles are labelled with a molecular Rhodamine B dye that is suspended in distilled water. As these particles have a limited take up of dye, only the particle surface is labelled. When the particles are suspended in a refractive index matched liquid consisting of distilled water, Glycerine and Ammoniumthiocyanate \citep{Bailey2003a}, they appear transparent with a fluorescent particle rim. In contrast to this, the material of commercially available particles is completely labelled, resulting in a plateau-shaped particle image.\\
Radial intensity distributions of ring- and plateau-shaped particle images are determined from approximately 500 recorded, individual particle images by fitting a smoo\-thing\--spline to all intensity distributions. With that, different synthetic particle image diameters are realized for synthetic data generation of ring-shaped and plateau-shaped particle images by scaling the smoothing-spline function. To create synthetic Gaussian particle images, the Gaussian density function is used:
\begin{equation}
I(x) = \frac{1}{\sqrt{2 \pi \sigma^2}} \cdot \exp\left(-0.5 \cdot\frac{\left(x-\mu\right)^2}{\sigma^2}\right),\label{eq:GaussFunc}
\end{equation}
\mbox{}\\
with \mbox{\(\mu = 0\)}. The parameter \(\sigma^2\) is adjusted iteratively, so that the width of the Gaussian density function corresponds to the prescribed particle image diameter. This point is defined to be where the Gaussian curve reaches \(0.5\%\) of its maximum value. Finally, this threshold value is subtracted from the resulting shape function to reach a zero intensity value at \mbox{\(R_{PI}/R_{PI,max} = 1\)}. The resulting radial intensity distributions \(I\), normalized by the corresponding maximum intensity \(I_{max}\) of all three particle image shapes are shown in Fig. \ref{fig:RadIntFuncs}.
\mbox{}\\
To create particle images of different discrete pixel sizes, radial intensity distributions are intersected and piecewise averaged intensity values are assigned to the corresponding pixel locations. Synthetic ring- and plateau-shaped particle images of \mbox{\(D_{PI} = \SI{60}{\pixel}\)} are shown in Fig. \ref{fig:CompSI_subfig:SynthRing} and \subref{fig:CompSI_subfig:SynthPlateau}, respectively. As experimental data is usually affected by image noise, both, synthetic particle images without (see Figs. \ref{fig:CompSI_subfig:SynthRing}, \subref{fig:CompSI_subfig:SynthPlateau}, \subref{fig:CompSI_subfig:SynthGauss}) and with \(8.5\%\) image noise (see Figs. \ref{fig:CompSI_subfig:SynthRingNoise}, \subref{fig:CompSI_subfig:SynthPlateauNoise}, \subref{fig:CompSI_subfig:SynthGaussNoise}), are analysed and compared in the course of this paper. Image noise is calculated as the ratio of the mean background intensity and the maximum intensity value of images with a particle image diameter of \mbox{\(D_{PI} = \SI{60}{\pixel}\)}. Noise levels in the order of magnitude as mentioned above are encountered, even after pre-processing of raw recordings of suspension flows with ring-shaped particle images.\\
It may be noted, that assigning a continuous intensity function (as given in Fig. \ref{fig:RadIntFuncs}) to discrete pixels leads to a reduction of the maximum particle intensity values for small particle image diameters. Fig. \ref{fig:MaxIntVal} shows the intensity maxima as a function of the particle image diameter.
\mbox{}\\
As can be seen, this discretisation error becomes prominent for small particle images of a few pixels diameter and is strongest for ring-shaped particle images due to a very sharp intensity peak at \mbox{\(R_{PI}/R_{PI,max} \approx 2/3\)}, evident in Fig. \ref{fig:RadIntFuncs}. However, in real experimental situations where e.g. noise-affected ring-shaped particle images give indeed a weaker fluorescence signal for identical laser energy input compared to noise-affected plateau-shaped particle images, the cross-correlation result may be significantly affected for particle image diameters smaller than \mbox{\(D_{PI} = \SI{10}{\pixel}\)}. Further details are discussed in Sect. \ref{sec:Results}.\\
To evaluate the size sensitivity of Gaussian, ring- and plateau-shaped particle images on the cross-cor\-re\-la\-tion result, 500 double-frame images with five randomly di\-stri\-bu\-ted particle images in each interrogation window and zero displacement between corresponding frames are created synthetically. Since neither an in-plane loss-of-pairs \(F_I\) nor an out-of-plane loss-of-pairs \(F_O\) is present (\mbox{\(F_I = F_O = 1\)}), this results in a constant effective number of particle images of \mbox{\(N_I F_I F_O = N_I = 5\)} for all particle image diameters, which is in agreement to the recommended value for PIV measurements \citep{Keane1992}. The displacement estimation error is evaluated based on (\ref{eq:EE}). In this way, errors resulting from a finite particle image displacement \citep{Westerweel1993} as well as in-plane and out-of-plane loss-of-pairs are excluded. For all investigations the interrogation window size was chosen to be \mbox{\(256 \times \SI{256}{\pixel}\)}. At first, overlapping particle images as well as particle images intersected at the interrogation window border, are suppressed to isolate the effect of the particle image size and shape on the cross-correlation result (see Sects. \ref{sec:PID} and \ref{sec:ResDetectability}).\\
The influence of particle images located on the interrogation window border is investigated separately in Sect. \ref{sec:PID-Cut}. For this, sets of double frame images with one, three and five out of five particle images are placed with their center point on the interrogation window border. Particle images inside the interrogation window and those on its border are distributed randomly whereas particle overlaps are excluded again. Examples of synthetically generated ring-shaped particle images of \mbox{\(D_{PI} = \SI{60}{\pixel}\)} with zero and three intersected particle images are shown in Figs. \ref{fig:SynthImages_subfig:PID60} and \subref{fig:SynthImages_subfig:PID60-Cut}.
\mbox{}

\section{Results}\label{sec:Results}
To evaluate the influence of the particle image diameter and shape on the displacement estimation error and the detectability, Monte Carlo simulations based on ensembles with 500 double-frame images are performed. For the cross-correlation of synthetically generated data (see Sect. \ref{sec:SIG}), a commercial PIV evaluation software (LaVision DaVis 8.4.0) is used.

\subsection{Influence of particle image shape and diameter on the estimation error}\label{sec:PID}
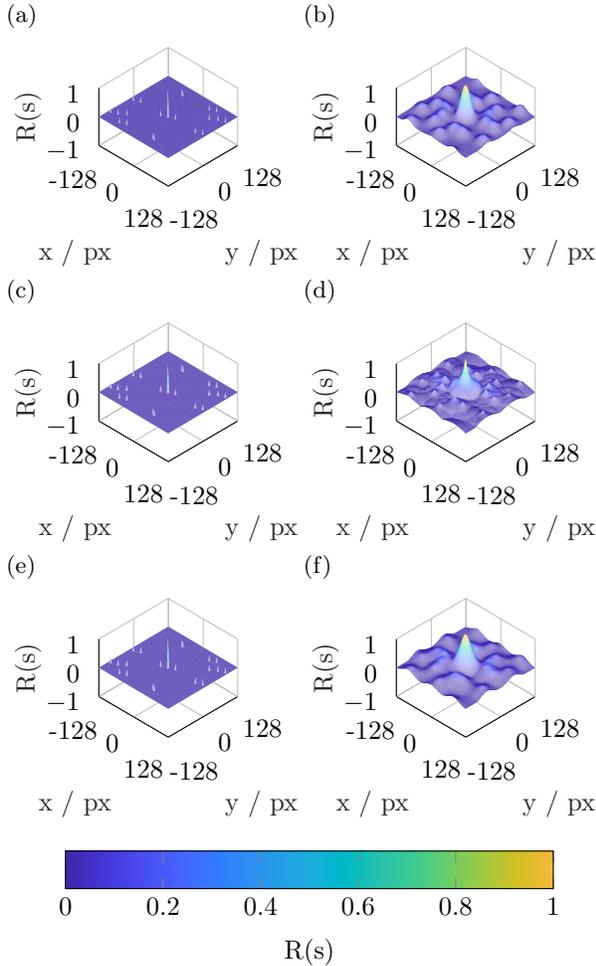
\begin{figure}[h!]
    \begin{subfigure}[c]{0.425\columnwidth}
        \subcaption{}%
        \input{Figures/Results_EE/CM_Gauss5.tikz}%
        \label{fig:CorrelationPeaks_subfig:CMGauss5}%
    \end{subfigure}\hspace{1.5em}%
    \begin{subfigure}[c]{0.425\columnwidth}
        \subcaption{}%
        \input{Figures/Results_EE/CM_Gauss60.tikz}%
        \label{fig:CorrelationPeaks_subfig:CMGauss60}%
    \end{subfigure}\hfill%
    \begin{subfigure}[c]{0.425\columnwidth}
        \subcaption{}%
        \input{Figures/Results_EE/CM_Halo5.tikz}%
        \label{fig:CorrelationPeaks_subfig:CMHalo5}%
    \end{subfigure}\hspace{1.5em}%
    \begin{subfigure}[c]{0.425\columnwidth}
        \subcaption{}%
        \input{Figures/Results_EE/CM_Halo60.tikz}%
        \label{fig:CorrelationPeaks_subfig:CMHalo60}%
    \end{subfigure}\hfill%
    \begin{subfigure}[c]{0.425\columnwidth}
        \subcaption{}%
        \input{Figures/Results_EE/CM_Plateau5.tikz}%
        \label{fig:CorrelationPeaks_subfig:CMPlateau5}%
    \end{subfigure}\hspace{1.5em}%
    \begin{subfigure}[c]{0.425\columnwidth}
        \subcaption{}%
        \input{Figures/Results_EE/CM_Plateau60.tikz}%
        \label{fig:CorrelationPeaks_subfig:CMPlateau60}%
    \end{subfigure}\hfill%
    \begin{subfigure}[c]{1\columnwidth}
        \centering
        \input{Figures/Results_EE/Colorbar.tikz}%
    \end{subfigure}
    \caption{Correlation maps resulting from the cross-correlation for a zero displacement of synthetically generated images with five particles in each image of \subref{fig:CorrelationPeaks_subfig:CMGauss5} \SI{5}{\pixel} Gaussian, \subref{fig:CorrelationPeaks_subfig:CMGauss60} \SI{60}{\pixel} Gaussian, \subref{fig:CorrelationPeaks_subfig:CMHalo5} \SI{5}{\pixel} ring-shaped, \subref{fig:CorrelationPeaks_subfig:CMHalo60} \SI{60}{\pixel} ring-shaped, \subref{fig:CorrelationPeaks_subfig:CMPlateau5} \SI{5}{\pixel} plateau-shaped and \subref{fig:CorrelationPeaks_subfig:CMPlateau60} \SI{60}{\pixel} plateau-shaped particle images.}
    \label{fig:CorrelationPeaks}
\end{figure}

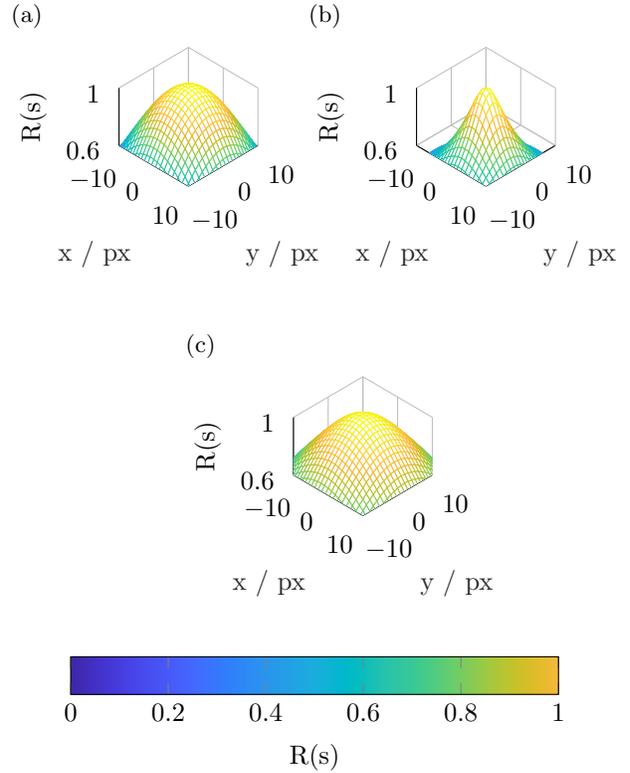
\begin{figure}[h!]
    \begin{subfigure}[c]{.425\columnwidth}
        \centering
        \subcaption{}%
        \input{Figures/Results_EE/CM_Gauss60Zoom_V2.tikz}
        \label{fig:CorrelationPeaksZoom_subfig:Gauss}%
    \end{subfigure}\hspace{1.5em}%
    \begin{subfigure}[c]{.425\columnwidth}
        \centering
        \subcaption{}%
        \input{Figures/Results_EE/CM_Halo60Zoom_V2.tikz}
        \label{fig:CorrelationPeaksZoom_subfig:Halo}%
    \end{subfigure}\hfill%
    \begin{center}
        \begin{subfigure}[c][][c]{.425\columnwidth}
            \subcaption{}%
            \input{Figures/Results_EE/CM_Plateau60Zoom_V2.tikz}%
            \label{fig:CorrelationPeaksZoom_subfig:Plateau}%
        \end{subfigure}\hfill%
    \end{center}
    \begin{center}
        \begin{subfigure}[c]{1\columnwidth}
            \centering
            \input{Figures/Results_EE/Colorbar.tikz}%
        \end{subfigure}
    \end{center}
    \caption{Zoom up of displacement correlation peak tops resulting from the cross-correlation of double-frame images each containing five particle images and zero displacement between corresponding frames, using \subref{fig:CorrelationPeaksZoom_subfig:Gauss} Gaussian, \subref{fig:CorrelationPeaksZoom_subfig:Halo} ring-shaped and \subref{fig:CorrelationPeaksZoom_subfig:Plateau} plateau-shaped particle images of \(D_{PI} = \SI{60}{\pixel}\).}
    \label{fig:CorrelationPeaksZoom}
\end{figure}

\begin{figure}[ht]
\centering
\input{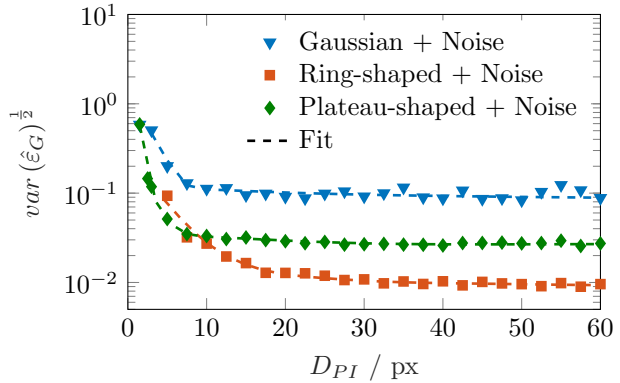}
\caption{Estimation error as a function of the particle image diameter (with noise, based on 500 samples per particle image diameter).}
\label{fig:EE-noise}
\end{figure}

In the present section, the influence of the particle image shape and diameter on the correlation peak shape, the resulting estimation error and the detectability is described. Excluded is the effect of particles intersected at interrogation window borders, which is discussed in Sect. \ref{sec:PID-Cut}.\\
Figs. \ref{fig:CorrelationPeaks_subfig:CMGauss5}-\subref{fig:CorrelationPeaks_subfig:CMPlateau60} show single cross-correlation results for two particle image sizes of synthetic image data with Gaussian, ring- and plateau-shaped particle images. 
\mbox{}\\
Cross-correlation results for particle image diameters of \mbox{\(D_{PI} = \SI{5}{\pixel}\)} (Figs. \ref{fig:CorrelationPeaks_subfig:CMGauss5}, \subref{fig:CorrelationPeaks_subfig:CMHalo5} and \subref{fig:CorrelationPeaks_subfig:CMPlateau5}), as well as \mbox{\(D_{PI} = \SI{60}{\pixel}\)} (Figs. \ref{fig:CorrelationPeaks_subfig:CMGauss60}, \subref{fig:CorrelationPeaks_subfig:CMHalo60} and \subref{fig:CorrelationPeaks_subfig:CMPlateau60}) and zero displacement are shown on the left and right hand side, respectively. It is evident that the correlation peak width increases with particle image diameter for all three types of particle image shapes. Furthermore, a comparison of displacement correlation peaks close to their maximum values reveal that correlation peaks resulting from ring-shaped particle images are most narrow compared to those of Gaussian and plateau-shaped particle images (see Figs. \ref{fig:CorrelationPeaksZoom_subfig:Gauss}-\subref{fig:CorrelationPeaksZoom_subfig:Plateau}).
\mbox{}\\
The scaling behaviour between particle image shape and displacement estimation error is evident in the correlation peak width (see (\ref{eq:EE})). Here, the first term \(\epsilon_1\) grows with an increasing displacement correlation peak width, while the second term \(\epsilon_2\) of the displacement estimation error denotes random errors. Thus, in comparison to Gaussian particle images a reduced estimation error is expected for ring-shaped particle images while a slightly increased estimation error is expected for plateau-shaped particle images. Fig. \ref{fig:EE} shows the displacement estimation error in a semi-logarithmic scale as a function of the particle image diameter derived from a Monte Carlo simulation for Gaussian, ring- and plateau-shaped particle images.
\begin{figure}[ht]
\centering
\input{Figures/Results_EE/Final_EE_Comparison_NEW_edit.tikz}
\caption{Estimation error as a function of the particle image diameter (without noise, based on 500 samples per particle image diameter).}
\label{fig:EE}
\end{figure}
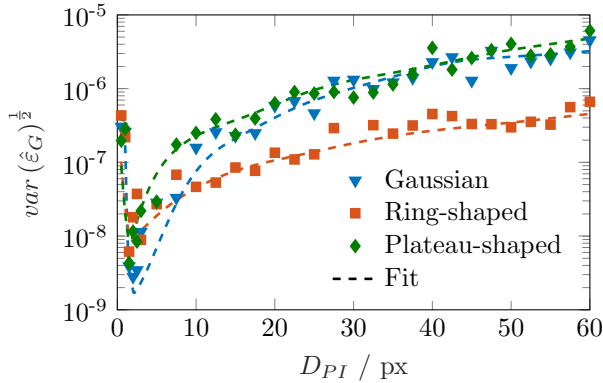\hfill%
\mbox{}\\
Every data point results from 500 cross-correlated double-frames, each containing five particle images and zero displacement between corresponding frames. All graphs resemble a non-monotonic relationship with a minimum estimation error at \mbox{\(D_{PI} \approx 2-\SI{3}{\pixel}\)} as it is also shown for Gaussian particle images by \citet{Westerweel1997b}. Obviously, this minimum value corresponds to the optimum particle image diameter for PIV measurements as also given in the common literature \citep{Raffel2007,adrian2011particle}. For smaller particle image diameters peak locking effects occur, which result in an increased estimation error. For larger particle image diameters random errors lead to an increased estimation error \citep{Westerweel1997b}. While the estimation error appears to stay in the same order of magnitude for Gaussian and plateau-shaped particle images with growing image diameter, a clear distinction has to be made for ring-shaped particle images for \mbox{\(D_{PI} \gtrsim \SI{10}{\pixel}\)}. Fig. \ref{fig:EE} shows a reduction of the displacement estimation error of up to one order of magnitude of ring-shaped particle images compared to Gaussian or plateau-shaped particle images (for \mbox{\(D_{PI} \gtrsim \SI{60}{\pixel}\)}). This can be understood from the fact that, for a constant particle image diameter, ring-shaped particle images decorrelate faster than Gaussian or plateau-shaped particle images, due to their large intensity gradient at the particle image border and a nearly transparent center region. Thus, the correlation peak width decreases (see also Fig. \ref{fig:CorrelationPeaks_subfig:CMHalo60}), resulting in an improved displacement estimation error.\\
It may be noted that Fig. \ref{fig:EE} displays results for noise-free data only, a situation that is hardly found in real experiments. To validate cross-correlation results of images that are closer to pre-processed experimental measurement data, image noise of 8.5\% is added (for details of the synthetic image generation see Sect. \ref{sec:SIG}) and the data set is analysed as well. The resulting estimation errors are shown in Fig. \ref{fig:EE-noise}.
\hfill%
\mbox{}\\
A clear difference in displacement estimation error for all three particle image shapes is evident in Fig. \ref{fig:EE-noise}. However, image noise leads to a strong increase in random errors in the correlation plane. Thus, in comparison to noise free data shown in Fig. \ref{fig:EE}, the overall level of displacement estimation error is now increased by approximately four orders of magnitude. This is due to an increase in the correlation peak asymmetry, represented by \(\epsilon_2\) in (\ref{eq:EE}), i.e. random errors in the correlation plane.\\
Furthermore, it is remarkable to see that noisy Gaussian particle images have a significantly higher estimation error compared to noisy plateau-shaped particle images while the corresponding noise free particle images lead to very similar displacement estimation errors (see Fig. \ref{fig:EE}). Figure \ref{fig:EE-parts} shows both \(\epsilon_{1}\) and \(\epsilon_{2}\) as a function of the particle image diameter \(D_{PI}\) for noise-free and noisy image data. It should be noted that results of \(\epsilon_1\) for images with and without noise coincide. Therefore, results of noisy image data are left out in Fig. \ref{fig:EE-parts} for a clearer presentation.
\begin{figure}[ht]
\input{Figures/Results_EE/EEParts_V3_Comparison_edit.tikz}
\caption{Factors \(\epsilon_1\) and \(\epsilon_2\) of (\ref{eq:EE}) for Gaussian, ring- and plateau-shaped particle images with and without noise. For a clearer presentation, results of \(\epsilon_1\) for images with noise are not plotted, as they coincide with corresponding results without image noise.}
\label{fig:EE-parts}
\end{figure}
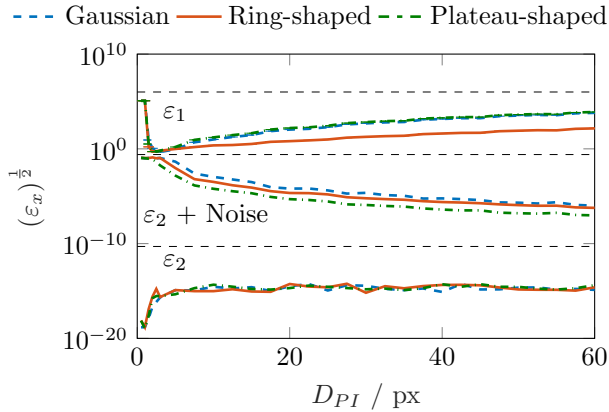\hfill%
\mbox{}\\
As can be seen from Fig. \ref{fig:EE-parts}, the increased estimation error of Gaussian particle images results from an increased amount of random errors in the correlation plane. This is obvious, since the values of \(\epsilon_1\) are the same for Gaussian and plateau-shaped particle images with and without noise, but values of \(\epsilon_2\) differ significantly.\\
For particle image diameters of \mbox{\(D_{PI} < \SI{3}{\pixel}\)} the particle image diameter is of the same length scale as the image noise, leading to an increased asymmetry of individual particle image peaks. This induces an increase in random errors represented by \(\epsilon_2\), as shown in Fig. \ref{fig:EE-parts}. Here, \(\epsilon_2\) for images with noise shows increased values for all three particle image shapes, compared to the values of \(\epsilon_2\) without noise. For larger particle images the random error decreases as the particle image diameter becomes significantly larger than the length scale of the image noise.

\subsection{Influence of the particle image size and shape on the detectability}\label{sec:ResDetectability}
\begin{figure}[ht]
\input{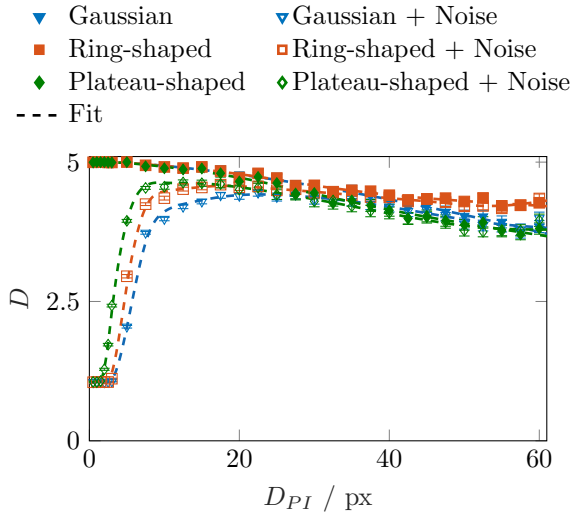}
\caption{Detectability with 95\% confidence interval as a function of the particle image diameter averaged over 500 samples per particle image diameter.}
\label{fig:Detectability}
\end{figure}

Figure \ref{fig:Detectability} shows the detectability as a function of the particle image diameter for Gaussian, ring- and plateau-shaped particle images without and with image noise of 8.5\%.
\mbox{}\\
A general decrease of the detectability with increasing particle image diameter can be associated with a significant increase of the particle image density \(N_{ppp}\), i.e. particles per pixel, within an interrogation window. While the effective number of particle images is constant (\mbox{\(N_I = 5\)} for all images), the particle image density increases from \mbox{\(N_{ppp} \approx 0.00006\)} for \mbox{\(D_{PI} = \SI{1}{\pixel}\)} to \mbox{\(N_{ppp} \approx 0.22\)} for \mbox{\(D_{PI} = \SI{60}{\pixel}\)}. Since the particle image density \(N_{ppp}\) mainly influences the height of the secondary correlation peak \citep{Scharnowski2018}, an increase in particle image diameter and \(N_{ppp}\), respectively, reduces the detectability for a constant \(N_I\). Furthermore, Fig. \ref{fig:Detectability} shows that the detectability decreases slower for growing ring-shaped particle images compared to those with Gaussian or plateau shape. We assume, that this is due to the characteristics of ring-shaped particle images, as slightly shifted ring-shaped particle image groups decorrelate faster than corresponding Gaussian or plateau-shaped particle image groups.\\
Fig. \ref{fig:Detectability} also shows the detectability as a function of the particle image diameter for images with noise (as described in Sect. \ref{sec:SIG}). Obviously, image noise affects especially the detectability for small particle image diameters (\mbox{\(D_{PI} < \SI{10}{\pixel}\)}). This is to be expected, as for small particle images the signal to noise ratio is decreased, leading to a lower detectability and thus to an increased probability of erroneous cross-correlations. A reduced signal to noise ratio originates from particle image discretization where very narrow particle image intensity peaks are averaged out over a full pixel, thereby being reduced in their maximum intensity value (see also Fig. \ref{fig:MaxIntVal}). This effect is well-known from classical PIV experiments where one strives for particle image diameters of two to three pixels.\\

\subsection{Influence of intersected particle images on the cross-correlation result}\label{sec:PID-Cut}
\begin{figure}[ht]
\input{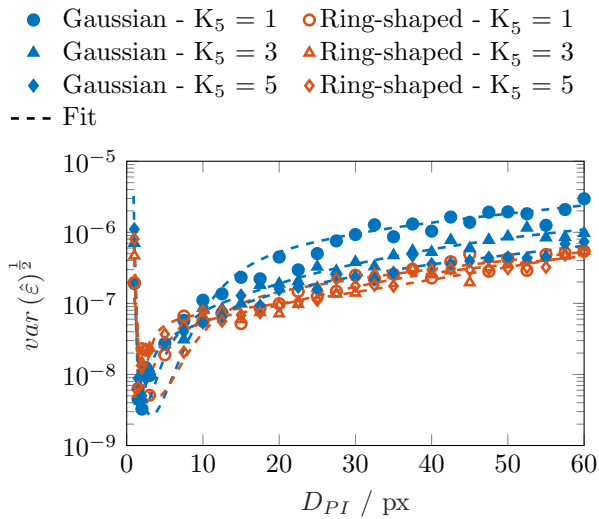}
\caption{Estimation error as a function of particle image diameter for one, three and five out of five particle images cut at the interrogation window edge.}
\label{fig:EE-Cut}
\end{figure}

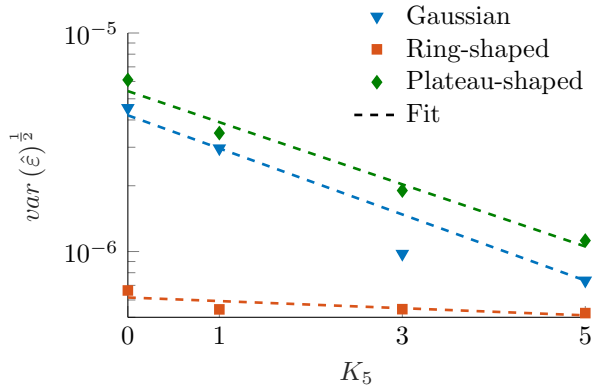
\begin{figure}[ht]
\input{Figures/Results_Cut/EEK-60_V3_edit.tikz}
\caption{Displacement estimation error as a function of the number of intersected particle images \(K_5\) for Gaussian, ring- and plateau-shaped particle images of \(D_{PI} = \SI{60}{\pixel}\). }
\label{fig:EE-K}
\end{figure}

\begin{figure}[ht]
\input{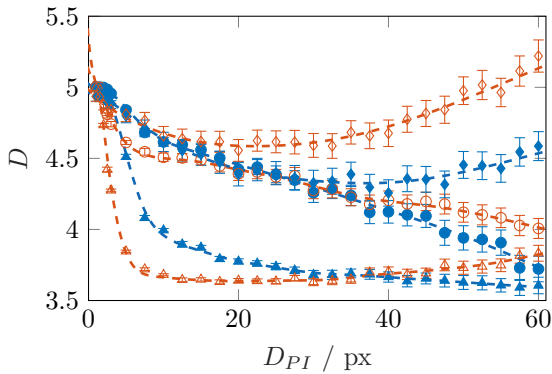}
\caption{Detectability with 95\% confidence interval for Gaussian and ring-shaped particle images as a function of the particle image diameter for \mbox{\(K_5 = 1\)}, 3 and 5.}
\label{fig:D-Cut}
\end{figure}

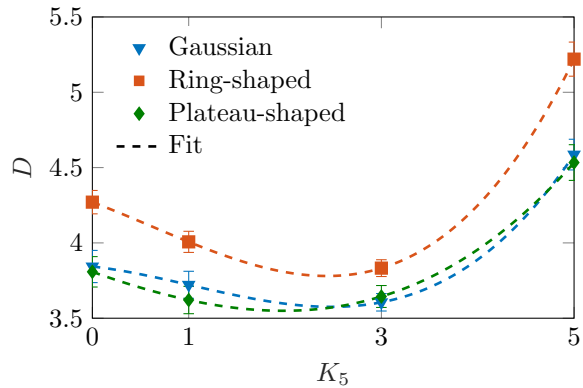
\begin{figure}[ht]
\input{Figures/Results_Cut/DK-60_V2_edit.tikz}
\caption{Detectability with 95\% confidence interval as a function of the number of intersected particle images \(K_5\) for Gaussian, ring- and plateau-shaped particle images of \mbox{\(D_{PI} = \SI{60}{\pixel}\)}.}
\label{fig:D-K}
\end{figure}

\begin{figure}[ht]
\input{Figures/Results_Cut/KlD.tikz}
\caption{Development of the ratio of the amount of intersected particle images and the total amount of particle images per interrogation window \mbox{\(K/N_I\)} as a function of the ratio of the corresponding interrogation window length and the particle image diameter \mbox{\(l/D_{PI}\)}.}
\label{fig:K-l}
\end{figure}
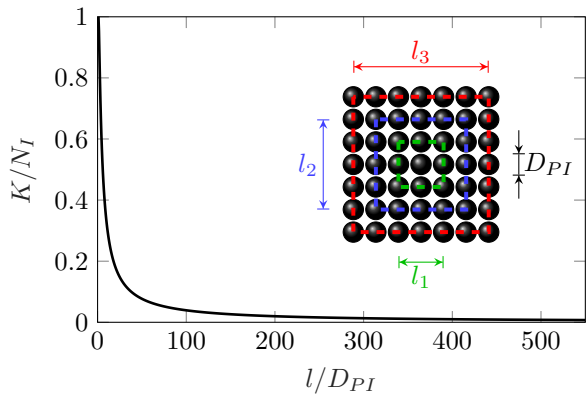

To study the influence of intersected particle images on the cross-correlation result, three cases are considered for one, three and five out of five particle images placed with their center points on the interrogation window border. These cases with particle images located only to 50\% inside the interrogation window are denoted with \mbox{\(K_{5}=1\)}, 3 and 5, respectively, where the index of \(K\) denotes the total amount of particle images inside the corresponding interrogation window.\\
Figure \ref{fig:EE-Cut} shows the displacement estimation error as a function of the particle image diameter for \mbox{\(K_5 = 1\)}, 3 and 5 for Gaussian and ring-shaped particle images. Since the results of plateau-shaped particle images strongly coincide with that of Gaussian particle images, results of plateau-shaped particle images are omitted here. It is obvious, that the amount of intersected particle images has a stronger influence on the displacement estimation error of Gaussian particle images compared to that of ring-shaped particle images.
\mbox{}\\
Fig. \ref{fig:EE-K} displays the displacement estimation error as a function of \(K_5\) for Gaussian, ring- and plateau-shaped particle images of \mbox{\(D_{PI} = \SI{60}{\pixel}\)}.
\mbox{}\\
The displacement estimation error reduces for intersected Gaussian and plateau-shaped particle images for increasing \(K_5\)-values. This is because intersected particle images deliver a sharp intensity jump at the interrogation window border. Such a sharp intensity jump leads to a reduction in the correlation peak width and hence an improved displacement estimation error. As more particles are intersected, their contribution to the cross-correlation result enhances the aforementioned effect. As already shown in Sect. \ref{sec:PID}, ring-shaped particle images decorrelate faster for slight image shifts due to the relatively small ring width and the transparent inner region. This is also the reason why the effect of intersected particle images on the estimation error is strongly reduced compared to Gaussian and plateau-shaped particle images.\\
The influence of intersected particle images at the interrogation window border on the detectability \(D\) is illustrated in Figs. \ref{fig:D-Cut} and \ref{fig:D-K}. Figure \ref{fig:D-Cut} shows how \(D\) alters as a function of the particle image diameter \(D_{PI}\) for Gaussian and ring-shaped particle images with \mbox{\(K_5=1\)}, 3 and 5. For a better conciseness, curves for plateau-shaped particle images are omitted as they show no significant differences to those of Gaussian particle images.
\mbox{}\\
Comparing different values of \(K_5\), a qualitative change in the detectability evolution can be recognised. While a mo\-no\-to\-nous decrease of detectability can be observed for \mbox{\(K_5=1\)} with growing particle image size, it appears to be non-monotonous for \mbox{\(K_5=3\)} and \mbox{\(K_5=5\)}. This qualitative behaviour is found for all particle image shapes and indicates that different competing effects come into play here. Firstly, a decrease in detectability with increasing particle image size is also observed when no particle images intersect the interrogation window border (see Fig. \ref{fig:Detectability}). It originates from the relative increase of secondary correlation peak values that are a result of an increase of the particle image density \(N_{ppp}\) for a constant effective number of particle images \(N_I\). A growing number of intersected particle images is in our case associated with a decreasing number of particle images located inside the interrogation window (as the total number of particle images is chosen to be constant here). Thus, an increasing number of intersected particle images leads to a decreased particle image density \(N_{ppp}\), which in turn results in an improved detectability due to a decrease in secondary correlation peak values. However, the overall detectability seems to be lowest for \mbox{\(K_5=3\)} while it is clearly enhanced for smaller or larger \(K_5\) values. Therefore, this effect alone does not explain the overall behaviour. We assume, that the detectability is decreased as the probability of erroneous correlations resulting from particle images inside the interrogation window correlating with intersected particle images increases for \mbox{\(K_5 \approx 1/2 N_I\)}. This is due to a similar number of particle images inside and on the interrogation window edge. This obviously leads to a relative increase of secondary correlation peak values and hence a reduced detectability.\\
Figure \ref{fig:D-K} shows detectability values of Gaussian, ring- and plateau-shaped particle images of \mbox{\(D_{PI} = \SI{60}{\pixel}\)} diameter as a function of \(K_5\).
\mbox{}\\
The detectability shows a non-monotonic behaviour for growing \(K_5\) values. Furthermore, the detectability of Gaussian and plateau-shaped particle images assumes similar values (see also Fig. \ref{fig:Detectability}), while it is slightly increased for ring-shaped particle images. It may be noted here that values of \mbox{\(K_5 = 5\)} are a hypothetical case. However, for high-resolution studies on dense particle systems such as suspension micro flows, large values of \(K\) can be expected as \(K\) scales with the particle image diameter or, respectively, decreasing interrogation window size.\\
Figure \ref{fig:K-l} illustrates the evolution of the ratio of \(K\) to the total amount of particle images per interrogation window as a function of the ratio between interrogation window edge length \(l\) to particle image diameter \(D_{PI}\), for in-plane particle images with simple cubic packing (see inset of Fig. \ref{fig:K-l}).
\mbox{}\\
As can be seen, values in the order of one can be expected for a ratio of interrogation window size and particle image diameter of \mbox{\(l/D_{PI} < 5\)} - a situation easily encountered in suspension flow studies.

\section{Experimental application}\label{sec:Experiments}
The previous study shows that the usage of particles with an image size larger than \SI{10}{\pixel} leads to a reduced displacement estimation error, if they assume a ring-shaped particle image instead of a plateau or Gaussian shape. Such particle images are encountered in a refractive index matched suspension with surface labelled suspension particles. We demonstrate in the following that such suspension systems allow to study in detail the bulk behaviour of suspensions of up to 5\% volume fraction by means of \textmu PIV. For this, PMMA suspension particles are used, which assume a particle image diameter of approximately \SI{270}{\pixel}. This is much larger than the commonly recommended particle image size of \mbox{\(2-\SI{3}{\pixel}\)}. Measurement results show that a spatial resolution beyond the particle image size can be reached through ensemble averaging to reveal the suspension bulk dynamics. Through refractive index matching and a labelling of only the suspension particle surfaces, an enhanced optical accessibility of the suspension flow could be reached. In this way, it was possible to measure velocity profiles of the suspension carrier liquid simultaneously with those of suspension particles. Measurement results of the suspension flow with a refractive index matched carrier liquid are compared to results obtained from the one phase flow in the same carrier liquid.

\subsection{Experimental set-up}\label{sec:ExSetup}
A straight microchannel with trapezoidal cross-section and a channel height of \mbox{\(h=\SI{533}{\micro\metre}\)} and a top and bottom width of \SI{578}{\micro\meter} and \SI{321}{\micro\meter}, respectively, is used (Micronit GmbH).\\
The suspension consists of a ternary carrier liquid of distilled water, Glycerine and Ammoniumthiocyanate \citep{Bailey2003a} and \SI{60}{\micro\meter} PMMA particles that are labelled as already described in Sect. \ref{sec:SIG}. The carrier liquid has a density of \mbox{\(\rho_l=\SI{1178.07+-0.37}{\kilo\gram\per\cubic\meter}\)}, a refractive index of \mbox{\(n_{D,l} = 1.489\)} and a dynamic viscosity of \SI[scientific-notation=true]{0.00596+-0.00011}{\pascal\second}. PMMA particles (Microbeads Spheromers CA60) assume a density of \mbox{\(\rho_p=\SI{1200}{\kilo\gram\per\cubic\meter}\)} and a refractive index of \mbox{\(n_{D,p} = 1.4895\pm 0.0035\)}. Thus, the deviations between carrier liquid and particle properties are less than \(2\%\) in density and \(0.3\%\) in refractive index. A nominal volumetric concentration of \mbox{\(\phi=5\%\)} is used throughout the experiments. Standard polystyrene tracer particles of \SI[multi-part-units = single]{1.19+-0.03}{\micro\metre} diameter (microParticles GmbH) are added to the suspension particle laden liquid. The same tracer particles are also used for reference measurements of the one phase flow. For both, reference and suspension measurements, the same liquid batch is used.\\
The experimental set-up consists of an EPI-fluorescent microscope (Nikon Eclipse LV100) that is used in combination with an infinity-corrected objective lens (20X Nikon CFI60 TU Plan Epi ELWD) of \mbox{\(M=20\)} magnification and a numerical aperture of \mbox{\(\mathit{NA}=0.4\)}. A double-pulsed Nd:YAG-laser (Litron Nano S 65-15 PIV) is coupled into the microscope for volume illumination. A double-frame CCD camera (LaVision Imager pro SX) with a resolution of \mbox{\(\SI{2058}{\pixel}\times\SI{2456}{\pixel}\)} is used for image recording. This results in a field of view of \mbox{\(\SI{0.57}{\milli\meter}\times\SI{0.48}{\milli\meter}\)}. For image acquisition and evaluation the commercial software DaVis 8.4 (LaVision GmbH) is used.

\subsection{Experimental procedure}\label{sec:Exex}
All measurements are performed at a bulk Reynolds number of \mbox{\(Re_b = \left(u_b \cdot h\right)/\nu = 0.81\)}. Here, \(u_b\) denotes the bulk fluid velocity calculated from the volume flow rate and the measured microchannel cross-section and \(\nu\) denotes the kinematic viscosity of the ternary carrier liquid, based on the measured dynamic viscosity and density (see Sect. \ref{sec:ExSetup}). Resulting velocity fields of the suspension flow are compared to velocity results of the one phase flow at the same bulk Reynolds number. As the depth of correlation \citep{Olsen2000} assumes \mbox{\(\delta_{DoC} \approx \SI{16}{\micro\meter}\)} for \SI{1.19}{\micro\meter} PIV tracer particles in combination with the prescribed optical configuration, measurements are performed at 27 different z-planes with a spatial distance of \mbox{\(dz = \SI{20}{\micro\metre}\)} from each other to prevent an overlapping of measurement volumes. At every measurement plane, 500 double-frame images are recorded with a frequency of \SI{4}{\hertz}. An inter-framing time of \mbox{\(dt=\SI{750}{\micro\second}\)} is used.

\subsection{Image pre-processing and evaluation}\label{sec:ExPrepEv}
Before cross-correlation, images are pre-processed and segmented to preserve solely suspension particle or PIV tracer signals, respectively, in individual images. Specifically, in a first step a minimum intensity image of the first 19 images of each recording sequence is generated and subtracted from each recorded image of the same recording sequence. This reduces stationary background noise and reflections. In a second step a sliding average filter of \SI{3}{\pixel} edge length including a Gaussian weighting function is applied to each image to reduce salt-and-pepper noise. In a third step, a sliding average filter of \SI{200}{\pixel} edge length is applied to each image. These spatially filtered images are subtracted from the original image to reduce large-scale intensity fluctuations, resulting in an increased intensity gradient at the ring-shaped suspension particle image rim.\\
Pre-processed images contain signals from both, PIV tracers and ring-shaped suspension particle images. To segment individual particle groups, an in-house Matlab code is used to detect the suspension particle images. The corresponding intensity information is extracted to generate two separate images containing either only the tracer or the suspension particle signals. Remaining masked regions inside the original images are filled utilizing a spectral random masking algorithm, as described by \citet{Anders2019a}, to decrease the velocity bias. Velocity fields are obtained by performing sum-of-correlation cross-correlations with the commercial DaVis software. A multi-pass method \citep{Willert1996a} with decreasing interrogation window sizes of \mbox{\(\SI{256}{\pixel}\times\SI{256}{\pixel}\)} during the first pass and \mbox{\(\SI{128}{\pixel}\times\SI{128}{\pixel}\)} during both, the second and third pass are utilized for the evaluation of the carrier liquid flow. As suspension particles assume \mbox{\(D_{PI}\approx\SI{270}{\pixel}\)} image diameter, interrogation window sizes are set to \mbox{\(\SI{512}{\pixel}\times\SI{512}{\pixel}\)} during the first pass and \mbox{\(\SI{256}{\pixel}\times\SI{256}{\pixel}\)} during the second and third pass. This is suitable, as we showed in Sect. \ref{sec:PID-Cut} that the estimation error of ring-shaped particle images is insensitive to intersection at the interrogation window borders. For all cross-correlation evaluations a 50\% overlap of interrogation windows is used. Vector post-processing is performed to eliminate erroneous velocity data. Specifically, vector results with a detectability below 2 are deleted and, additionally, an universal outlier detection is applied with a \mbox{\(5\times 5\)} median filter. This median filter defines velocity vectors as outliers if the value of a velocity component exceeds the median of the surrounding velocity data by an absolute value that corresponds to one times the median absolute deviation. Emerging empty vector spaces are filled with velocity information resulting from the 2nd, 3rd or 4th highest correlation peak, if they fulfill the median outlier criterion instead. In the end, all remaining empty vector spaces are filled up by interpolation. Overall, \mbox{\(4\%-6\%\)} of all vector information are replaced in an evaluation. 90\% of all replaced vectors are interpolated.

\subsection{Experimental results}\label{sec:ExResults}
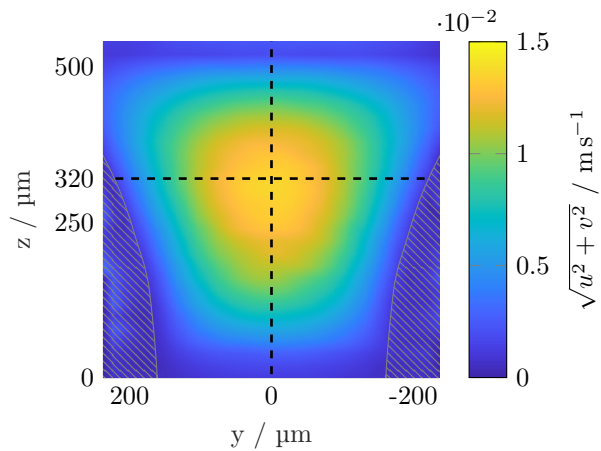
\begin{figure}[ht]
\input{Figures/Results_Exp/YZ-Projection.tikz}
\caption{Streamwise averaged velocity magnitudes in the microchannel cross-section obtained from \textmu PIV measurements of the one phase flow (OPF). Velocity profiles are analysed on xy- and xz-planes as indicated here by dashed lines. Channel walls are sketched as crosshatched regions.}
\label{fig:ExpYZ}
\end{figure}

\begin{figure}[h!]
    \begin{subfigure}{.95\columnwidth}
        \subcaption{}%
        \input{Figures/Results_Exp/2019-10-04_SuspensionPIV2/COMPARISON_V2DxxyMean_z=17.tikz}%
        \label{fig:ExpVelocities_subfig:XY}%
    \end{subfigure}
    \begin{subfigure}{.95\columnwidth}
        \subcaption{}%
        \input{Figures/Results_Exp/2019-10-04_SuspensionPIV2/COMPARISON_V2DxxzMean_y=17.tikz}%
        \label{fig:ExpVelocities_subfig:XZ}%
    \end{subfigure}%
    \caption{Velocity profiles obtained from \textmu PIV measurements of the one phase flow (OPF), the suspension carrier liquid (SCL) and the suspension particles (SP) \subref{fig:ExpVelocities_subfig:XY} on the xy-plane at \mbox{\(z=\SI{320}{\micro\metre}\)} above the channel bottom and \subref{fig:ExpVelocities_subfig:XZ} on the xz-plane in the channel bisector at \mbox{\(y=\SI{0}{\micro\metre}\)}.}
    \label{fig:ExpVelocities}
\end{figure}
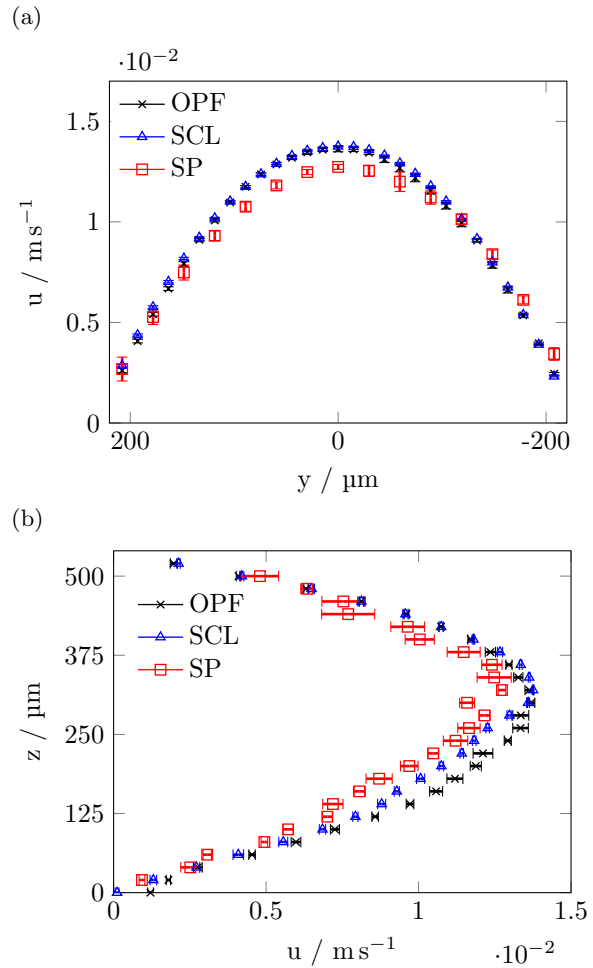

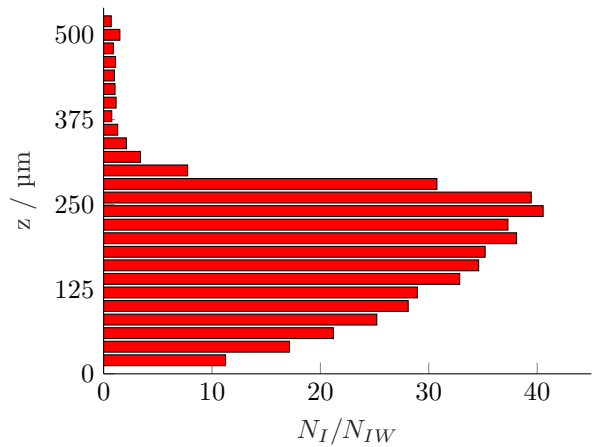
\begin{figure}[ht]
\input{Figures/Results_Exp/2019-10-04_SuspensionPIV2/NoP-z_y=9.tikz}
\caption{Average number of segmented suspension particle images \(N_I\) divided by the number of interrogation windows in streamwise direction \(N_{IW}\) along the microchannel height at the microchannel bisector.}
\label{fig:ExpParticleDistribution}
\end{figure}

PIV vector fields are derived through ensemble-averaging of 500 double-frame images. Measurements are performed at 27 equidistantly spaced measurement planes. Velocity results of the one phase flow (OPF) are compared against the suspension carrier liquid (SCL) and the suspension particles (SP) velocities. Velocity fields are averaged in streamwise direction. For the one phase flow this results in the streamwise averaged velocity field that is shown over the cross-section in Fig. \ref{fig:ExpYZ}. The microchannel walls are sketched as crosshatched regions, indicating the trapezoidal cross-sectional shape.
\mbox{}\\
The velocity profile assumes its maximum at \mbox{\(z\approx\SI{320}{\micro\meter}\)} (horizontal dashed line in Fig. \ref{fig:ExpYZ}). The corresponding velocity profile is shown in Fig. \ref{fig:ExpVelocities_subfig:XY} for the one phase flow (OPF), the suspension carrier liquid (SCL) and suspension particles (SP).
\mbox{}\\
The in-plane velocity profile of the one phase flow (OPF) as shown in Fig. \ref{fig:ExpVelocities_subfig:XY} assumes a parabolic shape with a maximum velocity of \mbox{\(u=\SI{0.0136+-0.0001}{\meter\per\second}\)}. At the microchannel side walls, velocities deviate from zero, which results from the presence of near-wall tracer particles that move with a non-zero velocity. Due to spatial averaging inside near-wall interrogation windows and a high velocity gradient in these regions, the presence of those tracer particles results in velocities that assume also non-zero values.\\
A comparison of the in-plane velocity profiles at \mbox{\(z=\SI{320}{\micro\meter}\)} between the one phase flow (OPF) and the suspension carrier liquid (SCL) shows no significant differences. Suspension particles (SP) generally tend to lag behind the suspension carrier liquid (SCL) up to 7.4\%. This is expected, due to a suspension particle size which assumes a similar order of magnitude as the microchannel dimensions \mbox{\(\left(D_P/h=0.113\right)\)} and is a well-known effect in Poiseuille flow for neutrally buoyant particles \citep{Brenner1966,Feng1994d,Guazzelli2011}. Numerical investigations of \citet{Loisel2015} showed similar trends of liquid-particle slip velocities compared to our results in suspension flows with homogeneously distributed particles and volume concentrations of up to 5\%. Their results indicate that the particle slip velocity is rather a function of the Reynolds number than of the particle volume concentration in the investigated particle volume concentration regime.\\
The velocity profiles in the xz-plane on the microchannel bisector (\mbox{\(y=\SI{0}{\micro\metre}\)}, see also vertical dashed line in Fig. \ref{fig:ExpYZ}), are shown in Fig. \ref{fig:ExpVelocities_subfig:XZ}. In the lower part of the microchannel for \mbox{\(z<\SI{300}{\micro\metre}\)}, the suspension carrier liquid (SCL) lags behind the one phase flow (OPF) up to 12\%. In contrast to this, similar flow velocities can be observed in the top region of the microchannel for \mbox{\(z\gtrsim\SI{300}{\micro\metre}\)}. Suspension particles (SP) in turn assume a generally lower velocity than the suspension carrier liquid (SCL) at the microchannel center, as already discussed above. This can be qualitatively understood when looking at the particle image density over the channel height, being an indication for the particle concentration distribution, see Fig. \ref{fig:ExpParticleDistribution}. The number of segmented suspension particle images \(N_I\) within the field of view during the whole time series is divided by the amount of interrogation windows \(N_{IW}\) of the resulting vector field in streamwise direction. These values are plotted for the microchannel bisector as a function of the microchannel height. A particle image is assigned to an interrogation window, if its center point is located inside the corresponding interrogation window borders. Therefore, the ratio \mbox{\(N_I/N_{IW}\)} indicates how many particle images contribute on average to the displacement correlation peak of a single interrogation window. Clearly, particles are distributed inhomogeneously over the microchannel height. We anticipate that a particle drift away from the channel top is induced at the inlet region where a sharp corner flow is induced, accelerating particles towards the channel bottom. It may be noted that the data of Fig. \ref{fig:ExpParticleDistribution} displays the particle image number density within each measurement volume.
\mbox{}\\
For \mbox{\(z\gtrsim\SI{300}{\micro\meter}\)}, the amount of particle images per interrogation window assumes on average \mbox{\(N_I/N_{IW}<5\)} which is below the effective number of particle images which ensures a high detectability in PIV measurements \citep{Keane1992}. This becomes evident through increasing standard deviations of suspension particle velocity data for \(z>\SI{300}{\micro\meter}\) in Fig. \ref{fig:ExpVelocities_subfig:XZ}. As the particle concentration is significantly reduced in this region, the velocity of the suspension carrier liquid assumes that of the one phase flow here.\\
Overall, we demonstrate that refractive index matched and surface labelled suspension particles open up the possibility to investigate both suspension and carrier bulk dynamics simultaneously by means of \textmu PIV. For studies of even higher volume fractions, a well defined temperature control will be required to achieve a matching of the refractive index between particles and carrier liquid up to the fourth digit \citep{Wiederseiner2011a}.

\section{Conclusion}\label{sec:Summary and Conclusion}
The size sensitivity of Gaussian, ring- and plateau-shaped particle images on cross-correlation based PIV evaluations is compared. When particle images become large compared to standard PIV tracer particle images, accuracy and reliability of cross-correlations on resulting standard Gaussian or plateau-shaped particle images are known to decrease. At higher particle volume fractions, the optical accessibility of such suspension flows is usually limited, as well. To overcome this, we perform a refractive index matching of a carrier fluid to PMMA suspension particles. Furthermore, a surface labelling of such suspension particles leads to ring-shaped particle images. The suitability of such particle images with regard to a PIV cross-correlation evaluation is investigated in this work. For all parameter studies, cross-correlation results of synthetic particle images are evaluated by means of displacement estimation error and detectability for particle image diameters between \mbox{\(\SI{1}{\pixel} \le D_{PI} \le \SI{60}{\pixel}\)}. The study is based on Monte Carlo simulations with 500 double-frame images for each particle image diameter. Particle images are cross-correlated using a commercial PIV evaluation algorithm.\\
It is shown that ring-shaped particle images assume a reduced displacement estimation error for particle image diameters beyond \mbox{\(D_{PI} = \SI{10}{\pixel}\)} compared to Gaussian and plateau-shaped particle images while there is no significant difference in the detectability for different particle image shapes. This reduction in displacement estimation error of ring-shaped particle images is related to increased intensity gradients at the particle image rim in combination with a nearly transparent center region. This leads to a faster decorrelation and thus reduces the correlation peak width. When image noise is added, the overall estimation error is increased by several orders of magnitude due to an increase in random errors (here an increase of four orders of magnitude was observed for 8.5\% image noise). Thus, special attention has to be spent on the elimination of image noise during pre-processing of experimental data, since image noise generally leads to an increased displacement estimation error for all investigated particle image shapes at various particle image diameters. However, also in this situation ring-shaped particle images assume a reduced displacement estimation error compared to Gaussian and plateau-shaped particle images.\\
Furthermore, the influence of particle images being intersected at the interrogation window edge is investigated. For this, synthetic images with one, three and five intersected particle images out of five within each interrogation window are evaluated for Gaussian, ring- and plateau-shaped particle images. While the displacement estimation error of ring-shaped particle images appears to be insensitive to particle image intersection at the interrogation window edge, a strong reduction in estimation error is observed for Gaussian and plateau-shaped particle images with increasing amount of intersected particle images. This can be explained by intensity jumps of particle images that occur at the intersection line along the interrogation window edge. These intensity jumps, which are much more prominent for Gaussian and plateau-shaped particle images, lead to a reduction of the correlation peak width and hence a reduced estimation error.\\
The quality of a PIV evaluation can be evaluated from the displacement estimation error and the detectability. The detectability is known to depend on the signal to noise ratio, the effective number of particle images, the correlated ensemble size and the particle image density. In the present study the first three parameters are fixed for all particle image shapes. Therefore, no significant difference in detectability is observed between different particle image shapes for particle images located inside the interrogation window. However, the detectability is found to decrease with increasing particle image size. This is due to a rise in the particle image density \(N_{ppp}\) associated with an increasing particle image diameter. Secondly, a decrease in detectability is observed for particle image diameters below \mbox{\(D_{PI} = \SI{10}{\pixel}\)} with image noise of 8.5\%. That is, because intensity values are averaged over a full pixel leading to a reduction in intensity peaks, especially for small particle images. This leads to a significant decrease in signal to noise ratio and hence a reduced detectability.\\
Furthermore, the detectability is investigated for a growing number of intersected particle images, keeping the total number of particle images within the interrogation window constant. Here a non-linear behaviour is observed with a minimum in detectability for a similar number of particle images located inside the interrogation window and on the interrogation window edge. We assume that for \(K_5\) being approximately equal to half the total amount of particle images per interrogation window the probability for particle images inside the interrogation window to correlate with intersected particle images is increased, leading to an increase in secondary correlation peak values.\\
Summed up, the usage of refractive index matched and surface labelled particles not only enhances optical accessibility, but also leads to a reduction of the displacement estimation error and a similar or even better detectability compared to Gaussian or plateau-shaped particle images. This holds true also for situations where intersected particle images at interrogation window borders occur.\\
Refractive index matching led to an enhanced optical accessibility of the system. Combined with a labelling, the accuracy and reliability of the \textmu PIV evaluation are enhanced compared to Gaussian particle images. We demonstrate both, theoretically and experimentally that such particle systems are suitable to study the bulk dynamics of suspension flows. Measurements are performed with surface labelled and refractive index matched suspension particles that assume ring-shaped particle images. Hence, the dynamics of a 5 Vol.-\% suspension particle volume fraction is measured together with the carrier liquid flow. By this, liquid-particle slip velocities are determined successfully.\\
Overall, we demonstrate that \textmu PIV measurements on ring-shaped particle images bring along advantageous properties, both from a theoretical and practical point of view. These are demonstrated to be an important step to perform \textmu PIV measurements in a variety of suspension flows.

\bibliographystyle{abbrvnat}
\bibliography{MyBibliography}

\end{document}

%% file: Figures/SIG/RadIntFunc.tikz
%
%
\definecolor{mycolor1}{rgb}{0.00000,0.49804,0.00000}%
\definecolor{mycolor2}{rgb}{0.85000,0.32500,0.09800}%
\begin{tikzpicture}

\begin{axis}[%
width=1\columnwidth,
height=3/5*1\columnwidth,
xmin=0,
xmax=5,
xtick={0, 1, 2, 3, 4, 5},
xticklabels={0, 0.2, 0.4, 0.6, 0.8, 1},
xlabel style={font=\color{white!15!black}},
xlabel={\(R_{PI}/R_{PI,max}\)},
ymin=0,
ymax=1,
ytick={0, 0.25,  0.5, 0.75, 1},
ylabel style={font=\color{white!15!black}},
ylabel={\(I/I_{max}\)},
axis background/.style={fill=white},
legend style={at={(0.4,1.03)}, anchor=south, legend columns=3, legend cell align=left, align=left, draw=white}
]
\addplot [color=blue, dashed, line width=1.0pt]
  table[row sep=crcr]{%
-5	0\\
-4.96031746031746	0.000439070404003656\\
-4.92063492063492	0.000912540096954157\\
-4.88095238095238	0.00142273054926445\\
-4.84126984126984	0.00197208688160297\\
-4.8015873015873	0.00256318168576078\\
-4.76190476190476	0.00319871872926004\\
-4.72222222222222	0.0038815365183539\\
-4.68253968253968	0.00461461169260692\\
-4.64285714285714	0.00540106222280952\\
-4.6031746031746	0.00624415038258156\\
-4.56349206349206	0.0071472854626699\\
-4.52380952380952	0.00811402619565548\\
-4.48412698412698	0.00914808285757077\\
-4.44444444444444	0.0102533190118006\\
-4.40476190476191	0.0114337528596136\\
-4.36507936507937	0.0126935581607625\\
-4.32539682539683	0.0140370646868121\\
-4.28571428571429	0.0154687581692222\\
-4.24603174603175	0.0169932797037421\\
-4.20634920634921	0.0186154245723782\\
-4.16666666666667	0.0203401404440952\\
-4.12698412698413	0.0221725249155172\\
-4.08730158730159	0.0241178223532213\\
-4.04761904761905	0.0261814199997795\\
-4.00793650793651	0.0283688433065225\\
-3.96825396825397	0.0306857504570707\\
-3.92857142857143	0.0331379260470366\\
-3.88888888888889	0.0357312738869383\\
-3.84920634920635	0.0384718088973029\\
-3.80952380952381	0.0413656480671795\\
-3.76984126984127	0.0444190004498378\\
-3.73015873015873	0.0476381561722986\\
-3.69047619047619	0.0510294744385404\\
-3.65079365079365	0.0545993705097416\\
-3.61111111111111	0.0583543016487615\\
-3.57142857142857	0.0623007520202233\\
-3.53174603174603	0.0664452165420405\\
-3.49206349206349	0.0707941836890136\\
-3.45238095238095	0.0753541172542045\\
-3.41269841269841	0.0801314370791679\\
-3.37301587301587	0.0851324987697513\\
-3.33333333333333	0.0903635724200689\\
-3.29365079365079	0.0958308203733659\\
-3.25396825396825	0.101540274054823\\
-3.21428571428571	0.107497809917839\\
-3.17460317460317	0.113709124552003\\
-3.13492063492063	0.12017970900771\\
-3.0952380952381	0.126914822399242\\
-3.05555555555556	0.133919464855027\\
-3.01587301587302	0.141198349890685\\
-2.97619047619048	0.14875587628732\\
-2.93650793650794	0.156596099564308\\
-2.8968253968254	0.164722703142472\\
-2.85714285714286	0.173138969300001\\
-2.81746031746032	0.181847750029721\\
-2.77777777777778	0.190851437912293\\
-2.73809523809524	0.200151937125547\\
-2.6984126984127	0.209750634715449\\
-2.65873015873016	0.219648372259005\\
-2.61904761904762	0.22984541805378\\
-2.57936507936508	0.240341439972547\\
-2.53968253968254	0.251135479124811\\
-2.5	0.262225924469597\\
-2.46031746031746	0.273610488525857\\
-2.42063492063492	0.285286184328068\\
-2.38095238095238	0.297249303775149\\
-2.34126984126984	0.309495397520478\\
-2.3015873015873	0.322019256549724\\
-2.26190476190476	0.334814895591272\\
-2.22222222222222	0.347875538501148\\
-2.18253968253968	0.361193605760724\\
-2.14285714285714	0.374760704220812\\
-2.1031746031746	0.388567619220291\\
-2.06349206349206	0.402604309200979\\
-2.02380952380952	0.41685990293314\\
-1.98412698412698	0.431322699457804\\
-1.94444444444444	0.445980170843006\\
-1.9047619047619	0.460818967841066\\
-1.86507936507937	0.475824928523344\\
-1.82539682539683	0.490983089957281\\
-1.78571428571429	0.506277702978324\\
-1.74603174603175	0.521692250096312\\
-1.70634920634921	0.537209466562331\\
-1.66666666666667	0.552811364607851\\
-1.62698412698413	0.568479260853291\\
-1.58730158730159	0.584193806868075\\
-1.54761904761905	0.599935022848753\\
-1.50793650793651	0.615682334366101\\
-1.46825396825397	0.631414612116186\\
-1.42857142857143	0.64711021459443\\
-1.38888888888889	0.662747033595746\\
-1.34920634920635	0.678302542427975\\
-1.30952380952381	0.693753846710184\\
-1.26984126984127	0.709077737612089\\
-1.23015873015873	0.724250747375905\\
-1.19047619047619	0.739249206947527\\
-1.15079365079365	0.754049305530134\\
-1.11111111111111	0.768627151860197\\
-1.07142857142857	0.782958836993565\\
-1.03174603174603	0.797020498377916\\
-0.992063492063492	0.810788384977373\\
-0.952380952380953	0.824238923205779\\
-0.912698412698413	0.837348783416822\\
-0.873015873015873	0.850094946692222\\
-0.833333333333333	0.862454771663416\\
-0.793650793650794	0.874406061097735\\
-0.753968253968254	0.885927127977057\\
-0.714285714285714	0.896996860795232\\
-0.674603174603175	0.907594787800424\\
-0.634920634920635	0.917701139909791\\
-0.595238095238095	0.927296912026652\\
-0.555555555555555	0.936363922494579\\
-0.515873015873016	0.944884870428535\\
-0.476190476190476	0.952843390670364\\
-0.436507936507937	0.96022410612455\\
-0.396825396825397	0.967012677240152\\
-0.357142857142857	0.97319584841616\\
-0.317460317460317	0.978761491120156\\
-0.277777777777778	0.983698643524004\\
-0.238095238095238	0.987997546475311\\
-0.198412698412699	0.991649675639443\\
-0.158730158730159	0.994647769663945\\
-0.119047619047619	0.99698585423511\\
-0.0793650793650791	0.998659261915154\\
-0.0396825396825395	0.999664647667802\\
0	1\\
0.0396825396825395	0.999664647667802\\
0.0793650793650791	0.998659261915154\\
0.119047619047619	0.99698585423511\\
0.158730158730159	0.994647769663945\\
0.198412698412699	0.991649675639443\\
0.238095238095238	0.987997546475311\\
0.277777777777778	0.983698643524004\\
0.317460317460317	0.978761491120156\\
0.357142857142857	0.97319584841616\\
0.396825396825397	0.967012677240152\\
0.436507936507937	0.96022410612455\\
0.476190476190476	0.952843390670364\\
0.515873015873016	0.944884870428535\\
0.555555555555555	0.936363922494579\\
0.595238095238095	0.927296912026652\\
0.634920634920635	0.917701139909791\\
0.674603174603175	0.907594787800424\\
0.714285714285714	0.896996860795232\\
0.753968253968254	0.885927127977057\\
0.793650793650794	0.874406061097735\\
0.833333333333333	0.862454771663416\\
0.873015873015873	0.850094946692222\\
0.912698412698413	0.837348783416822\\
0.952380952380953	0.824238923205779\\
0.992063492063492	0.810788384977373\\
1.03174603174603	0.797020498377916\\
1.07142857142857	0.782958836993565\\
1.11111111111111	0.768627151860197\\
1.15079365079365	0.754049305530134\\
1.19047619047619	0.739249206947527\\
1.23015873015873	0.724250747375905\\
1.26984126984127	0.709077737612089\\
1.30952380952381	0.693753846710184\\
1.34920634920635	0.678302542427975\\
1.38888888888889	0.662747033595746\\
1.42857142857143	0.64711021459443\\
1.46825396825397	0.631414612116186\\
1.50793650793651	0.615682334366101\\
1.54761904761905	0.599935022848753\\
1.58730158730159	0.584193806868075\\
1.62698412698413	0.568479260853291\\
1.66666666666667	0.552811364607851\\
1.70634920634921	0.537209466562331\\
1.74603174603175	0.521692250096312\\
1.78571428571429	0.506277702978324\\
1.82539682539683	0.490983089957281\\
1.86507936507936	0.475824928523344\\
1.90476190476191	0.460818967841066\\
1.94444444444444	0.445980170843006\\
1.98412698412698	0.431322699457804\\
2.02380952380952	0.41685990293314\\
2.06349206349206	0.402604309200979\\
2.1031746031746	0.388567619220291\\
2.14285714285714	0.374760704220812\\
2.18253968253968	0.361193605760724\\
2.22222222222222	0.347875538501148\\
2.26190476190476	0.334814895591272\\
2.3015873015873	0.322019256549724\\
2.34126984126984	0.309495397520478\\
2.38095238095238	0.297249303775149\\
2.42063492063492	0.285286184328068\\
2.46031746031746	0.273610488525857\\
2.5	0.262225924469597\\
2.53968253968254	0.251135479124811\\
2.57936507936508	0.240341439972547\\
2.61904761904762	0.22984541805378\\
2.65873015873016	0.219648372259005\\
2.6984126984127	0.209750634715449\\
2.73809523809524	0.200151937125547\\
2.77777777777778	0.190851437912293\\
2.81746031746032	0.181847750029721\\
2.85714285714286	0.173138969300001\\
2.8968253968254	0.164722703142472\\
2.93650793650794	0.156596099564308\\
2.97619047619048	0.14875587628732\\
3.01587301587302	0.141198349890685\\
3.05555555555556	0.133919464855027\\
3.09523809523809	0.126914822399242\\
3.13492063492063	0.12017970900771\\
3.17460317460317	0.113709124552003\\
3.21428571428571	0.107497809917839\\
3.25396825396825	0.101540274054823\\
3.29365079365079	0.0958308203733659\\
3.33333333333333	0.0903635724200689\\
3.37301587301587	0.0851324987697513\\
3.41269841269841	0.0801314370791679\\
3.45238095238095	0.0753541172542045\\
3.49206349206349	0.0707941836890136\\
3.53174603174603	0.0664452165420405\\
3.57142857142857	0.0623007520202233\\
3.61111111111111	0.0583543016487615\\
3.65079365079365	0.0545993705097416\\
3.69047619047619	0.0510294744385404\\
3.73015873015873	0.0476381561722986\\
3.76984126984127	0.0444190004498378\\
3.80952380952381	0.0413656480671795\\
3.84920634920635	0.0384718088973029\\
3.88888888888889	0.0357312738869383\\
3.92857142857143	0.0331379260470366\\
3.96825396825397	0.0306857504570707\\
4.00793650793651	0.0283688433065225\\
4.04761904761905	0.0261814199997795\\
4.08730158730159	0.0241178223532213\\
4.12698412698413	0.0221725249155172\\
4.16666666666667	0.0203401404440952\\
4.20634920634921	0.0186154245723782\\
4.24603174603175	0.0169932797037421\\
4.28571428571429	0.0154687581692222\\
4.32539682539683	0.0140370646868121\\
4.36507936507937	0.0126935581607625\\
4.40476190476191	0.0114337528596136\\
4.44444444444444	0.0102533190118006\\
4.48412698412698	0.00914808285757077\\
4.52380952380952	0.00811402619565548\\
4.56349206349206	0.0071472854626699\\
4.6031746031746	0.00624415038258156\\
4.64285714285714	0.00540106222280952\\
4.68253968253968	0.00461461169260692\\
4.72222222222222	0.0038815365183539\\
4.76190476190476	0.00319871872926004\\
4.8015873015873	0.00256318168576078\\
4.84126984126984	0.00197208688160297\\
4.88095238095238	0.00142273054926445\\
4.92063492063492	0.000912540096954157\\
4.96031746031746	0.000439070404003656\\
5	0\\
};
\addlegendentry{Gaussian}

\addplot [color=mycolor2, line width=1.0pt]
  table[row sep=crcr]{%
0.0197628458498024	0.252327388320769\\
0.0395256916996047	0.252447164898915\\
0.0592885375494071	0.252550944759613\\
0.0790513833992095	0.252642178265054\\
0.0988142292490119	0.252724315777424\\
0.118577075098814	0.252800807658913\\
0.138339920948617	0.252875104271708\\
0.158102766798419	0.252950655977998\\
0.177865612648221	0.253030913139971\\
0.197628458498024	0.253119326119815\\
0.217391304347826	0.253219345279718\\
0.237154150197628	0.253334420981869\\
0.256916996047431	0.253468003588457\\
0.276679841897233	0.253623283344217\\
0.296442687747036	0.253801441455264\\
0.316205533596838	0.254002723627655\\
0.33596837944664	0.254227370349488\\
0.355731225296443	0.254475622108862\\
0.375494071146245	0.254747678494354\\
0.395256916996047	0.255043190314824\\
0.41501976284585	0.255361416083169\\
0.434782608695652	0.255701605819213\\
0.454545454545455	0.256063021457754\\
0.474308300395257	0.256445576220386\\
0.494071146245059	0.256850157852351\\
0.513833992094862	0.257277733970749\\
0.533596837944664	0.257729269714035\\
0.553359683794466	0.258205620927891\\
0.573122529644269	0.258707493510142\\
0.592885375494071	0.259235582493161\\
0.612648221343874	0.259790611458866\\
0.632411067193676	0.26037399139288\\
0.652173913043478	0.260987838223056\\
0.671936758893281	0.261634216550236\\
0.691699604743083	0.26231406735492\\
0.711462450592885	0.263027526044527\\
0.731225296442688	0.263774752865566\\
0.75098814229249	0.264556035121208\\
0.770750988142292	0.265371676902236\\
0.790513833992095	0.266221565929886\\
0.810276679841897	0.267104943474321\\
0.8300395256917	0.26802099346543\\
0.849802371541502	0.268968472938808\\
0.869565217391304	0.269945139987324\\
0.889328063241107	0.270948638584145\\
0.909090909090909	0.271976709172554\\
0.928853754940711	0.273027116806826\\
0.948616600790514	0.274097916078583\\
0.968379446640316	0.275187631051057\\
0.988142292490119	0.276295281415111\\
1.00790513833992	0.277421797876403\\
1.02766798418972	0.278568607565583\\
1.04743083003953	0.279737032945758\\
1.06719367588933	0.280928304316369\\
1.08695652173913	0.282143206783392\\
1.10671936758893	0.283382116596436\\
1.12648221343874	0.284645836946336\\
1.14624505928854	0.285935206410815\\
1.16600790513834	0.287248767303323\\
1.18577075098814	0.28858335459948\\
1.20553359683794	0.289937364371164\\
1.22529644268775	0.291310750339842\\
1.24505928853755	0.292706196949843\\
1.26482213438735	0.294127643701114\\
1.28458498023715	0.295577978256093\\
1.30434782608696	0.297058660508607\\
1.32411067193676	0.298572039153562\\
1.34387351778656	0.30011886673436\\
1.36363636363636	0.301696541507798\\
1.38339920948617	0.303303993764634\\
1.40316205533597	0.304942415526772\\
1.42292490118577	0.306608536812077\\
1.44268774703557	0.308298640686728\\
1.46245059288538	0.310012695758569\\
1.48221343873518	0.311751359271785\\
1.50197628458498	0.313513743889006\\
1.52173913043478	0.315299637344279\\
1.54150197628459	0.317110088515137\\
1.56126482213439	0.318947071985984\\
1.58102766798419	0.320813062461476\\
1.60079051383399	0.322713581784755\\
1.62055335968379	0.324652834145901\\
1.6403162055336	0.326631652160172\\
1.6600790513834	0.328650470147698\\
1.6798418972332	0.330713786995796\\
1.699604743083	0.332828538500197\\
1.71936758893281	0.335002135003961\\
1.73913043478261	0.337238003843973\\
1.75889328063241	0.339536920611825\\
1.77865612648221	0.341898401320748\\
1.79841897233202	0.344324144143327\\
1.81818181818182	0.346816104357274\\
1.83794466403162	0.349374198758226\\
1.85770750988142	0.351996916717881\\
1.87747035573123	0.354681380059111\\
1.89723320158103	0.357422089331065\\
1.91699604743083	0.360212082802564\\
1.93675889328063	0.3630463725614\\
1.95652173913043	0.365922148148761\\
1.97628458498024	0.368838195431884\\
1.99604743083004	0.371793659851577\\
2.01581027667984	0.374788516476051\\
2.03557312252964	0.377826189759061\\
2.05533596837945	0.380912208929485\\
2.07509881422925	0.384053826712148\\
2.09486166007905	0.38725836883674\\
2.11462450592885	0.39052738468471\\
2.13438735177866	0.393867618320628\\
2.15415019762846	0.39729030684759\\
2.17391304347826	0.400805018563724\\
2.19367588932806	0.404418726230638\\
2.21343873517787	0.408136768156242\\
2.23320158102767	0.411966979601251\\
2.25296442687747	0.415916326564097\\
2.27272727272727	0.419985786810196\\
2.29249011857708	0.424174949152371\\
2.31225296442688	0.428482209553964\\
2.33201581027668	0.432905703590447\\
2.35177865612648	0.437448331238458\\
2.37154150197628	0.442114765950424\\
2.39130434782609	0.446908682289987\\
2.41106719367589	0.451831504885856\\
2.43083003952569	0.456881587285562\\
2.45059288537549	0.462060822322846\\
2.4703557312253	0.467373708580811\\
2.4901185770751	0.472825397688857\\
2.5098814229249	0.478418054785043\\
2.5296442687747	0.484150543034561\\
2.54940711462451	0.490024129964668\\
2.56916996047431	0.496046141805928\\
2.58893280632411	0.50222726619321\\
2.60869565217391	0.508578804633479\\
2.62845849802372	0.515112322862939\\
2.64822134387352	0.521840596500478\\
2.66798418972332	0.528776952719445\\
2.68774703557312	0.535937086556993\\
2.70750988142292	0.543341176215325\\
2.72727272727273	0.55100947586649\\
2.74703557312253	0.558963964534302\\
2.76679841897233	0.567227611890078\\
2.78656126482213	0.575824179146271\\
2.80632411067194	0.58478294455666\\
2.82608695652174	0.594140655107754\\
2.84584980237154	0.603939648478603\\
2.86561264822134	0.614221220345493\\
2.88537549407115	0.62503274426584\\
2.90513833992095	0.636424823983977\\
2.92490118577075	0.648450194623189\\
2.94466403162055	0.661166626322381\\
2.96442687747036	0.674624436902416\\
2.98418972332016	0.688868367923965\\
3.00395256916996	0.703932889917771\\
3.02371541501976	0.719846331607342\\
3.04347826086957	0.736632396304053\\
3.06324110671937	0.754293480691041\\
3.08300395256917	0.77279679192979\\
3.10276679841897	0.792084688860615\\
3.12252964426877	0.812068657498063\\
3.14229249011858	0.832624035415329\\
3.16205533596838	0.853574284669715\\
3.18181818181818	0.874668227886278\\
3.20158102766798	0.895586754219727\\
3.22134387351779	0.915943492588867\\
3.24110671936759	0.935296204734421\\
3.26086956521739	0.953156419686041\\
3.28063241106719	0.968976788239125\\
3.300395256917	0.982165499444745\\
3.3201581027668	0.992118681065139\\
3.3399209486166	0.998246945515084\\
3.3596837944664	1\\
3.37944664031621	0.996895347557424\\
3.39920948616601	0.988561560243809\\
3.41897233201581	0.974744214837295\\
3.43873517786561	0.955322030838304\\
3.45849802371542	0.930321282092426\\
3.47826086956522	0.8999527243079\\
3.49802371541502	0.864618491780636\\
3.51778656126482	0.824874130240333\\
3.53754940711462	0.781407226549736\\
3.55731225296443	0.734990827633008\\
3.57707509881423	0.686453258756476\\
3.59683794466403	0.636659247640111\\
3.61660079051383	0.586475630020249\\
3.63636363636364	0.536735252736955\\
3.65612648221344	0.488202390329822\\
3.67588932806324	0.441547275811873\\
3.69565217391304	0.397293142093062\\
3.71541501976285	0.355827932385105\\
3.73517786561265	0.317409408855299\\
3.75494071146245	0.28220495965162\\
3.77470355731225	0.250286543945922\\
3.79446640316206	0.221623935704537\\
3.81422924901186	0.196079652380489\\
3.83399209486166	0.173464032312526\\
3.85375494071146	0.153564217980404\\
3.87351778656127	0.136143055842376\\
3.89328063241107	0.120942470266755\\
3.91304347826087	0.10770730007982\\
3.93280632411067	0.0961921789367348\\
3.95256916996047	0.0861712893337306\\
3.97233201581028	0.0774410187028584\\
3.99209486166008	0.0698161410225846\\
4.01185770750988	0.0631384593073556\\
4.03162055335968	0.0572645021885174\\
4.05138339920949	0.0520707108314735\\
4.07114624505929	0.0474535087407068\\
4.09090909090909	0.043324613682328\\
4.11067193675889	0.0396083413028657\\
4.1304347826087	0.0362403974935927\\
4.1501976284585	0.0331748762082493\\
4.1699604743083	0.0303783138935241\\
4.1897233201581	0.0278268081215547\\
4.20948616600791	0.0254986052399997\\
4.22924901185771	0.0233680656966234\\
4.24901185770751	0.0214136977378218\\
4.26877470355731	0.0196187801729671\\
4.28853754940711	0.0179721235218655\\
4.30830039525692	0.0164610704390852\\
4.32806324110672	0.0150689205522382\\
4.34782608695652	0.0137816662747735\\
4.36758893280632	0.0125891034567796\\
4.38735177865613	0.0114875245479177\\
4.40711462450593	0.0104765869226642\\
4.42687747035573	0.00955371813034822\\
4.44664031620553	0.00871620563697543\\
4.46640316205534	0.00795977810480492\\
4.48616600790514	0.0072806758319852\\
4.50592885375494	0.00667838120146269\\
4.52569169960474	0.00615216205553749\\
4.54545454545455	0.00569193742176638\\
4.56521739130435	0.00528527626069625\\
4.58498023715415	0.00492240907928051\\
4.60474308300395	0.0045935953907203\\
4.62450592885375	0.00429181166512689\\
4.64426877470356	0.00401106464324434\\
4.66403162055336	0.00374682529253386\\
4.68379446640316	0.0034926564495119\\
4.70355731225296	0.00324246266449801\\
4.72332015810277	0.00298946158782635\\
4.74308300395257	0.00272968351173677\\
4.76284584980237	0.00246167430484076\\
4.78260869565217	0.00218216170397051\\
4.80237154150198	0.00189186647469782\\
4.82213438735178	0.0015956479918469\\
4.84189723320158	0.00130035859624236\\
4.86166007905138	0.00101646668045615\\
4.88142292490119	0.000755712006710899\\
4.90118577075099	0.000526802932556397\\
4.92094861660079	0.000334930522003142\\
4.94071146245059	0.000186041745968398\\
4.9604743083004	8.14969050137693e-05\\
4.9802371541502	2.02490136609152e-05\\
5	0\\
};
\addlegendentry{Ring-shaped}

\addplot [color=mycolor1, line width=1.0pt, dash pattern=on 4pt off 3pt on 2pt off 3pt]
  table[row sep=crcr]{%
0.0197628458498024	1\\
0.0395256916996047	0.999969719230833\\
0.0592885375494071	0.999941159470758\\
0.0790513833992095	0.999914021922994\\
0.0988142292490119	0.99988800779076\\
0.118577075098814	0.999862818277277\\
0.138339920948617	0.999838154585762\\
0.158102766798419	0.999813717919437\\
0.177865612648221	0.99978920948152\\
0.197628458498024	0.99976433047523\\
0.217391304347826	0.999738782103787\\
0.237154150197628	0.99971226557041\\
0.256916996047431	0.99968448207832\\
0.276679841897233	0.999655132830733\\
0.296442687747036	0.999623919030872\\
0.316205533596838	0.999590541933171\\
0.33596837944664	0.999554705495702\\
0.355731225296443	0.999516117667557\\
0.375494071146245	0.999474486718826\\
0.395256916996047	0.999429520919596\\
0.41501976284585	0.999380928539958\\
0.434782608695652	0.999328417941639\\
0.454545454545455	0.999271729320118\\
0.474308300395257	0.999210681252455\\
0.494071146245059	0.999145104163838\\
0.513833992094862	0.999074828479454\\
0.533596837944664	0.998999684624491\\
0.553359683794466	0.998919503076246\\
0.573122529644269	0.998834114502503\\
0.592885375494071	0.998743349614559\\
0.612648221343874	0.998647039123711\\
0.632411067193676	0.99854501399491\\
0.652173913043478	0.998437130101703\\
0.671936758893281	0.998323288858623\\
0.691699604743083	0.998203396537464\\
0.711462450592885	0.998077359410018\\
0.731225296442688	0.997945094973534\\
0.75098814229249	0.99780668360588\\
0.770750988142292	0.997662328519632\\
0.790513833992095	0.997512235698419\\
0.810276679841897	0.997356536454343\\
0.8300395256917	0.997195182613639\\
0.849802371541502	0.997028099665689\\
0.869565217391304	0.996855256274333\\
0.889328063241107	0.996676735488075\\
0.909090909090909	0.996492639083609\\
0.928853754940711	0.996303068618312\\
0.948616600790514	0.996108124147662\\
0.968379446640316	0.99590790509789\\
0.988142292490119	0.995702510893845\\
1.00790513833992	0.995492049626969\\
1.02766798418972	0.995276659556944\\
1.04743083003953	0.995056483191404\\
1.06719367588933	0.994831608872099\\
1.08695652173913	0.994602070222059\\
1.10671936758893	0.99436789842256\\
1.12648221343874	0.994129121092479\\
1.14624505928854	0.9938857581165\\
1.16600790513834	0.993637827446904\\
1.18577075098814	0.993385313375887\\
1.20553359683794	0.993128157614189\\
1.22529644268775	0.992866297489897\\
1.24505928853755	0.992599643986074\\
1.26482213438735	0.992328086916094\\
1.28458498023715	0.992051487109135\\
1.30434782608696	0.991769592076017\\
1.32411067193676	0.991482091778118\\
1.34387351778656	0.991188571874586\\
1.36363636363636	0.990888594958188\\
1.38339920948617	0.990581732046952\\
1.40316205533597	0.990267566853754\\
1.42292490118577	0.989945645736421\\
1.44268774703557	0.989615346551107\\
1.46245059288538	0.989275994109898\\
1.48221343873518	0.988926859252065\\
1.50197628458498	0.988567187237922\\
1.52173913043478	0.988196200872404\\
1.54150197628459	0.987812947513156\\
1.56126482213439	0.987416369438113\\
1.58102766798419	0.987005382201553\\
1.60079051383399	0.986578928567383\\
1.62055335968379	0.986136025342606\\
1.6403162055336	0.985675694737679\\
1.6600790513834	0.985196910255305\\
1.6798418972332	0.984698550682936\\
1.699604743083	0.984179332760031\\
1.71936758893281	0.983637953581022\\
1.73913043478261	0.983073125235228\\
1.75889328063241	0.982483565779475\\
1.77865612648221	0.981867978845976\\
1.79841897233202	0.981225003501711\\
1.81818181818182	0.980553114995012\\
1.83794466403162	0.979850744116037\\
1.85770750988142	0.979116320805697\\
1.87747035573123	0.978348274647922\\
1.89723320158103	0.977544982928521\\
1.91699604743083	0.976704703152315\\
1.93675889328063	0.975825625175355\\
1.95652173913043	0.974905876736449\\
1.97628458498024	0.973943515593679\\
1.99604743083004	0.972936443877325\\
2.01581027667984	0.9718824304837\\
2.03557312252964	0.970779127070351\\
2.05533596837945	0.969624107580904\\
2.07509881422925	0.968414822252095\\
2.09486166007905	0.967148634594241\\
2.11462450592885	0.96582276769258\\
2.13438735177866	0.964434276726406\\
2.15415019762846	0.962980028871452\\
2.17391304347826	0.961456740801615\\
2.19367588932806	0.959860979622857\\
2.21343873517787	0.958189258614612\\
2.23320158102767	0.956437943577845\\
2.25296442687747	0.954603252767788\\
2.27272727272727	0.952681227102973\\
2.29249011857708	0.950667791943592\\
2.31225296442688	0.94855879728481\\
2.33201581027668	0.946349963299446\\
2.35177865612648	0.944036831862849\\
2.37154150197628	0.941614836676736\\
2.39130434782609	0.939079354378193\\
2.41106719367589	0.936425598049917\\
2.43083003952569	0.933648590020798\\
2.45059288537549	0.930743207715885\\
2.4703557312253	0.927704227895772\\
2.4901185770751	0.92452612014836\\
2.5098814229249	0.921203598694906\\
2.5296442687747	0.917731864755494\\
2.54940711462451	0.914105770755319\\
2.56916996047431	0.910319561152616\\
2.58893280632411	0.906367463879572\\
2.60869565217391	0.902243818752994\\
2.62845849802372	0.897943291487107\\
2.64822134387352	0.89346053461676\\
2.66798418972332	0.88878934788905\\
2.68774703557312	0.88392276849798\\
2.70750988142292	0.878853918122682\\
2.72727272727273	0.873575966912141\\
2.74703557312253	0.868081694721072\\
2.76679841897233	0.862364080919354\\
2.78656126482213	0.856416292178191\\
2.80632411067194	0.850230921843315\\
2.82608695652174	0.843799883614635\\
2.84584980237154	0.837114529663878\\
2.86561264822134	0.830165922684873\\
2.88537549407115	0.822946090871552\\
2.90513833992095	0.815447093793328\\
2.92490118577075	0.807659523421881\\
2.94466403162055	0.799573043240381\\
2.96442687747036	0.791177761733227\\
2.98418972332016	0.782463974927878\\
3.00395256916996	0.773422433468912\\
3.02371541501976	0.764043744266039\\
3.04347826086957	0.754317391058637\\
3.06324110671937	0.744233024023065\\
3.08300395256917	0.733780498952515\\
3.10276679841897	0.722950210930617\\
3.12252964426877	0.711733614771742\\
3.14229249011858	0.700123851564066\\
3.16205533596838	0.688116724800488\\
3.18181818181818	0.675709635929075\\
3.20158102766798	0.66290178713441\\
3.22134387351779	0.649695139616869\\
3.24110671936759	0.636096376138988\\
3.26086956521739	0.622115933268303\\
3.28063241106719	0.607767827612606\\
3.300395256917	0.593069012544176\\
3.3201581027668	0.578040452533613\\
3.3399209486166	0.562707376647618\\
3.3596837944664	0.547099435904961\\
3.37944664031621	0.531249618711244\\
3.39920948616601	0.515195501944058\\
3.41897233201581	0.498978427199026\\
3.43873517786561	0.482641767307928\\
3.45849802371542	0.466230922606777\\
3.47826086956522	0.4497933484613\\
3.49802371541502	0.433378505022335\\
3.51778656126482	0.417037355876892\\
3.53754940711462	0.40081949160574\\
3.55731225296443	0.384770701537678\\
3.57707509881423	0.36893467319527\\
3.59683794466403	0.353353246191685\\
3.61660079051383	0.338066182411759\\
3.63636363636364	0.323110480320369\\
3.65612648221344	0.308519618196139\\
3.67588932806324	0.294322407535572\\
3.69565217391304	0.280543345819113\\
3.71541501976285	0.267201244898478\\
3.73517786561265	0.2543097305454\\
3.75494071146245	0.24188035242442\\
3.77470355731225	0.229922932496269\\
3.79446640316206	0.218441599820199\\
3.81422924901186	0.20743423801528\\
3.83399209486166	0.196895254908373\\
3.85375494071146	0.186817370933698\\
3.87351778656127	0.177192503466419\\
3.89328063241107	0.168011064738493\\
3.91304347826087	0.159260389731855\\
3.93280632411067	0.150926780176372\\
3.95256916996047	0.142994862816536\\
3.97233201581028	0.135447593505399\\
3.99209486166008	0.128267561907759\\
4.01185770750988	0.121438320813002\\
4.03162055335968	0.114946154238402\\
4.05138339920949	0.108776833220188\\
4.07114624505929	0.102914761958094\\
4.09090909090909	0.0973449472939414\\
4.11067193675889	0.0920542515003706\\
4.1304347826087	0.0870309031814065\\
4.1501976284585	0.0822626173095119\\
4.1699604743083	0.0777354458164879\\
4.1897233201581	0.0734356725621316\\
4.20948616600791	0.0693514184467267\\
4.22924901185771	0.065471606005257\\
4.24901185770751	0.0617856974375184\\
4.26877470355731	0.0582837776827924\\
4.28853754940711	0.054957029721282\\
4.30830039525692	0.051796929409218\\
4.32806324110672	0.048794668787643\\
4.34782608695652	0.0459417406995761\\
4.36758893280632	0.0432306811150404\\
4.38735177865613	0.0406548087337111\\
4.40711462450593	0.0382079005042418\\
4.42687747035573	0.0358832429191476\\
4.44664031620553	0.0336738673802643\\
4.46640316205534	0.0315739751436903\\
4.48616600790514	0.0295784376518646\\
4.50592885375494	0.0276814911015781\\
4.52569169960474	0.0258770763316781\\
4.54545454545455	0.0241599967531693\\
4.56521739130435	0.0225257597752495\\
4.58498023715415	0.0209699219735599\\
4.60474308300395	0.0194879480158323\\
4.62450592885375	0.0180752058671269\\
4.64426877470356	0.016727724603051\\
4.66403162055336	0.0154420827389265\\
4.68379446640316	0.0142146036610368\\
4.70355731225296	0.0130416800889675\\
4.72332015810277	0.0119199729135032\\
4.74308300395257	0.010846183469953\\
4.76284584980237	0.00981707214727702\\
4.78260869565217	0.00882941344532027\\
4.80237154150198	0.00788011707099172\\
4.82213438735178	0.00696671939629412\\
4.84189723320158	0.0060866363944767\\
4.86166007905138	0.00523702976492177\\
4.88142292490119	0.00441519994616436\\
4.90118577075099	0.00361895492728797\\
4.92094861660079	0.00284693961656121\\
4.94071146245059	0.00209854428796004\\
4.9604743083004	0.00137374580023672\\
4.9802371541502	0.000673497924118274\\
5	0\\
};
\addlegendentry{Plateau-shaped}

\end{axis}
\end{tikzpicture}%

%% file: Figures/SIG/Final_Halo_PID=60.tikz
%
%
\begin{tikzpicture}

\begin{axis}[%
width=1\linewidth,
height=1\linewidth,
point meta min=0,
point meta max=1973,
axis on top,
xmin=0,
xmax=256.5,
xtick={0,128,256},
tick align=outside,
xlabel style={font=\color{white!15!black}},
xlabel={x / \si{\pixel}},
xtick style={draw=none},
y dir=reverse,
ymin=0,
ymax=256.5,
ytick={0,128,256},
ylabel={y / \si{\pixel}},
ytick style={draw=none},
axis background/.style={fill=white},
]
\addplot [forget plot] graphics [xmin=0.5, xmax=256.5, ymin=0.5, ymax=256.5] {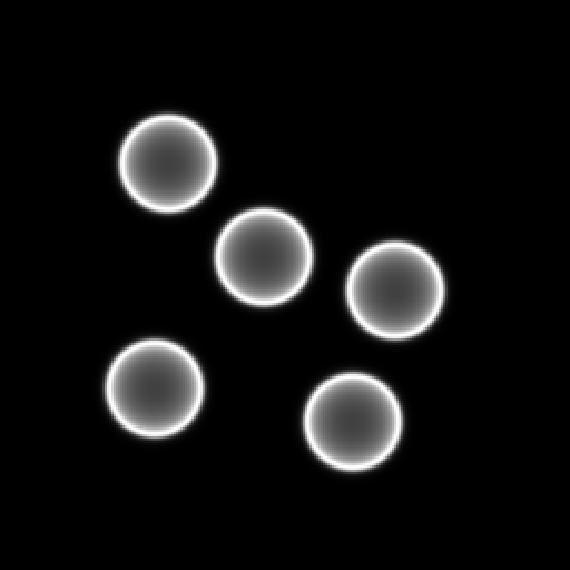};
\end{axis}
\end{tikzpicture}%

%% file: Figures/SIG/Cut50-3_Halo_PID=60.tikz
%
%
\begin{tikzpicture}

\begin{axis}[%
width=1\linewidth,
height=1\linewidth,
point meta min=0,
point meta max=1973,
axis on top,
xmin=0,
xmax=256.5,
xtick={0,128,256},
tick align=outside,
xlabel style={font=\color{white!15!black}},
xlabel={x / \si{\pixel}},
xtick style={draw=none},
y dir=reverse,
ymin=0,
ymax=256.5,
ytick={0,128,256},
ylabel={y / \si{\pixel}},
ytick style={draw=none},
axis background/.style={fill=white},
]
\addplot [forget plot] graphics [xmin=0.5, xmax=256.5, ymin=0.5, ymax=256.5] {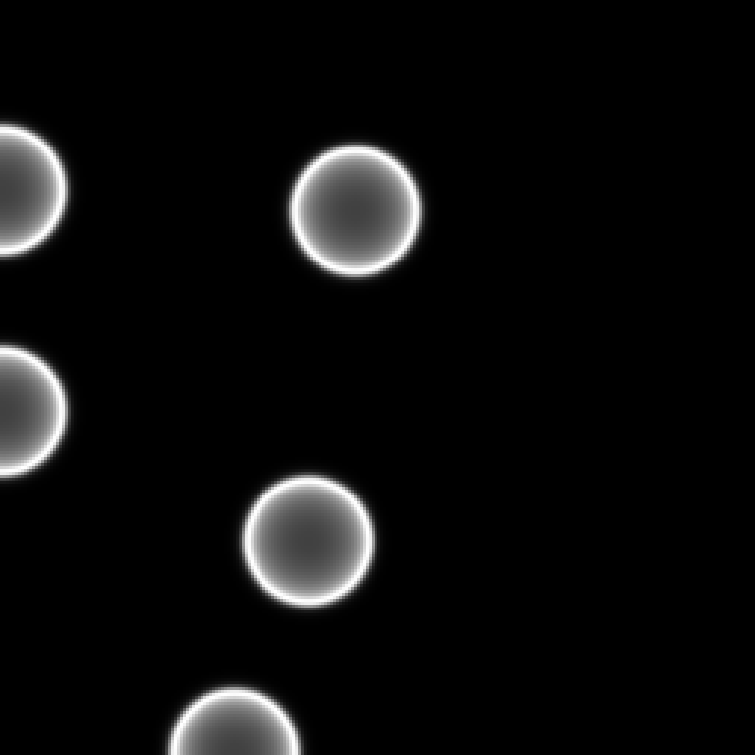};
\end{axis}
\end{tikzpicture}%

%% file: Figures/SIG/Colorbar_BlackWhite.tikz
%
%
\begin{tikzpicture}
\begin{axis}[hide axis,
width=1\columnwidth,
height=1\columnwidth,
point meta min=0,
point meta max=2000,
colormap = {whiteblack}{color(0cm)  = (black);color(1cm) = (white)},
colorbar horizontal,
colorbar style={title=counts,  title style={at={(0.5,-1.5)}, anchor=north}, xtick={0,500,1000,1500,2000}},
]
\end{axis}
\end{tikzpicture}%

%% file: Figures/Results_EE/CM_Gauss5.tikz
%
%
\begin{tikzpicture}

\begin{axis}[%
width=1\linewidth,
height=1\linewidth,
xmin=0,
xmax=256,
xtick={1,129,256},
xticklabels={{-128},{0},{128}},
xtick style={draw=none},
xlabel style={font=\color{white!15!black},yshift=2mm,xshift=0mm},
xlabel={x / \si{\pixel}},
ymin=0,
ymax=256,
ytick={1,129,256},
yticklabels={{-128},{0},{128}},
ytick style={draw=none},
ylabel style={font=\color{white!15!black},yshift=2mm,xshift=0mm},
ylabel={y / \si{\pixel}},
zmin=-1,
zmax=1,
ztick={-1,0,1},
ztick style={draw=none},
zlabel style={font=\color{white!15!black},yshift=-2mm},
zlabel={R(s)},
view={45}{45},
axis background/.style={fill=white},
axis x line*=bottom,
axis y line*=left,
axis z line*=left,
xmajorgrids,
ymajorgrids,
zmajorgrids,
enlargelimits=false,
tick align = outside,
]

\addplot3 graphics [xmin=0, xmax=256, ymin=0, ymax=256, zmin=-1.425, zmax=1.405] {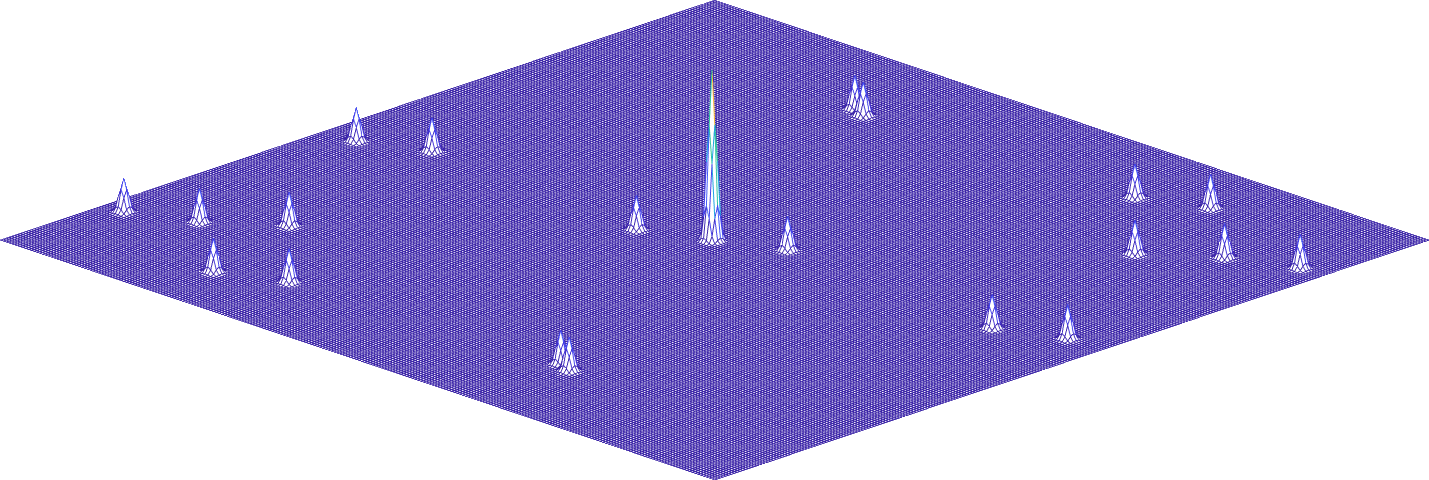};
\end{axis}
\end{tikzpicture}%

%% file: Figures/Results_EE/CM_Gauss60.tikz
%
%
\begin{tikzpicture}

\begin{axis}[%
width=1\linewidth,
height=1\linewidth,
xmin=0,
xmax=256,
xtick={1,129,256},
xticklabels={{-128},{0},{128}},
xtick style={draw=none},
xlabel style={font=\color{white!15!black},yshift=2mm,xshift=0mm},
xlabel={x / \si{\pixel}},
ymin=0,
ymax=256,
ytick={1,129,256},
yticklabels={{-128},{0},{128}},
ytick style={draw=none},
ylabel style={font=\color{white!15!black},yshift=2mm,xshift=0mm},
ylabel={y / \si{\pixel}},
zmin=-1,
zmax=1,
ztick={-1,0,1},
ztick style={draw=none},
zlabel style={font=\color{white!15!black},yshift=-2mm},
zlabel={R(s)},
view={45}{45},
axis background/.style={fill=white},
axis x line*=bottom,
axis y line*=left,
axis z line*=left,
xmajorgrids,
ymajorgrids,
zmajorgrids,
enlargelimits=false,
tick align = outside,
]

\addplot3 graphics [xmin=0, xmax=256, ymin=0, ymax=256, zmin=-1.425, zmax=1.405] {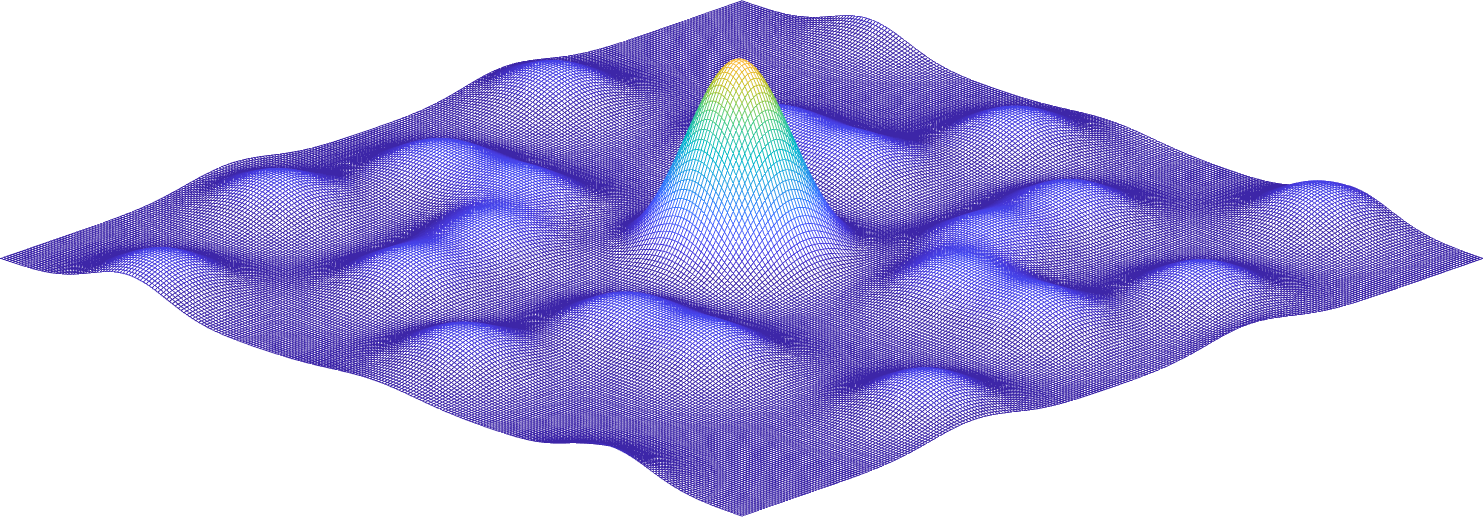};
\end{axis}
\end{tikzpicture}%

%% file: Figures/Results_EE/CM_Halo5.tikz
%
%
\begin{tikzpicture}

\begin{axis}[%
width=1\linewidth,
height=1\linewidth,
xmin=0,
xmax=256,
xtick={1,129,256},
xticklabels={{-128},{0},{128}},
xlabel style={font=\color{white!15!black},yshift=2mm,xshift=0mm},
xlabel={x / \si{\pixel}},
ymin=0,
ymax=256,
ytick={1,129,256},
yticklabels={{-128},{0},{128}},
ylabel style={font=\color{white!15!black},yshift=2mm,xshift=0mm},
ylabel={y / \si{\pixel}},
zmin=-1,
zmax=1,
ztick={-1,0,1},
zlabel style={font=\color{white!15!black},yshift=-2mm},
zlabel={R(s)},
view={45}{45},
axis background/.style={fill=white},
axis x line*=bottom,
axis y line*=left,
axis z line*=left,
xtick style={draw=none},
ytick style={draw=none},
ztick style={draw=none},
xmajorgrids,
ymajorgrids,
zmajorgrids,
enlargelimits=false,
tick align = outside,
]

\addplot3 graphics [xmin=0, xmax=256, ymin=0, ymax=256, zmin=-1.425, zmax=1.405] {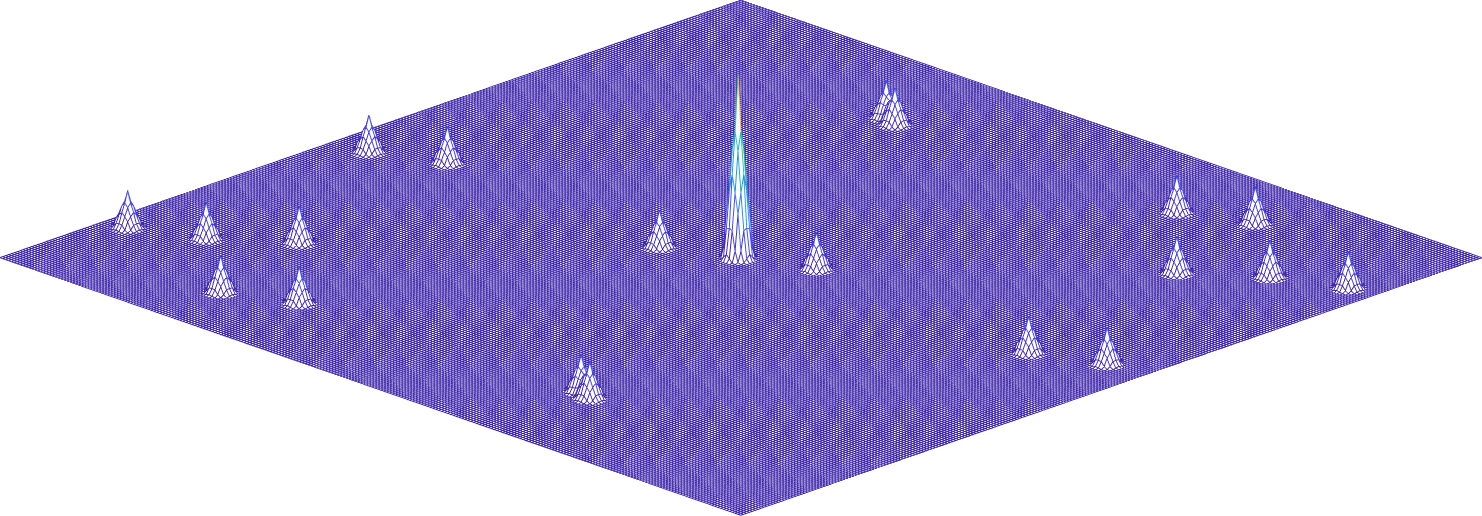};
\end{axis}
\end{tikzpicture}%

%% file: Figures/Results_EE/CM_Halo60.tikz
%
%
\begin{tikzpicture}

\begin{axis}[%
width=1\linewidth,
height=1\linewidth,
xmin=0,
xmax=256,
xtick={1,129,256},
xticklabels={{-128},{0},{128}},
xlabel style={font=\color{white!15!black},yshift=2mm,xshift=0mm},
xlabel={x / \si{\pixel}},
ymin=0,
ymax=256,
ytick={1,129,256},
yticklabels={{-128},{0},{128}},
ylabel style={font=\color{white!15!black},yshift=2mm,xshift=0mm},
ylabel={y / \si{\pixel}},
zmin=-1,
zmax=1,
ztick={-1,0,1},
zlabel style={font=\color{white!15!black},yshift=-2mm},
zlabel={R(s)},
view={45}{45},
axis background/.style={fill=white},
axis x line*=bottom,
axis y line*=left,
axis z line*=left,
xtick style={draw=none},
ytick style={draw=none},
ztick style={draw=none},
xmajorgrids,
ymajorgrids,
zmajorgrids,
enlargelimits=false,
tick align = outside,
]

\addplot3 graphics [xmin=0, xmax=256, ymin=0, ymax=256, zmin=-1.425, zmax=1.405] {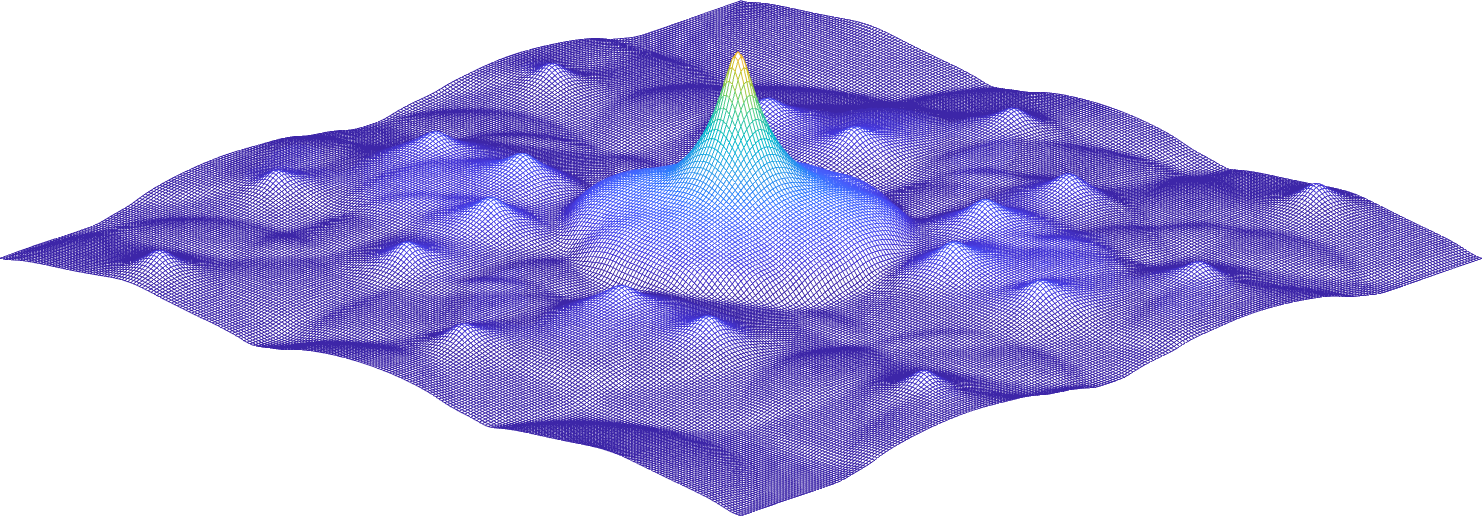};
\end{axis}
\end{tikzpicture}%

%% file: Figures/Results_EE/CM_Plateau5.tikz
%
%
\begin{tikzpicture}

\begin{axis}[%
width=1\linewidth,
height=1\linewidth,
xmin=0,
xmax=256,
xtick={1,129,256},
xticklabels={{-128},{0},{128}},
xlabel style={font=\color{white!15!black},yshift=2mm,xshift=0mm},
xlabel={x / \si{\pixel}},
ymin=0,
ymax=256,
ytick={1,129,256},
yticklabels={{-128},{0},{128}},
ylabel style={font=\color{white!15!black},yshift=2mm,xshift=0mm},
ylabel={y / \si{\pixel}},
zmin=-1,
zmax=1,
ztick={-1,0,1},
zlabel style={font=\color{white!15!black},yshift=-2mm},
zlabel={R(s)},
view={45}{45},
axis background/.style={fill=white},
axis x line*=bottom,
axis y line*=left,
axis z line*=left,
xtick style={draw=none},
ytick style={draw=none},
ztick style={draw=none},
xmajorgrids,
ymajorgrids,
zmajorgrids,
enlargelimits=false,
tick align = outside,
]

\addplot3 graphics [xmin=0, xmax=256, ymin=0, ymax=256, zmin=-1.425, zmax=1.405] {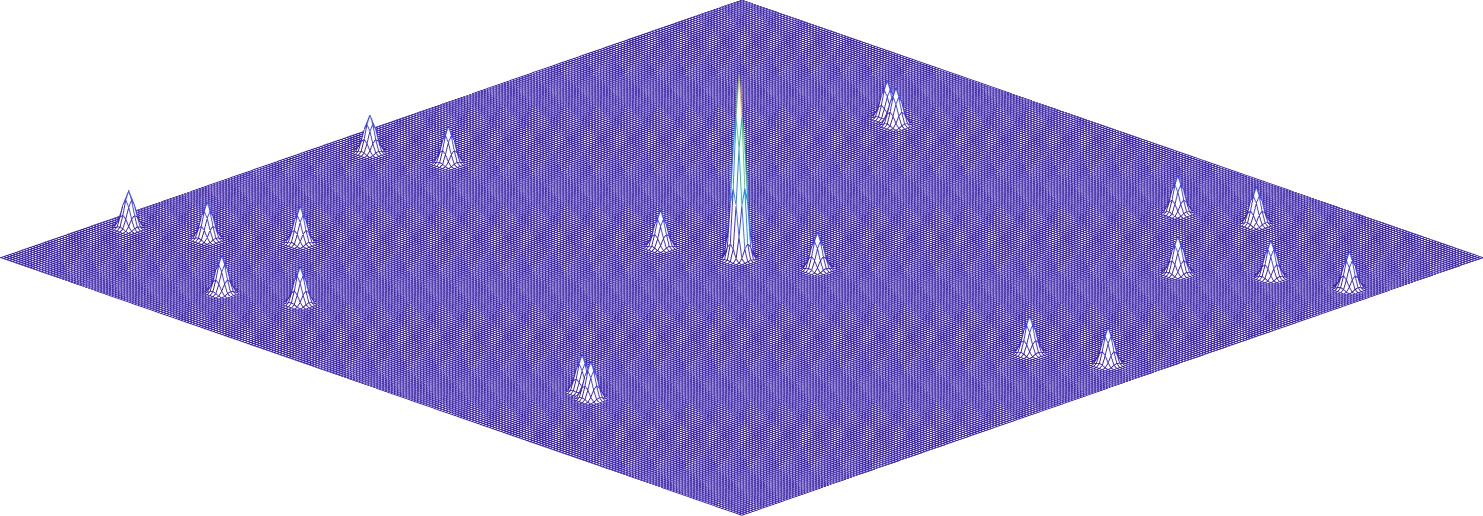};
\end{axis}
\end{tikzpicture}%

%% file: Figures/Results_EE/CM_Plateau60.tikz
%
%
\begin{tikzpicture}

\begin{axis}[%
width=1\linewidth,
height=1\linewidth,
xmin=0,
xmax=256,
xtick={1,129,256},
xticklabels={{-128},{0},{128}},
xlabel style={font=\color{white!15!black},yshift=2mm,xshift=0mm},
xlabel={x / \si{\pixel}},
ymin=0,
ymax=256,
ytick={1,129,256},
yticklabels={{-128},{0},{128}},
ylabel style={font=\color{white!15!black},yshift=2mm,xshift=0mm},
ylabel={y / \si{\pixel}},
zmin=-1,
zmax=1,
ztick={-1,0,1},
zlabel style={font=\color{white!15!black},yshift=-2mm},
zlabel={R(s)},
view={45}{45},
axis background/.style={fill=white},
axis x line*=bottom,
axis y line*=left,
axis z line*=left,
xtick style={draw=none},
ytick style={draw=none},
ztick style={draw=none},
xmajorgrids,
ymajorgrids,
zmajorgrids,
enlargelimits=false,
tick align = outside,
]

\addplot3 graphics [xmin=0, xmax=256, ymin=0, ymax=256, zmin=-1.425, zmax=1.405] {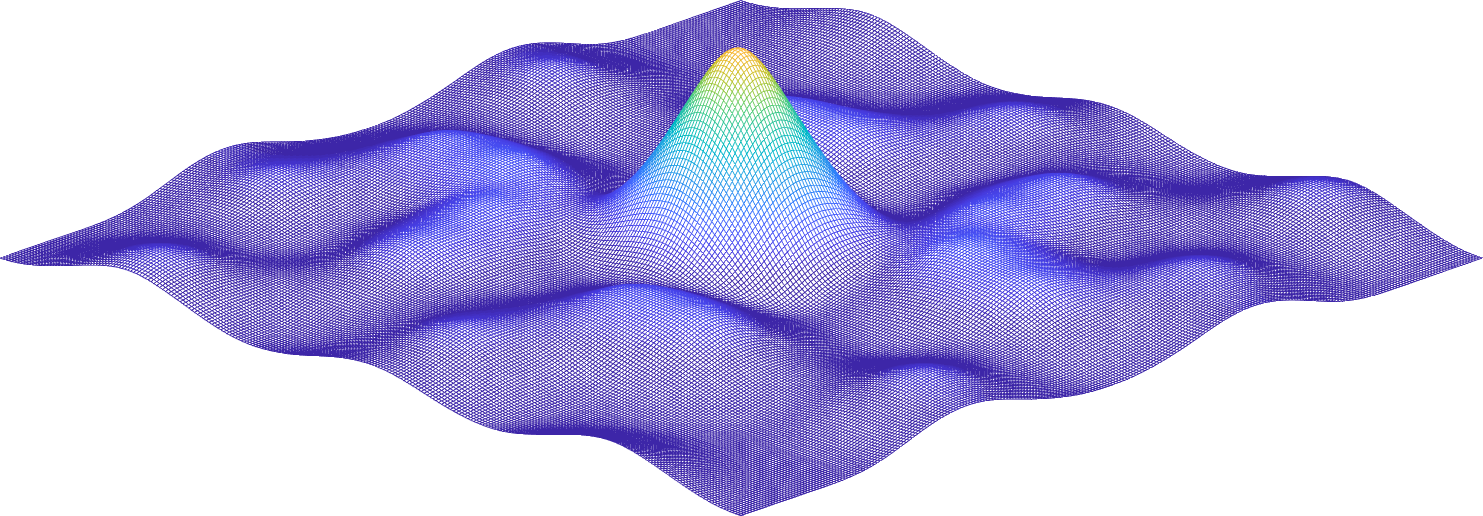};
\end{axis}
\end{tikzpicture}%

%% file: Figures/Results_EE/Colorbar.tikz
%
%
\begin{tikzpicture}

\begin{axis}[hide axis,
width=1\columnwidth,
height=1\columnwidth,
colormap={mymap}{[1pt] rgb(0pt)=(0.2422,0.1504,0.6603); rgb(1pt)=(0.25039,0.164995,0.707614); rgb(2pt)=(0.257771,0.181781,0.751138); rgb(3pt)=(0.264729,0.197757,0.795214); rgb(4pt)=(0.270648,0.214676,0.836371); rgb(5pt)=(0.275114,0.234238,0.870986); rgb(6pt)=(0.2783,0.255871,0.899071); rgb(7pt)=(0.280333,0.278233,0.9221); rgb(8pt)=(0.281338,0.300595,0.941376); rgb(9pt)=(0.281014,0.322757,0.957886); rgb(10pt)=(0.279467,0.344671,0.971676); rgb(11pt)=(0.275971,0.366681,0.982905); rgb(12pt)=(0.269914,0.3892,0.9906); rgb(13pt)=(0.260243,0.412329,0.995157); rgb(14pt)=(0.244033,0.435833,0.998833); rgb(15pt)=(0.220643,0.460257,0.997286); rgb(16pt)=(0.196333,0.484719,0.989152); rgb(17pt)=(0.183405,0.507371,0.979795); rgb(18pt)=(0.178643,0.528857,0.968157); rgb(19pt)=(0.176438,0.549905,0.952019); rgb(20pt)=(0.168743,0.570262,0.935871); rgb(21pt)=(0.154,0.5902,0.9218); rgb(22pt)=(0.146029,0.609119,0.907857); rgb(23pt)=(0.138024,0.627629,0.89729); rgb(24pt)=(0.124814,0.645929,0.888343); rgb(25pt)=(0.111252,0.6635,0.876314); rgb(26pt)=(0.0952095,0.679829,0.859781); rgb(27pt)=(0.0688714,0.694771,0.839357); rgb(28pt)=(0.0296667,0.708167,0.816333); rgb(29pt)=(0.00357143,0.720267,0.7917); rgb(30pt)=(0.00665714,0.731214,0.766014); rgb(31pt)=(0.0433286,0.741095,0.73941); rgb(32pt)=(0.0963952,0.75,0.712038); rgb(33pt)=(0.140771,0.7584,0.684157); rgb(34pt)=(0.1717,0.766962,0.655443); rgb(35pt)=(0.193767,0.775767,0.6251); rgb(36pt)=(0.216086,0.7843,0.5923); rgb(37pt)=(0.246957,0.791795,0.556743); rgb(38pt)=(0.290614,0.79729,0.518829); rgb(39pt)=(0.340643,0.8008,0.478857); rgb(40pt)=(0.3909,0.802871,0.435448); rgb(41pt)=(0.445629,0.802419,0.390919); rgb(42pt)=(0.5044,0.7993,0.348); rgb(43pt)=(0.561562,0.794233,0.304481); rgb(44pt)=(0.617395,0.787619,0.261238); rgb(45pt)=(0.671986,0.779271,0.2227); rgb(46pt)=(0.7242,0.769843,0.191029); rgb(47pt)=(0.773833,0.759805,0.16461); rgb(48pt)=(0.820314,0.749814,0.153529); rgb(49pt)=(0.863433,0.7406,0.159633); rgb(50pt)=(0.903543,0.733029,0.177414); rgb(51pt)=(0.939257,0.728786,0.209957); rgb(52pt)=(0.972757,0.729771,0.239443)},
colorbar horizontal,
colorbar style={title=R(s),  title style={at={(0.5,-1.5)}, anchor=north}}
]
\end{axis}
\end{tikzpicture}%

%% file: Figures/Results_EE/CM_Gauss60Zoom_V2.tikz
%
%
\begin{tikzpicture}

\begin{axis}[%
width=1\linewidth,
height=1\linewidth,
xmin=-10,
xmax=10,
xlabel={x / \si{\pixel}},
tick align=outside,
ymin=-10,
ymax=10,
ylabel={y / \si{\pixel}},
zmin=0.6,
zmax=1,
ztick={0.6,1},
zlabel={R(s)},
view={45}{45},
axis background/.style={fill=white},
axis x line*=bottom,
axis y line*=left,
axis z line*=left,
xmajorgrids,
ymajorgrids,
zmajorgrids,
xlabel style={font=\color{white!15!black},yshift=2mm,xshift=0mm},
ylabel style={font=\color{white!15!black},yshift=2mm,xshift=0mm},
xtick style={draw=none},
ytick style={draw=none},
ztick style={draw=none},
]

\addplot3[%
surf,
shader=flat corner, fill=white, z buffer=sort, colormap={mymap}{[1pt] rgb(0pt)=(0.2422,0.1504,0.6603); rgb(1pt)=(0.25039,0.164995,0.707614); rgb(2pt)=(0.257771,0.181781,0.751138); rgb(3pt)=(0.264729,0.197757,0.795214); rgb(4pt)=(0.270648,0.214676,0.836371); rgb(5pt)=(0.275114,0.234238,0.870986); rgb(6pt)=(0.2783,0.255871,0.899071); rgb(7pt)=(0.280333,0.278233,0.9221); rgb(8pt)=(0.281338,0.300595,0.941376); rgb(9pt)=(0.281014,0.322757,0.957886); rgb(10pt)=(0.279467,0.344671,0.971676); rgb(11pt)=(0.275971,0.366681,0.982905); rgb(12pt)=(0.269914,0.3892,0.9906); rgb(13pt)=(0.260243,0.412329,0.995157); rgb(14pt)=(0.244033,0.435833,0.998833); rgb(15pt)=(0.220643,0.460257,0.997286); rgb(16pt)=(0.196333,0.484719,0.989152); rgb(17pt)=(0.183405,0.507371,0.979795); rgb(18pt)=(0.178643,0.528857,0.968157); rgb(19pt)=(0.176438,0.549905,0.952019); rgb(20pt)=(0.168743,0.570262,0.935871); rgb(21pt)=(0.154,0.5902,0.9218); rgb(22pt)=(0.146029,0.609119,0.907857); rgb(23pt)=(0.138024,0.627629,0.89729); rgb(24pt)=(0.124814,0.645929,0.888343); rgb(25pt)=(0.111252,0.6635,0.876314); rgb(26pt)=(0.0952095,0.679829,0.859781); rgb(27pt)=(0.0688714,0.694771,0.839357); rgb(28pt)=(0.0296667,0.708167,0.816333); rgb(29pt)=(0.00357143,0.720267,0.7917); rgb(30pt)=(0.00665714,0.731214,0.766014); rgb(31pt)=(0.0433286,0.741095,0.73941); rgb(32pt)=(0.0963952,0.75,0.712038); rgb(33pt)=(0.140771,0.7584,0.684157); rgb(34pt)=(0.1717,0.766962,0.655443); rgb(35pt)=(0.193767,0.775767,0.6251); rgb(36pt)=(0.216086,0.7843,0.5923); rgb(37pt)=(0.246957,0.791795,0.556743); rgb(38pt)=(0.290614,0.79729,0.518829); rgb(39pt)=(0.340643,0.8008,0.478857); rgb(40pt)=(0.3909,0.802871,0.435448); rgb(41pt)=(0.445629,0.802419,0.390919); rgb(42pt)=(0.5044,0.7993,0.348); rgb(43pt)=(0.561562,0.794233,0.304481); rgb(44pt)=(0.617395,0.787619,0.261238); rgb(45pt)=(0.671986,0.779271,0.2227); rgb(46pt)=(0.7242,0.769843,0.191029); rgb(47pt)=(0.773833,0.759805,0.16461); rgb(48pt)=(0.820314,0.749814,0.153529); rgb(49pt)=(0.863433,0.7406,0.159633); rgb(50pt)=(0.903543,0.733029,0.177414); rgb(51pt)=(0.939257,0.728786,0.209957); rgb(52pt)=(0.972757,0.729771,0.239443); rgb(53pt)=(0.995648,0.743371,0.237148); rgb(54pt)=(0.996986,0.765857,0.219943); rgb(55pt)=(0.995205,0.789252,0.202762); rgb(56pt)=(0.9892,0.813567,0.188533); rgb(57pt)=(0.978629,0.838629,0.176557); rgb(58pt)=(0.967648,0.8639,0.16429); rgb(59pt)=(0.96101,0.889019,0.153676); rgb(60pt)=(0.959671,0.913457,0.142257); rgb(61pt)=(0.962795,0.937338,0.12651); rgb(62pt)=(0.969114,0.960629,0.106362); rgb(63pt)=(0.9769,0.9839,0.0805)}, mesh/rows=31]
table[row sep=crcr, point meta=\thisrow{c}] {%
x	y	z	c\\
-15	-15	0.203708182017719	0.203708182017719\\
-15	-14	0.229539793003409	0.229539793003409\\
-15	-13	0.255673856311582	0.255673856311582\\
-15	-12	0.281805696081471	0.281805696081471\\
-15	-11	0.307608988901426	0.307608988901426\\
-15	-10	0.332741769918073	0.332741769918073\\
-15	-9	0.356853307914457	0.356853307914457\\
-15	-8	0.379592130494197	0.379592130494197\\
-15	-7	0.40061376812861	0.40061376812861\\
-15	-6	0.419589111593713	0.419589111593713\\
-15	-5	0.436212820523044	0.436212820523044\\
-15	-4	0.450210888549412	0.450210888549412\\
-15	-3	0.461348106214999	0.461348106214999\\
-15	-2	0.469434271543811	0.469434271543811\\
-15	-1	0.474329531645141	0.474329531645141\\
-15	0	0.475947960821153	0.475947960821153\\
-15	1	0.474259963010546	0.474259963010546\\
-15	2	0.469293498568297	0.469293498568297\\
-15	3	0.461132601906804	0.461132601906804\\
-15	4	0.449915541326274	0.449915541326274\\
-15	5	0.435830729748908	0.435830729748908\\
-15	6	0.41911153646291	0.41911153646291\\
-15	7	0.400029974318388	0.400029974318388\\
-15	8	0.378889696859659	0.378889696859659\\
-15	9	0.356018024256911	0.356018024256911\\
-15	10	0.331758045911619	0.331758045911619\\
-15	11	0.306460237461621	0.306460237461621\\
-15	12	0.280474767849824	0.280474767849824\\
-15	13	0.25414395974036	0.25414395974036\\
-15	14	0.227795618903735	0.227795618903735\\
-15	15	0.201737309244695	0.201737309244695\\
-14	-15	0.229408859823751	0.229408859823751\\
-14	-14	0.25759829035975	0.25759829035975\\
-14	-13	0.286102517689397	0.286102517689397\\
-14	-12	0.314591589177635	0.314591589177635\\
-14	-11	0.342711808889798	0.342711808889798\\
-14	-10	0.370092663785574	0.370092663785574\\
-14	-9	0.396354184397468	0.396354184397468\\
-14	-8	0.421115455615262	0.421115455615262\\
-14	-7	0.444003229702137	0.444003229702137\\
-14	-6	0.464660820447801	0.464660820447801\\
-14	-5	0.482757252900746	0.482757252900746\\
-14	-4	0.497995493015687	0.497995493015687\\
-14	-3	0.510120779532652	0.510120779532652\\
-14	-2	0.518927115686454	0.518927115686454\\
-14	-1	0.524262636368069	0.524262636368069\\
-14	0	0.526034464127785	0.526034464127785\\
-14	1	0.524210907155571	0.524210907155571\\
-14	2	0.518822226018424	0.518822226018424\\
-14	3	0.509960020270491	0.509960020270491\\
-14	4	0.497774365967004	0.497774365967004\\
-14	5	0.482470033093406	0.482470033093406\\
-14	6	0.464300300550814	0.464300300550814\\
-14	7	0.443560617794011	0.443560617794011\\
-14	8	0.420580630763117	0.420580630763117\\
-14	9	0.395715645541213	0.395715645541213\\
-14	10	0.369337989780311	0.369337989780311\\
-14	11	0.341828092990313	0.341828092990313\\
-14	12	0.313565796849159	0.313565796849159\\
-14	13	0.284922457808233	0.284922457808233\\
-14	14	0.256253535298082	0.256253535298082\\
-14	15	0.227892432271791	0.227892432271791\\
-13	-15	0.255450965767018	0.255450965767018\\
-13	-14	0.28600897573399	0.28600897573399\\
-13	-13	0.316895047050581	0.316895047050581\\
-13	-12	0.347753515822954	0.347753515822954\\
-13	-11	0.378203160245957	0.378203160245957\\
-13	-10	0.407844586839495	0.407844586839495\\
-13	-9	0.436268383422239	0.436268383422239\\
-13	-8	0.46306380880147	0.46306380880147\\
-13	-7	0.487828453663558	0.487828453663558\\
-13	-6	0.510177978138172	0.510177978138172\\
-13	-5	0.52975549155173	0.52975549155173\\
-13	-4	0.546240753275476	0.546240753275476\\
-13	-3	0.559359271341904	0.559359271341904\\
-13	-2	0.568888691922586	0.568888691922586\\
-13	-1	0.574665699964229	0.574665699964229\\
-13	0	0.576590057338663	0.576590057338663\\
-13	1	0.574627823139663	0.574627823139663\\
-13	2	0.568811762609532	0.568811762609532\\
-13	3	0.559240835980877	0.559240835980877\\
-13	4	0.546077387106361	0.546077387106361\\
-13	5	0.529542338571376	0.529542338571376\\
-13	6	0.509909160026977	0.509909160026977\\
-13	7	0.487496916438	0.487496916438\\
-13	8	0.462661373929929	0.462661373929929\\
-13	9	0.435786105635755	0.435786105635755\\
-13	10	0.407272830805503	0.407272830805503\\
-13	11	0.3775320861686	0.3775320861686\\
-13	12	0.346973692832958	0.346973692832958\\
-13	13	0.315998219942593	0.315998219942593\\
-13	14	0.284988959493471	0.284988959493471\\
-13	15	0.254305102371191	0.254305102371191\\
-12	-15	0.281521926592158	0.281521926592158\\
-12	-14	0.314434024654388	0.314434024654388\\
-12	-13	0.347688164243726	0.347688164243726\\
-12	-12	0.380902842417995	0.380902842417995\\
-12	-11	0.413669362617384	0.413669362617384\\
-12	-10	0.445559655385302	0.445559655385302\\
-12	-9	0.476134584689564	0.476134584689564\\
-12	-8	0.504953941065706	0.504953941065706\\
-12	-7	0.531586204739111	0.531586204739111\\
-12	-6	0.555619279947764	0.555619279947764\\
-12	-5	0.576670309180187	0.576670309180187\\
-12	-4	0.594396254150733	0.594396254150733\\
-12	-3	0.608502381026134	0.608502381026134\\
-12	-2	0.61875069453624	0.61875069453624\\
-12	-1	0.624965765177796	0.624965765177796\\
-12	0	0.62704055641822	0.62704055641822\\
-12	1	0.624938469328513	0.624938469328513\\
-12	2	0.618694876057931	0.618694876057931\\
-12	3	0.608416404212475	0.608416404212475\\
-12	4	0.594277103168189	0.594277103168189\\
-12	5	0.576514201457885	0.576514201457885\\
-12	6	0.555421461714198	0.555421461714198\\
-12	7	0.531341104369616	0.531341104369616\\
-12	8	0.504655117966629	0.504655117966629\\
-12	9	0.475775189340674	0.475775189340674\\
-12	10	0.445132429339787	0.445132429339787\\
-12	11	0.413167200775897	0.413167200775897\\
-12	12	0.380319253071064	0.380319253071064\\
-12	13	0.347018189156555	0.347018189156555\\
-12	14	0.313675005804202	0.313675005804202\\
-12	15	0.280674681832118	0.280674681832118\\
-11	-15	0.307288646041023	0.307288646041023\\
-11	-14	0.342513377266361	0.342513377266361\\
-11	-13	0.378094334659584	0.378094334659584\\
-11	-12	0.41362473850423	0.41362473850423\\
-11	-11	0.448668826412404	0.448668826412404\\
-11	-10	0.482770031856403	0.482770031856403\\
-11	-9	0.515460133900964	0.515460133900964\\
-11	-8	0.546269735946906	0.546269735946906\\
-11	-7	0.574738540011472	0.574738540011472\\
-11	-6	0.600427052251718	0.600427052251718\\
-11	-5	0.622927163939791	0.622927163939791\\
-11	-4	0.641872936901504	0.641872936901504\\
-11	-3	0.656950111059348	0.656950111059348\\
-11	-2	0.667904691890689	0.667904691890689\\
-11	-1	0.67454990217965	0.67454990217965\\
-11	0	0.676771446946837	0.676771446946837\\
-11	1	0.674530427051229	0.674530427051229\\
-11	2	0.667864974896508	0.667864974896508\\
-11	3	0.656888618724728	0.656888618724728\\
-11	4	0.641787471246072	0.641787471246072\\
-11	5	0.622814606898367	0.622814606898367\\
-11	6	0.600283570137605	0.600283570137605\\
-11	7	0.574559941327214	0.574559941327214\\
-11	8	0.546051062457708	0.546051062457708\\
-11	9	0.515196325140387	0.515196325140387\\
-11	10	0.482455771778893	0.482455771778893\\
-11	11	0.448299361246457	0.448299361246457\\
-11	12	0.413196004541948	0.413196004541948\\
-11	13	0.377603980573125	0.377603980573125\\
-11	14	0.341961479729458	0.341961479729458\\
-11	15	0.306678885392907	0.306678885392907\\
-10	-15	0.332403996562154	0.332403996562154\\
-10	-14	0.369871383389423	0.369871383389423\\
-10	-13	0.407709334372825	0.407709334372825\\
-10	-12	0.445486355295655	0.445486355295655\\
-10	-11	0.482740486910925	0.482740486910925\\
-10	-10	0.518987227893864	0.518987227893864\\
-10	-9	0.553730243606765	0.553730243606765\\
-10	-8	0.586471563705599	0.586471563705599\\
-10	-7	0.616723032084277	0.616723032084277\\
-10	-6	0.644018574672089	0.644018574672089\\
-10	-5	0.667925291567208	0.667925291567208\\
-10	-4	0.688055009212605	0.688055009212605\\
-10	-3	0.704074503560575	0.704074503560575\\
-10	-2	0.715714240878407	0.715714240878407\\
-10	-1	0.722776402932535	0.722776402932535\\
-10	0	0.725139691875866	0.725139691875866\\
-10	1	0.722762806123716	0.722762806123716\\
-10	2	0.715686587218366	0.715686587218366\\
-10	3	0.704031361806278	0.704031361806278\\
-10	4	0.687994743657728	0.687994743657728\\
-10	5	0.667845448652265	0.667845448652265\\
-10	6	0.643916343026836	0.643916343026836\\
-10	7	0.61659503806442	0.61659503806442\\
-10	8	0.586314229203553	0.586314229203553\\
-10	9	0.553539939399125	0.553539939399125\\
-10	10	0.518760324757223	0.518760324757223\\
-10	11	0.482473866780103	0.482473866780103\\
-10	12	0.445178178001038	0.445178178001038\\
-10	13	0.407359139872008	0.407359139872008\\
-10	14	0.36948136966278	0.36948136966278\\
-10	15	0.331979939697642	0.331979939697642\\
-9	-15	0.356512978767407	0.356512978767407\\
-9	-14	0.396123856500711	0.396123856500711\\
-9	-13	0.436119483030927	0.436119483030927\\
-9	-12	0.476044723073386	0.476044723073386\\
-9	-11	0.515412238375163	0.515412238375163\\
-9	-10	0.553711177404925	0.553711177404925\\
-9	-9	0.590417807444038	0.590417807444038\\
-9	-8	0.625007066762478	0.625007066762478\\
-9	-7	0.65696396344728	0.65696396344728\\
-9	-6	0.685796865516418	0.685796865516418\\
-9	-5	0.711049206442189	0.711049206442189\\
-9	-4	0.732311548485351	0.732311548485351\\
-9	-3	0.749232419249524	0.749232419249524\\
-9	-2	0.761527461413432	0.761527461413432\\
-9	-1	0.768987969073292	0.768987969073292\\
-9	0	0.771486203788357	0.771486203788357\\
-9	1	0.768978768225219	0.768978768225219\\
-9	2	0.76150854855906	0.76150854855906\\
-9	3	0.7492027720724	0.7492027720724\\
-9	4	0.73226988908991	0.73226988908991\\
-9	5	0.710993694658816	0.710993694658816\\
-9	6	0.685725354480563	0.685725354480563\\
-9	7	0.656873999599456	0.656873999599456\\
-9	8	0.624896043195733	0.624896043195733\\
-9	9	0.590283321714707	0.590283321714707\\
-9	10	0.553550775953522	0.553550775953522\\
-9	11	0.515224387727009	0.515224387727009\\
-9	12	0.475828912070255	0.475828912070255\\
-9	13	0.43587688733674	0.43587688733674\\
-9	14	0.395858156454697	0.395858156454697\\
-9	15	0.356231279468911	0.356231279468911\\
-8	-15	0.379260542152816	0.379260542152816\\
-8	-14	0.420886482287805	0.420886482287805\\
-8	-13	0.462910972491816	0.462910972491816\\
-8	-12	0.504856463191956	0.504856463191956\\
-8	-11	0.54621182171987	0.54621182171987\\
-8	-10	0.586441149791136	0.586441149791136\\
-8	-9	0.624995105659984	0.624995105659984\\
-8	-8	0.661322814105963	0.661322814105963\\
-8	-7	0.69488438981049	0.69488438981049\\
-8	-6	0.725163767428323	0.725163767428323\\
-8	-5	0.751682196164739	0.751682196164739\\
-8	-4	0.774010200878027	0.774010200878027\\
-8	-3	0.791778980907934	0.791778980907934\\
-8	-2	0.804690479892722	0.804690479892722\\
-8	-1	0.812525615416614	0.812525615416614\\
-8	0	0.815150361792681	0.815150361792681\\
-8	1	0.812519532633721	0.812519532633721\\
-8	2	0.804677854284533	0.804677854284533\\
-8	3	0.791759147968754	0.791759147968754\\
-8	4	0.773982087175582	0.773982087175582\\
-8	5	0.751644523803463	0.751644523803463\\
-8	6	0.725114849586069	0.725114849586069\\
-8	7	0.694822488549289	0.694822488549289\\
-8	8	0.661246344835314	0.661246344835314\\
-8	9	0.624902279326093	0.624902279326093\\
-8	10	0.586330688498439	0.586330688498439\\
-8	11	0.54608321431014	0.54608321431014\\
-8	12	0.504710271939243	0.504710271939243\\
-8	13	0.462749395376492	0.462749395376492\\
-8	14	0.42071417084973	0.42071417084973\\
-8	15	0.379085265997029	0.379085265997029\\
-7	-15	0.400299559166923	0.400299559166923\\
-7	-14	0.443783431664841	0.443783431664841\\
-7	-13	0.487678888766552	0.487678888766552\\
-7	-12	0.531487806780555	0.531487806780555\\
-7	-11	0.574676792097739	0.574676792097739\\
-7	-10	0.616687046545148	0.616687046545148\\
-7	-9	0.656945357287844	0.656945357287844\\
-7	-8	0.694877233595322	0.694877233595322\\
-7	-7	0.729919327986414	0.729919327986414\\
-7	-6	0.761533493080502	0.761533493080502\\
-7	-5	0.789220327290523	0.789220327290523\\
-7	-4	0.812531595967861	0.812531595967861\\
-7	-3	0.831082805894718	0.831082805894718\\
-7	-2	0.844563019522053	0.844563019522053\\
-7	-1	0.852743595775257	0.852743595775257\\
-7	0	0.855484886226913	0.855484886226913\\
-7	1	0.852739710972737	0.852739710972737\\
-7	2	0.844554840990432	0.844554840990432\\
-7	3	0.831069924707416	0.831069924707416\\
-7	4	0.812513194271716	0.812513194271716\\
-7	5	0.789195331653259	0.789195331653259\\
-7	6	0.761500881185666	0.761500881185666\\
-7	7	0.729877821938441	0.729877821938441\\
-7	8	0.69482581107776	0.69482581107776\\
-7	9	0.656883047100061	0.656883047100061\\
-7	10	0.616613184181452	0.616613184181452\\
-7	11	0.574591735368888	0.574591735368888\\
-7	12	0.531392731350469	0.531392731350469\\
-7	13	0.487576657121298	0.487576657121298\\
-7	14	0.44367910427086	0.44367910427086\\
-7	15	0.400201135650454	0.400201135650454\\
-6	-15	0.419298799279103	0.419298799279103\\
-6	-14	0.464456050462358	0.464456050462358\\
-6	-13	0.510036949583544	0.510036949583544\\
-6	-12	0.555524715675904	0.555524715675904\\
-6	-11	0.600365968843679	0.600365968843679\\
-6	-10	0.643980953426636	0.643980953426636\\
-6	-9	0.685775345755092	0.685775345755092\\
-6	-8	0.725152879758104	0.725152879758104\\
-6	-7	0.761529454930515	0.761529454930515\\
-6	-6	0.794346784257594	0.794346784257594\\
-6	-5	0.823086604413729	0.823086604413729\\
-6	-4	0.847284323687062	0.847284323687062\\
-6	-3	0.866540778618701	0.866540778618701\\
-6	-2	0.880533428367871	0.880533428367871\\
-6	-1	0.889025402212518	0.889025402212518\\
-6	0	0.891871531216382	0.891871531216382\\
-6	1	0.889022948653032	0.889022948653032\\
-6	2	0.880528316785608	0.880528316785608\\
-6	3	0.866532600087081	0.866532600087081\\
-6	4	0.847272567047858	0.847272567047858\\
-6	5	0.823070656277069	0.823070656277069\\
-6	6	0.794325673422849	0.794325673422849\\
-6	7	0.76150251689199	0.76150251689199\\
-6	8	0.725119398894283	0.725119398894283\\
-6	9	0.685734810907748	0.685734810907748\\
-6	10	0.643933415711593	0.643933415711593\\
-6	11	0.600311990534985	0.600311990534985\\
-6	12	0.555466034711528	0.555466034711528\\
-6	13	0.509976684028667	0.509976684028667\\
-6	14	0.464399311899242	0.464399311899242\\
-6	15	0.41925320396532	0.41925320396532\\
-5	-15	0.435951209742839	0.435951209742839\\
-5	-14	0.482571753580433	0.482571753580433\\
-5	-13	0.529626781910356	0.529626781910356\\
-5	-12	0.57658300335514	0.57658300335514\\
-5	-11	0.622869709755158	0.622869709755158\\
-5	-10	0.667888845985675	0.667888845985675\\
-5	-9	0.711027226638459	0.711027226638459\\
-5	-8	0.751669928367309	0.751669928367309\\
-5	-7	0.78921424450763	0.78921424450763\\
-5	-6	0.823084457549178	0.823084457549178\\
-5	-5	0.85274599821892	0.85274599821892\\
-5	-4	0.877719195637326	0.877719195637326\\
-5	-3	0.897592823011362	0.897592823011362\\
-5	-2	0.912034065219905	0.912034065219905\\
-5	-1	0.920798384167508	0.920798384167508\\
-5	0	0.923736112725519	0.923736112725519\\
-5	1	0.920796952924474	0.920796952924474\\
-5	2	0.912031100502192	0.912031100502192\\
-5	3	0.897588018124035	0.897588018124035\\
-5	4	0.877712039422158	0.877712039422158\\
-5	5	0.852736030633508	0.852736030633508\\
-5	6	0.823071269666941	0.823071269666941\\
-5	7	0.789197325170341	0.789197325170341\\
-5	8	0.751648919764209	0.751648919764209\\
-5	9	0.711001975422081	0.711001975422081\\
-5	10	0.667859607735133	0.667859607735133\\
-5	11	0.622837353439435	0.622837353439435\\
-5	12	0.576549266912207	0.576549266912207\\
-5	13	0.529594783405391	0.529594783405391\\
-5	14	0.482546349016588	0.482546349016588\\
-5	15	0.43593935087199	0.43593935087199\\
-4	-15	0.44998163408493	0.44998163408493\\
-4	-14	0.49783248465733	0.49783248465733\\
-4	-13	0.546127122801777	0.546127122801777\\
-4	-12	0.59431855810034	0.59431855810034\\
-4	-11	0.641821309920651	0.641821309920651\\
-4	-10	0.688021783927897	0.688021783927897\\
-4	-9	0.732290948808833	0.732290948808833\\
-4	-8	0.773998137543887	0.773998137543887\\
-4	-7	0.81252515537421	0.81252515537421\\
-4	-6	0.847281256737704	0.847281256737704\\
-4	-5	0.877718173320873	0.877718173320873\\
-4	-4	0.903344375373337	0.903344375373337\\
-4	-3	0.923737339505262	0.923737339505262\\
-4	-2	0.938556020948082	0.938556020948082\\
-4	-1	0.947549441012698	0.947549441012698\\
-4	0	0.950564149999586	0.950564149999586\\
-4	1	0.947548725391182	0.947548725391182\\
-4	2	0.938554283010113	0.938554283010113\\
-4	3	0.923734681482486	0.923734681482486\\
-4	4	0.903340286107527	0.903340286107527\\
-4	5	0.877712346117094	0.877712346117094\\
-4	6	0.847273333785197	0.847273333785197\\
-4	7	0.812514932209685	0.812514932209685\\
-4	8	0.773985460819875	0.773985460819875\\
-4	9	0.732275971872803	0.732275971872803\\
-4	10	0.68800471124314	0.68800471124314\\
-4	11	0.641803163803618	0.641803163803618\\
-4	12	0.594301076489001	0.594301076489001\\
-4	13	0.546112963718909	0.546112963718909\\
-4	14	0.497825890716211	0.497825890716211\\
-4	15	0.449988534720985	0.449988534720985\\
-3	-15	0.461153712741549	0.461153712741549\\
-3	-14	0.509982357884979	0.509982357884979\\
-3	-13	0.559262509089671	0.559262509089671\\
-3	-12	0.608436032688364	0.608436032688364\\
-3	-11	0.656905691409485	0.656905691409485\\
-3	-10	0.704045623120791	0.704045623120791\\
-3	-9	0.749214222016668	0.749214222016668\\
-3	-8	0.791768042121891	0.791768042121891\\
-3	-7	0.83107662088018	0.83107662088018\\
-3	-6	0.866537507206053	0.866537507206053\\
-3	-5	0.897591596231619	0.897591596231619\\
-3	-4	0.923736930578681	0.923736930578681\\
-3	-3	0.944543012789077	0.944543012789077\\
-3	-2	0.959661948574007	0.959661948574007\\
-3	-1	0.968837647662111	0.968837647662111\\
-3	0	0.971913797867795	0.971913797867795\\
-3	1	0.96883723873553	0.96883723873553\\
-3	2	0.959661130720845	0.959661130720845\\
-3	3	0.944541683777688	0.944541683777688\\
-3	4	0.92373457925084	0.92373457925084\\
-3	5	0.897588222587326	0.897588222587326\\
-3	6	0.866533009013662	0.866533009013662\\
-3	7	0.8310707936764	0.8310707936764\\
-3	8	0.791760834790901	0.791760834790901\\
-3	9	0.749205787905935	0.749205787905935\\
-3	10	0.704036320041073	0.704036320041073\\
-3	11	0.65689643944559	0.65689643944559\\
-3	12	0.608428263083325	0.608428263083325\\
-3	13	0.559258215360571	0.559258215360571\\
-3	14	0.5099851181394	0.5099851181394\\
-3	15	0.461168229635175	0.461168229635175\\
-2	-15	0.469276783694298	0.469276783694298\\
-2	-14	0.518814660876675	0.518814660876675\\
-2	-13	0.568810024671563	0.568810024671563\\
-2	-12	0.618696562880078	0.618696562880078\\
-2	-11	0.667868348540802	0.667868348540802\\
-2	-10	0.715690574252531	0.715690574252531\\
-2	-9	0.76151243336158	0.76151243336158\\
-2	-8	0.804681330160472	0.804681330160472\\
-2	-7	0.8445577034765	0.8445577034765\\
-2	-6	0.880530565881804	0.880530565881804\\
-2	-5	0.912032736208517	0.912032736208517\\
-2	-4	0.93855540755821	0.93855540755821\\
-2	-3	0.959661744110716	0.959661744110716\\
-2	-2	0.974998944458263	0.974998944458263\\
-2	-1	0.98430703352697	0.98430703352697\\
-2	0	0.987427552266695	0.987427552266695\\
-2	1	0.984306931295325	0.984306931295325\\
-2	2	0.974998433300036	0.974998433300036\\
-2	3	0.959661130720845	0.959661130720845\\
-2	4	0.938554180778467	0.938554180778467\\
-2	5	0.912031100502192	0.912031100502192\\
-2	6	0.880528214553963	0.880528214553963\\
-2	7	0.844554636527142	0.844554636527142\\
-2	8	0.804677547589598	0.804677547589598\\
-2	9	0.761508088516657	0.761508088516657\\
-2	10	0.715686024944317	0.715686024944317\\
-2	11	0.667864105927524	0.667864105927524\\
-2	12	0.618693751509834	0.618693751509834\\
-2	13	0.568810178019031	0.568810178019031\\
-2	14	0.518820079153874	0.518820079153874\\
-2	15	0.46929063608223	0.46929063608223\\
-1	-15	0.474210073967662	0.474210073967662\\
-1	-14	0.524177324060105	0.524177324060105\\
-1	-13	0.574605945567578	0.574605945567578\\
-1	-12	0.624924668056404	0.624924668056404\\
-1	-11	0.674522248519609	0.674522248519609\\
-1	-10	0.722758359047148	0.722758359047148\\
-1	-9	0.768976519129023	0.768976519129023\\
-1	-8	0.812518612548914	0.812518612548914\\
-1	-7	0.852739506509447	0.852739506509447\\
-1	-6	0.889023153116322	0.889023153116322\\
-1	-5	0.92079725961941	0.92079725961941\\
-1	-4	0.947548929854472	0.947548929854472\\
-1	-3	0.96883744319882	0.96883744319882\\
-1	-2	0.984307135758616	0.984307135758616\\
-1	-1	0.993695681132138	0.993695681132138\\
-1	0	0.99684318902621	0.99684318902621\\
-1	1	0.993695681132138	0.993695681132138\\
-1	2	0.984306931295325	0.984306931295325\\
-1	3	0.96883723873553	0.96883723873553\\
-1	4	0.947548418696246	0.947548418696246\\
-1	5	0.920796543997893	0.920796543997893\\
-1	6	0.88902213079987	0.88902213079987\\
-1	7	0.852738177498058	0.852738177498058\\
-1	8	0.81251697684259	0.81251697684259\\
-1	9	0.768974678959409	0.768974678959409\\
-1	10	0.722756467761711	0.722756467761711\\
-1	11	0.67452071504493	0.67452071504493\\
-1	12	0.624923901319064	0.624923901319064\\
-1	13	0.574606814536563	0.574606814536563\\
-1	14	0.524181004399334	0.524181004399334\\
-1	15	0.474218252499283	0.474218252499283\\
0	-15	0.47586714670558	0.47586714670558\\
0	-14	0.525976805479862	0.525976805479862\\
0	-13	0.576549675838788	0.576549675838788\\
0	-12	0.627012647179066	0.627012647179066\\
0	-11	0.67675263632411	0.67675263632411\\
0	-10	0.725127372962613	0.725127372962613\\
0	-9	0.771478434183318	0.771478434183318\\
0	-8	0.815145556905354	0.815145556905354\\
0	-7	0.855482074856669	0.855482074856669\\
0	-6	0.891869997741703	0.891869997741703\\
0	-5	0.923735397104002	0.923735397104002\\
0	-4	0.950563741073005	0.950563741073005\\
0	-3	0.971913593404505	0.971913593404505\\
0	-2	0.98742745003505	0.98742745003505\\
0	-1	0.99684318902621	0.99684318902621\\
0	0	1	1\\
0	1	0.99684318902621	0.99684318902621\\
0	2	0.987427552266695	0.987427552266695\\
0	3	0.971913593404505	0.971913593404505\\
0	4	0.95056384330465	0.95056384330465\\
0	5	0.923735397104002	0.923735397104002\\
0	6	0.891869895510058	0.891869895510058\\
0	7	0.855482074856669	0.855482074856669\\
0	8	0.815145505789531	0.815145505789531\\
0	9	0.771478383067495	0.771478383067495\\
0	10	0.725127424078435	0.725127424078435\\
0	11	0.676752687439933	0.676752687439933\\
0	12	0.627012698294889	0.627012698294889\\
0	13	0.576549675838788	0.576549675838788\\
0	14	0.525976805479862	0.525976805479862\\
0	15	0.47586714670558	0.47586714670558\\
1	-15	0.474218252499283	0.474218252499283\\
1	-14	0.524181055515157	0.524181055515157\\
1	-13	0.574606865652386	0.574606865652386\\
1	-12	0.624923952434887	0.624923952434887\\
1	-11	0.67452071504493	0.67452071504493\\
1	-10	0.722756518877533	0.722756518877533\\
1	-9	0.768974678959409	0.768974678959409\\
1	-8	0.81251697684259	0.81251697684259\\
1	-7	0.852738279729704	0.852738279729704\\
1	-6	0.88902213079987	0.88902213079987\\
1	-5	0.920796543997893	0.920796543997893\\
1	-4	0.947548316464601	0.947548316464601\\
1	-3	0.968837136503884	0.968837136503884\\
1	-2	0.984306931295325	0.984306931295325\\
1	-1	0.993695681132138	0.993695681132138\\
1	0	0.99684318902621	0.99684318902621\\
1	1	0.993695681132138	0.993695681132138\\
1	2	0.98430703352697	0.98430703352697\\
1	3	0.96883744319882	0.96883744319882\\
1	4	0.947548929854472	0.947548929854472\\
1	5	0.92079725961941	0.92079725961941\\
1	6	0.889023050884677	0.889023050884677\\
1	7	0.852739506509447	0.852739506509447\\
1	8	0.812518510317269	0.812518510317269\\
1	9	0.768976468013201	0.768976468013201\\
1	10	0.722758307931325	0.722758307931325\\
1	11	0.674522197403786	0.674522197403786\\
1	12	0.624924565824758	0.624924565824758\\
1	13	0.574605843335933	0.574605843335933\\
1	14	0.524177272944283	0.524177272944283\\
1	15	0.474209971736017	0.474209971736017\\
2	-15	0.469290687198052	0.469290687198052\\
2	-14	0.518820181385519	0.518820181385519\\
2	-13	0.568810178019031	0.568810178019031\\
2	-12	0.618693751509834	0.618693751509834\\
2	-11	0.667864208159169	0.667864208159169\\
2	-10	0.715686024944317	0.715686024944317\\
2	-9	0.761508088516657	0.761508088516657\\
2	-8	0.80467759870542	0.80467759870542\\
2	-7	0.844554636527142	0.844554636527142\\
2	-6	0.880528316785608	0.880528316785608\\
2	-5	0.912030998270547	0.912030998270547\\
2	-4	0.938554283010113	0.938554283010113\\
2	-3	0.959661028489199	0.959661028489199\\
2	-2	0.974998535531682	0.974998535531682\\
2	-1	0.984306931295325	0.984306931295325\\
2	0	0.987427552266695	0.987427552266695\\
2	1	0.984307135758616	0.984307135758616\\
2	2	0.974998944458263	0.974998944458263\\
2	3	0.959661744110716	0.959661744110716\\
2	4	0.93855540755821	0.93855540755821\\
2	5	0.912032838440162	0.912032838440162\\
2	6	0.880530565881804	0.880530565881804\\
2	7	0.844557601244854	0.844557601244854\\
2	8	0.804681279044649	0.804681279044649\\
2	9	0.761512331129935	0.761512331129935\\
2	10	0.715690523136708	0.715690523136708\\
2	11	0.667868348540802	0.667868348540802\\
2	12	0.618696511764255	0.618696511764255\\
2	13	0.56880997355574	0.56880997355574\\
2	14	0.518814660876675	0.518814660876675\\
2	15	0.469276681462653	0.469276681462653\\
3	-15	0.461168280750998	0.461168280750998\\
3	-14	0.509985220371046	0.509985220371046\\
3	-13	0.559258317592216	0.559258317592216\\
3	-12	0.608428263083325	0.608428263083325\\
3	-11	0.656896490561412	0.656896490561412\\
3	-10	0.704036422272718	0.704036422272718\\
3	-9	0.749205839021758	0.749205839021758\\
3	-8	0.791760937022546	0.791760937022546\\
3	-7	0.831070895908046	0.831070895908046\\
3	-6	0.866533111245307	0.866533111245307\\
3	-5	0.897588222587326	0.897588222587326\\
3	-4	0.923734681482486	0.923734681482486\\
3	-3	0.944541683777688	0.944541683777688\\
3	-2	0.959661130720845	0.959661130720845\\
3	-1	0.96883723873553	0.96883723873553\\
3	0	0.971913797867795	0.971913797867795\\
3	1	0.968837647662111	0.968837647662111\\
3	2	0.959661948574007	0.959661948574007\\
3	3	0.944543115020722	0.944543115020722\\
3	4	0.923736930578681	0.923736930578681\\
3	5	0.897591493999974	0.897591493999974\\
3	6	0.866537507206053	0.866537507206053\\
3	7	0.83107662088018	0.83107662088018\\
3	8	0.791768042121891	0.791768042121891\\
3	9	0.749214170900846	0.749214170900846\\
3	10	0.704045520889145	0.704045520889145\\
3	11	0.656905640293663	0.656905640293663\\
3	12	0.608436032688364	0.608436032688364\\
3	13	0.559262406858026	0.559262406858026\\
3	14	0.509982306769156	0.509982306769156\\
3	15	0.461153712741549	0.461153712741549\\
4	-15	0.44998863695263	0.44998863695263\\
4	-14	0.497825941832034	0.497825941832034\\
4	-13	0.546113065950554	0.546113065950554\\
4	-12	0.594301178720647	0.594301178720647\\
4	-11	0.641803266035264	0.641803266035264\\
4	-10	0.688004762358962	0.688004762358962\\
4	-9	0.732276022988626	0.732276022988626\\
4	-8	0.773985563051521	0.773985563051521\\
4	-7	0.812514983325508	0.812514983325508\\
4	-6	0.84727338490102	0.84727338490102\\
4	-5	0.877712448348739	0.877712448348739\\
4	-4	0.903340286107527	0.903340286107527\\
4	-3	0.923734681482486	0.923734681482486\\
4	-2	0.938554385241758	0.938554385241758\\
4	-1	0.947548725391182	0.947548725391182\\
4	0	0.950564252231231	0.950564252231231\\
4	1	0.947549441012698	0.947549441012698\\
4	2	0.938556020948082	0.938556020948082\\
4	3	0.923737339505262	0.923737339505262\\
4	4	0.903344375373337	0.903344375373337\\
4	5	0.877718275552519	0.877718275552519\\
4	6	0.847281256737704	0.847281256737704\\
4	7	0.812525104258388	0.812525104258388\\
4	8	0.773998086428064	0.773998086428064\\
4	9	0.732290948808833	0.732290948808833\\
4	10	0.688021732812075	0.688021732812075\\
4	11	0.641821207689006	0.641821207689006\\
4	12	0.59431855810034	0.59431855810034\\
4	13	0.546127071685954	0.546127071685954\\
4	14	0.497832433541508	0.497832433541508\\
4	15	0.449981582969108	0.449981582969108\\
5	-15	0.435939453103635	0.435939453103635\\
5	-14	0.482546451248233	0.482546451248233\\
5	-13	0.529594885637037	0.529594885637037\\
5	-12	0.576549369143852	0.576549369143852\\
5	-11	0.622837404555258	0.622837404555258\\
5	-10	0.667859709966778	0.667859709966778\\
5	-9	0.711002077653726	0.711002077653726\\
5	-8	0.751648970880031	0.751648970880031\\
5	-7	0.789197427401986	0.789197427401986\\
5	-6	0.823071320782763	0.823071320782763\\
5	-5	0.852736030633508	0.852736030633508\\
5	-4	0.877712141653803	0.877712141653803\\
5	-3	0.897588018124035	0.897588018124035\\
5	-2	0.912031202733838	0.912031202733838\\
5	-1	0.920796952924474	0.920796952924474\\
5	0	0.923736112725519	0.923736112725519\\
5	1	0.920798384167508	0.920798384167508\\
5	2	0.91203416745155	0.91203416745155\\
5	3	0.897592925243007	0.897592925243007\\
5	4	0.877719297868971	0.877719297868971\\
5	5	0.852745947103098	0.852745947103098\\
5	6	0.823084457549178	0.823084457549178\\
5	7	0.789214295623453	0.789214295623453\\
5	8	0.751669928367309	0.751669928367309\\
5	9	0.711027226638459	0.711027226638459\\
5	10	0.667888845985675	0.667888845985675\\
5	11	0.622869658639336	0.622869658639336\\
5	12	0.576582952239318	0.576582952239318\\
5	13	0.52962667967871	0.52962667967871\\
5	14	0.482571702464611	0.482571702464611\\
5	15	0.435951209742839	0.435951209742839\\
6	-15	0.419253280639054	0.419253280639054\\
6	-14	0.464399414130887	0.464399414130887\\
6	-13	0.50997673514449	0.50997673514449\\
6	-12	0.555466136943174	0.555466136943174\\
6	-11	0.600312092766631	0.600312092766631\\
6	-10	0.643933517943238	0.643933517943238\\
6	-9	0.685734913139394	0.685734913139394\\
6	-8	0.725119450010106	0.725119450010106\\
6	-7	0.761502568007813	0.761502568007813\\
6	-6	0.794325724538672	0.794325724538672\\
6	-5	0.823070605161246	0.823070605161246\\
6	-4	0.84727261816368	0.84727261816368\\
6	-3	0.866532702318726	0.866532702318726\\
6	-2	0.880528419017253	0.880528419017253\\
6	-1	0.889022948653032	0.889022948653032\\
6	0	0.891871531216382	0.891871531216382\\
6	1	0.889025402212518	0.889025402212518\\
6	2	0.880533428367871	0.880533428367871\\
6	3	0.866540778618701	0.866540778618701\\
6	4	0.847284323687062	0.847284323687062\\
6	5	0.823086706645374	0.823086706645374\\
6	6	0.794346784257594	0.794346784257594\\
6	7	0.761529506046337	0.761529506046337\\
6	8	0.725152879758104	0.725152879758104\\
6	9	0.685775345755092	0.685775345755092\\
6	10	0.643980953426636	0.643980953426636\\
6	11	0.600365866612034	0.600365866612034\\
6	12	0.555524664560081	0.555524664560081\\
6	13	0.510036898467721	0.510036898467721\\
6	14	0.464456050462358	0.464456050462358\\
6	15	0.419298748163281	0.419298748163281\\
7	-15	0.4002012378821	0.4002012378821\\
7	-14	0.443679206502505	0.443679206502505\\
7	-13	0.487576759352943	0.487576759352943\\
7	-12	0.531392833582114	0.531392833582114\\
7	-11	0.574591837600533	0.574591837600533\\
7	-10	0.616613235297275	0.616613235297275\\
7	-9	0.656883098215884	0.656883098215884\\
7	-8	0.694825862193582	0.694825862193582\\
7	-7	0.729877924170086	0.729877924170086\\
7	-6	0.761501034533134	0.761501034533134\\
7	-5	0.789195382769081	0.789195382769081\\
7	-4	0.812513245387538	0.812513245387538\\
7	-3	0.831069975823238	0.831069975823238\\
7	-2	0.844554892106255	0.844554892106255\\
7	-1	0.852739710972737	0.852739710972737\\
7	0	0.855484988458558	0.855484988458558\\
7	1	0.852743698006902	0.852743698006902\\
7	2	0.844563019522053	0.844563019522053\\
7	3	0.83108285701054	0.83108285701054\\
7	4	0.812531698199506	0.812531698199506\\
7	5	0.789220327290523	0.789220327290523\\
7	6	0.761533544196325	0.761533544196325\\
7	7	0.729919327986414	0.729919327986414\\
7	8	0.6948771824795	0.6948771824795\\
7	9	0.656945408403666	0.656945408403666\\
7	10	0.616687046545148	0.616687046545148\\
7	11	0.574676792097739	0.574676792097739\\
7	12	0.531487857896377	0.531487857896377\\
7	13	0.487678888766552	0.487678888766552\\
7	14	0.443783431664841	0.443783431664841\\
7	15	0.400299559166923	0.400299559166923\\
8	-15	0.37908529155494	0.37908529155494\\
8	-14	0.420714196407641	0.420714196407641\\
8	-13	0.462749395376492	0.462749395376492\\
8	-12	0.504710271939243	0.504710271939243\\
8	-11	0.54608321431014	0.54608321431014\\
8	-10	0.586330688498439	0.586330688498439\\
8	-9	0.624902279326093	0.624902279326093\\
8	-8	0.661246344835314	0.661246344835314\\
8	-7	0.694822488549289	0.694822488549289\\
8	-6	0.725114849586069	0.725114849586069\\
8	-5	0.75164447268764	0.75164447268764\\
8	-4	0.773982138291405	0.773982138291405\\
8	-3	0.791759096852932	0.791759096852932\\
8	-2	0.804677854284533	0.804677854284533\\
8	-1	0.812519532633721	0.812519532633721\\
8	0	0.815150361792681	0.815150361792681\\
8	1	0.812525666532437	0.812525666532437\\
8	2	0.804690479892722	0.804690479892722\\
8	3	0.791778980907934	0.791778980907934\\
8	4	0.774010200878027	0.774010200878027\\
8	5	0.751682247280562	0.751682247280562\\
8	6	0.7251637163125	0.7251637163125\\
8	7	0.694884338694667	0.694884338694667\\
8	8	0.661322814105963	0.661322814105963\\
8	9	0.624995105659984	0.624995105659984\\
8	10	0.586441098675313	0.586441098675313\\
8	11	0.54621182171987	0.54621182171987\\
8	12	0.504856463191956	0.504856463191956\\
8	13	0.462910921375993	0.462910921375993\\
8	14	0.420886482287805	0.420886482287805\\
8	15	0.379260542152816	0.379260542152816\\
9	-15	0.356231330584733	0.356231330584733\\
9	-14	0.39585823312843	0.39585823312843\\
9	-13	0.435876989568385	0.435876989568385\\
9	-12	0.475829014301901	0.475829014301901\\
9	-11	0.515224438842832	0.515224438842832\\
9	-10	0.553550827069345	0.553550827069345\\
9	-9	0.590283372830529	0.590283372830529\\
9	-8	0.624896094311555	0.624896094311555\\
9	-7	0.656874050715279	0.656874050715279\\
9	-6	0.685725405596385	0.685725405596385\\
9	-5	0.710993745774638	0.710993745774638\\
9	-4	0.732269940205733	0.732269940205733\\
9	-3	0.7492027720724	0.7492027720724\\
9	-2	0.76150854855906	0.76150854855906\\
9	-1	0.768978768225219	0.768978768225219\\
9	0	0.771486203788357	0.771486203788357\\
9	1	0.768988020189114	0.768988020189114\\
9	2	0.761527512529255	0.761527512529255\\
9	3	0.749232419249524	0.749232419249524\\
9	4	0.732311599601174	0.732311599601174\\
9	5	0.711049308673834	0.711049308673834\\
9	6	0.685796865516418	0.685796865516418\\
9	7	0.65696396344728	0.65696396344728\\
9	8	0.625007015646656	0.625007015646656\\
9	9	0.590417909675683	0.590417909675683\\
9	10	0.553711126289103	0.553711126289103\\
9	11	0.51541218725934	0.51541218725934\\
9	12	0.476044723073386	0.476044723073386\\
9	13	0.436119431915104	0.436119431915104\\
9	14	0.396123805384888	0.396123805384888\\
9	15	0.356512953209496	0.356512953209496\\
10	-15	0.331980016371376	0.331980016371376\\
10	-14	0.369481446336514	0.369481446336514\\
10	-13	0.407359190987831	0.407359190987831\\
10	-12	0.445178229116861	0.445178229116861\\
10	-11	0.482473917895926	0.482473917895926\\
10	-10	0.518760375873045	0.518760375873045\\
10	-9	0.553539990514948	0.553539990514948\\
10	-8	0.586314229203553	0.586314229203553\\
10	-7	0.616595089180242	0.616595089180242\\
10	-6	0.643916394142658	0.643916394142658\\
10	-5	0.667845499768087	0.667845499768087\\
10	-4	0.687994743657728	0.687994743657728\\
10	-3	0.7040314129221	0.7040314129221\\
10	-2	0.715686587218366	0.715686587218366\\
10	-1	0.722762806123716	0.722762806123716\\
10	0	0.725139691875866	0.725139691875866\\
10	1	0.722776454048358	0.722776454048358\\
10	2	0.715714343110052	0.715714343110052\\
10	3	0.704074503560575	0.704074503560575\\
10	4	0.688055060328427	0.688055060328427\\
10	5	0.667925342683031	0.667925342683031\\
10	6	0.644018574672089	0.644018574672089\\
10	7	0.616723032084277	0.616723032084277\\
10	8	0.586471461473953	0.586471461473953\\
10	9	0.553730192490943	0.553730192490943\\
10	10	0.518987227893864	0.518987227893864\\
10	11	0.482740384679279	0.482740384679279\\
10	12	0.445486355295655	0.445486355295655\\
10	13	0.407709283257002	0.407709283257002\\
10	14	0.3698713322736	0.3698713322736\\
10	15	0.332403945446332	0.332403945446332\\
11	-15	0.30667893650873	0.30667893650873\\
11	-14	0.341961530845281	0.341961530845281\\
11	-13	0.377604031688948	0.377604031688948\\
11	-12	0.41319605565777	0.41319605565777\\
11	-11	0.44829941236228	0.44829941236228\\
11	-10	0.482455874010538	0.482455874010538\\
11	-9	0.51519637625621	0.51519637625621\\
11	-8	0.546051113573531	0.546051113573531\\
11	-7	0.574559992443037	0.574559992443037\\
11	-6	0.60028367236925	0.60028367236925\\
11	-5	0.622814606898367	0.622814606898367\\
11	-4	0.641787471246072	0.641787471246072\\
11	-3	0.65688866984055	0.65688866984055\\
11	-2	0.667865026012331	0.667865026012331\\
11	-1	0.674530427051229	0.674530427051229\\
11	0	0.676771446946837	0.676771446946837\\
11	1	0.67454990217965	0.67454990217965\\
11	2	0.667904691890689	0.667904691890689\\
11	3	0.656950111059348	0.656950111059348\\
11	4	0.641872936901504	0.641872936901504\\
11	5	0.622927215055614	0.622927215055614\\
11	6	0.600427001135896	0.600427001135896\\
11	7	0.57473848889565	0.57473848889565\\
11	8	0.546269684831083	0.546269684831083\\
11	9	0.515460133900964	0.515460133900964\\
11	10	0.482769929624758	0.482769929624758\\
11	11	0.448668826412404	0.448668826412404\\
11	12	0.413624687388408	0.413624687388408\\
11	13	0.378094334659584	0.378094334659584\\
11	14	0.342513326150538	0.342513326150538\\
11	15	0.307288594925201	0.307288594925201\\
12	-15	0.28067473294794	0.28067473294794\\
12	-14	0.313675056920024	0.313675056920024\\
12	-13	0.347018240272378	0.347018240272378\\
12	-12	0.380319304186887	0.380319304186887\\
12	-11	0.413167277449631	0.413167277449631\\
12	-10	0.445132429339787	0.445132429339787\\
12	-9	0.475775189340674	0.475775189340674\\
12	-8	0.504655117966629	0.504655117966629\\
12	-7	0.531341104369616	0.531341104369616\\
12	-6	0.555421461714198	0.555421461714198\\
12	-5	0.576514201457885	0.576514201457885\\
12	-4	0.594277154284012	0.594277154284012\\
12	-3	0.608416506444121	0.608416506444121\\
12	-2	0.618694876057931	0.618694876057931\\
12	-1	0.624938469328513	0.624938469328513\\
12	0	0.62704055641822	0.62704055641822\\
12	1	0.624965816293618	0.624965816293618\\
12	2	0.61875069453624	0.61875069453624\\
12	3	0.608502483257779	0.608502483257779\\
12	4	0.594396254150733	0.594396254150733\\
12	5	0.576670309180187	0.576670309180187\\
12	6	0.555619279947764	0.555619279947764\\
12	7	0.531586204739111	0.531586204739111\\
12	8	0.504953889949883	0.504953889949883\\
12	9	0.476134533573741	0.476134533573741\\
12	10	0.445559553153657	0.445559553153657\\
12	11	0.413669311501561	0.413669311501561\\
12	12	0.380902791302173	0.380902791302173\\
12	13	0.347688138685814	0.347688138685814\\
12	14	0.314433973538565	0.314433973538565\\
12	15	0.281521849918424	0.281521849918424\\
13	-15	0.254305153487014	0.254305153487014\\
13	-14	0.284989010609293	0.284989010609293\\
13	-13	0.315998296616327	0.315998296616327\\
13	-12	0.346973795064604	0.346973795064604\\
13	-11	0.377532162842334	0.377532162842334\\
13	-10	0.407272907479237	0.407272907479237\\
13	-9	0.4357862078674	0.4357862078674\\
13	-8	0.462661476161574	0.462661476161574\\
13	-7	0.487497018669645	0.487497018669645\\
13	-6	0.509909262258622	0.509909262258622\\
13	-5	0.529542389687199	0.529542389687199\\
13	-4	0.546077489338006	0.546077489338006\\
13	-3	0.559240938212523	0.559240938212523\\
13	-2	0.568811762609532	0.568811762609532\\
13	-1	0.574627925371308	0.574627925371308\\
13	0	0.576590159570308	0.576590159570308\\
13	1	0.574665751080052	0.574665751080052\\
13	2	0.568888794154231	0.568888794154231\\
13	3	0.559359322457726	0.559359322457726\\
13	4	0.546240855507122	0.546240855507122\\
13	5	0.52975549155173	0.52975549155173\\
13	6	0.510178029253994	0.510178029253994\\
13	7	0.487828453663558	0.487828453663558\\
13	8	0.46306380880147	0.46306380880147\\
13	9	0.436268383422239	0.436268383422239\\
13	10	0.407844586839495	0.407844586839495\\
13	11	0.378203160245957	0.378203160245957\\
13	12	0.347753515822954	0.347753515822954\\
13	13	0.316895047050581	0.316895047050581\\
13	14	0.286008924618167	0.286008924618167\\
13	15	0.255450940209107	0.255450940209107\\
14	-15	0.227892483387613	0.227892483387613\\
14	-14	0.256253586413905	0.256253586413905\\
14	-13	0.284922508924056	0.284922508924056\\
14	-12	0.313565847964982	0.313565847964982\\
14	-11	0.341828169664047	0.341828169664047\\
14	-10	0.369338092011957	0.369338092011957\\
14	-9	0.395715696657035	0.395715696657035\\
14	-8	0.42058068187894	0.42058068187894\\
14	-7	0.443560668909833	0.443560668909833\\
14	-6	0.464300351666637	0.464300351666637\\
14	-5	0.482470084209228	0.482470084209228\\
14	-4	0.497774468198649	0.497774468198649\\
14	-3	0.509960071386313	0.509960071386313\\
14	-2	0.518822277134247	0.518822277134247\\
14	-1	0.524210958271394	0.524210958271394\\
14	0	0.52603456635943	0.52603456635943\\
14	1	0.524262687483892	0.524262687483892\\
14	2	0.518927115686454	0.518927115686454\\
14	3	0.510120830648475	0.510120830648475\\
14	4	0.497995493015687	0.497995493015687\\
14	5	0.482757201784924	0.482757201784924\\
14	6	0.464660820447801	0.464660820447801\\
14	7	0.444003229702137	0.444003229702137\\
14	8	0.421115455615262	0.421115455615262\\
14	9	0.396354184397468	0.396354184397468\\
14	10	0.370092612669752	0.370092612669752\\
14	11	0.342711757773975	0.342711757773975\\
14	12	0.314591563619724	0.314591563619724\\
14	13	0.286102492131486	0.286102492131486\\
14	14	0.257598239243927	0.257598239243927\\
14	15	0.229408834265839	0.229408834265839\\
15	-15	0.201737360360518	0.201737360360518\\
15	-14	0.227795670019558	0.227795670019558\\
15	-13	0.254144010856183	0.254144010856183\\
15	-12	0.280474818965647	0.280474818965647\\
15	-11	0.306460314135355	0.306460314135355\\
15	-10	0.331758148143264	0.331758148143264\\
15	-9	0.356018100930645	0.356018100930645\\
15	-8	0.37888972241757	0.37888972241757\\
15	-7	0.400030025434211	0.400030025434211\\
15	-6	0.419111562020821	0.419111562020821\\
15	-5	0.435830831980553	0.435830831980553\\
15	-4	0.449915592442096	0.449915592442096\\
15	-3	0.461132653022627	0.461132653022627\\
15	-2	0.46929354968412	0.46929354968412\\
15	-1	0.474260065242191	0.474260065242191\\
15	0	0.475947960821153	0.475947960821153\\
15	1	0.474329582760964	0.474329582760964\\
15	2	0.469434373775457	0.469434373775457\\
15	3	0.461348157330822	0.461348157330822\\
15	4	0.450210939665234	0.450210939665234\\
15	5	0.436212820523044	0.436212820523044\\
15	6	0.419589188267447	0.419589188267447\\
15	7	0.400613819244433	0.400613819244433\\
15	8	0.379592156052108	0.379592156052108\\
15	9	0.356853333472368	0.356853333472368\\
15	10	0.332741744360161	0.332741744360161\\
15	11	0.307608988901426	0.307608988901426\\
15	12	0.281805721639383	0.281805721639383\\
15	13	0.255673830753671	0.255673830753671\\
15	14	0.229539793003409	0.229539793003409\\
15	15	0.203708207575631	0.203708207575631\\
};
\end{axis}
\end{tikzpicture}%

%% file: Figures/Results_EE/CM_Halo60Zoom_V2.tikz
%
%
\begin{tikzpicture}

\begin{axis}[%
width=1\linewidth,
height=1\linewidth,
point meta min=0,
point meta max=1,
xmin=-10,
xmax=10,
xlabel={x / \si{\pixel}},
tick align=outside,
ymin=-10,
ymax=10,
ylabel={y / \si{\pixel}},
zmin=0.6,
zmax=1,
ztick={0.6,1},
zlabel={R(s)},
view={45}{45},
axis background/.style={fill=white},
axis x line*=bottom,
axis y line*=left,
axis z line*=left,
xmajorgrids,
ymajorgrids,
zmajorgrids,
xlabel style={font=\color{white!15!black},yshift=2mm,xshift=0mm},
ylabel style={font=\color{white!15!black},yshift=2mm,xshift=0mm},
xtick style={draw=none},
ytick style={draw=none},
ztick style={draw=none},
]

\addplot3[%
surf,
shader=flat corner, fill=white, z buffer=sort, colormap={mymap}{[1pt] rgb(0pt)=(0.2422,0.1504,0.6603); rgb(1pt)=(0.25039,0.164995,0.707614); rgb(2pt)=(0.257771,0.181781,0.751138); rgb(3pt)=(0.264729,0.197757,0.795214); rgb(4pt)=(0.270648,0.214676,0.836371); rgb(5pt)=(0.275114,0.234238,0.870986); rgb(6pt)=(0.2783,0.255871,0.899071); rgb(7pt)=(0.280333,0.278233,0.9221); rgb(8pt)=(0.281338,0.300595,0.941376); rgb(9pt)=(0.281014,0.322757,0.957886); rgb(10pt)=(0.279467,0.344671,0.971676); rgb(11pt)=(0.275971,0.366681,0.982905); rgb(12pt)=(0.269914,0.3892,0.9906); rgb(13pt)=(0.260243,0.412329,0.995157); rgb(14pt)=(0.244033,0.435833,0.998833); rgb(15pt)=(0.220643,0.460257,0.997286); rgb(16pt)=(0.196333,0.484719,0.989152); rgb(17pt)=(0.183405,0.507371,0.979795); rgb(18pt)=(0.178643,0.528857,0.968157); rgb(19pt)=(0.176438,0.549905,0.952019); rgb(20pt)=(0.168743,0.570262,0.935871); rgb(21pt)=(0.154,0.5902,0.9218); rgb(22pt)=(0.146029,0.609119,0.907857); rgb(23pt)=(0.138024,0.627629,0.89729); rgb(24pt)=(0.124814,0.645929,0.888343); rgb(25pt)=(0.111252,0.6635,0.876314); rgb(26pt)=(0.0952095,0.679829,0.859781); rgb(27pt)=(0.0688714,0.694771,0.839357); rgb(28pt)=(0.0296667,0.708167,0.816333); rgb(29pt)=(0.00357143,0.720267,0.7917); rgb(30pt)=(0.00665714,0.731214,0.766014); rgb(31pt)=(0.0433286,0.741095,0.73941); rgb(32pt)=(0.0963952,0.75,0.712038); rgb(33pt)=(0.140771,0.7584,0.684157); rgb(34pt)=(0.1717,0.766962,0.655443); rgb(35pt)=(0.193767,0.775767,0.6251); rgb(36pt)=(0.216086,0.7843,0.5923); rgb(37pt)=(0.246957,0.791795,0.556743); rgb(38pt)=(0.290614,0.79729,0.518829); rgb(39pt)=(0.340643,0.8008,0.478857); rgb(40pt)=(0.3909,0.802871,0.435448); rgb(41pt)=(0.445629,0.802419,0.390919); rgb(42pt)=(0.5044,0.7993,0.348); rgb(43pt)=(0.561562,0.794233,0.304481); rgb(44pt)=(0.617395,0.787619,0.261238); rgb(45pt)=(0.671986,0.779271,0.2227); rgb(46pt)=(0.7242,0.769843,0.191029); rgb(47pt)=(0.773833,0.759805,0.16461); rgb(48pt)=(0.820314,0.749814,0.153529); rgb(49pt)=(0.863433,0.7406,0.159633); rgb(50pt)=(0.903543,0.733029,0.177414); rgb(51pt)=(0.939257,0.728786,0.209957); rgb(52pt)=(0.972757,0.729771,0.239443); rgb(53pt)=(0.995648,0.743371,0.237148); rgb(54pt)=(0.996986,0.765857,0.219943); rgb(55pt)=(0.995205,0.789252,0.202762); rgb(56pt)=(0.9892,0.813567,0.188533); rgb(57pt)=(0.978629,0.838629,0.176557); rgb(58pt)=(0.967648,0.8639,0.16429); rgb(59pt)=(0.96101,0.889019,0.153676); rgb(60pt)=(0.959671,0.913457,0.142257); rgb(61pt)=(0.962795,0.937338,0.12651); rgb(62pt)=(0.969114,0.960629,0.106362); rgb(63pt)=(0.9769,0.9839,0.0805)}, mesh/rows=31]
table[row sep=crcr, point meta=\thisrow{c}] {%
x	y	z	c\\
-15	-15	0.280507444739691	0.280507444739691\\
-15	-14	0.287228485876209	0.287228485876209\\
-15	-13	0.293560299668651	0.293560299668651\\
-15	-12	0.299505603976113	0.299505603976113\\
-15	-11	0.305170420237867	0.305170420237867\\
-15	-10	0.310766716528992	0.310766716528992\\
-15	-9	0.316529300914776	0.316529300914776\\
-15	-8	0.322575814305494	0.322575814305494\\
-15	-7	0.328818886258245	0.328818886258245\\
-15	-6	0.334998026185999	0.334998026185999\\
-15	-5	0.340785171041605	0.340785171041605\\
-15	-4	0.345876793313249	0.345876793313249\\
-15	-3	0.350030579655014	0.350030579655014\\
-15	-2	0.353063361324456	0.353063361324456\\
-15	-1	0.354843085277741	0.354843085277741\\
-15	0	0.355286021507092	0.355286021507092\\
-15	1	0.354356364088993	0.354356364088993\\
-15	2	0.352065408346111	0.352065408346111\\
-15	3	0.348466140063597	0.348466140063597\\
-15	4	0.343648049115753	0.343648049115753\\
-15	5	0.337729050692211	0.337729050692211\\
-15	6	0.330847705737944	0.330847705737944\\
-15	7	0.323155092310463	0.323155092310463\\
-15	8	0.31480637772317	0.31480637772317\\
-15	9	0.305954136102803	0.305954136102803\\
-15	10	0.296747675158289	0.296747675158289\\
-15	11	0.287336202860604	0.287336202860604\\
-15	12	0.277882167393391	0.277882167393391\\
-15	13	0.268591926572124	0.268591926572124\\
-15	14	0.259769631270799	0.259769631270799\\
-15	15	0.251841611350338	0.251841611350338\\
-14	-15	0.284968723176725	0.284968723176725\\
-14	-14	0.292259068523362	0.292259068523362\\
-14	-13	0.299315952267014	0.299315952267014\\
-14	-12	0.306237591352481	0.306237591352481\\
-14	-11	0.313216704430779	0.313216704430779\\
-14	-10	0.320470246474413	0.320470246474413\\
-14	-9	0.328117005380064	0.328117005380064\\
-14	-8	0.336085867990442	0.336085867990442\\
-14	-7	0.344126392316827	0.344126392316827\\
-14	-6	0.351898920775554	0.351898920775554\\
-14	-5	0.359063895520908	0.359063895520908\\
-14	-4	0.365326092225287	0.365326092225287\\
-14	-3	0.370440230338808	0.370440230338808\\
-14	-2	0.374207257850723	0.374207257850723\\
-14	-1	0.37647819120044	0.37647819120044\\
-14	0	0.37715827933687	0.37715827933687\\
-14	1	0.376209422363682	0.376209422363682\\
-14	2	0.373652415643139	0.373652415643139\\
-14	3	0.36956196289868	0.36956196289868\\
-14	4	0.364059420279159	0.364059420279159\\
-14	5	0.35730182518451	0.35730182518451\\
-14	6	0.349467085180397	0.349467085180397\\
-14	7	0.340743779792972	0.340743779792972\\
-14	8	0.331322832390178	0.331322832390178\\
-14	9	0.32138933166997	0.32138933166997\\
-14	10	0.311118168125068	0.311118168125068\\
-14	11	0.300676402819233	0.300676402819233\\
-14	12	0.290236931486213	0.290236931486213\\
-14	13	0.280006884910586	0.280006884910586\\
-14	14	0.270280140957648	0.270280140957648\\
-14	15	0.261472357527831	0.261472357527831\\
-13	-15	0.290060519989779	0.290060519989779\\
-13	-14	0.298047335430889	0.298047335430889\\
-13	-13	0.306056267762082	0.306056267762082\\
-13	-12	0.314272480484774	0.314272480484774\\
-13	-11	0.322903253990465	0.322903253990465\\
-13	-10	0.332068772507115	0.332068772507115\\
-13	-9	0.341715277280473	0.341715277280473\\
-13	-8	0.351612722732293	0.351612722732293\\
-13	-7	0.361425690141722	0.361425690141722\\
-13	-6	0.370794125514623	0.370794125514623\\
-13	-5	0.379379967074508	0.379379967074508\\
-13	-4	0.386878814602513	0.386878814602513\\
-13	-3	0.393019829699448	0.393019829699448\\
-13	-2	0.397570672814248	0.397570672814248\\
-13	-1	0.400350618784206	0.400350618784206\\
-13	0	0.401242824602639	0.401242824602639\\
-13	1	0.400202258585792	0.400202258585792\\
-13	2	0.397261285697957	0.397261285697957\\
-13	3	0.392523758078026	0.392523758078026\\
-13	4	0.386153196092586	0.386153196092586\\
-13	5	0.378355708212243	0.378355708212243\\
-13	6	0.369359868880503	0.369359868880503\\
-13	7	0.359402156773294	0.359402156773294\\
-13	8	0.34871645737988	0.34871645737988\\
-13	9	0.337524089208015	0.337524089208015\\
-13	10	0.326026847062029	0.326026847062029\\
-13	11	0.314410542740005	0.314410542740005\\
-13	12	0.302858419787849	0.302858419787849\\
-13	13	0.291577234972832	0.291577234972832\\
-13	14	0.280847975030405	0.280847975030405\\
-13	15	0.271069492016675	0.271069492016675\\
-12	-15	0.295393732507211	0.295393732507211\\
-12	-14	0.30430821060745	0.30430821060745\\
-12	-13	0.31359216793796	0.31359216793796\\
-12	-12	0.323449369127554	0.323449369127554\\
-12	-11	0.334004037791108	0.334004037791108\\
-12	-10	0.345222263026125	0.345222263026125\\
-12	-9	0.356903072866351	0.356903072866351\\
-12	-8	0.368735434519344	0.368735434519344\\
-12	-7	0.380364280888394	0.380364280888394\\
-12	-6	0.39143020627603	0.39143020627603\\
-12	-5	0.401582382448355	0.401582382448355\\
-12	-4	0.410481475970036	0.410481475970036\\
-12	-3	0.417805308332886	0.417805308332886\\
-12	-2	0.423265013503521	0.423265013503521\\
-12	-1	0.426628376603312	0.426628376603312\\
-12	0	0.427741601715652	0.427741601715652\\
-12	1	0.426547339520145	0.426547339520145\\
-12	2	0.423094012790793	0.423094012790793\\
-12	3	0.417526565701276	0.417526565701276\\
-12	4	0.410066466366269	0.410066466366269\\
-12	5	0.400986173926625	0.400986173926625\\
-12	6	0.390581511137637	0.390581511137637\\
-12	7	0.379150569792926	0.379150569792926\\
-12	8	0.36697914898396	0.36697914898396\\
-12	9	0.354329534580842	0.354329534580842\\
-12	10	0.341432270851565	0.341432270851565\\
-12	11	0.328490573866224	0.328490573866224\\
-12	12	0.315696489044684	0.315696489044684\\
-12	13	0.30325437944345	0.30325437944345\\
-12	14	0.291428750101982	0.291428750101982\\
-12	15	0.280595313872332	0.280595313872332\\
-11	-15	0.300826807645592	0.300826807645592\\
-11	-14	0.310990902502764	0.310990902502764\\
-11	-13	0.321890514861254	0.321890514861254\\
-11	-12	0.333655403792077	0.333655403792077\\
-11	-11	0.346273949824232	0.346273949824232\\
-11	-10	0.359582607656643	0.359582607656643\\
-11	-9	0.373310863038453	0.373310863038453\\
-11	-8	0.387136886544294	0.387136886544294\\
-11	-7	0.400714836837693	0.400714836837693\\
-11	-6	0.413678600844142	0.413678600844142\\
-11	-5	0.425641669078663	0.425641669078663\\
-11	-4	0.436201324639642	0.436201324639642\\
-11	-3	0.444954327001237	0.444954327001237\\
-11	-2	0.451524265145756	0.451524265145756\\
-11	-1	0.455599059032299	0.455599059032299\\
-11	0	0.45696861067232	0.45696861067232\\
-11	1	0.455556271452386	0.455556271452386\\
-11	2	0.451432855315943	0.451432855315943\\
-11	3	0.444802276498727	0.444802276498727\\
-11	4	0.43596923443345	0.43596923443345\\
-11	5	0.425299667653208	0.425299667653208\\
-11	6	0.413180808743108	0.413180808743108\\
-11	7	0.399990265576225	0.399990265576225\\
-11	8	0.38607674695505	0.38607674695505\\
-11	9	0.371747844712802	0.371747844712802\\
-11	10	0.357263201663948	0.357263201663948\\
-11	11	0.342840869898429	0.342840869898429\\
-11	12	0.328676684878151	0.328676684878151\\
-11	13	0.3149688508413	0.3149688508413\\
-11	14	0.30196080318571	0.30196080318571\\
-11	15	0.289997460671831	0.289997460671831\\
-10	-15	0.306456316673572	0.306456316673572\\
-10	-14	0.318188192343437	0.318188192343437\\
-10	-13	0.330939763864423	0.330939763864423\\
-10	-12	0.344721578524585	0.344721578524585\\
-10	-11	0.359414449475448	0.359414449475448\\
-10	-10	0.374803566110402	0.374803566110402\\
-10	-9	0.390622254089605	0.390622254089605\\
-10	-8	0.406572621095334	0.406572621095334\\
-10	-7	0.422318699848019	0.422318699848019\\
-10	-6	0.437473083221145	0.437473083221145\\
-10	-5	0.451590890095363	0.451590890095363\\
-10	-4	0.464175724700921	0.464175724700921\\
-10	-3	0.474704960187603	0.474704960187603\\
-10	-2	0.482672426466701	0.482672426466701\\
-10	-1	0.487646457697047	0.487646457697047\\
-10	0	0.489332777061548	0.489332777061548\\
-10	1	0.487625313251962	0.487625313251962\\
-10	2	0.482626796355256	0.482626796355256\\
-10	3	0.474627064849814	0.474627064849814\\
-10	4	0.464052149382712	0.464052149382712\\
-10	5	0.451401338124212	0.451401338124212\\
-10	6	0.437187533474549	0.437187533474549\\
-10	7	0.421892320118122	0.421892320118122\\
-10	8	0.405938985908424	0.405938985908424\\
-10	9	0.389682897156012	0.389682897156012\\
-10	10	0.37341075059389	0.37341075059389\\
-10	11	0.35735019808954	0.35735019808954\\
-10	12	0.341693584276671	0.341693584276671\\
-10	13	0.32662487580132	0.32662487580132\\
-10	14	0.312360304535863	0.312360304535863\\
-10	15	0.299206290392622	0.299206290392622\\
-9	-15	0.312453360041826	0.312453360041826\\
-9	-14	0.325939825701951	0.325939825701951\\
-9	-13	0.340601179461084	0.340601179461084\\
-9	-12	0.356362592923353	0.356362592923353\\
-9	-11	0.373074633707397	0.373074633707397\\
-9	-10	0.39054388499656	0.39054388499656\\
-9	-9	0.408546235620531	0.408546235620531\\
-9	-8	0.426816781588056	0.426816781588056\\
-9	-7	0.445024193434171	0.445024193434171\\
-9	-6	0.462748474657684	0.462748474657684\\
-9	-5	0.479466948539408	0.479466948539408\\
-9	-4	0.494557599363433	0.494557599363433\\
-9	-3	0.507330440001939	0.507330440001939\\
-9	-2	0.517091044988398	0.517091044988398\\
-9	-1	0.52323016506431	0.52323016506431\\
-9	0	0.52532436276926	0.52532436276926\\
-9	1	0.523220540352278	0.523220540352278\\
-9	2	0.517070050150235	0.517070050150235\\
-9	3	0.50729343722304	0.50729343722304\\
-9	4	0.494495961311251	0.494495961311251\\
-9	5	0.479367011114996	0.479367011114996\\
-9	6	0.462590240402367	0.462590240402367\\
-9	7	0.444779187163646	0.444779187163646\\
-9	8	0.426444559564201	0.426444559564201\\
-9	9	0.407989847474745	0.407989847474745\\
-9	10	0.38972164534901	0.38972164534901\\
-9	11	0.371867704792428	0.371867704792428\\
-9	12	0.354606182715533	0.354606182715533\\
-9	13	0.338096410491076	0.338096410491076\\
-9	14	0.322519063412789	0.322519063412789\\
-9	15	0.308133185867452	0.308133185867452\\
-8	-15	0.318871895980501	0.318871895980501\\
-8	-14	0.334106767878077	0.334106767878077\\
-8	-13	0.350582778806963	0.350582778806963\\
-8	-12	0.36821385491759	0.36821385491759\\
-8	-11	0.386886170276216	0.386886170276216\\
-8	-10	0.406464953979913	0.406464953979913\\
-8	-9	0.426782521602741	0.426782521602741\\
-8	-8	0.447611395818626	0.447611395818626\\
-8	-7	0.468628026722589	0.468628026722589\\
-8	-6	0.489377010841714	0.489377010841714\\
-8	-5	0.5092460070411	0.5092460070411\\
-8	-4	0.527457657750028	0.527457657750028\\
-8	-3	0.54309921113267	0.54309921113267\\
-8	-2	0.555205004371514	0.555205004371514\\
-8	-1	0.562892955049902	0.562892955049902\\
-8	0	0.565531372870919	0.565531372870919\\
-8	1	0.56288871618709	0.56288871618709\\
-8	2	0.555195678873328	0.555195678873328\\
-8	3	0.543082205812448	0.543082205812448\\
-8	4	0.527428334793164	0.527428334793164\\
-8	5	0.509195788984023	0.509195788984023\\
-8	6	0.489292782144194	0.489292782144194\\
-8	7	0.468491435602099	0.468491435602099\\
-8	8	0.447397707263935	0.447397707263935\\
-8	9	0.426459370649555	0.426459370649555\\
-8	10	0.405989278768962	0.405989278768962\\
-8	11	0.386199200221338	0.386199200221338\\
-8	12	0.367237944025866	0.367237944025866\\
-8	13	0.349227689169476	0.349227689169476\\
-8	14	0.332304004458685	0.332304004458685\\
-8	15	0.316669582339362	0.316669582339362\\
-7	-15	0.325554463203379	0.325554463203379\\
-7	-14	0.342383172452698	0.342383172452698\\
-7	-13	0.360508749312058	0.360508749312058\\
-7	-12	0.379889478384493	0.379889478384493\\
-7	-11	0.400477236109845	0.400477236109845\\
-7	-10	0.422207990725169	0.422207990725169\\
-7	-9	0.44497966044016	0.44497966044016\\
-7	-8	0.468615060789282	0.468615060789282\\
-7	-7	0.492813232512896	0.492813232512896\\
-7	-6	0.517098176251716	0.517098176251716\\
-7	-5	0.540773870732039	0.540773870732039\\
-7	-4	0.562890162387343	0.562890162387343\\
-7	-3	0.582247004221109	0.582247004221109\\
-7	-2	0.597484718650218	0.597484718650218\\
-7	-1	0.607288859251203	0.607288859251203\\
-7	0	0.610678702776326	0.610678702776326\\
-7	1	0.607286914361207	0.607286914361207\\
-7	2	0.597480380049458	0.597480380049458\\
-7	3	0.582239075054203	0.582239075054203\\
-7	4	0.5628762988125	0.5628762988125\\
-7	5	0.54074928532773	0.54074928532773\\
-7	6	0.517055039588984	0.517055039588984\\
-7	7	0.492739775513815	0.492739775513815\\
-7	8	0.468495973678756	0.468495973678756\\
-7	9	0.444796641304636	0.444796641304636\\
-7	10	0.421940144464434	0.421940144464434\\
-7	11	0.400099503564335	0.400099503564335\\
-7	12	0.379374032666572	0.379374032666572\\
-7	13	0.359834071959334	0.359834071959334\\
-7	14	0.341559835687714	0.341559835687714\\
-7	15	0.324690059338095	0.324690059338095\\
-6	-15	0.33218369555829	0.33218369555829\\
-6	-14	0.350394772774014	0.350394772774014\\
-6	-13	0.369997019830098	0.369997019830098\\
-6	-12	0.391012802961499	0.391012802961499\\
-6	-11	0.413466632769063	0.413466632769063\\
-6	-10	0.437371749465454	0.437371749465454\\
-6	-9	0.462704939043158	0.462704939043158\\
-6	-8	0.489361202376875	0.489361202376875\\
-6	-7	0.517093987257879	0.517093987257879\\
-6	-6	0.545448488640945	0.545448488640945\\
-6	-5	0.573685099768867	0.573685099768867\\
-6	-4	0.600687453783928	0.600687453783928\\
-6	-3	0.624899538951359	0.624899538951359\\
-6	-2	0.644394617581766	0.644394617581766\\
-6	-1	0.657163269226434	0.657163269226434\\
-6	0	0.661622552904499	0.661622552904499\\
-6	1	0.657162321715924	0.657162321715924\\
-6	2	0.644392473215873	0.644392473215873\\
-6	3	0.624895848647264	0.624895848647264\\
-6	4	0.600680771341378	0.600680771341378\\
-6	5	0.573673330690943	0.573673330690943\\
-6	6	0.545427244457911	0.545427244457911\\
-6	7	0.517056386051289	0.517056386051289\\
-6	8	0.489297818910594	0.489297818910594\\
-6	9	0.462605550177464	0.462605550177464\\
-6	10	0.437227179309083	0.437227179309083\\
-6	11	0.413270024338054	0.413270024338054\\
-6	12	0.390762161496883	0.390762161496883\\
-6	13	0.369705136723772	0.369705136723772\\
-6	14	0.350113711235101	0.350113711235101\\
-6	15	0.332056504739448	0.332056504739448\\
-5	-15	0.33840203249981	0.33840203249981\\
-5	-14	0.357787997814542	0.357787997814542\\
-5	-13	0.378699928807052	0.378699928807052\\
-5	-12	0.401224472820112	0.401224472820112\\
-5	-11	0.425459148632882	0.425459148632882\\
-5	-10	0.451503320176568	0.451503320176568\\
-5	-9	0.479428848643075	0.479428848643075\\
-5	-8	0.50923124582472	0.50923124582472\\
-5	-7	0.540768784096664	0.540768784096664\\
-5	-6	0.573683753306562	0.573683753306562\\
-5	-5	0.607277189911227	0.607277189911227\\
-5	-4	0.640320621598811	0.640320621598811\\
-5	-3	0.670866615055603	0.670866615055603\\
-5	-2	0.696207683931239	0.696207683931239\\
-5	-1	0.7132189884298	0.7132189884298\\
-5	0	0.719244307506458	0.719244307506458\\
-5	1	0.713218539609031	0.713218539609031\\
-5	2	0.696206686551754	0.696206686551754\\
-5	3	0.670864769903555	0.670864769903555\\
-5	4	0.640317429984459	0.640317429984459\\
-5	5	0.607271504848162	0.607271504848162\\
-5	6	0.573673729642737	0.573673729642737\\
-5	7	0.540750631790035	0.540750631790035\\
-5	8	0.509199928108887	0.509199928108887\\
-5	9	0.479378680454973	0.479378680454973\\
-5	10	0.451431109901844	0.451431109901844\\
-5	11	0.425366841161532	0.425366841161532\\
-5	12	0.401121667929682	0.401121667929682\\
-5	13	0.378613131857358	0.378613131857358\\
-5	14	0.357781215634043	0.357781215634043\\
-5	15	0.338621356248592	0.338621356248592\\
-4	-15	0.343897842807807	0.343897842807807\\
-4	-14	0.364264282025406	0.364264282025406\\
-4	-13	0.386310308295983	0.386310308295983\\
-4	-12	0.410181439286419	0.410181439286419\\
-4	-11	0.436048675709441	0.436048675709441\\
-4	-10	0.464103115474403	0.464103115474403\\
-4	-9	0.494526181909651	0.494526181909651\\
-4	-8	0.527445290244412	0.527445290244412\\
-4	-7	0.562885574441712	0.562885574441712\\
-4	-6	0.600685857976752	0.600685857976752\\
-4	-5	0.640320222647017	0.640320222647017\\
-4	-4	0.680594206781103	0.680594206781103\\
-4	-3	0.719239719560826	0.719239719560826\\
-4	-2	0.752565010690911	0.752565010690911\\
-4	-1	0.775689702874667	0.775689702874667\\
-4	0	0.784040312482982	0.784040312482982\\
-4	1	0.77568950339877	0.77568950339877\\
-4	2	0.752564512001168	0.752564512001168\\
-4	3	0.719238821919289	0.719238821919289\\
-4	4	0.680592610973927	0.680592610973927\\
-4	5	0.640317529722407	0.640317529722407\\
-4	6	0.600681120424197	0.600681120424197\\
-4	7	0.562877296191985	0.562877296191985\\
-4	8	0.527431127455723	0.527431127455723\\
-4	9	0.494503691002261	0.494503691002261\\
-4	10	0.464071847627543	0.464071847627543\\
-4	11	0.43601386716541	0.43601386716541\\
-4	12	0.410156604537239	0.410156604537239\\
-4	13	0.38632362331211	0.38632362331211\\
-4	14	0.364375589575948	0.364375589575948\\
-4	15	0.344234807466862	0.344234807466862\\
-3	-15	0.348429261957134	0.348429261957134\\
-3	-14	0.369578968218902	0.369578968218902\\
-3	-13	0.392557020683855	0.392557020683855\\
-3	-12	0.417560551407233	0.417560551407233\\
-3	-11	0.444830352731234	0.444830352731234\\
-3	-10	0.474646962570542	0.474646962570542\\
-3	-9	0.50730585459763	0.50730585459763\\
-3	-8	0.543089436813716	0.543089436813716\\
-3	-7	0.582243214179066	0.582243214179066\\
-3	-6	0.62489814262008	0.62489814262008\\
-3	-5	0.670866066496886	0.670866066496886\\
-3	-4	0.719239569953903	0.719239569953903\\
-3	-3	0.767674013297462	0.767674013297462\\
-3	-2	0.811396935691165	0.811396935691165\\
-3	-1	0.842967387486119	0.842967387486119\\
-3	0	0.85463573008278	0.85463573008278\\
-3	1	0.842967188010222	0.842967188010222\\
-3	2	0.811396736215268	0.811396736215268\\
-3	3	0.767673564476694	0.767673564476694\\
-3	4	0.719238821919289	0.719238821919289\\
-3	5	0.67086481977253	0.67086481977253\\
-3	6	0.624896048123161	0.624896048123161\\
-3	7	0.582239673481894	0.582239673481894\\
-3	8	0.543083701881676	0.543083701881676\\
-3	9	0.507297277134058	0.507297277134058\\
-3	10	0.474636789299794	0.474636789299794\\
-3	11	0.444824418323297	0.444824418323297\\
-3	12	0.417571372974646	0.417571372974646\\
-3	13	0.392608236120417	0.392608236120417\\
-3	14	0.36971835200195	0.36971835200195\\
-3	15	0.348755928673005	0.348755928673005\\
-2	-15	0.3518153403747	0.3518153403747\\
-2	-14	0.373534425650047	0.373534425650047\\
-2	-13	0.397208025633451	0.397208025633451\\
-2	-12	0.423073067821606	0.423073067821606\\
-2	-11	0.451427768680569	0.451427768680569\\
-2	-10	0.482627943341664	0.482627943341664\\
-2	-9	0.517072543598948	0.517072543598948\\
-2	-8	0.555197673632299	0.555197673632299\\
-2	-7	0.597481826249711	0.597481826249711\\
-2	-6	0.644393420726384	0.644393420726384\\
-2	-5	0.696207235110471	0.696207235110471\\
-2	-4	0.752564861083988	0.752564861083988\\
-2	-3	0.811396935691165	0.811396935691165\\
-2	-2	0.866969923223722	0.866969923223722\\
-2	-1	0.908692501322027	0.908692501322027\\
-2	0	0.924462866265175	0.924462866265175\\
-2	1	0.908692301846129	0.908692301846129\\
-2	2	0.866969923223722	0.866969923223722\\
-2	3	0.811396736215268	0.811396736215268\\
-2	4	0.752564462132194	0.752564462132194\\
-2	5	0.69620663668278	0.69620663668278\\
-2	6	0.644392423346898	0.644392423346898\\
-2	7	0.597480330180484	0.597480330180484\\
-2	8	0.555195529266405	0.555195529266405\\
-2	9	0.517069950412287	0.517069950412287\\
-2	10	0.482626547010385	0.482626547010385\\
-2	11	0.451432506233123	0.451432506233123\\
-2	12	0.423093414363102	0.423093414363102\\
-2	13	0.397260188580524	0.397260188580524\\
-2	14	0.373650470753143	0.373650470753143\\
-2	15	0.352062017255862	0.352062017255862\\
-1	-15	0.353927291434485	0.353927291434485\\
-1	-14	0.375983540844784	0.375983540844784\\
-1	-13	0.400083595361547	0.400083595361547\\
-1	-12	0.426486998061294	0.426486998061294\\
-1	-11	0.455528444564751	0.455528444564751\\
-1	-10	0.487614441815574	0.487614441815574\\
-1	-9	0.523216999655106	0.523216999655106\\
-1	-8	0.562887768676579	0.562887768676579\\
-1	-7	0.607286764754285	0.607286764754285\\
-1	-6	0.657162371584898	0.657162371584898\\
-1	-5	0.713218739084928	0.713218739084928\\
-1	-4	0.775689653005693	0.775689653005693\\
-1	-3	0.842967387486119	0.842967387486119\\
-1	-2	0.908692401584078	0.908692401584078\\
-1	-1	0.959454728645956	0.959454728645956\\
-1	0	0.978951502821487	0.978951502821487\\
-1	1	0.959454728645956	0.959454728645956\\
-1	2	0.908692401584078	0.908692401584078\\
-1	3	0.84296728774817	0.84296728774817\\
-1	4	0.775689453529796	0.775689453529796\\
-1	5	0.713218489740057	0.713218489740057\\
-1	6	0.657162072371052	0.657162072371052\\
-1	7	0.607286166326593	0.607286166326593\\
-1	8	0.56288707051094	0.56288707051094\\
-1	9	0.523216451096389	0.523216451096389\\
-1	10	0.487615139981213	0.487615139981213\\
-1	11	0.455533231986279	0.455533231986279\\
-1	12	0.426500961374086	0.426500961374086\\
-1	13	0.400114938011868	0.400114938011868\\
-1	14	0.376047996494012	0.376047996494012\\
-1	15	0.354058097753961	0.354058097753961\\
0	-15	0.354685274908705	0.354685274908705\\
0	-14	0.376833482707534	0.376833482707534\\
0	-13	0.401067360616716	0.401067360616716\\
0	-12	0.427648546209689	0.427648546209689\\
0	-11	0.456922382133183	0.456922382133183\\
0	-10	0.489312081437232	0.489312081437232\\
0	-9	0.525315984781585	0.525315984781585\\
0	-8	0.565528031649644	0.565528031649644\\
0	-7	0.610677356314021	0.610677356314021\\
0	-6	0.661622004345782	0.661622004345782\\
0	-5	0.719244058161586	0.719244058161586\\
0	-4	0.784040262614008	0.784040262614008\\
0	-3	0.85463573008278	0.85463573008278\\
0	-2	0.924462866265175	0.924462866265175\\
0	-1	0.978951502821487	0.978951502821487\\
0	0	1	1\\
0	1	0.978951502821487	0.978951502821487\\
0	2	0.924462866265175	0.924462866265175\\
0	3	0.85463573008278	0.85463573008278\\
0	4	0.784040212745034	0.784040212745034\\
0	5	0.719244157899535	0.719244157899535\\
0	6	0.661622004345782	0.661622004345782\\
0	7	0.610677356314021	0.610677356314021\\
0	8	0.565528031649644	0.565528031649644\\
0	9	0.525315984781585	0.525315984781585\\
0	10	0.489312131306206	0.489312131306206\\
0	11	0.456922332264209	0.456922332264209\\
0	12	0.427648546209689	0.427648546209689\\
0	13	0.401067335682229	0.401067335682229\\
0	14	0.376833457773047	0.376833457773047\\
0	15	0.354685249974218	0.354685249974218\\
1	-15	0.354058147622936	0.354058147622936\\
1	-14	0.376048021428499	0.376048021428499\\
1	-13	0.400114987880842	0.400114987880842\\
1	-12	0.42650101124306	0.42650101124306\\
1	-11	0.455533331724228	0.455533331724228\\
1	-10	0.487615139981213	0.487615139981213\\
1	-9	0.523216500965364	0.523216500965364\\
1	-8	0.56288707051094	0.56288707051094\\
1	-7	0.607286216195568	0.607286216195568\\
1	-6	0.657162072371052	0.657162072371052\\
1	-5	0.713218489740057	0.713218489740057\\
1	-4	0.77568950339877	0.77568950339877\\
1	-3	0.842967188010222	0.842967188010222\\
1	-2	0.908692501322027	0.908692501322027\\
1	-1	0.959454728645956	0.959454728645956\\
1	0	0.978951502821487	0.978951502821487\\
1	1	0.959454728645956	0.959454728645956\\
1	2	0.908692401584078	0.908692401584078\\
1	3	0.842967387486119	0.842967387486119\\
1	4	0.775689653005693	0.775689653005693\\
1	5	0.713218739084928	0.713218739084928\\
1	6	0.657162421453872	0.657162421453872\\
1	7	0.60728671488531	0.60728671488531\\
1	8	0.562887718807605	0.562887718807605\\
1	9	0.523216999655106	0.523216999655106\\
1	10	0.487614441815574	0.487614441815574\\
1	11	0.455528444564751	0.455528444564751\\
1	12	0.42648694819232	0.42648694819232\\
1	13	0.400083520558086	0.400083520558086\\
1	14	0.37598349097581	0.37598349097581\\
1	15	0.353927266499998	0.353927266499998\\
2	-15	0.35206211699381	0.35206211699381\\
2	-14	0.373650545556605	0.373650545556605\\
2	-13	0.397260238449498	0.397260238449498\\
2	-12	0.423093464232077	0.423093464232077\\
2	-11	0.451432605971072	0.451432605971072\\
2	-10	0.482626696617308	0.482626696617308\\
2	-9	0.517070000281261	0.517070000281261\\
2	-8	0.55519557913538	0.55519557913538\\
2	-7	0.597480380049458	0.597480380049458\\
2	-6	0.644392523084847	0.644392523084847\\
2	-5	0.696206686551754	0.696206686551754\\
2	-4	0.752564611739117	0.752564611739117\\
2	-3	0.811396736215268	0.811396736215268\\
2	-2	0.866969923223722	0.866969923223722\\
2	-1	0.908692501322027	0.908692501322027\\
2	0	0.924462866265175	0.924462866265175\\
2	1	0.908692501322027	0.908692501322027\\
2	2	0.86697002296167	0.86697002296167\\
2	3	0.811396885822191	0.811396885822191\\
2	4	0.752564861083988	0.752564861083988\\
2	5	0.696207284979445	0.696207284979445\\
2	6	0.644393420726384	0.644393420726384\\
2	7	0.597481826249711	0.597481826249711\\
2	8	0.555197673632299	0.555197673632299\\
2	9	0.517072543598948	0.517072543598948\\
2	10	0.482627943341664	0.482627943341664\\
2	11	0.451427768680569	0.451427768680569\\
2	12	0.423073067821606	0.423073067821606\\
2	13	0.397208050567938	0.397208050567938\\
2	14	0.373534375781073	0.373534375781073\\
2	15	0.351815315440212	0.351815315440212\\
3	-15	0.348756003476466	0.348756003476466\\
3	-14	0.369718401870924	0.369718401870924\\
3	-13	0.392608261054904	0.392608261054904\\
3	-12	0.417571422843621	0.417571422843621\\
3	-11	0.444824468192272	0.444824468192272\\
3	-10	0.474636839168768	0.474636839168768\\
3	-9	0.507297327003032	0.507297327003032\\
3	-8	0.543083801619625	0.543083801619625\\
3	-7	0.582239723350868	0.582239723350868\\
3	-6	0.624896048123161	0.624896048123161\\
3	-5	0.670864869641504	0.670864869641504\\
3	-4	0.719238971526212	0.719238971526212\\
3	-3	0.767673614345668	0.767673614345668\\
3	-2	0.811396736215268	0.811396736215268\\
3	-1	0.84296728774817	0.84296728774817\\
3	0	0.85463573008278	0.85463573008278\\
3	1	0.842967387486119	0.842967387486119\\
3	2	0.811397035429114	0.811397035429114\\
3	3	0.767674013297462	0.767674013297462\\
3	4	0.719239569953903	0.719239569953903\\
3	5	0.670866166234835	0.670866166234835\\
3	6	0.62489814262008	0.62489814262008\\
3	7	0.582243214179066	0.582243214179066\\
3	8	0.54308948668269	0.54308948668269\\
3	9	0.507305804728656	0.507305804728656\\
3	10	0.474646912701568	0.474646912701568\\
3	11	0.444830402600208	0.444830402600208\\
3	12	0.41756057634172	0.41756057634172\\
3	13	0.392556995749368	0.392556995749368\\
3	14	0.369578943284415	0.369578943284415\\
3	15	0.348429261957134	0.348429261957134\\
4	-15	0.344234857335836	0.344234857335836\\
4	-14	0.364375689313897	0.364375689313897\\
4	-13	0.386323673181084	0.386323673181084\\
4	-12	0.4101566793407	0.4101566793407\\
4	-11	0.436013917034384	0.436013917034384\\
4	-10	0.464071897496518	0.464071897496518\\
4	-9	0.494503740871235	0.494503740871235\\
4	-8	0.527431227193671	0.527431227193671\\
4	-7	0.562877395929934	0.562877395929934\\
4	-6	0.600681170293172	0.600681170293172\\
4	-5	0.640317579591382	0.640317579591382\\
4	-4	0.680592710711876	0.680592710711876\\
4	-3	0.719238871788264	0.719238871788264\\
4	-2	0.752564561870142	0.752564561870142\\
4	-1	0.775689553267744	0.775689553267744\\
4	0	0.784040362351957	0.784040362351957\\
4	1	0.775689752743641	0.775689752743641\\
4	2	0.752565060559885	0.752565060559885\\
4	3	0.7192397694298	0.7192397694298\\
4	4	0.680594256650078	0.680594256650078\\
4	5	0.640320322384966	0.640320322384966\\
4	6	0.600685857976752	0.600685857976752\\
4	7	0.562885524572738	0.562885524572738\\
4	8	0.527445290244412	0.527445290244412\\
4	9	0.494526181909651	0.494526181909651\\
4	10	0.464103065605428	0.464103065605428\\
4	11	0.436048675709441	0.436048675709441\\
4	12	0.410181439286419	0.410181439286419\\
4	13	0.386310308295983	0.386310308295983\\
4	14	0.364264257090919	0.364264257090919\\
4	15	0.343897842807807	0.343897842807807\\
5	-15	0.338621406117567	0.338621406117567\\
5	-14	0.357781265503017	0.357781265503017\\
5	-13	0.378613206660819	0.378613206660819\\
5	-12	0.401121717798656	0.401121717798656\\
5	-11	0.425366891030506	0.425366891030506\\
5	-10	0.451431209639792	0.451431209639792\\
5	-9	0.479378780192921	0.479378780192921\\
5	-8	0.509200027846835	0.509200027846835\\
5	-7	0.540750731527983	0.540750731527983\\
5	-6	0.573673729642737	0.573673729642737\\
5	-5	0.607271554717136	0.607271554717136\\
5	-4	0.640317429984459	0.640317429984459\\
5	-3	0.670864869641504	0.670864869641504\\
5	-2	0.696206686551754	0.696206686551754\\
5	-1	0.713218589478006	0.713218589478006\\
5	0	0.719244307506458	0.719244307506458\\
5	1	0.7132189884298	0.7132189884298\\
5	2	0.696207683931239	0.696207683931239\\
5	3	0.670866615055603	0.670866615055603\\
5	4	0.640320621598811	0.640320621598811\\
5	5	0.607277239780202	0.607277239780202\\
5	6	0.573683803175537	0.573683803175537\\
5	7	0.540768784096664	0.540768784096664\\
5	8	0.50923124582472	0.50923124582472\\
5	9	0.479428848643075	0.479428848643075\\
5	10	0.451503320176568	0.451503320176568\\
5	11	0.425459148632882	0.425459148632882\\
5	12	0.401224422951138	0.401224422951138\\
5	13	0.378699903872565	0.378699903872565\\
5	14	0.357787947945568	0.357787947945568\\
5	15	0.338401982630836	0.338401982630836\\
6	-15	0.332056579542909	0.332056579542909\\
6	-14	0.350113786038562	0.350113786038562\\
6	-13	0.369705211527233	0.369705211527233\\
6	-12	0.390762211365857	0.390762211365857\\
6	-11	0.413270074207028	0.413270074207028\\
6	-10	0.437227229178058	0.437227229178058\\
6	-9	0.462605649915412	0.462605649915412\\
6	-8	0.489297918648543	0.489297918648543\\
6	-7	0.517056386051289	0.517056386051289\\
6	-6	0.54542734419586	0.54542734419586\\
6	-5	0.573673380559917	0.573673380559917\\
6	-4	0.600680821210352	0.600680821210352\\
6	-3	0.624895848647264	0.624895848647264\\
6	-2	0.644392523084847	0.644392523084847\\
6	-1	0.657162371584898	0.657162371584898\\
6	0	0.661622652642447	0.661622652642447\\
6	1	0.657163319095409	0.657163319095409\\
6	2	0.644394617581766	0.644394617581766\\
6	3	0.624899638689308	0.624899638689308\\
6	4	0.600687503652902	0.600687503652902\\
6	5	0.573685199506816	0.573685199506816\\
6	6	0.545448488640945	0.545448488640945\\
6	7	0.517093987257879	0.517093987257879\\
6	8	0.489361202376875	0.489361202376875\\
6	9	0.462704988912132	0.462704988912132\\
6	10	0.437371749465454	0.437371749465454\\
6	11	0.413466607834575	0.413466607834575\\
6	12	0.391012802961499	0.391012802961499\\
6	13	0.369996969961124	0.369996969961124\\
6	14	0.350394797708501	0.350394797708501\\
6	15	0.33218369555829	0.33218369555829\\
7	-15	0.324690109207069	0.324690109207069\\
7	-14	0.341559885556688	0.341559885556688\\
7	-13	0.359834096893821	0.359834096893821\\
7	-12	0.379374107470033	0.379374107470033\\
7	-11	0.400099578367796	0.400099578367796\\
7	-10	0.421940194333408	0.421940194333408\\
7	-9	0.444796741042585	0.444796741042585\\
7	-8	0.468495973678756	0.468495973678756\\
7	-7	0.492739875251763	0.492739875251763\\
7	-6	0.517055089457958	0.517055089457958\\
7	-5	0.540749335196704	0.540749335196704\\
7	-4	0.562876348681474	0.562876348681474\\
7	-3	0.582239124923177	0.582239124923177\\
7	-2	0.597480429918432	0.597480429918432\\
7	-1	0.607286914361207	0.607286914361207\\
7	0	0.610678702776326	0.610678702776326\\
7	1	0.607288909120178	0.607288909120178\\
7	2	0.597484768519193	0.597484768519193\\
7	3	0.582247004221109	0.582247004221109\\
7	4	0.562890162387343	0.562890162387343\\
7	5	0.540773920601013	0.540773920601013\\
7	6	0.517098176251716	0.517098176251716\\
7	7	0.492813232512896	0.492813232512896\\
7	8	0.468615060789282	0.468615060789282\\
7	9	0.444979710309134	0.444979710309134\\
7	10	0.422208040594143	0.422208040594143\\
7	11	0.400477261044332	0.400477261044332\\
7	12	0.379889528253467	0.379889528253467\\
7	13	0.360508749312058	0.360508749312058\\
7	14	0.342383172452698	0.342383172452698\\
7	15	0.325554463203379	0.325554463203379\\
8	-15	0.316669557404874	0.316669557404874\\
8	-14	0.332304004458685	0.332304004458685\\
8	-13	0.349227714103963	0.349227714103963\\
8	-12	0.367237919091379	0.367237919091379\\
8	-11	0.386199250090313	0.386199250090313\\
8	-10	0.405989278768962	0.405989278768962\\
8	-9	0.426459320780581	0.426459320780581\\
8	-8	0.447397707263935	0.447397707263935\\
8	-7	0.468491385733125	0.468491385733125\\
8	-6	0.489292782144194	0.489292782144194\\
8	-5	0.509195739115049	0.509195739115049\\
8	-4	0.527428334793164	0.527428334793164\\
8	-3	0.543082305550397	0.543082305550397\\
8	-2	0.555195629004354	0.555195629004354\\
8	-1	0.562888766056064	0.562888766056064\\
8	0	0.565531372870919	0.565531372870919\\
8	1	0.562892955049902	0.562892955049902\\
8	2	0.555205054240488	0.555205054240488\\
8	3	0.54309921113267	0.54309921113267\\
8	4	0.527457657750028	0.527457657750028\\
8	5	0.5092460070411	0.5092460070411\\
8	6	0.489377010841714	0.489377010841714\\
8	7	0.468628026722589	0.468628026722589\\
8	8	0.4476114456876	0.4476114456876\\
8	9	0.426782571471715	0.426782571471715\\
8	10	0.406464953979913	0.406464953979913\\
8	11	0.386886170276216	0.386886170276216\\
8	12	0.36821385491759	0.36821385491759\\
8	13	0.350582778806963	0.350582778806963\\
8	14	0.334106792812564	0.334106792812564\\
8	15	0.318871895980501	0.318871895980501\\
9	-15	0.308133185867452	0.308133185867452\\
9	-14	0.322519063412789	0.322519063412789\\
9	-13	0.338096435425563	0.338096435425563\\
9	-12	0.354606182715533	0.354606182715533\\
9	-11	0.371867729726916	0.371867729726916\\
9	-10	0.389721670283497	0.389721670283497\\
9	-9	0.407989847474745	0.407989847474745\\
9	-8	0.426444559564201	0.426444559564201\\
9	-7	0.444779187163646	0.444779187163646\\
9	-6	0.462590290271341	0.462590290271341\\
9	-5	0.479367011114996	0.479367011114996\\
9	-4	0.494495961311251	0.494495961311251\\
9	-3	0.50729343722304	0.50729343722304\\
9	-2	0.517070050150235	0.517070050150235\\
9	-1	0.523220590221253	0.523220590221253\\
9	0	0.52532436276926	0.52532436276926\\
9	1	0.52323016506431	0.52323016506431\\
9	2	0.517091094857372	0.517091094857372\\
9	3	0.507330489870913	0.507330489870913\\
9	4	0.494557649232407	0.494557649232407\\
9	5	0.479467048277356	0.479467048277356\\
9	6	0.462748474657684	0.462748474657684\\
9	7	0.445024193434171	0.445024193434171\\
9	8	0.426816781588056	0.426816781588056\\
9	9	0.408546235620531	0.408546235620531\\
9	10	0.390543860062073	0.390543860062073\\
9	11	0.373074633707397	0.373074633707397\\
9	12	0.356362592923353	0.356362592923353\\
9	13	0.340601179461084	0.340601179461084\\
9	14	0.325939800767464	0.325939800767464\\
9	15	0.312453360041826	0.312453360041826\\
10	-15	0.299206290392622	0.299206290392622\\
10	-14	0.312360354404837	0.312360354404837\\
10	-13	0.326624900735807	0.326624900735807\\
10	-12	0.341693584276671	0.341693584276671\\
10	-11	0.35735019808954	0.35735019808954\\
10	-10	0.37341075059389	0.37341075059389\\
10	-9	0.389682897156012	0.389682897156012\\
10	-8	0.405938985908424	0.405938985908424\\
10	-7	0.421892320118122	0.421892320118122\\
10	-6	0.437187533474549	0.437187533474549\\
10	-5	0.451401338124212	0.451401338124212\\
10	-4	0.464052149382712	0.464052149382712\\
10	-3	0.474627114718788	0.474627114718788\\
10	-2	0.48262684622423	0.48262684622423\\
10	-1	0.487625313251962	0.487625313251962\\
10	0	0.489332826930523	0.489332826930523\\
10	1	0.487646457697047	0.487646457697047\\
10	2	0.482672426466701	0.482672426466701\\
10	3	0.474705059925552	0.474705059925552\\
10	4	0.464175774569895	0.464175774569895\\
10	5	0.451590890095363	0.451590890095363\\
10	6	0.437473083221145	0.437473083221145\\
10	7	0.422318699848019	0.422318699848019\\
10	8	0.406572621095334	0.406572621095334\\
10	9	0.390622229155118	0.390622229155118\\
10	10	0.374803566110402	0.374803566110402\\
10	11	0.359414449475448	0.359414449475448\\
10	12	0.344721578524585	0.344721578524585\\
10	13	0.330939763864423	0.330939763864423\\
10	14	0.318188142474463	0.318188142474463\\
10	15	0.306456291739085	0.306456291739085\\
11	-15	0.289997510540805	0.289997510540805\\
11	-14	0.301960828120197	0.301960828120197\\
11	-13	0.314968875775787	0.314968875775787\\
11	-12	0.328676734747125	0.328676734747125\\
11	-11	0.342840869898429	0.342840869898429\\
11	-10	0.35726327646741	0.35726327646741\\
11	-9	0.371747894581776	0.371747894581776\\
11	-8	0.386076796824025	0.386076796824025\\
11	-7	0.399990340379686	0.399990340379686\\
11	-6	0.413180833677596	0.413180833677596\\
11	-5	0.425299667653208	0.425299667653208\\
11	-4	0.435969284302424	0.435969284302424\\
11	-3	0.444802326367702	0.444802326367702\\
11	-2	0.451432905184917	0.451432905184917\\
11	-1	0.455556271452386	0.455556271452386\\
11	0	0.456968660541294	0.456968660541294\\
11	1	0.455599108901273	0.455599108901273\\
11	2	0.45152431501473	0.45152431501473\\
11	3	0.444954327001237	0.444954327001237\\
11	4	0.436201374508616	0.436201374508616\\
11	5	0.425641669078663	0.425641669078663\\
11	6	0.413678650713116	0.413678650713116\\
11	7	0.40071486177218	0.40071486177218\\
11	8	0.387136911478781	0.387136911478781\\
11	9	0.37331088797294	0.37331088797294\\
11	10	0.35958263259113	0.35958263259113\\
11	11	0.346273949824232	0.346273949824232\\
11	12	0.333655428726564	0.333655428726564\\
11	13	0.321890514861254	0.321890514861254\\
11	14	0.31099085263379	0.31099085263379\\
11	15	0.300826832580079	0.300826832580079\\
12	-15	0.280595363741306	0.280595363741306\\
12	-14	0.291428775036469	0.291428775036469\\
12	-13	0.303254404377937	0.303254404377937\\
12	-12	0.315696513979171	0.315696513979171\\
12	-11	0.328490648669685	0.328490648669685\\
12	-10	0.341432320720539	0.341432320720539\\
12	-9	0.354329584449817	0.354329584449817\\
12	-8	0.366979173918447	0.366979173918447\\
12	-7	0.3791506196619	0.3791506196619\\
12	-6	0.390581536072124	0.390581536072124\\
12	-5	0.400986173926625	0.400986173926625\\
12	-4	0.410066516235244	0.410066516235244\\
12	-3	0.417526615570251	0.417526615570251\\
12	-2	0.423094062659768	0.423094062659768\\
12	-1	0.426547389389119	0.426547389389119\\
12	0	0.427741651584627	0.427741651584627\\
12	1	0.426628426472287	0.426628426472287\\
12	2	0.423265013503521	0.423265013503521\\
12	3	0.41780535820186	0.41780535820186\\
12	4	0.410481550773498	0.410481550773498\\
12	5	0.401582457251817	0.401582457251817\\
12	6	0.391430256145005	0.391430256145005\\
12	7	0.380364355691855	0.380364355691855\\
12	8	0.368735459453831	0.368735459453831\\
12	9	0.356903147669812	0.356903147669812\\
12	10	0.345222263026125	0.345222263026125\\
12	11	0.334004087660082	0.334004087660082\\
12	12	0.323449418996529	0.323449418996529\\
12	13	0.313592217806934	0.313592217806934\\
12	14	0.30430821060745	0.30430821060745\\
12	15	0.295393757441698	0.295393757441698\\
13	-15	0.271069516951162	0.271069516951162\\
13	-14	0.280847999964892	0.280847999964892\\
13	-13	0.29157725990732	0.29157725990732\\
13	-12	0.302858444722337	0.302858444722337\\
13	-11	0.314410592608979	0.314410592608979\\
13	-10	0.326026896931004	0.326026896931004\\
13	-9	0.337524089208015	0.337524089208015\\
13	-8	0.348716432445393	0.348716432445393\\
13	-7	0.359402156773294	0.359402156773294\\
13	-6	0.369359868880503	0.369359868880503\\
13	-5	0.37835573314673	0.37835573314673\\
13	-4	0.386153196092586	0.386153196092586\\
13	-3	0.392523758078026	0.392523758078026\\
13	-2	0.397261285697957	0.397261285697957\\
13	-1	0.400202258585792	0.400202258585792\\
13	0	0.401242849537126	0.401242849537126\\
13	1	0.400350643718694	0.400350643718694\\
13	2	0.397570672814248	0.397570672814248\\
13	3	0.393019879568422	0.393019879568422\\
13	4	0.386878789668026	0.386878789668026\\
13	5	0.379379967074508	0.379379967074508\\
13	6	0.37079415044911	0.37079415044911\\
13	7	0.361425690141722	0.361425690141722\\
13	8	0.351612722732293	0.351612722732293\\
13	9	0.341715277280473	0.341715277280473\\
13	10	0.332068772507115	0.332068772507115\\
13	11	0.322903253990465	0.322903253990465\\
13	12	0.3142724306158	0.3142724306158\\
13	13	0.306056242827595	0.306056242827595\\
13	14	0.298047310496402	0.298047310496402\\
13	15	0.290060495055292	0.290060495055292\\
14	-15	0.261472357527831	0.261472357527831\\
14	-14	0.270280140957648	0.270280140957648\\
14	-13	0.280006884910586	0.280006884910586\\
14	-12	0.290236981355187	0.290236981355187\\
14	-11	0.300676452688207	0.300676452688207\\
14	-10	0.311118193059555	0.311118193059555\\
14	-9	0.32138933166997	0.32138933166997\\
14	-8	0.331322832390178	0.331322832390178\\
14	-7	0.340743779792972	0.340743779792972\\
14	-6	0.34946706024591	0.34946706024591\\
14	-5	0.357301850118997	0.357301850118997\\
14	-4	0.364059470148133	0.364059470148133\\
14	-3	0.36956196289868	0.36956196289868\\
14	-2	0.373652415643139	0.373652415643139\\
14	-1	0.376209422363682	0.376209422363682\\
14	0	0.37715827933687	0.37715827933687\\
14	1	0.37647819120044	0.37647819120044\\
14	2	0.374207257850723	0.374207257850723\\
14	3	0.370440205404321	0.370440205404321\\
14	4	0.365326092225287	0.365326092225287\\
14	5	0.359063895520908	0.359063895520908\\
14	6	0.35189887090658	0.35189887090658\\
14	7	0.344126392316827	0.344126392316827\\
14	8	0.336085867990442	0.336085867990442\\
14	9	0.328116980445577	0.328116980445577\\
14	10	0.320470246474413	0.320470246474413\\
14	11	0.313216679496291	0.313216679496291\\
14	12	0.306237591352481	0.306237591352481\\
14	13	0.299315927332527	0.299315927332527\\
14	14	0.292259043588875	0.292259043588875\\
14	15	0.284968723176725	0.284968723176725\\
15	-15	0.251841611350338	0.251841611350338\\
15	-14	0.259769631270799	0.259769631270799\\
15	-13	0.268591926572124	0.268591926572124\\
15	-12	0.277882167393391	0.277882167393391\\
15	-11	0.287336227795091	0.287336227795091\\
15	-10	0.296747675158289	0.296747675158289\\
15	-9	0.305954136102803	0.305954136102803\\
15	-8	0.31480637772317	0.31480637772317\\
15	-7	0.323155142179438	0.323155142179438\\
15	-6	0.330847730672431	0.330847730672431\\
15	-5	0.337729050692211	0.337729050692211\\
15	-4	0.343648049115753	0.343648049115753\\
15	-3	0.348466214867058	0.348466214867058\\
15	-2	0.352065433280599	0.352065433280599\\
15	-1	0.354356364088993	0.354356364088993\\
15	0	0.355286021507092	0.355286021507092\\
15	1	0.354843110212228	0.354843110212228\\
15	2	0.353063386258943	0.353063386258943\\
15	3	0.350030604589502	0.350030604589502\\
15	4	0.345876843182223	0.345876843182223\\
15	5	0.34078522091058	0.34078522091058\\
15	6	0.334998051120486	0.334998051120486\\
15	7	0.328818911192732	0.328818911192732\\
15	8	0.322575839239981	0.322575839239981\\
15	9	0.316529300914776	0.316529300914776\\
15	10	0.310766741463479	0.310766741463479\\
15	11	0.305170420237867	0.305170420237867\\
15	12	0.2995056289106	0.2995056289106\\
15	13	0.293560299668651	0.293560299668651\\
15	14	0.287228460941722	0.287228460941722\\
15	15	0.280507494608665	0.280507494608665\\
};
\end{axis}
\end{tikzpicture}%

%% file: Figures/Results_EE/CM_Plateau60Zoom_V2.tikz
%
%
\begin{tikzpicture}

\begin{axis}[%
width=1\linewidth,
height=1\linewidth,
point meta min=0,
point meta max=1,
xmin=-10,
xmax=10,
xlabel={x / \si{\pixel}},
tick align=outside,
ymin=-10,
ymax=10,
ylabel={y / \si{\pixel}},
zmin=0.6,
zmax=1,
ztick={0.6,1},
zlabel={R(s)},
view={45}{45},
axis background/.style={fill=white},
axis x line*=bottom,
axis y line*=left,
axis z line*=left,
xmajorgrids,
ymajorgrids,
zmajorgrids,
xlabel style={font=\color{white!15!black},yshift=2mm,xshift=0mm},
ylabel style={font=\color{white!15!black},yshift=2mm,xshift=0mm},
xtick style={draw=none},
ytick style={draw=none},
ztick style={draw=none},
]

\addplot3[%
surf,
shader=flat corner, fill=white, z buffer=sort, colormap={mymap}{[1pt] rgb(0pt)=(0.2422,0.1504,0.6603); rgb(1pt)=(0.25039,0.164995,0.707614); rgb(2pt)=(0.257771,0.181781,0.751138); rgb(3pt)=(0.264729,0.197757,0.795214); rgb(4pt)=(0.270648,0.214676,0.836371); rgb(5pt)=(0.275114,0.234238,0.870986); rgb(6pt)=(0.2783,0.255871,0.899071); rgb(7pt)=(0.280333,0.278233,0.9221); rgb(8pt)=(0.281338,0.300595,0.941376); rgb(9pt)=(0.281014,0.322757,0.957886); rgb(10pt)=(0.279467,0.344671,0.971676); rgb(11pt)=(0.275971,0.366681,0.982905); rgb(12pt)=(0.269914,0.3892,0.9906); rgb(13pt)=(0.260243,0.412329,0.995157); rgb(14pt)=(0.244033,0.435833,0.998833); rgb(15pt)=(0.220643,0.460257,0.997286); rgb(16pt)=(0.196333,0.484719,0.989152); rgb(17pt)=(0.183405,0.507371,0.979795); rgb(18pt)=(0.178643,0.528857,0.968157); rgb(19pt)=(0.176438,0.549905,0.952019); rgb(20pt)=(0.168743,0.570262,0.935871); rgb(21pt)=(0.154,0.5902,0.9218); rgb(22pt)=(0.146029,0.609119,0.907857); rgb(23pt)=(0.138024,0.627629,0.89729); rgb(24pt)=(0.124814,0.645929,0.888343); rgb(25pt)=(0.111252,0.6635,0.876314); rgb(26pt)=(0.0952095,0.679829,0.859781); rgb(27pt)=(0.0688714,0.694771,0.839357); rgb(28pt)=(0.0296667,0.708167,0.816333); rgb(29pt)=(0.00357143,0.720267,0.7917); rgb(30pt)=(0.00665714,0.731214,0.766014); rgb(31pt)=(0.0433286,0.741095,0.73941); rgb(32pt)=(0.0963952,0.75,0.712038); rgb(33pt)=(0.140771,0.7584,0.684157); rgb(34pt)=(0.1717,0.766962,0.655443); rgb(35pt)=(0.193767,0.775767,0.6251); rgb(36pt)=(0.216086,0.7843,0.5923); rgb(37pt)=(0.246957,0.791795,0.556743); rgb(38pt)=(0.290614,0.79729,0.518829); rgb(39pt)=(0.340643,0.8008,0.478857); rgb(40pt)=(0.3909,0.802871,0.435448); rgb(41pt)=(0.445629,0.802419,0.390919); rgb(42pt)=(0.5044,0.7993,0.348); rgb(43pt)=(0.561562,0.794233,0.304481); rgb(44pt)=(0.617395,0.787619,0.261238); rgb(45pt)=(0.671986,0.779271,0.2227); rgb(46pt)=(0.7242,0.769843,0.191029); rgb(47pt)=(0.773833,0.759805,0.16461); rgb(48pt)=(0.820314,0.749814,0.153529); rgb(49pt)=(0.863433,0.7406,0.159633); rgb(50pt)=(0.903543,0.733029,0.177414); rgb(51pt)=(0.939257,0.728786,0.209957); rgb(52pt)=(0.972757,0.729771,0.239443); rgb(53pt)=(0.995648,0.743371,0.237148); rgb(54pt)=(0.996986,0.765857,0.219943); rgb(55pt)=(0.995205,0.789252,0.202762); rgb(56pt)=(0.9892,0.813567,0.188533); rgb(57pt)=(0.978629,0.838629,0.176557); rgb(58pt)=(0.967648,0.8639,0.16429); rgb(59pt)=(0.96101,0.889019,0.153676); rgb(60pt)=(0.959671,0.913457,0.142257); rgb(61pt)=(0.962795,0.937338,0.12651); rgb(62pt)=(0.969114,0.960629,0.106362); rgb(63pt)=(0.9769,0.9839,0.0805)}, mesh/rows=31]
table[row sep=crcr, point meta=\thisrow{c}] {%
x	y	z	c\\
-15	-15	0.391344706549839	0.391344706549839\\
-15	-14	0.414004668332403	0.414004668332403\\
-15	-13	0.436109944723482	0.436109944723482\\
-15	-12	0.457526492499945	0.457526492499945\\
-15	-11	0.478110057389808	0.478110057389808\\
-15	-10	0.49770572319475	0.49770572319475\\
-15	-9	0.516149476600201	0.516149476600201\\
-15	-8	0.533270249385109	0.533270249385109\\
-15	-7	0.548893313264158	0.548893313264158\\
-15	-6	0.562845106929043	0.562845106929043\\
-15	-5	0.574958010045338	0.574958010045338\\
-15	-4	0.58507702684811	0.58507702684811\\
-15	-3	0.593064082739095	0.593064082739095\\
-15	-2	0.598804548749086	0.598804548749086\\
-15	-1	0.602210954646605	0.602210954646605\\
-15	0	0.60322675509099	0.60322675509099\\
-15	1	0.601828398364365	0.601828398364365\\
-15	2	0.598026387259836	0.598026387259836\\
-15	3	0.591864324282118	0.591864324282118\\
-15	4	0.583416153338233	0.583416153338233\\
-15	5	0.57278340142821	0.57278340142821\\
-15	6	0.560090298559399	0.560090298559399\\
-15	7	0.545479852460157	0.545479852460157\\
-15	8	0.529109498938263	0.529109498938263\\
-15	9	0.511146911372552	0.511146911372552\\
-15	10	0.491766977181568	0.491766977181568\\
-15	11	0.471149145603085	0.471149145603085\\
-15	12	0.449475836361815	0.449475836361815\\
-15	13	0.426931140081374	0.426931140081374\\
-15	14	0.403700473495618	0.403700473495618\\
-15	15	0.379969147249584	0.379969147249584\\
-14	-15	0.41319133839957	0.41319133839957\\
-14	-14	0.437176004745989	0.437176004745989\\
-14	-13	0.46061516238379	0.46061516238379\\
-14	-12	0.483365167051698	0.483365167051698\\
-14	-11	0.50526953774131	0.50526953774131\\
-14	-10	0.526159911496463	0.526159911496463\\
-14	-9	0.545857051257025	0.545857051257025\\
-14	-8	0.564173445034958	0.564173445034958\\
-14	-7	0.580917284245047	0.580917284245047\\
-14	-6	0.59589713161294	0.59589713161294\\
-14	-5	0.608927862149037	0.608927862149037\\
-14	-4	0.619836657166767	0.619836657166767\\
-14	-3	0.628469422656158	0.628469422656158\\
-14	-2	0.634697791145912	0.634697791145912\\
-14	-1	0.638423312211757	0.638423312211757\\
-14	0	0.639583181272692	0.639583181272692\\
-14	1	0.638152361367375	0.638152361367375\\
-14	2	0.634144750131126	0.634144750131126\\
-14	3	0.627612702396249	0.627612702396249\\
-14	4	0.618642945772488	0.618642945772488\\
-14	5	0.607353185804807	0.607353185804807\\
-14	6	0.593885740644239	0.593885740644239\\
-14	7	0.578402820095429	0.578402820095429\\
-14	8	0.561080690731572	0.561080690731572\\
-14	9	0.542105061030776	0.542105061030776\\
-14	10	0.521666943912114	0.521666943912114\\
-14	11	0.499960110603959	0.499960110603959\\
-14	12	0.477178756689961	0.477178756689961\\
-14	13	0.453516653398492	0.453516653398492\\
-14	14	0.429166643680756	0.429166643680756\\
-14	15	0.404319395667163	0.404319395667163\\
-13	-15	0.434727156531082	0.434727156531082\\
-13	-14	0.460037827029394	0.460037827029394\\
-13	-13	0.484815825566217	0.484815825566217\\
-13	-12	0.508907853797725	0.508907853797725\\
-13	-11	0.532146079211853	0.532146079211853\\
-13	-10	0.554348983838854	0.554348983838854\\
-13	-9	0.575322425139486	0.575322425139486\\
-13	-8	0.594862394314318	0.594862394314318\\
-13	-7	0.612759153767681	0.612759153767681\\
-13	-6	0.62880190501571	0.62880190501571\\
-13	-5	0.642785154015504	0.642785154015504\\
-13	-4	0.654516031293656	0.654516031293656\\
-13	-3	0.663821081630679	0.663821081630679\\
-13	-2	0.670554645077733	0.670554645077733\\
-13	-1	0.67460543446342	0.67460543446342\\
-13	0	0.675902211145606	0.675902211145606\\
-13	1	0.674417869430973	0.674417869430973\\
-13	2	0.670170815729662	0.670170815729662\\
-13	3	0.663222846778889	0.663222846778889\\
-13	4	0.653676444378056	0.653676444378056\\
-13	5	0.641668038748729	0.641668038748729\\
-13	6	0.627361218850201	0.627361218850201\\
-13	7	0.61094047313916	0.61094047313916\\
-13	8	0.5926035511747	0.5926035511747\\
-13	9	0.572556159177362	0.572556159177362\\
-13	10	0.551007504298719	0.551007504298719\\
-13	11	0.528166157157431	0.528166157157431\\
-13	12	0.504238831817819	0.504238831817819\\
-13	13	0.47942895359081	0.47942895359081\\
-13	14	0.453936287723062	0.453936287723062\\
-13	15	0.427956674174926	0.427956674174926\\
-12	-15	0.455772207784713	0.455772207784713\\
-12	-14	0.482400289240557	0.482400289240557\\
-12	-13	0.508512885123632	0.508512885123632\\
-12	-12	0.533947573451655	0.533947573451655\\
-12	-11	0.558526337185913	0.558526337185913\\
-12	-10	0.582055192920396	0.582055192920396\\
-12	-9	0.604325623080853	0.604325623080853\\
-12	-8	0.62511712205646	0.62511712205646\\
-12	-7	0.644200909308488	0.644200909308488\\
-12	-6	0.661344968589222	0.661344968589222\\
-12	-5	0.676320890670804	0.676320890670804\\
-12	-4	0.688911670873442	0.688911670873442\\
-12	-3	0.698921097925921	0.698921097925921\\
-12	-2	0.706182877604056	0.706182877604056\\
-12	-1	0.710569278969462	0.710569278969462\\
-12	0	0.711999303208635	0.711999303208635\\
-12	1	0.710442821096907	0.710442821096907\\
-12	2	0.705922853908058	0.705922853908058\\
-12	3	0.698513398593517	0.698513398593517\\
-12	4	0.688334600741094	0.688334600741094\\
-12	5	0.675545328357626	0.675545328357626\\
-12	6	0.660334154319343	0.660334154319343\\
-12	7	0.642910763177504	0.642910763177504\\
-12	8	0.623497517097274	0.623497517097274\\
-12	9	0.602322719217693	0.602322719217693\\
-12	10	0.57961477876662	0.57961477876662\\
-12	11	0.555598604040881	0.555598604040881\\
-12	12	0.53049337454106	0.53049337454106\\
-12	13	0.504511798349767	0.504511798349767\\
-12	14	0.477860430398096	0.477860430398096\\
-12	15	0.450739035932707	0.450739035932707\\
-11	-15	0.4761410489044	0.4761410489044\\
-11	-14	0.504066490530936	0.504066490530936\\
-11	-13	0.531498300881471	0.531498300881471\\
-11	-12	0.558266737845192	0.558266737845192\\
-11	-11	0.584184554656111	0.584184554656111\\
-11	-10	0.609046628582195	0.609046628582195\\
-11	-9	0.63263086268033	0.63263086268033\\
-11	-8	0.654699936261837	0.654699936261837\\
-11	-7	0.675004911912332	0.675004911912332\\
-11	-6	0.693290327755042	0.693290327755042\\
-11	-5	0.709301517579342	0.709301517579342\\
-11	-4	0.722793946656844	0.722793946656844\\
-11	-3	0.733544138889785	0.733544138889785\\
-11	-2	0.741361028314682	0.741361028314682\\
-11	-1	0.746097416694816	0.746097416694816\\
-11	0	0.747659839780422	0.747659839780422\\
-11	1	0.746014455238157	0.746014455238157\\
-11	2	0.741189535738353	0.741189535738353\\
-11	3	0.733273294134222	0.733273294134222\\
-11	4	0.722406934644195	0.722406934644195\\
-11	5	0.708775423124647	0.708775423124647\\
-11	6	0.692596135566234	0.692596135566234\\
-11	7	0.674108037034334	0.674108037034334\\
-11	8	0.653561231920343	0.653561231920343\\
-11	9	0.631208529880552	0.631208529880552\\
-11	10	0.607299345729034	0.607299345729034\\
-11	11	0.582075296751645	0.582075296751645\\
-11	12	0.555768240062873	0.555768240062873\\
-11	13	0.528599901294963	0.528599901294963\\
-11	14	0.500782458086429	0.500782458086429\\
-11	15	0.472519176614969	0.472519176614969\\
-10	-15	0.495643886992492	0.495643886992492\\
-10	-14	0.524833058634017	0.524833058634017\\
-10	-13	0.553556288181366	0.553556288181366\\
-10	-12	0.581637998638668	0.581637998638668\\
-10	-11	0.608883357889359	0.608883357889359\\
-10	-10	0.635077536074442	0.635077536074442\\
-10	-9	0.659985440370445	0.659985440370445\\
-10	-8	0.683352828922019	0.683352828922019\\
-10	-7	0.704909599595335	0.704909599595335\\
-10	-6	0.724374298752905	0.724374298752905\\
-10	-5	0.741463138803216	0.741463138803216\\
-10	-4	0.755899758372105	0.755899758372105\\
-10	-3	0.76742880167162	0.76742880167162\\
-10	-2	0.775830028312984	0.775830028312984\\
-10	-1	0.78093305965247	0.78093305965247\\
-10	0	0.782629260739159	0.782629260739159\\
-10	1	0.780880015242842	0.780880015242842\\
-10	2	0.775719908118595	0.775719908118595\\
-10	3	0.767253436853389	0.767253436853389\\
-10	4	0.755646471316129	0.755646471316129\\
-10	5	0.741114530943138	0.741114530943138\\
-10	6	0.723908250569909	0.723908250569909\\
-10	7	0.704299429751379	0.704299429751379\\
-10	8	0.682568991680939	0.682568991680939\\
-10	9	0.658996586486152	0.658996586486152\\
-10	10	0.633853695455494	0.633853695455494\\
-10	11	0.607399228352364	0.607399228352364\\
-10	12	0.579877560771919	0.579877560771919\\
-10	13	0.55151848109667	0.55151848109667\\
-10	14	0.522538940961996	0.522538940961996\\
-10	15	0.49314432832198	0.49314432832198\\
-9	-15	0.514087481264714	0.514087481264714\\
-9	-14	0.544492112508466	0.544492112508466\\
-9	-13	0.574464803124613	0.574464803124613\\
-9	-12	0.603825732564515	0.603825732564515\\
-9	-11	0.632374446004186	0.632374446004186\\
-9	-10	0.65988773256791	0.65988773256791\\
-9	-9	0.686118299217998	0.686118299217998\\
-9	-8	0.71079503597684	0.71079503597684\\
-9	-7	0.733625509014155	0.733625509014155\\
-9	-6	0.754300840732675	0.754300840732675\\
-9	-5	0.772504356006919	0.772504356006919\\
-9	-4	0.78792425979709	0.78792425979709\\
-9	-3	0.800268701761414	0.800268701761414\\
-9	-2	0.809283599177769	0.809283599177769\\
-9	-1	0.814770831176194	0.814770831176194\\
-9	0	0.816603674662086	0.816603674662086\\
-9	1	0.814737731464586	0.814737731464586\\
-9	2	0.809214588400843	0.809214588400843\\
-9	3	0.80015794503411	0.80015794503411\\
-9	4	0.787762527392133	0.787762527392133\\
-9	5	0.772278758132769	0.772278758132769\\
-9	6	0.753995092755576	0.753995092755576\\
-9	7	0.733219878413726	0.733219878413726\\
-9	8	0.710267721500724	0.710267721500724\\
-9	9	0.685446863080921	0.685446863080921\\
-9	10	0.659051805716575	0.659051805716575\\
-9	11	0.631358592515391	0.631358592515391\\
-9	12	0.602623534064695	0.602623534064695\\
-9	13	0.573083791919936	0.573083791919936\\
-9	14	0.542959924736348	0.542959924736348\\
-9	15	0.5124577978677	0.5124577978677\\
-8	-15	0.531276999604501	0.531276999604501\\
-8	-14	0.562833012803647	0.562833012803647\\
-8	-13	0.593997133904459	0.593997133904459\\
-8	-12	0.624587526670729	0.624587526670729\\
-8	-11	0.65439975794775	0.65439975794775\\
-8	-10	0.683203296731276	0.683203296731276\\
-8	-9	0.710739710657598	0.710739710657598\\
-8	-8	0.736721499026529	0.736721499026529\\
-8	-7	0.760833312822828	0.760833312822828\\
-8	-6	0.782737153068343	0.782737153068343\\
-8	-5	0.802081547504876	0.802081547504876\\
-8	-4	0.818515554318324	0.818515554318324\\
-8	-3	0.831706478971499	0.831706478971499\\
-8	-2	0.84136146327884	0.84136146327884\\
-8	-1	0.847249764058466	0.847249764058466\\
-8	0	0.849223069141054	0.849223069141054\\
-8	1	0.847229660227217	0.847229660227217\\
-8	2	0.841319346017596	0.841319346017596\\
-8	3	0.831638210816307	0.831638210816307\\
-8	4	0.818414716895621	0.818414716895621\\
-8	5	0.801939070220614	0.801939070220614\\
-8	6	0.782541366152405	0.782541366152405\\
-8	7	0.760570159506662	0.760570159506662\\
-8	8	0.736375649475752	0.736375649475752\\
-8	9	0.710295994171056	0.710295994171056\\
-8	10	0.68264887656184	0.68264887656184\\
-8	11	0.653727154833661	0.653727154833661\\
-8	12	0.623798172811048	0.623798172811048\\
-8	13	0.593105245200966	0.593105245200966\\
-8	14	0.561870840257397	0.561870840257397\\
-8	15	0.530300452023241	0.530300452023241\\
-7	-15	0.547019095138756	0.547019095138756\\
-7	-14	0.579645809746253	0.579645809746253\\
-7	-13	0.611926303492104	0.611926303492104\\
-7	-12	0.64367810360719	0.64367810360719\\
-7	-11	0.674694336893957	0.674694336893957\\
-7	-10	0.70473890268515	0.70473890268515\\
-7	-9	0.733543237134821	0.733543237134821\\
-7	-8	0.760803873175484	0.760803873175484\\
-7	-7	0.786183077051036	0.786183077051036\\
-7	-6	0.8093135162248	0.8093135162248\\
-7	-5	0.82980754215121	0.82980754215121\\
-7	-4	0.847272414021377	0.847272414021377\\
-7	-3	0.861330879994025	0.861330879994025\\
-7	-2	0.871645948934783	0.871645948934783\\
-7	-1	0.877949056997703	0.877949056997703\\
-7	0	0.880066324608023	0.880066324608023\\
-7	1	0.877937175049947	0.877937175049947\\
-7	2	0.87162075284021	0.87162075284021\\
-7	3	0.861289717532153	0.861289717532153\\
-7	4	0.847210988595028	0.847210988595028\\
-7	5	0.829719594520046	0.829719594520046\\
-7	6	0.809191195816196	0.809191195816196\\
-7	7	0.78601694196008	0.78601694196008\\
-7	8	0.760583526697888	0.760583526697888\\
-7	9	0.733259184321261	0.733259184321261\\
-7	10	0.704384141673555	0.704384141673555\\
-7	11	0.674267329396448	0.674267329396448\\
-7	12	0.643185745397018	0.643185745397018\\
-7	13	0.61138731924587	0.61138731924587\\
-7	14	0.579095367907538	0.579095367907538\\
-7	15	0.54651315755972	0.54651315755972\\
-6	-15	0.561125619346527	0.561125619346527\\
-6	-14	0.594725221471019	0.594725221471019\\
-6	-13	0.628028252301279	0.628028252301279\\
-6	-12	0.66085282255669	0.66085282255669\\
-6	-11	0.692990998151508	0.692990998151508\\
-6	-10	0.72420227573248	0.72420227573248\\
-6	-9	0.754209498259295	0.754209498259295\\
-6	-8	0.782694558407412	0.782694558407412\\
-6	-7	0.809298769878923	0.809298769878923\\
-6	-6	0.833626209200362	0.833626209200362\\
-6	-5	0.855253475894074	0.855253475894074\\
-6	-4	0.873745234490239	0.873745234490239\\
-6	-3	0.888675803601588	0.888675803601588\\
-6	-2	0.899659497325713	0.899659497325713\\
-6	-1	0.906384891933685	0.906384891933685\\
-6	0	0.908648190803712	0.908648190803712\\
-6	1	0.906378102249253	0.906378102249253\\
-6	2	0.899644857068656	0.899644857068656\\
-6	3	0.888651615350798	0.888651615350798\\
-6	4	0.873708686892005	0.873708686892005\\
-6	5	0.855200643662084	0.855200643662084\\
-6	6	0.833551840938063	0.833551840938063\\
-6	7	0.809196977656846	0.809196977656846\\
-6	8	0.782558923851992	0.782558923851992\\
-6	9	0.75403455779634	0.75403455779634\\
-6	10	0.723985430185919	0.723985430185919\\
-6	11	0.692734528430955	0.692734528430955\\
-6	12	0.66056717841084	0.66056717841084\\
-6	13	0.627734280183118	0.627734280183118\\
-6	14	0.594458567223817	0.594458567223817\\
-6	15	0.560939751735189	0.560939751735189\\
-5	-15	0.573418343011463	0.573418343011463\\
-5	-14	0.607876203683743	0.607876203683743\\
-5	-13	0.642088362650625	0.642088362650625\\
-5	-12	0.675874681097009	0.675874681097009\\
-5	-11	0.709026906670695	0.709026906670695\\
-5	-10	0.741301830753536	0.741301830753536\\
-5	-9	0.772414817043466	0.772414817043466\\
-5	-8	0.802035611046138	0.802035611046138\\
-5	-7	0.829786748742635	0.829786748742635\\
-5	-6	0.855246474032003	0.855246474032003\\
-5	-5	0.877957756280882	0.877957756280882\\
-5	-4	0.897442241003244	0.897442241003244\\
-5	-3	0.913224650288816	0.913224650288816\\
-5	-2	0.924867367758165	0.924867367758165\\
-5	-1	0.932011388846925	0.932011388846925\\
-5	0	0.934419286777599	0.934419286777599\\
-5	1	0.932007569649432	0.932007569649432\\
-5	2	0.924859198919082	0.924859198919082\\
-5	3	0.913211070919951	0.913211070919951\\
-5	4	0.897421235417031	0.897421235417031\\
-5	5	0.877926937478888	0.877926937478888\\
-5	6	0.855202871527289	0.855202871527289\\
-5	7	0.829726596382117	0.829726596382117\\
-5	8	0.801955513987599	0.801955513987599\\
-5	9	0.772312282199654	0.772312282199654\\
-5	10	0.741176805080042	0.741176805080042\\
-5	11	0.708884376342024	0.708884376342024\\
-5	12	0.675726262838869	0.675726262838869\\
-5	13	0.641955168138048	0.641955168138048\\
-5	14	0.607791756983615	0.607791756983615\\
-5	15	0.57343155106946	0.57343155106946\\
-4	-15	0.583734472840414	0.583734472840414\\
-4	-14	0.618919041924604	0.618919041924604\\
-4	-13	0.653907081471121	0.653907081471121\\
-4	-12	0.68852078661889	0.68852078661889\\
-4	-11	0.722552117193348	0.722552117193348\\
-4	-10	0.755756114110831	0.755756114110831\\
-4	-9	0.787842783583901	0.787842783583901\\
-4	-8	0.81847184572479	0.81847184572479\\
-4	-7	0.847250718857839	0.847250718857839\\
-4	-6	0.873735739540915	0.873735739540915\\
-4	-5	0.897439058338666	0.897439058338666\\
-4	-4	0.91784259050225	0.91784259050225\\
-4	-3	0.934422469442177	0.934422469442177\\
-4	-2	0.946687716102925	0.946687716102925\\
-4	-1	0.954229676352717	0.954229676352717\\
-4	0	0.956775171481967	0.956775171481967\\
-4	1	0.954227660665151	0.954227660665151\\
-4	2	0.946683366461335	0.946683366461335\\
-4	3	0.934415043224828	0.934415043224828\\
-4	4	0.917831026820951	0.917831026820951\\
-4	5	0.897421871949946	0.897421871949946\\
-4	6	0.873711073890438	0.873711073890438\\
-4	7	0.847216717391267	0.847216717391267\\
-4	8	0.818426864065425	0.818426864065425\\
-4	9	0.787786185198827	0.787786185198827\\
-4	10	0.755689596421157	0.755689596421157\\
-4	11	0.722481355950904	0.722481355950904\\
-4	12	0.688457239416155	0.688457239416155\\
-4	13	0.653870268650839	0.653870268650839\\
-4	14	0.618938084867661	0.618938084867661\\
-4	15	0.583851488808054	0.583851488808054\\
-3	-15	0.591930948060612	0.591930948060612\\
-3	-14	0.627695080364402	0.627695080364402\\
-3	-13	0.663307187390198	0.663307187390198\\
-3	-12	0.698590949520394	0.698590949520394\\
-3	-11	0.733339493557438	0.733339493557438\\
-3	-10	0.767306693440656	0.767306693440656\\
-3	-9	0.800198630096295	0.800198630096295\\
-3	-8	0.831668127863337	0.831668127863337\\
-3	-7	0.861310935296005	0.861310935296005\\
-3	-6	0.888666202563445	0.888666202563445\\
-3	-5	0.913220725002504	0.913220725002504\\
-3	-4	0.934421302465165	0.934421302465165\\
-3	-3	0.951699882368717	0.951699882368717\\
-3	-2	0.964515836090228	0.964515836090228\\
-3	-1	0.972412451262796	0.972412451262796\\
-3	0	0.975080903333565	0.975080903333565\\
-3	1	0.972411496463423	0.972411496463423\\
-3	2	0.964513714313843	0.964513714313843\\
-3	3	0.951696169260043	0.951696169260043\\
-3	4	0.934415043224828	0.934415043224828\\
-3	5	0.913211495275228	0.913211495275228\\
-3	6	0.888652994505448	0.888652994505448\\
-3	7	0.861292794107912	0.861292794107912\\
-3	8	0.831644417012233	0.831644417012233\\
-3	9	0.800169773937457	0.800169773937457\\
-3	10	0.767274760706059	0.767274760706059\\
-3	11	0.733310213043323	0.733310213043323\\
-3	12	0.698574081398132	0.698574081398132\\
-3	13	0.663318591938268	0.663318591938268\\
-3	14	0.627758415389499	0.627758415389499\\
-3	15	0.592079154141113	0.592079154141113\\
-2	-15	0.597889161372127	0.597889161372127\\
-2	-14	0.634072609734031	0.634072609734031\\
-2	-13	0.670139943883258	0.670139943883258\\
-2	-12	0.705916488578902	0.705916488578902\\
-2	-11	0.741196007156327	0.741196007156327\\
-2	-10	0.775731418755485	0.775731418755485\\
-2	-9	0.809226841659467	0.809226841659467\\
-2	-8	0.84133006098834	0.84133006098834\\
-2	-7	0.87162923994575	0.87162923994575\\
-2	-6	0.899651116308992	0.899651116308992\\
-2	-5	0.924863654649491	0.924863654649491\\
-2	-4	0.946686336948275	0.946686336948275\\
-2	-3	0.964515411734951	0.964515411734951\\
-2	-2	0.977768981835897	0.977768981835897\\
-2	-1	0.985949278511158	0.985949278511158\\
-2	0	0.988716393183836	0.988716393183836\\
-2	1	0.9859489602447	0.9859489602447\\
-2	2	0.977768027036523	0.977768027036523\\
-2	3	0.964513714313843	0.964513714313843\\
-2	4	0.946683366461335	0.946683366461335\\
-2	5	0.924859198919082	0.924859198919082\\
-2	6	0.899644644891017	0.899644644891017\\
-2	7	0.871620275440523	0.871620275440523\\
-2	8	0.841318603395861	0.841318603395861\\
-2	9	0.80921352751265	0.80921352751265\\
-2	10	0.775718263741897	0.775718263741897\\
-2	11	0.741187201784329	0.741187201784329\\
-2	12	0.705919406021432	0.705919406021432\\
-2	13	0.670165670421928	0.670165670421928\\
-2	14	0.634137536091416	0.634137536091416\\
-2	15	0.598016308822007	0.598016308822007\\
-1	-15	0.601518937278593	0.601518937278593\\
-1	-14	0.637951004788425	0.637951004788425\\
-1	-13	0.674292366357793	0.674292366357793\\
-1	-12	0.710368452834608	0.710368452834608\\
-1	-11	0.745973080598647	0.745973080598647\\
-1	-10	0.78085879747899	0.78085879747899\\
-1	-9	0.814728024337624	0.814728024337624\\
-1	-8	0.847226000162952	0.847226000162952\\
-1	-7	0.877936326339393	0.877936326339393\\
-1	-6	0.906378314426891	0.906378314426891\\
-1	-5	0.932008312271166	0.932008312271166\\
-1	-4	0.954228297198067	0.954228297198067\\
-1	-3	0.972412026907519	0.972412026907519\\
-1	-2	0.985949172422339	0.985949172422339\\
-1	-1	0.994314275820746	0.994314275820746\\
-1	0	0.997145998584351	0.997145998584351\\
-1	1	0.994314275820746	0.994314275820746\\
-1	2	0.985948854155881	0.985948854155881\\
-1	3	0.972411390374604	0.972411390374604\\
-1	4	0.954227130221055	0.954227130221055\\
-1	5	0.932006508761239	0.932006508761239\\
-1	6	0.906375768295229	0.906375768295229\\
-1	7	0.877932719319538	0.877932719319538\\
-1	8	0.847221438343724	0.847221438343724\\
-1	9	0.814722985118709	0.814722985118709\\
-1	10	0.78085455392622	0.78085455392622\\
-1	11	0.745971807532815	0.745971807532815\\
-1	12	0.7103739164088	0.7103739164088\\
-1	13	0.674310507545886	0.674310507545886\\
-1	14	0.63799025765155	0.63799025765155\\
-1	15	0.60159075940923	0.60159075940923\\
0	-15	0.602761237352091	0.602761237352091\\
0	-14	0.639265710481066	0.639265710481066\\
0	-13	0.67569189006143	0.67569189006143\\
0	-12	0.71186446431936	0.71186446431936\\
0	-11	0.747576400924076	0.747576400924076\\
0	-10	0.782579452038518	0.782579452038518\\
0	-9	0.816574977636477	0.816574977636477\\
0	-8	0.849207102773756	0.849207102773756\\
0	-7	0.880057731413663	0.880057731413663\\
0	-6	0.908643735073303	0.908643735073303\\
0	-5	0.934417165001214	0.934417165001214\\
0	-4	0.956774322771413	0.956774322771413\\
0	-3	0.975080585067108	0.975080585067108\\
0	-2	0.988716393183836	0.988716393183836\\
0	-1	0.997145892495532	0.997145892495532\\
0	0	1	1\\
0	1	0.997145786406712	0.997145786406712\\
0	2	0.988716393183836	0.988716393183836\\
0	3	0.975080585067108	0.975080585067108\\
0	4	0.956774322771413	0.956774322771413\\
0	5	0.934417165001214	0.934417165001214\\
0	6	0.908643735073303	0.908643735073303\\
0	7	0.880057784458073	0.880057784458073\\
0	8	0.849207102773756	0.849207102773756\\
0	9	0.816574977636477	0.816574977636477\\
0	10	0.782579452038518	0.782579452038518\\
0	11	0.747576453968486	0.747576453968486\\
0	12	0.71186446431936	0.71186446431936\\
0	13	0.67569189006143	0.67569189006143\\
0	14	0.639265710481066	0.639265710481066\\
0	15	0.602761237352091	0.602761237352091\\
1	-15	0.601590812453639	0.601590812453639\\
1	-14	0.63799031069596	0.63799031069596\\
1	-13	0.674310613634705	0.674310613634705\\
1	-12	0.71037396945321	0.71037396945321\\
1	-11	0.745971860577225	0.745971860577225\\
1	-10	0.780854606970629	0.780854606970629\\
1	-9	0.814723091207528	0.814723091207528\\
1	-8	0.847221491388134	0.847221491388134\\
1	-7	0.877932719319538	0.877932719319538\\
1	-6	0.906375768295229	0.906375768295229\\
1	-5	0.93200640267242	0.93200640267242\\
1	-4	0.954227130221055	0.954227130221055\\
1	-3	0.972411390374604	0.972411390374604\\
1	-2	0.9859489602447	0.9859489602447\\
1	-1	0.994314275820746	0.994314275820746\\
1	0	0.997145998584351	0.997145998584351\\
1	1	0.994314381909565	0.994314381909565\\
1	2	0.985949172422339	0.985949172422339\\
1	3	0.972412026907519	0.972412026907519\\
1	4	0.954228403286886	0.954228403286886\\
1	5	0.932008312271166	0.932008312271166\\
1	6	0.90637842051571	0.90637842051571\\
1	7	0.877936326339393	0.877936326339393\\
1	8	0.847225947118543	0.847225947118543\\
1	9	0.814727971293214	0.814727971293214\\
1	10	0.78085879747899	0.78085879747899\\
1	11	0.745973080598647	0.745973080598647\\
1	12	0.710368399790199	0.710368399790199\\
1	13	0.674292313313383	0.674292313313383\\
1	14	0.637951004788425	0.637951004788425\\
1	15	0.601518937278593	0.601518937278593\\
2	-15	0.598016361866416	0.598016361866416\\
2	-14	0.634137642180236	0.634137642180236\\
2	-13	0.670165723466338	0.670165723466338\\
2	-12	0.705919459065841	0.705919459065841\\
2	-11	0.741187201784329	0.741187201784329\\
2	-10	0.775718369830716	0.775718369830716\\
2	-9	0.80921358055706	0.80921358055706\\
2	-8	0.84131870948468	0.84131870948468\\
2	-7	0.871620328484933	0.871620328484933\\
2	-6	0.899644644891017	0.899644644891017\\
2	-5	0.924859198919082	0.924859198919082\\
2	-4	0.946683366461335	0.946683366461335\\
2	-3	0.964513714313843	0.964513714313843\\
2	-2	0.977768027036523	0.977768027036523\\
2	-1	0.9859489602447	0.9859489602447\\
2	0	0.988716393183836	0.988716393183836\\
2	1	0.985949384599977	0.985949384599977\\
2	2	0.977769087924716	0.977769087924716\\
2	3	0.964515411734951	0.964515411734951\\
2	4	0.946686336948275	0.946686336948275\\
2	5	0.924863654649491	0.924863654649491\\
2	6	0.899651222397811	0.899651222397811\\
2	7	0.87162923994575	0.87162923994575\\
2	8	0.84133011403275	0.84133011403275\\
2	9	0.809226841659467	0.809226841659467\\
2	10	0.775731418755485	0.775731418755485\\
2	11	0.741196007156327	0.741196007156327\\
2	12	0.705916541623312	0.705916541623312\\
2	13	0.670139890838849	0.670139890838849\\
2	14	0.634072609734031	0.634072609734031\\
2	15	0.597889161372127	0.597889161372127\\
3	-15	0.592079260229933	0.592079260229933\\
3	-14	0.627758468433908	0.627758468433908\\
3	-13	0.663318751071497	0.663318751071497\\
3	-12	0.698574081398132	0.698574081398132\\
3	-11	0.733310319132143	0.733310319132143\\
3	-10	0.767274866794879	0.767274866794879\\
3	-9	0.800169880026276	0.800169880026276\\
3	-8	0.831644470056643	0.831644470056643\\
3	-7	0.861292794107912	0.861292794107912\\
3	-6	0.888653100594267	0.888653100594267\\
3	-5	0.913211495275228	0.913211495275228\\
3	-4	0.934415255402467	0.934415255402467\\
3	-3	0.951696275348862	0.951696275348862\\
3	-2	0.964513714313843	0.964513714313843\\
3	-1	0.972411602552242	0.972411602552242\\
3	0	0.975081009422385	0.975081009422385\\
3	1	0.972412557351616	0.972412557351616\\
3	2	0.964515942179047	0.964515942179047\\
3	3	0.951700094546356	0.951700094546356\\
3	4	0.934421302465165	0.934421302465165\\
3	5	0.913220831091323	0.913220831091323\\
3	6	0.888666255607855	0.888666255607855\\
3	7	0.861310988340414	0.861310988340414\\
3	8	0.831668180907747	0.831668180907747\\
3	9	0.800198683140704	0.800198683140704\\
3	10	0.767306693440656	0.767306693440656\\
3	11	0.733339493557438	0.733339493557438\\
3	12	0.698590949520394	0.698590949520394\\
3	13	0.663307134345788	0.663307134345788\\
3	14	0.627695133408812	0.627695133408812\\
3	15	0.591930948060612	0.591930948060612\\
4	-15	0.583851541852464	0.583851541852464\\
4	-14	0.61893813791207	0.61893813791207\\
4	-13	0.653870321695248	0.653870321695248\\
4	-12	0.688457345504974	0.688457345504974\\
4	-11	0.722481408995313	0.722481408995313\\
4	-10	0.755689649465567	0.755689649465567\\
4	-9	0.787786185198827	0.787786185198827\\
4	-8	0.818426970154245	0.818426970154245\\
4	-7	0.847216823480087	0.847216823480087\\
4	-6	0.873711126934848	0.873711126934848\\
4	-5	0.897421871949946	0.897421871949946\\
4	-4	0.917831026820951	0.917831026820951\\
4	-3	0.934415043224828	0.934415043224828\\
4	-2	0.946683366461335	0.946683366461335\\
4	-1	0.954227766753971	0.954227766753971\\
4	0	0.956775171481967	0.956775171481967\\
4	1	0.954229676352717	0.954229676352717\\
4	2	0.946687822191744	0.946687822191744\\
4	3	0.934422575530996	0.934422575530996\\
4	4	0.917842696591069	0.917842696591069\\
4	5	0.897439058338666	0.897439058338666\\
4	6	0.873735739540915	0.873735739540915\\
4	7	0.84725066581343	0.84725066581343\\
4	8	0.81847184572479	0.81847184572479\\
4	9	0.787842836628311	0.787842836628311\\
4	10	0.755756114110831	0.755756114110831\\
4	11	0.722552064148938	0.722552064148938\\
4	12	0.688520680530071	0.688520680530071\\
4	13	0.653907081471121	0.653907081471121\\
4	14	0.618918988880194	0.618918988880194\\
4	15	0.583734419796004	0.583734419796004\\
5	-15	0.57343160411387	0.57343160411387\\
5	-14	0.607791863072434	0.607791863072434\\
5	-13	0.641955274226867	0.641955274226867\\
5	-12	0.675726368927688	0.675726368927688\\
5	-11	0.708884429386433	0.708884429386433\\
5	-10	0.741176911168861	0.741176911168861\\
5	-9	0.772312335244064	0.772312335244064\\
5	-8	0.801955620076418	0.801955620076418\\
5	-7	0.829726702470936	0.829726702470936\\
5	-6	0.855202977616108	0.855202977616108\\
5	-5	0.877926990523298	0.877926990523298\\
5	-4	0.897421235417031	0.897421235417031\\
5	-3	0.913211070919951	0.913211070919951\\
5	-2	0.924859411096721	0.924859411096721\\
5	-1	0.932007675738251	0.932007675738251\\
5	0	0.934419286777599	0.934419286777599\\
5	1	0.932011494935744	0.932011494935744\\
5	2	0.924867473846984	0.924867473846984\\
5	3	0.913224862466455	0.913224862466455\\
5	4	0.897442241003244	0.897442241003244\\
5	5	0.877957756280882	0.877957756280882\\
5	6	0.855246580120822	0.855246580120822\\
5	7	0.829786748742635	0.829786748742635\\
5	8	0.802035664090547	0.802035664090547\\
5	9	0.772414870087875	0.772414870087875\\
5	10	0.741301883797946	0.741301883797946\\
5	11	0.709026906670695	0.709026906670695\\
5	12	0.675874681097009	0.675874681097009\\
5	13	0.642088362650625	0.642088362650625\\
5	14	0.607876203683743	0.607876203683743\\
5	15	0.573418343011463	0.573418343011463\\
6	-15	0.560939857824008	0.560939857824008\\
6	-14	0.594458620268226	0.594458620268226\\
6	-13	0.627734386271937	0.627734386271937\\
6	-12	0.66056723145525	0.66056723145525\\
6	-11	0.692734634519774	0.692734634519774\\
6	-10	0.723985536274738	0.723985536274738\\
6	-9	0.754034663885159	0.754034663885159\\
6	-8	0.782558976896401	0.782558976896401\\
6	-7	0.809197030701256	0.809197030701256\\
6	-6	0.833551947026882	0.833551947026882\\
6	-5	0.855200643662084	0.855200643662084\\
6	-4	0.873708792980824	0.873708792980824\\
6	-3	0.888651668395207	0.888651668395207\\
6	-2	0.899644857068656	0.899644857068656\\
6	-1	0.906378102249253	0.906378102249253\\
6	0	0.908648190803712	0.908648190803712\\
6	1	0.906384891933685	0.906384891933685\\
6	2	0.899659497325713	0.899659497325713\\
6	3	0.888675803601588	0.888675803601588\\
6	4	0.873745340579058	0.873745340579058\\
6	5	0.855253581982893	0.855253581982893\\
6	6	0.833626209200362	0.833626209200362\\
6	7	0.809298769878923	0.809298769878923\\
6	8	0.782694558407412	0.782694558407412\\
6	9	0.754209498259295	0.754209498259295\\
6	10	0.72420227573248	0.72420227573248\\
6	11	0.692990945107099	0.692990945107099\\
6	12	0.660852928645509	0.660852928645509\\
6	13	0.628028252301279	0.628028252301279\\
6	14	0.594725221471019	0.594725221471019\\
6	15	0.561125566302117	0.561125566302117\\
7	-15	0.546513210604129	0.546513210604129\\
7	-14	0.579095473996357	0.579095473996357\\
7	-13	0.611387425334689	0.611387425334689\\
7	-12	0.643185798441428	0.643185798441428\\
7	-11	0.674267382440858	0.674267382440858\\
7	-10	0.704384247762375	0.704384247762375\\
7	-9	0.73325923736567	0.73325923736567\\
7	-8	0.760583632786707	0.760583632786707\\
7	-7	0.786017048048899	0.786017048048899\\
7	-6	0.809191248860606	0.809191248860606\\
7	-5	0.829719594520046	0.829719594520046\\
7	-4	0.847211041639437	0.847211041639437\\
7	-3	0.861289823620973	0.861289823620973\\
7	-2	0.87162080588462	0.87162080588462\\
7	-1	0.877937175049947	0.877937175049947\\
7	0	0.880066430696842	0.880066430696842\\
7	1	0.877949110042113	0.877949110042113\\
7	2	0.871646001979193	0.871646001979193\\
7	3	0.861330986082844	0.861330986082844\\
7	4	0.847272520110197	0.847272520110197\\
7	5	0.82980754215121	0.82980754215121\\
7	6	0.8093135162248	0.8093135162248\\
7	7	0.786183130095446	0.786183130095446\\
7	8	0.760803873175484	0.760803873175484\\
7	9	0.733543237134821	0.733543237134821\\
7	10	0.70473890268515	0.70473890268515\\
7	11	0.674694336893957	0.674694336893957\\
7	12	0.643678156651599	0.643678156651599\\
7	13	0.611926250447695	0.611926250447695\\
7	14	0.579645756701843	0.579645756701843\\
7	15	0.547019095138756	0.547019095138756\\
8	-15	0.530300452023241	0.530300452023241\\
8	-14	0.561870893301807	0.561870893301807\\
8	-13	0.593105245200966	0.593105245200966\\
8	-12	0.623798119766639	0.623798119766639\\
8	-11	0.653727154833661	0.653727154833661\\
8	-10	0.68264887656184	0.68264887656184\\
8	-9	0.710295941126646	0.710295941126646\\
8	-8	0.736375649475752	0.736375649475752\\
8	-7	0.760570159506662	0.760570159506662\\
8	-6	0.782541260063585	0.782541260063585\\
8	-5	0.801939070220614	0.801939070220614\\
8	-4	0.818414663851211	0.818414663851211\\
8	-3	0.831638210816307	0.831638210816307\\
8	-2	0.841319346017596	0.841319346017596\\
8	-1	0.847229660227217	0.847229660227217\\
8	0	0.849223069141054	0.849223069141054\\
8	1	0.847249764058466	0.847249764058466\\
8	2	0.84136146327884	0.84136146327884\\
8	3	0.831706532015908	0.831706532015908\\
8	4	0.818515554318324	0.818515554318324\\
8	5	0.802081600549285	0.802081600549285\\
8	6	0.782737153068343	0.782737153068343\\
8	7	0.760833312822828	0.760833312822828\\
8	8	0.736721499026529	0.736721499026529\\
8	9	0.710739763702008	0.710739763702008\\
8	10	0.683203243686867	0.683203243686867\\
8	11	0.65439970490334	0.65439970490334\\
8	12	0.624587526670729	0.624587526670729\\
8	13	0.593997133904459	0.593997133904459\\
8	14	0.562833012803647	0.562833012803647\\
8	15	0.531276999604501	0.531276999604501\\
9	-15	0.5124577978677	0.5124577978677\\
9	-14	0.542959924736348	0.542959924736348\\
9	-13	0.573083844964346	0.573083844964346\\
9	-12	0.602623534064695	0.602623534064695\\
9	-11	0.631358645559801	0.631358645559801\\
9	-10	0.659051858760984	0.659051858760984\\
9	-9	0.685446863080921	0.685446863080921\\
9	-8	0.710267721500724	0.710267721500724\\
9	-7	0.733219931458136	0.733219931458136\\
9	-6	0.753995092755576	0.753995092755576\\
9	-5	0.772278811177179	0.772278811177179\\
9	-4	0.787762527392133	0.787762527392133\\
9	-3	0.80015794503411	0.80015794503411\\
9	-2	0.809214588400843	0.809214588400843\\
9	-1	0.814737731464586	0.814737731464586\\
9	0	0.816603727706495	0.816603727706495\\
9	1	0.814770831176194	0.814770831176194\\
9	2	0.809283599177769	0.809283599177769\\
9	3	0.800268595672595	0.800268595672595\\
9	4	0.78792425979709	0.78792425979709\\
9	5	0.772504409051328	0.772504409051328\\
9	6	0.754300840732675	0.754300840732675\\
9	7	0.733625509014155	0.733625509014155\\
9	8	0.710794982932431	0.710794982932431\\
9	9	0.686118299217998	0.686118299217998\\
9	10	0.65988773256791	0.65988773256791\\
9	11	0.632374446004186	0.632374446004186\\
9	12	0.603825732564515	0.603825732564515\\
9	13	0.574464750080203	0.574464750080203\\
9	14	0.544492059464056	0.544492059464056\\
9	15	0.514087428220305	0.514087428220305\\
10	-15	0.49314432832198	0.49314432832198\\
10	-14	0.522538994006406	0.522538994006406\\
10	-13	0.551518534141079	0.551518534141079\\
10	-12	0.579877613816329	0.579877613816329\\
10	-11	0.607399334441184	0.607399334441184\\
10	-10	0.633853801544314	0.633853801544314\\
10	-9	0.658996586486152	0.658996586486152\\
10	-8	0.682568991680939	0.682568991680939\\
10	-7	0.704299482795788	0.704299482795788\\
10	-6	0.723908303614319	0.723908303614319\\
10	-5	0.741114583987548	0.741114583987548\\
10	-4	0.755646471316129	0.755646471316129\\
10	-3	0.767253436853389	0.767253436853389\\
10	-2	0.775719961163005	0.775719961163005\\
10	-1	0.780880015242842	0.780880015242842\\
10	0	0.782629313783569	0.782629313783569\\
10	1	0.780933165741289	0.780933165741289\\
10	2	0.775830081357394	0.775830081357394\\
10	3	0.76742880167162	0.76742880167162\\
10	4	0.755899811416515	0.755899811416515\\
10	5	0.741463138803216	0.741463138803216\\
10	6	0.724374351797314	0.724374351797314\\
10	7	0.704909493506516	0.704909493506516\\
10	8	0.683352775877609	0.683352775877609\\
10	9	0.659985387326036	0.659985387326036\\
10	10	0.635077536074442	0.635077536074442\\
10	11	0.608883357889359	0.608883357889359\\
10	12	0.581637998638668	0.581637998638668\\
10	13	0.553556288181366	0.553556288181366\\
10	14	0.524833058634017	0.524833058634017\\
10	15	0.495643833948083	0.495643833948083\\
11	-15	0.472519176614969	0.472519176614969\\
11	-14	0.500782564175249	0.500782564175249\\
11	-13	0.528599954339372	0.528599954339372\\
11	-12	0.555768346151692	0.555768346151692\\
11	-11	0.582075402840464	0.582075402840464\\
11	-10	0.607299398773444	0.607299398773444\\
11	-9	0.631208635969372	0.631208635969372\\
11	-8	0.653561284964753	0.653561284964753\\
11	-7	0.674108090078743	0.674108090078743\\
11	-6	0.692596135566234	0.692596135566234\\
11	-5	0.708775423124647	0.708775423124647\\
11	-4	0.722407040733014	0.722407040733014\\
11	-3	0.733273347178631	0.733273347178631\\
11	-2	0.741189641827172	0.741189641827172\\
11	-1	0.746014455238157	0.746014455238157\\
11	0	0.747659892824831	0.747659892824831\\
11	1	0.746097522783635	0.746097522783635\\
11	2	0.741361081359091	0.741361081359091\\
11	3	0.733544244978604	0.733544244978604\\
11	4	0.722794052745664	0.722794052745664\\
11	5	0.709301570623751	0.709301570623751\\
11	6	0.693290327755042	0.693290327755042\\
11	7	0.675004964956741	0.675004964956741\\
11	8	0.654699936261837	0.654699936261837\\
11	9	0.63263086268033	0.63263086268033\\
11	10	0.609046628582195	0.609046628582195\\
11	11	0.584184607700521	0.584184607700521\\
11	12	0.558266790889601	0.558266790889601\\
11	13	0.531498300881471	0.531498300881471\\
11	14	0.504066490530936	0.504066490530936\\
11	15	0.4761410489044	0.4761410489044\\
12	-15	0.450739088977117	0.450739088977117\\
12	-14	0.477860430398096	0.477860430398096\\
12	-13	0.504511904438586	0.504511904438586\\
12	-12	0.530493427585469	0.530493427585469\\
12	-11	0.5555987101297	0.5555987101297\\
12	-10	0.579614831811029	0.579614831811029\\
12	-9	0.602322772262102	0.602322772262102\\
12	-8	0.623497517097274	0.623497517097274\\
12	-7	0.642910763177504	0.642910763177504\\
12	-6	0.660334101274933	0.660334101274933\\
12	-5	0.675545275313217	0.675545275313217\\
12	-4	0.688334653785504	0.688334653785504\\
12	-3	0.698513398593517	0.698513398593517\\
12	-2	0.705922853908058	0.705922853908058\\
12	-1	0.710442874141317	0.710442874141317\\
12	0	0.711999303208635	0.711999303208635\\
12	1	0.710569332013871	0.710569332013871\\
12	2	0.706182877604056	0.706182877604056\\
12	3	0.698921150970331	0.698921150970331\\
12	4	0.688911670873442	0.688911670873442\\
12	5	0.676320890670804	0.676320890670804\\
12	6	0.661344968589222	0.661344968589222\\
12	7	0.644200962352897	0.644200962352897\\
12	8	0.625117175100869	0.625117175100869\\
12	9	0.604325623080853	0.604325623080853\\
12	10	0.582055139875986	0.582055139875986\\
12	11	0.558526337185913	0.558526337185913\\
12	12	0.533947573451655	0.533947573451655\\
12	13	0.508512885123632	0.508512885123632\\
12	14	0.482400342284966	0.482400342284966\\
12	15	0.455772207784713	0.455772207784713\\
13	-15	0.427956700697131	0.427956700697131\\
13	-14	0.453936340767472	0.453936340767472\\
13	-13	0.479429006635219	0.479429006635219\\
13	-12	0.504238884862229	0.504238884862229\\
13	-11	0.52816626324625	0.52816626324625\\
13	-10	0.551007557343129	0.551007557343129\\
13	-9	0.572556265266181	0.572556265266181\\
13	-8	0.5926035511747	0.5926035511747\\
13	-7	0.610940579227979	0.610940579227979\\
13	-6	0.62736132493902	0.62736132493902\\
13	-5	0.641668038748729	0.641668038748729\\
13	-4	0.653676497422466	0.653676497422466\\
13	-3	0.663222899823299	0.663222899823299\\
13	-2	0.670170815729662	0.670170815729662\\
13	-1	0.674417975519793	0.674417975519793\\
13	0	0.675902264190016	0.675902264190016\\
13	1	0.67460543446342	0.67460543446342\\
13	2	0.670554698122143	0.670554698122143\\
13	3	0.663821081630679	0.663821081630679\\
13	4	0.654516031293656	0.654516031293656\\
13	5	0.642785207059914	0.642785207059914\\
13	6	0.62880190501571	0.62880190501571\\
13	7	0.612759153767681	0.612759153767681\\
13	8	0.594862447358728	0.594862447358728\\
13	9	0.575322425139486	0.575322425139486\\
13	10	0.554348983838854	0.554348983838854\\
13	11	0.532146079211853	0.532146079211853\\
13	12	0.508907853797725	0.508907853797725\\
13	13	0.484815825566217	0.484815825566217\\
13	14	0.460037773984985	0.460037773984985\\
13	15	0.434727156531082	0.434727156531082\\
14	-15	0.404319448711572	0.404319448711572\\
14	-14	0.429166696725165	0.429166696725165\\
14	-13	0.453516706442901	0.453516706442901\\
14	-12	0.477178809734371	0.477178809734371\\
14	-11	0.499960216692779	0.499960216692779\\
14	-10	0.521667050000934	0.521667050000934\\
14	-9	0.542105061030776	0.542105061030776\\
14	-8	0.561080743775981	0.561080743775981\\
14	-7	0.578402926184248	0.578402926184248\\
14	-6	0.593885793688649	0.593885793688649\\
14	-5	0.607353185804807	0.607353185804807\\
14	-4	0.618642998816897	0.618642998816897\\
14	-3	0.627612702396249	0.627612702396249\\
14	-2	0.634144856219945	0.634144856219945\\
14	-1	0.638152308322965	0.638152308322965\\
14	0	0.639583234317102	0.639583234317102\\
14	1	0.638423365256166	0.638423365256166\\
14	2	0.634697791145912	0.634697791145912\\
14	3	0.628469422656158	0.628469422656158\\
14	4	0.619836604122357	0.619836604122357\\
14	5	0.608927862149037	0.608927862149037\\
14	6	0.59589718465735	0.59589718465735\\
14	7	0.580917284245047	0.580917284245047\\
14	8	0.564173498079368	0.564173498079368\\
14	9	0.545857051257025	0.545857051257025\\
14	10	0.526159911496463	0.526159911496463\\
14	11	0.50526953774131	0.50526953774131\\
14	12	0.483365167051698	0.483365167051698\\
14	13	0.46061516238379	0.46061516238379\\
14	14	0.43717595170158	0.43717595170158\\
14	15	0.41319133839957	0.41319133839957\\
15	-15	0.379969226816198	0.379969226816198\\
15	-14	0.403700553062233	0.403700553062233\\
15	-13	0.426931193125784	0.426931193125784\\
15	-12	0.449475942450635	0.449475942450635\\
15	-11	0.471149198647495	0.471149198647495\\
15	-10	0.491767030225978	0.491767030225978\\
15	-9	0.511146964416961	0.511146964416961\\
15	-8	0.529109498938263	0.529109498938263\\
15	-7	0.545479905504567	0.545479905504567\\
15	-6	0.560090298559399	0.560090298559399\\
15	-5	0.57278340142821	0.57278340142821\\
15	-4	0.583416206382643	0.583416206382643\\
15	-3	0.591864377326528	0.591864377326528\\
15	-2	0.598026440304246	0.598026440304246\\
15	-1	0.601828398364365	0.601828398364365\\
15	0	0.60322675509099	0.60322675509099\\
15	1	0.602211060735425	0.602211060735425\\
15	2	0.598804601793495	0.598804601793495\\
15	3	0.593064135783504	0.593064135783504\\
15	4	0.58507702684811	0.58507702684811\\
15	5	0.574958116134157	0.574958116134157\\
15	6	0.562845106929043	0.562845106929043\\
15	7	0.548893313264158	0.548893313264158\\
15	8	0.533270249385109	0.533270249385109\\
15	9	0.516149476600201	0.516149476600201\\
15	10	0.497705670150341	0.497705670150341\\
15	11	0.478110057389808	0.478110057389808\\
15	12	0.457526492499945	0.457526492499945\\
15	13	0.436109971245686	0.436109971245686\\
15	14	0.414004668332403	0.414004668332403\\
15	15	0.391344706549839	0.391344706549839\\
};
\end{axis}
\end{tikzpicture}%

%% file: Figures/Results_EE/Final_EE_Comparison_NEW_edit.tikz
%
%
\definecolor{mycolor1}{rgb}{0.00000,0.45000,0.74000}%
\definecolor{mycolor2}{rgb}{0.85000,0.33000,0.10000}%
\begin{tikzpicture}

\begin{axis}[%
width=.975\columnwidth,
height=3.5/5*.975\columnwidth,
xmin=0,
xmax=60,
xlabel style={font=\color{white!15!black}},
xlabel={$D_{PI}$ / \si{\pixel}},
ymode=log,
ymin=1e-09,
ymax=1e-05,
ytick={1e-09,1e-08,1e-07,1e-06,1e-05},
yticklabels={{$\text{10}^{\text{-9}}$},{$\text{10}^{\text{-8}}$},{$\text{10}^{\text{-7}}$},{$\text{10}^{\text{-6}}$},{$\text{10}^{\text{-5}}$}},
yminorticks=true,
ylabel style={font=\color{white!15!black}},
ylabel={$var\left(\hat{\varepsilon}_G\right)^\frac{1}{2}$},
axis background/.style={fill=white},
legend style={at={(0.97,0.03)}, anchor=south east, legend cell align=left, align=left, draw=white}
]
\addlegendimage{only marks, color=mycolor1, line width=1.0pt, draw=none, mark size=2pt, mark=triangle*, mark options={solid, rotate=180, fill=mycolor1, mycolor1}},
\addlegendimage{only marks, color=mycolor2, line width=1.0pt, draw=none, mark size=1.5pt, mark=square*, mark options={solid, fill=mycolor2, mycolor2}},
\addlegendimage{only marks, color=black!50!green, line width=1.0pt, draw=none, mark size=2pt, mark=diamond*, mark options={solid, fill=black!50!green, black!50!green}},
\addlegendimage{no markers, dashed, color=black, line width=1.0pt},

\addplot [color=mycolor1, line width=1.0pt, draw=none, mark size=2pt, mark=triangle*, mark options={solid, rotate=180, fill=mycolor1, mycolor1}]
  table[row sep=crcr]{%
0.5	3.05792389026424e-07\\
1	2.60102755932102e-07\\
1.5	3.89126020945025e-09\\
2	2.80416428771174e-09\\
2.5	3.45286596144783e-09\\
3	1.13996839095855e-08\\
5	2.84884288853656e-08\\
7.5	3.30442026912913e-08\\
10	1.58148031741414e-07\\
12.5	2.56649467632019e-07\\
15	2.28693981040332e-07\\
17.5	2.49084213653064e-07\\
20	5.84932595554233e-07\\
22.5	6.77048521308057e-07\\
25	4.63540881157296e-07\\
27.5	1.28303869280692e-06\\
30	1.3239323348282e-06\\
32.5	9.91075471407787e-07\\
35	1.21641925929422e-06\\
37.5	1.39607459734253e-06\\
40	2.28987438202775e-06\\
42.5	2.67002479586915e-06\\
45	1.28167925137092e-06\\
47.5	3.26627258692328e-06\\
50	1.90311376240918e-06\\
52.5	2.33131083496465e-06\\
55	2.55117686617952e-06\\
57.5	3.0991718792784e-06\\
60	4.53747405607136e-06\\
};
\addlegendentry{Gaussian}

\addplot [color=mycolor2, line width=1.0pt, draw=none, mark size=1.5pt, mark=square*, mark options={solid, fill=mycolor2, mycolor2}]
  table[row sep=crcr]{%
0.5	4.32455736211561e-07\\
1	2.19790269506006e-07\\
1.5	6.16001299479348e-09\\
2	1.79942679989175e-08\\
2.5	3.73842869064392e-08\\
3	8.89471452628929e-09\\
5	2.71191912867339e-08\\
7.5	6.79534039668293e-08\\
10	4.65482365148984e-08\\
12.5	5.31813462303338e-08\\
15	8.50885408830729e-08\\
17.5	7.67873327094737e-08\\
20	1.35093324223658e-07\\
22.5	1.09585124293228e-07\\
25	1.28460121649726e-07\\
27.5	2.91636001346397e-07\\
30	9.36113144912651e-08\\
32.5	3.1962439305388e-07\\
35	2.45551656619955e-07\\
37.5	3.16506495173077e-07\\
40	4.53194020878436e-07\\
42.5	4.24181386981471e-07\\
45	3.31613415994043e-07\\
47.5	3.29592200128889e-07\\
50	2.98946916201095e-07\\
52.5	3.55265723605815e-07\\
55	3.2533032436582e-07\\
57.5	5.61424613589505e-07\\
60	6.63430604033796e-07\\
};
\addlegendentry{Ring-shaped}

\addplot [color=black!50!green, line width=1.0pt, draw=none, mark size=2pt, mark=diamond*, mark options={solid, fill=black!50!green, black!50!green}]
  table[row sep=crcr]{%
0.5	1.96706185126319e-07\\
1	2.86387764019186e-07\\
1.5	4.32143861457243e-09\\
2	1.15348958578106e-08\\
2.5	8.4270884514101e-09\\
3	2.19351184399128e-08\\
5	2.93754488440177e-08\\
7.5	1.74580728433584e-07\\
10	2.50987198554069e-07\\
12.5	3.83790585548792e-07\\
15	2.37111056384282e-07\\
17.5	3.94396810438334e-07\\
20	6.34762530943354e-07\\
22.5	8.99428365939899e-07\\
25	8.58870351519607e-07\\
27.5	9.01075442761004e-07\\
30	7.57602119364027e-07\\
32.5	8.8699293009953e-07\\
35	1.14178352146396e-06\\
37.5	1.56093354015593e-06\\
40	3.57421553365012e-06\\
42.5	1.81335763049497e-06\\
45	2.604532545743e-06\\
47.5	3.32837275553681e-06\\
50	4.04536142458901e-06\\
52.5	2.79822701450236e-06\\
55	2.91279394369855e-06\\
57.5	3.69573661078306e-06\\
60	6.10093911311509e-06\\
};
\addlegendentry{Plateau-shaped}

\addlegendentry{Fit}

\addplot [color=mycolor1, dashed, line width=1.0pt, forget plot, each nth point={1}, filter discard warning=false, unbounded coords=discard]
  table[row sep=crcr]{%
0.5	3.65862405557555e-07\\
0.530000000000001	3.7707358381001e-07\\
0.560000000000002	3.86851189289863e-07\\
0.590000000000003	3.94868371802401e-07\\
0.619999999999997	4.00804350424411e-07\\
0.649999999999999	4.04357264774555e-07\\
0.68	4.05258282231154e-07\\
0.710000000000001	4.03286286561158e-07\\
0.729999999999997	4.00294025571837e-07\\
0.75	3.95923651770163e-07\\
0.770000000000003	3.90094992671791e-07\\
0.789999999999999	3.82485423527538e-07\\
0.810000000000002	3.72759526135267e-07\\
0.829999999999998	3.60658130101559e-07\\
0.850000000000001	3.46018708489727e-07\\
0.869999999999997	3.28794147178114e-07\\
0.890000000000001	3.0906724679201e-07\\
0.909999999999997	2.87058268873732e-07\\
0.93	2.63123160537263e-07\\
0.950000000000003	2.37740841194855e-07\\
0.969999999999999	2.11489095632487e-07\\
0.990000000000002	1.85010082598646e-07\\
1.01	1.58980667540401e-07\\
1.03	1.34290915649789e-07\\
1.06	1.01523564090654e-07\\
1.09	7.49319884886881e-08\\
1.14	4.3866170071835e-08\\
1.2	2.31242259696261e-08\\
1.23	1.71559839408137e-08\\
1.26	1.30652605020135e-08\\
1.28	1.10850160293078e-08\\
1.3	9.53143895136724e-09\\
1.32	8.30081879678171e-09\\
1.34	7.31745632663994e-09\\
1.36	6.52548073881655e-09\\
1.38	5.88321171617784e-09\\
1.4	5.35923078857512e-09\\
1.42	4.92961365642602e-09\\
1.44	4.5759595324846e-09\\
1.46	4.28397332239653e-09\\
1.48	4.04243510407526e-09\\
1.5	3.84244355706131e-09\\
1.52	3.67672389925759e-09\\
1.54	3.53885380591071e-09\\
1.56	3.42338563719124e-09\\
1.59	3.28227577605616e-09\\
1.62	3.16825542775027e-09\\
1.67	3.00928210357132e-09\\
1.72	2.8531260656275e-09\\
1.75	2.74626232156974e-09\\
1.78	2.62440194836086e-09\\
1.82	2.44879337783719e-09\\
1.94	1.96779145285362e-09\\
1.97	1.87981603157815e-09\\
2	1.81006612216438e-09\\
2.02	1.77441261843525e-09\\
2.04	1.746676549907e-09\\
2.07	1.71785310562005e-09\\
2.1	1.70194545904576e-09\\
2.13	1.69658682838852e-09\\
2.16	1.6996603413183e-09\\
2.2	1.71345545879423e-09\\
2.26	1.74557389474191e-09\\
2.41	1.83946554628523e-09\\
2.57	1.95530035207741e-09\\
2.75	2.10583887239387e-09\\
2.94	2.28943153474716e-09\\
3.15	2.52459054340053e-09\\
3.37	2.81179112515115e-09\\
3.61	3.17974330029903e-09\\
3.87	3.65274514631836e-09\\
4.16	4.28688721302425e-09\\
4.51	5.22937263002606e-09\\
5.06	7.19067485053543e-09\\
5.8	1.10240233845107e-08\\
6.24	1.41362940574892e-08\\
6.61	1.73316753655331e-08\\
6.94	2.06754608012994e-08\\
7.24	2.41443619293966e-08\\
7.52	2.77596716160786e-08\\
7.79	3.15914451270188e-08\\
8.06	3.57671089192618e-08\\
8.34	4.0463777484875e-08\\
8.62	4.55320835039823e-08\\
8.9	5.09655891305057e-08\\
9.18	5.67524887021928e-08\\
9.46	6.28754377871572e-08\\
9.75	6.95468089566459e-08\\
10.04	7.65222994271247e-08\\
10.33	8.37658018714951e-08\\
10.63	9.15025632451109e-08\\
10.93	9.94451135261843e-08\\
11.24	1.07821562178652e-07\\
11.56	1.16596098475146e-07\\
11.89	1.25723339179892e-07\\
12.22	1.34872129448709e-07\\
12.56	1.44261859280397e-07\\
12.92	1.5411615542182e-07\\
13.31	1.64685617602315e-07\\
13.74	1.76247419999489e-07\\
14.23	1.89413899915713e-07\\
14.86	2.06657112406305e-07\\
15.95	2.38839681786782e-07\\
19.4	3.74343970794368e-07\\
20.5	4.30136524297136e-07\\
21.42	4.806579510214e-07\\
22.25	5.28623252544362e-07\\
23.03	5.75095417837984e-07\\
23.81	6.22404048984057e-07\\
24.59	6.70142944779041e-07\\
25.38	7.18506135597642e-07\\
26.28	7.73706186750803e-07\\
27.52	8.51943304074477e-07\\
29.39	9.80121196501199e-07\\
30.84	1.0870348538622e-06\\
37.25	1.69721544604863e-06\\
38.69	1.86128906988266e-06\\
39.83	1.99231709146161e-06\\
40.83	2.1041225261334e-06\\
41.76	2.20240531913189e-06\\
42.63	2.28678718780944e-06\\
43.5	2.36194086265888e-06\\
44.41	2.43044240487569e-06\\
45.37	2.49186167646676e-06\\
46.5	2.55224816150113e-06\\
48.21	2.63092575524098e-06\\
50.85	2.74316073480232e-06\\
54.35	2.8995619010044e-06\\
56.85	3.03472318580942e-06\\
60	3.2187461246646e-06\\
};
\addplot [color=mycolor2, dashed, line width=1.0pt, forget plot, each nth point={1}, filter discard warning=false, unbounded coords=discard]
  table[row sep=crcr]{%
0.5	1.02402299960245e-07\\
0.609999999999999	7.38193103049165e-08\\
0.700000000000003	5.69157801097821e-08\\
0.780000000000001	4.55042954375144e-08\\
0.859999999999999	3.66824846741191e-08\\
0.93	3.06152768888871e-08\\
1	2.57603806557772e-08\\
1.06	2.23758806495705e-08\\
1.12	1.95777213770517e-08\\
1.18	1.72657358731041e-08\\
1.23	1.56509111206361e-08\\
1.28	1.42774489173437e-08\\
1.33	1.31133131431018e-08\\
1.38	1.21322105529518e-08\\
1.42	1.14642588161309e-08\\
1.46	1.08899577543609e-08\\
1.5	1.04013774562196e-08\\
1.54	9.99151080617115e-09\\
1.58	9.65299250928419e-09\\
1.62	9.37948622738988e-09\\
1.66	9.16596464857422e-09\\
1.7	9.00855482165009e-09\\
1.74	8.90442061854124e-09\\
1.78	8.85111719353239e-09\\
1.82	8.84222359835699e-09\\
1.87	8.88112997818684e-09\\
1.93	8.98120685138676e-09\\
2.04	9.23410928787415e-09\\
2.89	1.14863768638217e-08\\
3.49	1.33278440253888e-08\\
4.06	1.52711735438223e-08\\
4.61	1.73247544534819e-08\\
5.14	1.94637173857969e-08\\
5.66	2.17071451293687e-08\\
6.19	2.41349228442801e-08\\
6.72	2.66969571353207e-08\\
7.26	2.94343183071674e-08\\
7.8	3.22862863484125e-08\\
8.36	3.53513032736125e-08\\
8.93	3.85710711019403e-08\\
9.52	4.19970872740831e-08\\
10.14	4.56858342014813e-08\\
10.78	4.95759649526502e-08\\
11.46	5.37910937915675e-08\\
12.18	5.8340500338124e-08\\
12.95	6.3303594779667e-08\\
13.72	6.83366975543088e-08\\
14.47	7.32431586611245e-08\\
15.2	7.7955211006558e-08\\
15.95	8.26801362527515e-08\\
16.76	8.76468090545041e-08\\
17.65	9.29597794503691e-08\\
18.66	9.88556879902097e-08\\
19.87	1.05843707886159e-07\\
21.72	1.16809279050192e-07\\
24.33	1.34359535676432e-07\\
29.06	1.73493011688099e-07\\
30.67	1.88116177868835e-07\\
32.13	2.01401968414479e-07\\
33.55	2.14110667818522e-07\\
35.11	2.27791168613793e-07\\
36.84	2.4272302333929e-07\\
38.63	2.57873808309665e-07\\
40.25	2.71037152969435e-07\\
41.95	2.84029800785628e-07\\
44.27	3.01092151706442e-07\\
51.68	3.61344856132276e-07\\
55.41	4.0009035435178e-07\\
60	4.55012762830549e-07\\
};
\addplot [color=black!50!green, dashed, line width=1.0pt, forget plot, each nth point={1}, filter discard warning=false, unbounded coords=discard]
  table[row sep=crcr]{%
0.5	8.98502634185503e-08\\
0.600000000000001	6.51520571721002e-08\\
0.68	5.07827663134986e-08\\
0.759999999999998	3.99353977470929e-08\\
0.829999999999998	3.26415728763288e-08\\
0.890000000000001	2.76600190775504e-08\\
0.950000000000003	2.36141002862736e-08\\
1.01	2.03252179820778e-08\\
1.06	1.80590208128338e-08\\
1.11	1.61527465216648e-08\\
1.16	1.45513449283712e-08\\
1.21	1.3209179941042e-08\\
1.25	1.22965584256006e-08\\
1.29	1.15096625199591e-08\\
1.33	1.08354989336213e-08\\
1.37	1.02632316029829e-08\\
1.41	9.78385069041686e-09\\
1.45	9.39004242047423e-09\\
1.49	9.07609269248755e-09\\
1.52	8.89038811000083e-09\\
1.55	8.74493825486327e-09\\
1.58	8.63747669479711e-09\\
1.61	8.56624582995425e-09\\
1.64	8.52996150493669e-09\\
1.67	8.52777810599687e-09\\
1.7	8.55926324089599e-09\\
1.74	8.65361062088248e-09\\
1.78	8.80814436978015e-09\\
1.82	9.01870888799328e-09\\
1.87	9.35050288668047e-09\\
1.93	9.83083857180859e-09\\
2.33	1.39530526831957e-08\\
2.5	1.60386892216775e-08\\
2.67	1.8331419957454e-08\\
2.84	2.08361471977521e-08\\
3.01	2.3555999910526e-08\\
3.18	2.64924148152369e-08\\
3.36	2.9837173489892e-08\\
3.54	3.3421635521968e-08\\
3.72	3.72409133516474e-08\\
3.91	4.15193925343239e-08\\
4.1	4.60412795579398e-08\\
4.29	5.0794167735509e-08\\
4.49	5.60314101995803e-08\\
4.69	6.14915918854393e-08\\
4.9	6.74447989867641e-08\\
5.12	7.39004784347073e-08\\
5.34	8.05527402659322e-08\\
5.57	8.76832655304938e-08\\
5.8	9.49581081485012e-08\\
6.04	1.02665631982385e-07\\
6.29	1.10781852299238e-07\\
6.55	1.19278706089863e-07\\
6.82	1.28126711158963e-07\\
7.11	1.37625858261393e-07\\
7.41	1.47426604868961e-07\\
7.73	1.57842908688985e-07\\
8.07	1.688344652358e-07\\
8.42	1.80019012021783e-07\\
8.79	1.91658897734345e-07\\
9.19	2.04004598431977e-07\\
9.62	2.17001401322404e-07\\
10.09	2.30928734488374e-07\\
10.6	2.45782924903602e-07\\
11.15	2.61511591107832e-07\\
11.75	2.78370682583106e-07\\
12.44	2.97517007220611e-07\\
13.29	3.2115912859473e-07\\
16.33	4.20097863904208e-07\\
17.44	4.6736844005765e-07\\
18.69	5.29870445942189e-07\\
21.98	7.39293138255939e-07\\
22.88	8.04190471737681e-07\\
23.74	8.66995013613684e-07\\
24.6	9.2993084940438e-07\\
25.47	9.93104405026535e-07\\
26.41	1.0605500506305e-06\\
27.46	1.13531662360343e-06\\
28.59	1.21547661152657e-06\\
29.59	1.28454531269012e-06\\
30.52	1.34533801431095e-06\\
31.71	1.41927922330208e-06\\
35.11	1.64961165680389e-06\\
37.8	1.8698348847742e-06\\
40.51	2.131964971124e-06\\
42.79	2.39271334919203e-06\\
46.15	2.84048420896313e-06\\
47.62	3.04250463827269e-06\\
49.05	3.23572059797775e-06\\
50.61	3.44238489447718e-06\\
52.44	3.68183304765832e-06\\
55.4	4.0797427558586e-06\\
60	4.78352855818785e-06\\
};
\end{axis}
\end{tikzpicture}%

%% file: Figures/Results_EE/EEParts_V3_Comparison_edit.tikz
%
%
\definecolor{mycolor1}{rgb}{0.00000,0.44700,0.74100}%
\definecolor{mycolor2}{rgb}{0.00000,0.44706,0.74118}%
\definecolor{mycolor3}{rgb}{0.85098,0.32549,0.09804}%
\definecolor{mycolor4}{rgb}{0.00000,0.49804,0.00000}%
\begin{tikzpicture}

\begin{axis}[%
width=0.95\columnwidth,
height=3.5/5*0.95\columnwidth,
xmin=0,
xmax=60,
xlabel style={font=\color{white!15!black}},
xlabel={\(D_{PI}\) / \si{\pixel}},
ymode=log,
ymin=1e-20,
ymax=10000000000,
yminorticks=true,
ylabel style={font=\color{white!15!black}},
ylabel={\(\left(\varepsilon_x\right)^{\frac{1}{2}}\)},
axis background/.style={fill=white},
legend style={draw=white!15!black, legend columns=3, legend style={draw=none}},
legend style={at={(axis cs:23,2*10E15)},anchor=north}
]

\addlegendimage{color=mycolor1, line width = 1pt, draw=none, dashed},
\addlegendimage{color=mycolor3, line width = 1pt, draw=none},
\addlegendimage{color=mycolor4, line width = 1pt, draw=none,  dash pattern=on 4pt off 3pt on 2pt off 3pt},

\addplot [color=mycolor1, dashed, line width=1.0pt]
 plot [error bars/.cd, y dir = both, y explicit]
 table[row sep=crcr, y error plus index=2, y error minus index=3]{%
0.5	119437.494517564	6.17532117692627	6.17532117692627\\
1	119436.960428694	6.08261377785929	6.08261377785929\\
1.5	8.58251944648139	1.87635235320999e-06	1.87635235320999e-06\\
2	1.13968651556085	0.0010656820143581	0.0010656820143581\\
2.5	0.565464191312471	3.65925900593862e-05	3.65925900593862e-05\\
3	0.463624216679652	1.94219478829676e-09	1.94219478829676e-09\\
5	0.902756537892284	2.9608355607878e-05	2.9608355607878e-05\\
7.5	4.89708573095835	0.000636813201825102	0.000636813201825102\\
10	9.78172376807106	0.000152537332464298	0.000152537332464298\\
12.5	17.8480639804739	0.000134430935935985	0.000134430935935985\\
15	28.6728192281276	7.55837530394073e-05	7.55837530394073e-05\\
17.5	74.0270244793637	0.00192656790460035	0.00192656790460035\\
20	108.46051673025	0.000830345068071997	0.000830345068071997\\
22.5	129.713639222918	0.000880011778363465	0.000880011778363465\\
25	193.401166803668	0.000889157747543554	0.000889157747543554\\
27.5	375.963956835396	0.00589312073565027	0.00589312073565027\\
30	486.305240552494	0.00720262706227132	0.00720262706227132\\
32.5	618.656388388113	0.00469357982636869	0.00469357982636869\\
35	681.516630261221	0.00576145357285224	0.00576145357285224\\
37.5	1175.40695349678	0.0196724488532277	0.0196724488532277\\
40	1423.24847264635	0.0145892070076013	0.0145892070076013\\
42.5	1706.59882413357	0.0453682774980607	0.0453682774980607\\
45	1725.48621685123	0.0223046403110726	0.0223046403110726\\
47.5	2796.71300206741	0.0513304443549282	0.0513304443549282\\
50	3248.56203522256	0.0853641394307572	0.0853641394307572\\
52.5	3749.67475866898	0.107200687237279	0.107200687237279\\
55	3557.10294009229	0.0677561599833708	0.0677561599833708\\
57.5	5570.05698764927	0.199305370507753	0.199305370507753\\
60	6291.40120095946	0.266320690131588	0.266320690131588\\
};
\addlegendentry{Gaussian}

\addplot [color=mycolor2, dashed, line width=1.0pt]
  table[row sep=crcr]{%
0.5	1.49802272347004e-19\\
1	1.49619564150896e-19\\
1.5	1.36903293288215e-18\\
2	2.23200824750368e-17\\
2.5	6.92726926592532e-17\\
3	2.08920257697256e-16\\
5	1.21909877252471e-15\\
7.5	1.14822629851163e-15\\
10	2.78865412950701e-15\\
12.5	1.83572768326646e-15\\
15	2.53593823478434e-15\\
17.5	9.95591286806009e-16\\
20	3.16239570768199e-15\\
22.5	5.47505850153955e-15\\
25	1.12646157983337e-15\\
27.5	4.07286958363139e-15\\
30	3.56796429265907e-15\\
32.5	1.68208549812909e-15\\
35	1.24157860388465e-15\\
37.5	1.84541333886225e-15\\
40	7.93339160408215e-16\\
42.5	5.59550875071118e-15\\
45	1.45226151528934e-15\\
47.5	1.48749931149473e-15\\
50	2.26277210815294e-15\\
52.5	2.40036974173361e-15\\
55	1.39384603619904e-15\\
57.5	1.65569895747662e-15\\
60	1.31632034274103e-15\\
};

\addplot [color=mycolor2, dashed, line width=1.0pt]
  table[row sep=crcr]{%
0.5	0.113737346122436\\
1	0.0982405201317306\\
1.5	0.109759449735225\\
2	0.0967815618103575\\
2.5	0.102371521743979\\
3	0.109169769116401\\
5	0.0445628932374052\\
7.5	0.00248102198082403\\
10	0.0011030078597742\\
12.5	0.000587533869870893\\
15	0.000312579730763213\\
17.5	9.05913144827149e-05\\
20	6.39934178709325e-05\\
22.5	6.53212306110828e-05\\
25	4.09965202711169e-05\\
27.5	1.90709644212039e-05\\
30	2.27584485630643e-05\\
32.5	1.32293071871952e-05\\
35	1.11750047188486e-05\\
37.5	5.70028454225027e-06\\
40	5.91500339784405e-06\\
42.5	4.96906268226091e-06\\
45	3.92282517025245e-06\\
47.5	2.23509033272513e-06\\
50	2.08856558699291e-06\\
52.5	1.71272793258407e-06\\
55	2.33590710941231e-06\\
57.5	1.21536296535532e-06\\
60	1.00054370750502e-06\\
};

\addplot [color=mycolor3, line width=1.0pt]
 plot [error bars/.cd, y dir = both, y explicit]
 table[row sep=crcr, y error plus index=2, y error minus index=3]{%
0.5	119436.946637323	8.61451007544333	8.61451007544333\\
1	119437.100821024	5.4706328080472	5.4706328080472\\
1.5	1.62745444667511	1.51746268585226e-07	1.51746268585226e-07\\
2	0.504451108976156	0.00157762615012959	0.00157762615012959\\
2.5	0.578485913626461	0.000734242953502694	0.000734242953502694\\
3	0.635405946833102	3.31314000211105e-08	3.31314000211105e-08\\
5	0.971267584171311	0.000911849591427324	0.000911849591427324\\
7.5	1.54060835787747	0.000700120033533462	0.000700120033533462\\
10	2.30904595572803	4.06168761111771e-05	4.06168761111771e-05\\
12.5	2.54071253144231	3.5034961245351e-06	3.5034961245351e-06\\
15	3.27572765183369	2.63432073778772e-06	2.63432073778772e-06\\
17.5	5.50096378287633	2.55708251125061e-05	2.55708251125061e-05\\
20	6.46622103145024	8.62782122185088e-06	8.62782122185088e-06\\
22.5	7.81439943693707	9.65454511882552e-06	9.65454511882552e-06\\
25	9.75540917859562	1.19643616806032e-05	1.19643616806032e-05\\
27.5	15.2060742410483	4.02014302218072e-05	4.02014302218072e-05\\
30	18.5164679879507	1.88832999430052e-05	1.88832999430052e-05\\
32.5	21.7755056955463	4.73303217678104e-05	4.73303217678104e-05\\
35	23.9960736495105	5.35059832787444e-05	5.35059832787444e-05\\
37.5	35.8172536703176	0.000121435996598912	0.000121435996598912\\
40	41.6031993809905	0.000174539794973359	0.000174539794973359\\
42.5	48.2877360112901	0.000191042163879581	0.000191042163879581\\
45	49.3519744572763	0.000163290662679962	0.000163290662679962\\
47.5	72.483400370443	0.000297979509833976	0.000297979509833976\\
50	82.258319733705	0.000264840471422151	0.000264840471422151\\
52.5	92.7838113997344	0.000349749625220066	0.000349749625220066\\
55	89.7577741537612	0.000367319090197971	0.000367319090197971\\
57.5	129.870962240263	0.000723287586408347	0.000723287586408347\\
60	144.109767165409	0.000931409010807694	0.000931409010807694\\
};
\addlegendentry{Ring-shaped}

\addplot [color=mycolor3, line width=1.0pt]
  table[row sep=crcr]{%
0.5	7.84035205474145e-19\\
1	1.50990816687101e-19\\
1.5	1.25078169110789e-17\\
2	2.5490270701492e-16\\
2.5	1.64767527641602e-15\\
3	2.27978998275648e-16\\
5	1.24551315189875e-15\\
7.5	9.61087450651335e-16\\
10	9.59213655869369e-16\\
12.5	1.52215474852561e-15\\
15	8.85610268278027e-16\\
17.5	8.18685373460984e-16\\
20	5.46861378382022e-15\\
22.5	2.94569169455316e-15\\
25	2.44950017715328e-15\\
27.5	5.81737751085417e-15\\
30	6.50265928198333e-16\\
32.5	3.21082737296158e-15\\
35	1.92285544318837e-15\\
37.5	3.30580780037921e-15\\
40	4.34756006910413e-15\\
42.5	4.50706595109659e-15\\
45	3.97534815282327e-15\\
47.5	2.35356713865419e-15\\
50	1.4490480299936e-15\\
52.5	1.35007503700594e-15\\
55	1.33653019603048e-15\\
57.5	1.09062539844858e-15\\
60	2.42667073170405e-15\\
};

\addplot [color=mycolor3, line width=1.0pt]
  table[row sep=crcr]{%
0.5	0.112806760064107\\
1	0.108291353567799\\
1.5	0.107934904288982\\
2	0.125528956397773\\
2.5	0.101409982659081\\
3	0.109040627669442\\
5	0.00941937647731067\\
7.5	0.000637689650087193\\
10	0.000324952400163101\\
12.5	0.000143681972340592\\
15	8.13952714943038e-05\\
17.5	3.4372892378583e-05\\
20	2.19628914607174e-05\\
22.5	2.11432602076009e-05\\
25	1.55788411461494e-05\\
27.5	7.83818524124106e-06\\
30	6.08021059149277e-06\\
32.5	4.97205676075169e-06\\
35	4.24614402811726e-06\\
37.5	3.00557465468651e-06\\
40	2.42495069330592e-06\\
42.5	2.16239718103567e-06\\
45	1.82117664528717e-06\\
47.5	1.38137973981597e-06\\
50	1.14940932759481e-06\\
52.5	9.87759810308512e-07\\
55	1.00296540345431e-06\\
57.5	6.6534824598071e-07\\
60	5.90399026436937e-07\\
};

\addplot [color=mycolor4, dashdotted, line width=1.0pt]
 plot [error bars/.cd, y dir = both, y explicit]
 table[row sep=crcr, y error plus index=2, y error minus index=3]{%
0.5	119437.800650782	7.34677996865195	7.34677996865195\\
1	119437.045518149	7.27689340431879	7.27689340431879\\
1.5	3.25524793466449	3.52907335066619e-07	3.52907335066619e-07\\
2	0.515985858266046	0.00040914283750758	0.00040914283750758\\
2.5	0.49552562727807	9.35396619351465e-05	9.35396619351465e-05\\
3	0.656432924633339	2.95913337789369e-08	2.95913337789369e-08\\
5	1.56954344766252	0.000519644665367327	0.000519644665367327\\
7.5	8.84860883762602	0.00865203020517054	0.00865203020517054\\
10	16.2400958789717	0.00134920311959901	0.00134920311959901\\
12.5	28.3594812876161	0.000784942964435487	0.000784942964435487\\
15	44.8211067008188	0.000237280168358103	0.000237280168358103\\
17.5	108.939127005912	0.010479459364244	0.010479459364244\\
20	157.747444803351	0.00339133788630092	0.00339133788630092\\
22.5	186.905742760961	0.00140768975700045	0.00140768975700045\\
25	274.18571074179	0.00233146948095862	0.00233146948095862\\
27.5	519.487220575625	0.0182443957059618	0.0182443957059618\\
30	665.987730080996	0.0120276728352406	0.0120276728352406\\
32.5	840.587654807287	0.00862227110429805	0.00862227110429805\\
35	923.900335134982	0.0082376082649225	0.0082376082649225\\
37.5	1565.87969433874	0.0427867551558404	0.0427867551558404\\
40	1877.7120260239	0.0535718885637595	0.0535718885637595\\
42.5	2241.06626881034	0.0495394411476593	0.0495394411476593\\
45	2261.1503048585	0.0624599410058982	0.0624599410058982\\
47.5	3583.61272155913	0.112877297605284	0.112877297605284\\
50	4146.0265950529	0.133517494981667	0.133517494981667\\
52.5	4735.12197700742	0.147529517797923	0.147529517797923\\
55	4510.57717542755	0.12927526293874	0.12927526293874\\
57.5	6870.30581695778	0.352279374690053	0.352279374690053\\
60	7694.35775306392	0.415040810851205	0.415040810851205\\
};
\addlegendentry{Plateau-shaped}

\addplot [color=mycolor4, dashdotted, line width=1.0pt]
  table[row sep=crcr]{%
0.5	7.38371175071666e-19\\
1	1.15673518745549e-19\\
1.5	2.30019411216666e-18\\
2	1.73290500211817e-16\\
2.5	2.93351193469463e-16\\
3	2.86657400413877e-16\\
5	4.51245577808816e-16\\
7.5	1.99431823184036e-15\\
10	4.60280281137466e-15\\
12.5	2.83484597819947e-15\\
15	1.64261734065241e-15\\
17.5	1.41074182597418e-15\\
20	2.03322858508148e-15\\
22.5	3.70801203363302e-15\\
25	2.75780376944698e-15\\
27.5	2.04723529707033e-15\\
30	1.50851334513148e-15\\
32.5	1.79744506571698e-15\\
35	1.23973138879472e-15\\
37.5	1.99430181024655e-15\\
40	5.18466194834549e-15\\
42.5	2.79784006707743e-15\\
45	5.18815323881828e-15\\
47.5	2.77661685214647e-15\\
50	2.74369267380542e-15\\
52.5	1.77959505994476e-15\\
55	1.41276187445141e-15\\
57.5	1.96143087656162e-15\\
60	4.09793499017891e-15\\
};

\addplot [color=mycolor4, dashdotted, line width=1.0pt]
  table[row sep=crcr]{%
0.5	0.110424625852782\\
1	0.103897936366805\\
1.5	0.0905951755929935\\
2	0.0800990002685981\\
2.5	0.0423878722499585\\
3	0.0179873885932816\\
5	0.0015092983014954\\
7.5	0.000138555623230126\\
10	6.10009002916739e-05\\
12.5	3.37775496450263e-05\\
15	2.07621840781543e-05\\
17.5	8.49784688830262e-06\\
20	4.91378412914334e-06\\
22.5	4.02887536429268e-06\\
25	2.7708635079857e-06\\
27.5	1.3890136823588e-06\\
30	1.19329931844756e-06\\
32.5	9.42201010647518e-07\\
35	8.11340766778084e-07\\
37.5	5.12597542407469e-07\\
40	3.92142703587275e-07\\
42.5	3.20526204908704e-07\\
45	3.22208958316265e-07\\
47.5	1.84258902951236e-07\\
50	1.60338622479338e-07\\
52.5	1.46719863543308e-07\\
55	1.75963557445026e-07\\
57.5	1.06470311399463e-07\\
60	9.47478357970481e-08\\
};

\draw[dashed, color=black, line width=0.05pt] (axis cs:0,10E5) -- (axis cs:60,10E5);
\draw[dashed, color=black, line width=0.05pt] (axis cs:0,2.5*10E-2) -- (axis cs:60,2.5*10E-2);
\draw[dashed, color=black, line width=0.05pt] (axis cs:0,5*10E-12) -- (axis cs:60,5*10E-12);
\node at (axis cs:5,5*10E2) {\(\varepsilon_1\)};
\node at (axis cs:9,1*10E-8) {\(\varepsilon_2\) + Noise};
\node at (axis cs:5,10E-13) {\(\varepsilon_2\)};

\end{axis}

\end{tikzpicture}%

%% file: Figures/Results_Cut/EEK-60_V3_edit.tikz
%
%
\definecolor{mycolor1}{rgb}{0.00000,0.45000,0.74000}%
\definecolor{mycolor2}{rgb}{0.85000,0.33000,0.10000}%
\definecolor{mycolor3}{rgb}{0.00000,0.44700,0.74100}%
\definecolor{mycolor4}{rgb}{0.85098,0.32549,0.09804}%
\definecolor{mycolor5}{rgb}{0.00000,0.49804,0.00000}%
\begin{tikzpicture}

\begin{axis}[%
width=0.95\columnwidth,
height=3.5/5*0.95\columnwidth,
xmin=0,
xmax=1,
xtick={0,0.2,0.6,1},
xticklabels={{0},{1},{3},{5}},
xlabel style={font=\color{white!15!black}},
xlabel={$K_5$},
ymode=log,
ymin=5E-7,
ymax=1E-5,
yminorticks=true,
ylabel style={font=\color{white!15!black}},
ylabel={$var\left(\hat\varepsilon\right)^\frac{1}{2}$},
axis background/.style={fill=white},
axis x line*=bottom,
axis y line*=left,
legend style={at={(axis cs:0.75,1.5E-5)}, anchor=north, legend cell align=left, align=left, draw=white},
]

\addlegendimage{only marks, color=mycolor1, line width=1.0pt, draw=none, mark size=2pt, mark=triangle*, mark options={solid, rotate=180, fill=mycolor1, mycolor1}},
\addlegendimage{only marks, color=mycolor2, line width=1.0pt, draw=none, mark size=1.5pt, mark=square*, mark options={solid, fill=mycolor2, mycolor2}},
\addlegendimage{only marks, color=black!50!green, line width=1.0pt, draw=none, mark size=2pt, mark=diamond*, mark options={solid, fill=black!50!green, black!50!green}},
\addlegendimage{no markers, dashed, color=black, line width=1.0pt},

\addplot [color=mycolor1, line width=1.0pt, draw=none, mark size=2pt, mark=triangle*, mark options={solid, rotate=180, fill=mycolor1, mycolor1}]
  table[row sep=crcr]{%
0	4.53747405607136e-06\\
0.2	2.96524330669553e-06\\
0.6	9.76132906100159e-07\\
1	7.36423760804585e-07\\
};
\addlegendentry{Gaussian}

\addplot [color=mycolor2, line width=1.0pt, draw=none, mark size=1.5pt, mark=square*, mark options={solid, fill=mycolor2, mycolor2}]
  table[row sep=crcr]{%
0	6.63430604033796e-07\\
0.2	5.43489770025257e-07\\
0.6	5.44668278808001e-07\\
1	5.2263455941996e-07\\
};
\addlegendentry{Ring-shaped}

\addplot [color=black!50!green, line width=1.0pt, draw=none, mark size=2pt, mark=diamond*, mark options={solid, fill=black!50!green, black!50!green}]
  table[row sep=crcr]{%
0	6.10093911311509e-06\\
0.2	3.4834625913898e-06\\
0.6	1.90129330101966e-06\\
1	1.12217207417917e-06\\
};
\addlegendentry{Plateau-shaped}

\addplot [color=mycolor1, line width=1pt, dashed]
  table[row sep=crcr]{%
0	4.2004321719101e-06\\
0.2	2.96524386571015e-06\\
0.6	1.47772389538651e-06\\
1	7.36421019615974e-07\\
};

\addplot [color=mycolor2, line width=1pt, dashed]
  table[row sep=crcr]{%
0	6.15991145485587e-07\\
0.2	5.93257544676468e-07\\
0.6	5.50276379320455e-07\\
1	5.10409174489576e-07\\
};

\addplot [color=black!50!green, line width=1pt, dashed]
  table[row sep=crcr]{%
0	5.42148396529287e-06\\
0.2	3.90751641985338e-06\\
0.6	2.0298599726917e-06\\
1	1.05446300565784e-06\\
};
\addlegendentry{Fit}

\end{axis}
\end{tikzpicture}%

%% file: Figures/Results_Cut/DK-60_V2_edit.tikz
%
%
\definecolor{mycolor1}{rgb}{0.00000,0.45000,0.74000}%
\definecolor{mycolor2}{rgb}{0.85000,0.33000,0.10000}%
\definecolor{mycolor3}{rgb}{0.00000,0.44700,0.74100}%
\definecolor{mycolor4}{rgb}{0.85098,0.32549,0.09804}%
\definecolor{mycolor5}{rgb}{0.00000,0.49804,0.00000}%
\begin{tikzpicture}

\begin{axis}[%
width=0.99\columnwidth,
height=3.5/5*0.99\columnwidth,
xmin=0,
xmax=1,
xtick={0,0.2,0.6,1},
xticklabels={{0},{1},{3},{5}},
xlabel style={font=\color{white!15!black}},
xlabel={$K_5$},
ymin=3.5,
ymax=5.5,
ylabel style={font=\color{white!15!black}},
ylabel={\(D\)},
axis background/.style={fill=white},
legend style={legend cell align=left, align=left, draw=white},
legend pos=north west
]

\addlegendimage{only marks, color=mycolor1, line width=1.0pt, draw=none, mark size=2pt, mark=triangle*, mark options={solid, rotate=180, fill=mycolor1, mycolor1}},
\addlegendimage{only marks, color=mycolor2, line width=1.0pt, draw=none, mark size=1.5pt, mark=square*, mark options={solid, fill=mycolor2, mycolor2}},
\addlegendimage{only marks, color=black!50!green, line width=1.0pt, draw=none, mark size=2pt, mark=diamond*, mark options={solid, fill=black!50!green, black!50!green}},
\addlegendimage{no markers, dashed, color=black, line width=1.0pt},

\addplot [color=mycolor1, line width=1.0pt, draw=none, mark size=2pt, mark=triangle*, mark options={solid, rotate=180, fill=mycolor1, mycolor1}]
 plot [error bars/.cd, y dir = both, y explicit]
 table[row sep=crcr, y error plus index=3, y error minus index=2]{%
0	3.84331432914734	0.10691504354	0.10691504354\\
0.2	3.72188874459267	0.08976131077	0.08976131077\\
0.6	3.60579150104523	0.05799201886	0.05799201886\\
1	4.58630038642883	0.1018322006	0.1018322006\\
};
\addlegendentry{Gaussian}

\addplot [color=mycolor2, line width=1.0pt, draw=none, mark size=2pt, mark=square*, mark options={solid, fill=mycolor2, mycolor2}]
 plot [error bars/.cd, y dir = both, y explicit]
 table[row sep=crcr, y error plus index=3, y error minus index=2]{%
0	4.27056998491287	0.07759671716	0.07759671716\\
0.2	4.00698041534424	0.07016951254	0.07016951254\\
0.6	3.83254039573669	0.05574620689	0.05574620689\\
1	5.22000195837021	0.1132190456	0.1132190456\\
};
\addlegendentry{Ring-shaped}

\addplot [color=black!50!green, line width=1.0pt, draw=none, mark size=2pt, mark=diamond*, mark options={solid, fill=black!50!green, black!50!green}]
 plot [error bars/.cd, y dir = both, y explicit]
 table[row sep=crcr, y error plus index=3, y error minus index=2]{%
0	3.80806147813797	0.10092746067	0.10092746067\\
0.2	3.62074344587326	0.09074344587	0.09074344587\\
0.6	3.64428399443626	0.07271600556	0.07271600556\\
1	4.5334847972393	0.1185152028	0.1185152028\\
};
\addlegendentry{Plateau-shaped}

\addplot [color=mycolor1, dashed, line width=1pt]
  table[row sep=crcr]{%
0	3.84331432914734\\
0.0101010101010101	3.83944867660765\\
0.0202020202020202	3.83523550445021\\
0.0303030303030303	3.83069273757938\\
0.0404040404040404	3.82583830089953\\
0.0505050505050505	3.82069011931502\\
0.0606060606060606	3.81526611773021\\
0.0707070707070707	3.80958422104946\\
0.0808080808080808	3.80366235417714\\
0.0909090909090909	3.79751844201761\\
0.101010101010101	3.79117040947524\\
0.111111111111111	3.78463618145438\\
0.121212121212121	3.7779336828594\\
0.131313131313131	3.77108083859466\\
0.141414141414141	3.76409557356453\\
0.151515151515152	3.75699581267337\\
0.161616161616162	3.74979948082553\\
0.171717171717172	3.74252450292539\\
0.181818181818182	3.73518880387731\\
0.191919191919192	3.72781030858564\\
0.202020202020202	3.72040694195476\\
0.212121212121212	3.71299662888902\\
0.222222222222222	3.7055972942928\\
0.232323232323232	3.69822686307044\\
0.242424242424242	3.69090326012631\\
0.252525252525253	3.68364441036478\\
0.262626262626263	3.67646823869022\\
0.272727272727273	3.66939267000697\\
0.282828282828283	3.66243562921941\\
0.292929292929293	3.6556150412319\\
0.303030303030303	3.6489488309488\\
0.313131313131313	3.64245492327447\\
0.323232323232323	3.63615124311328\\
0.333333333333333	3.63005571536959\\
0.343434343434343	3.62418626494776\\
0.353535353535354	3.61856081675216\\
0.363636363636364	3.61319729568715\\
0.373737373737374	3.60811362665708\\
0.383838383838384	3.60332773456634\\
0.393939393939394	3.59885754431927\\
0.404040404040404	3.59472098082024\\
0.414141414141414	3.59093596897361\\
0.424242424242424	3.58752043368375\\
0.434343434343434	3.58449229985501\\
0.444444444444444	3.58186949239177\\
0.454545454545455	3.57966993619839\\
0.464646464646465	3.57791155617922\\
0.474747474747475	3.57661227723863\\
0.484848484848485	3.57579002428098\\
0.494949494949495	3.57546272221064\\
0.505050505050505	3.57564829593197\\
0.515151515151515	3.57636467034933\\
0.525252525252525	3.57762977036709\\
0.535353535353535	3.5794615208896\\
0.545454545454545	3.58187784682124\\
0.555555555555556	3.58489667306635\\
0.565656565656566	3.58853592452932\\
0.575757575757576	3.59281352611449\\
0.585858585858586	3.59774740272624\\
0.595959595959596	3.60335547926892\\
0.606060606060606	3.60965568064689\\
0.616161616161616	3.61666593176453\\
0.626262626262626	3.6244041575262\\
0.636363636363636	3.63288828283625\\
0.646464646464647	3.64213623259904\\
0.656565656565657	3.65216593171895\\
0.666666666666667	3.66299530510034\\
0.676767676767677	3.67464227764757\\
0.686868686868687	3.68712477426499\\
0.696969696969697	3.70046071985698\\
0.707070707070707	3.7146680393279\\
0.717171717171717	3.7297646575821\\
0.727272727272727	3.74576849952396\\
0.737373737373737	3.76269749005784\\
0.747474747474748	3.78056955408809\\
0.757575757575758	3.79940261651908\\
0.767676767676768	3.81921460225518\\
0.777777777777778	3.84002343620074\\
0.787878787878788	3.86184704326014\\
0.797979797979798	3.88470334833773\\
0.808080808080808	3.90861027633787\\
0.818181818181818	3.93358575216493\\
0.828282828282828	3.95964770072328\\
0.838383838383838	3.98681404691726\\
0.848484848484849	4.01510271565126\\
0.858585858585859	4.04453163182963\\
0.868686868686869	4.07511872035672\\
0.878787878787879	4.10688190613692\\
0.888888888888889	4.13983911407457\\
0.898989898989899	4.17400826907405\\
0.909090909090909	4.20940729603971\\
0.919191919191919	4.24605411987592\\
0.929292929292929	4.28396666548703\\
0.939393939393939	4.32316285777742\\
0.94949494949495	4.36366062165145\\
0.95959595959596	4.40547788201348\\
0.96969696969697	4.44863256376787\\
0.97979797979798	4.49314259181898\\
0.98989898989899	4.53902589107118\\
1	4.58630038642883\\
};

\addplot [color=mycolor2, dashed, line width=1pt]
  table[row sep=crcr]{%
0	4.27056998491287\\
0.0101010101010101	4.25829801689875\\
0.0202020202020202	4.24579018703788\\
0.0303030303030303	4.23306758901035\\
0.0404040404040404	4.22015131649622\\
0.0505050505050505	4.20706246317557\\
0.0606060606060606	4.19382212272845\\
0.0707070707070707	4.18045138883495\\
0.0808080808080808	4.16697135517514\\
0.0909090909090909	4.15340311542908\\
0.101010101010101	4.13976776327685\\
0.111111111111111	4.12608639239851\\
0.121212121212121	4.11238009647413\\
0.131313131313131	4.0986699691838\\
0.141414141414141	4.08497710420757\\
0.151515151515152	4.07132259522552\\
0.161616161616162	4.05772753591772\\
0.171717171717172	4.04421301996424\\
0.181818181818182	4.03080014104515\\
0.191919191919192	4.01750999284052\\
0.202020202020202	4.00436366903042\\
0.212121212121212	3.99138226329493\\
0.222222222222222	3.9785868693141\\
0.232323232323232	3.96599858076802\\
0.242424242424242	3.95363849133675\\
0.252525252525253	3.94152769470036\\
0.262626262626263	3.92968728453893\\
0.272727272727273	3.91813835453253\\
0.282828282828283	3.90690199836122\\
0.292929292929293	3.89599930970507\\
0.303030303030303	3.88545138224416\\
0.313131313131313	3.87527930965856\\
0.323232323232323	3.86550418562834\\
0.333333333333333	3.85614710383356\\
0.343434343434343	3.84722915795431\\
0.353535353535354	3.83877144167064\\
0.363636363636364	3.83079504866263\\
0.373737373737374	3.82332107261036\\
0.383838383838384	3.81637060719388\\
0.393939393939394	3.80996474609328\\
0.404040404040404	3.80412458298861\\
0.414141414141414	3.79887121155996\\
0.424242424242424	3.7942257254874\\
0.434343434343434	3.79020921845099\\
0.444444444444444	3.7868427841308\\
0.454545454545455	3.7841475162069\\
0.464646464646465	3.78214450835938\\
0.474747474747475	3.78085485426828\\
0.484848484848485	3.7802996476137\\
0.494949494949495	3.78049998207569\\
0.505050505050505	3.78147695133433\\
0.515151515151515	3.78325164906969\\
0.525252525252525	3.78584516896183\\
0.535353535353535	3.78927860469084\\
0.545454545454545	3.79357304993677\\
0.555555555555556	3.79874959837971\\
0.565656565656566	3.80482934369971\\
0.575757575757576	3.81183337957686\\
0.585858585858586	3.81978279969122\\
0.595959595959596	3.82869869772286\\
0.606060606060606	3.83860216735186\\
0.616161616161616	3.84951430225828\\
0.626262626262626	3.86145619612219\\
0.636363636363636	3.87444894262367\\
0.646464646464647	3.88851363544278\\
0.656565656565657	3.9036713682596\\
0.666666666666667	3.9199432347542\\
0.676767676767677	3.93735032860664\\
0.686868686868687	3.955913743497\\
0.696969696969697	3.97565457310535\\
0.707070707070707	3.99659391111176\\
0.717171717171717	4.01875285119629\\
0.727272727272727	4.04215248703903\\
0.737373737373737	4.06681391232004\\
0.747474747474748	4.09275822071938\\
0.757575757575758	4.12000650591714\\
0.767676767676768	4.14857986159338\\
0.777777777777778	4.17849938142817\\
0.787878787878788	4.20978615910159\\
0.797979797979798	4.2424612882937\\
0.808080808080808	4.27654586268457\\
0.818181818181818	4.31206097595428\\
0.828282828282828	4.3490277217829\\
0.838383838383838	4.38746719385049\\
0.848484848484849	4.42740048583712\\
0.858585858585859	4.46884869142288\\
0.868686868686869	4.51183290428782\\
0.878787878787879	4.55637421811202\\
0.888888888888889	4.60249372657554\\
0.898989898989899	4.65021252335847\\
0.909090909090909	4.69955170214087\\
0.919191919191919	4.7505323566028\\
0.929292929292929	4.80317558042435\\
0.939393939393939	4.85750246728558\\
0.94949494949495	4.91353411086656\\
0.95959595959596	4.97129160484736\\
0.96969696969697	5.03079604290806\\
0.97979797979798	5.09206851872872\\
0.98989898989899	5.15513012598941\\
1	5.22000195837021\\
};

\addplot [color=green!50!black, dashed, line width=1pt]
  table[row sep=crcr]{%
0	3.80806147813797\\
0.0101010101010101	3.79660230446807\\
0.0202020202020202	3.78531737312867\\
0.0303030303030303	3.77421315296538\\
0.0404040404040404	3.76329611282383\\
0.0505050505050505	3.75257272154964\\
0.0606060606060606	3.74204944798843\\
0.0707070707070707	3.73173276098583\\
0.0808080808080808	3.72162912938747\\
0.0909090909090909	3.71174502203896\\
0.101010101010101	3.70208690778593\\
0.111111111111111	3.69266125547401\\
0.121212121212121	3.68347453394881\\
0.131313131313131	3.67453321205596\\
0.141414141414141	3.66584375864108\\
0.151515151515152	3.65741264254981\\
0.161616161616162	3.64924633262775\\
0.171717171717172	3.64135129772054\\
0.181818181818182	3.63373400667379\\
0.191919191919192	3.62640092833314\\
0.202020202020202	3.6193585315442\\
0.212121212121212	3.6126132851526\\
0.222222222222222	3.60617165800396\\
0.232323232323232	3.60004011894391\\
0.242424242424242	3.59422513681807\\
0.252525252525253	3.58873318047206\\
0.262626262626263	3.5835707187515\\
0.272727272727273	3.57874422050203\\
0.282828282828283	3.57426015456926\\
0.292929292929293	3.57012498979882\\
0.303030303030303	3.56634519503633\\
0.313131313131313	3.56292723912741\\
0.323232323232323	3.55987759091769\\
0.333333333333333	3.55720271925279\\
0.343434343434343	3.55490909297833\\
0.353535353535354	3.55300318093995\\
0.363636363636364	3.55149145198325\\
0.373737373737374	3.55038037495387\\
0.383838383838384	3.54967641869743\\
0.393939393939394	3.54938605205956\\
0.404040404040404	3.54951574388586\\
0.414141414141414	3.55007196302198\\
0.424242424242424	3.55106117831353\\
0.434343434343434	3.55248985860614\\
0.444444444444444	3.55436447274543\\
0.454545454545455	3.55669148957703\\
0.464646464646465	3.55947737794655\\
0.474747474747475	3.56272860669962\\
0.484848484848485	3.56645164468186\\
0.494949494949495	3.57065296073891\\
0.505050505050505	3.57533902371637\\
0.515151515151515	3.58051630245988\\
0.525252525252525	3.58619126581506\\
0.535353535353535	3.59237038262754\\
0.545454545454545	3.59906012174292\\
0.555555555555556	3.60626695200685\\
0.565656565656566	3.61399734226494\\
0.575757575757576	3.62225776136282\\
0.585858585858586	3.63105467814611\\
0.595959595959596	3.64039456146043\\
0.606060606060606	3.65028388015141\\
0.616161616161616	3.66072910306466\\
0.626262626262626	3.67173669904583\\
0.636363636363636	3.68331313694052\\
0.646464646464647	3.69546488559436\\
0.656565656565657	3.70819841385298\\
0.666666666666667	3.721520190562\\
0.676767676767677	3.73543668456703\\
0.686868686868687	3.74995436471372\\
0.696969696969697	3.76507969984767\\
0.707070707070707	3.78081915881452\\
0.717171717171717	3.79717921045988\\
0.727272727272727	3.81416632362938\\
0.737373737373737	3.83178696716865\\
0.747474747474748	3.8500476099233\\
0.757575757575758	3.86895472073896\\
0.767676767676768	3.88851476846126\\
0.777777777777778	3.90873422193582\\
0.787878787878788	3.92961955000825\\
0.797979797979798	3.95117722152419\\
0.808080808080808	3.97341370532926\\
0.818181818181818	3.99633547026909\\
0.828282828282828	4.01994898518928\\
0.838383838383838	4.04426071893548\\
0.848484848484849	4.0692771403533\\
0.858585858585859	4.09500471828836\\
0.868686868686869	4.1214499215863\\
0.878787878787879	4.14861921909272\\
0.888888888888889	4.17651907965327\\
0.898989898989899	4.20515597211355\\
0.909090909090909	4.2345363653192\\
0.919191919191919	4.26466672811584\\
0.929292929292929	4.29555352934909\\
0.939393939393939	4.32720323786457\\
0.94949494949495	4.35962232250791\\
0.95959595959596	4.39281725212473\\
0.96969696969697	4.42679449556066\\
0.97979797979798	4.46156052166131\\
0.98989898989899	4.49712179927232\\
1	4.5334847972393\\
};

\addlegendentry{Fit}

\end{axis}
\end{tikzpicture}%

%% file: Figures/Results_Cut/KlD.tikz
%
%
\definecolor{mycolor1}{rgb}{0.00000,0.44700,0.74100}%
\begin{tikzpicture}

\begin{axis}[%
width=1\columnwidth,
height=3.5/5*1\columnwidth,
xmin=0,
xmax=550,
xlabel style={font=\color{white!15!black}},
xlabel={$l/D_{PI}$},
ymin=0,
ymax=1,
ylabel style={font=\color{white!15!black}},
ylabel={$K/N_I$},
axis background/.style={fill=white},
]
\addplot [color=black, line width=1pt]
  table[row sep=crcr]{%
1	1\\
1.54954954954955	0.953539187653735\\
2.0990990990991	0.874222552731206\\
2.64864864864865	0.795829903978052\\
3.1981981981982	0.725837646668754\\
3.74774774774775	0.665051182987733\\
4.2972972972973	0.612557267805081\\
4.84684684684685	0.567121160681005\\
5.3963963963964	0.527585796468955\\
5.94594594594595	0.492967342427592\\
6.4954954954955	0.462457239275148\\
7.04504504504505	0.435398376573298\\
7.59459459459459	0.411257466081247\\
8.14414414414414	0.389600330024995\\
8.69369369369369	0.370071585522588\\
9.24324324324324	0.352378499175027\\
9.79279279279279	0.336278326983481\\
10.3423423423423	0.321568424578933\\
10.8918918918919	0.308078512396694\\
11.4414414414414	0.295664603040855\\
11.990990990991	0.284204208594551\\
12.5405405405405	0.273592535487906\\
13.0900900900901	0.263739444404471\\
13.6396396396396	0.25456700591716\\
14.1891891891892	0.24600752270108\\
14.7387387387387	0.238001919391457\\
15.2882882882883	0.2304984239173\\
15.8378378378378	0.22345148133739\\
16.3873873873874	0.216820854251121\\
16.9369369369369	0.210570873804042\\
17.4864864864865	0.204669812933894\\
18.036036036036	0.199089359382311\\
18.5855855855856	0.193804170556269\\
19.1351351351351	0.188791495878564\\
19.6846846846847	0.184030855054693\\
20.2342342342342	0.179503762882172\\
20.7837837837838	0.175193492971449\\
21.3333333333333	0.171084874136779\\
21.8828828828829	0.167164114328229\\
22.4324324324324	0.16341864787166\\
22.981981981982	0.159837002507958\\
23.5315315315315	0.156408683311285\\
24.0810810810811	0.153124071046373\\
24.6306306306306	0.149974332918418\\
25.1801801801802	0.146951343992944\\
25.7297297297297	0.144047617830514\\
26.2792792792793	0.141256245102949\\
26.8288288288288	0.138570839142349\\
27.3783783783784	0.135985487528345\\
27.9279279279279	0.133494708948268\\
28.4774774774775	0.131093414673513\\
29.027027027027	0.128776874087049\\
29.5765765765766	0.126540683774514\\
30.1261261261261	0.124380739757184\\
30.6756756756757	0.122293212501019\\
31.2252252252252	0.12027452438384\\
31.7747747747748	0.118321329343522\\
32.3243243243243	0.116430494465197\\
32.8738738738739	0.114599083295609\\
33.4234234234234	0.112824340698769\\
33.972972972973	0.111103679089556\\
34.5225225225225	0.109434665901344\\
35.0720720720721	0.107815012160667\\
35.6216216216216	0.106242562056617\\
36.1711711711712	0.104715283405495\\
36.7207207207207	0.103231258922425\\
37.2702702702703	0.101788678221456\\
37.8198198198198	0.100385830474258\\
38.3693693693694	0.0990210976650661\\
38.9189189189189	0.0976929483862007\\
39.4684684684685	0.0963999321243108\\
40.018018018018	0.095140673992703\\
40.5675675675676	0.0939138698696735\\
41.1171171171171	0.0927182819068318\\
41.6666666666667	0.091552734375\\
42.2162162162162	0.0904161098184802\\
42.7657657657658	0.0893073454913279\\
43.3153153153153	0.088225430051816\\
43.8648648648649	0.0871694004935404\\
44.4144144144144	0.08613833929365\\
44.963963963964	0.0851313717604998\\
45.5135135135135	0.0841476635646545\\
46.0630630630631	0.0831864184386352\\
46.6126126126126	0.082246876032111\\
47.1621621621622	0.0813283099104274\\
47.7117117117117	0.0804300256854226\\
48.2612612612613	0.0795513592684497\\
48.8108108108108	0.0786916752363874\\
49.3603603603604	0.0778503653022104\\
49.9099099099099	0.077026846882399\\
50.4594594594595	0.0762205617541134\\
51.009009009009	0.0754309747956412\\
51.5585585585586	0.0746575728041581\\
52.1081081081081	0.0738998633853246\\
52.6576576576577	0.0731573739096784\\
53.2072072072072	0.0724296505311843\\
53.7567567567568	0.0717162572636671\\
54.3063063063063	0.0710167751111839\\
54.8558558558559	0.0703308012486993\\
55.4054054054054	0.0696579482497006\\
55.954954954955	0.0689978433576486\\
56.5045045045045	0.0683501277983886\\
57.0540540540541	0.0677144561308601\\
57.6036036036036	0.0670904956336431\\
58.1531531531532	0.066477925725052\\
58.7027027027027	0.0658764374146589\\
59.2522522522523	0.0652857327842769\\
59.8018018018018	0.0647055244965746\\
60.3513513513514	0.0641355353296202\\
60.9009009009009	0.0635754977357753\\
61.4504504504505	0.0630251534234634\\
62	0.0624842529604434\\
62.5495495495495	0.0619525553973081\\
63.0990990990991	0.0614298279100156\\
63.6486486486486	0.0609158454603416\\
64.1981981981982	0.0604103904732121\\
64.7477477477478	0.0599132525299468\\
65.2972972972973	0.0594242280765052\\
65.8468468468468	0.0589431201458868\\
66.3963963963964	0.0584697380938899\\
66.945945945946	0.0580038973474873\\
67.4954954954955	0.0575454191651194\\
68.045045045045	0.0570941304082531\\
68.5945945945946	0.0566498633235932\\
69.1441441441442	0.0562124553353715\\
69.6936936936937	0.0557817488471737\\
70.2432432432432	0.0553575910527976\\
70.7927927927928	0.0549398337556652\\
71.3423423423423	0.0545283331963419\\
71.8918918918919	0.0541229498877411\\
72.4414414414414	0.0537235484576173\\
72.990990990991	0.0533299974979752\\
73.5405405405405	0.0529421694210421\\
74.0900900900901	0.0525599403214738\\
74.6396396396396	0.05218318984448\\
75.1891891891892	0.0518118010595763\\
75.7387387387387	0.0514456603396831\\
76.2882882882883	0.0510846572453105\\
76.8378378378378	0.0507286844135802\\
77.3873873873874	0.0503776374518523\\
77.9369369369369	0.0500314148357324\\
78.4864864864865	0.0496899178112536\\
79.036036036036	0.0493530503010334\\
79.5855855855856	0.0490207188142181\\
80.1351351351351	0.0486928323600389\\
80.6846846846847	0.0483693023648114\\
81.2342342342342	0.0480500425922185\\
81.7837837837838	0.0477349690667279\\
82.3333333333333	0.047424\\
82.8828828828829	0.0471170557201516\\
83.4324324324324	0.0468140586037468\\
83.981981981982	0.0465149330103928\\
84.5315315315315	0.0462196052198257\\
85.0810810810811	0.0459280033713763\\
85.6306306306306	0.0456400574057104\\
86.1801801801802	0.0453556990087462\\
86.7297297297297	0.0450748615576534\\
87.2792792792793	0.044797480068845\\
87.8288288288288	0.0445234911478755\\
88.3783783783784	0.0442528329411666\\
88.9279279279279	0.04398544508948\\
89.4774774774775	0.0437212686830677\\
90.027027027027	0.0434602462184258\\
90.5765765765766	0.0432023215565888\\
91.1261261261261	0.0429474398828984\\
91.6756756756757	0.0426955476681872\\
92.2252252252252	0.0424465926313199\\
92.7747747747748	0.042200523703036\\
93.3243243243243	0.0419572909910428\\
93.8738738738739	0.0417168457463074\\
94.4234234234234	0.0414791403305008\\
94.972972972973	0.0412441281845476\\
95.5225225225225	0.0410117637982387\\
96.0720720720721	0.0407820026808641\\
96.6216216216216	0.0405548013328281\\
97.1711711711712	0.0403301172182057\\
97.7207207207207	0.0401079087382074\\
98.2702702702703	0.0398881352055148\\
98.8198198198198	0.0396707568194555\\
99.3693693693694	0.0394557346419847\\
99.9189189189189	0.0392430305744431\\
100.468468468468	0.0390326073350626\\
101.018018018018	0.0388244284371908\\
101.567567567568	0.038618458168209\\
102.117117117117	0.0384146615691169\\
102.666666666667	0.03821300441476\\
103.216216216216	0.0380134531946764\\
103.765765765766	0.0378159750945403\\
104.315315315315	0.0376205379781803\\
104.864864864865	0.0374271103701526\\
105.414414414414	0.0372356614388485\\
105.963963963964	0.0370461609801177\\
106.513513513514	0.0368585794013884\\
107.063063063063	0.0366728877062678\\
107.612612612613	0.0364890574796052\\
108.162162162162	0.0363070608730022\\
108.711711711712	0.0361268705907535\\
109.261261261261	0.0359484598762049\\
109.810810810811	0.0357718024985128\\
110.36036036036	0.0355968727397925\\
110.90990990991	0.0354236453826409\\
111.459459459459	0.0352520956980224\\
112.009009009009	0.0350821994335043\\
112.558558558559	0.0349139328018307\\
113.108108108108	0.0347472724698243\\
113.657657657658	0.0345821955476031\\
114.207207207207	0.0344186795781038\\
114.756756756757	0.0342567025269012\\
115.306306306306	0.0340962427723122\\
115.855855855856	0.0339372790957781\\
116.405405405405	0.0337797906725137\\
116.954954954955	0.0336237570624157\\
117.504504504505	0.0334691582012214\\
118.054054054054	0.033315974391911\\
118.603603603604	0.0331641862963439\\
119.153153153153	0.0330137749271234\\
119.702702702703	0.0328647216396803\\
120.252252252252	0.032717008124572\\
120.801801801802	0.032570616399986\\
121.351351351351	0.0324255288044455\\
121.900900900901	0.032281727989708\\
122.45045045045	0.0321391969138521\\
123	0.0319979188345473\\
123.54954954955	0.0318578773024993\\
124.099099099099	0.0317190561550673\\
124.648648648649	0.0315814395100476\\
125.198198198198	0.0314450117596178\\
125.747747747748	0.0313097575644375\\
126.297297297297	0.0311756618479001\\
126.846846846847	0.0310427097905325\\
127.396396396396	0.0309108868245366\\
127.945945945946	0.0307801786284696\\
128.495495495496	0.0306505711220588\\
129.045045045045	0.0305220504611466\\
129.594594594595	0.0303946030327617\\
130.144144144144	0.0302682154503145\\
130.693693693694	0.0301428745489104\\
131.243243243243	0.0300185673807797\\
131.792792792793	0.0298952812108202\\
132.342342342342	0.0297730035122491\\
132.891891891892	0.0296517219623605\\
133.441441441441	0.0295314244383875\\
133.990990990991	0.0294120990134627\\
134.540540540541	0.0292937339526784\\
135.09009009009	0.0291763177092393\\
135.63963963964	0.0290598389207093\\
136.189189189189	0.0289442864053464\\
136.738738738739	0.0288296491585254\\
137.288288288288	0.0287159163492451\\
137.837837837838	0.0286030773167176\\
138.387387387387	0.028491121567038\\
138.936936936937	0.0283800387699319\\
139.486486486486	0.0282698187555784\\
140.036036036036	0.0281604515115074\\
140.585585585586	0.0280519271795676\\
141.135135135135	0.027944236052966\\
141.684684684685	0.0278373685733731\\
142.234234234234	0.0277313153280967\\
142.783783783784	0.0276260670473175\\
143.333333333333	0.0275216146013899\\
143.882882882883	0.0274179489982018\\
144.432432432432	0.0273150613805946\\
144.981981981982	0.0272129430238417\\
145.531531531532	0.027111585333182\\
146.081081081081	0.0270109798414093\\
146.630630630631	0.026911118206515\\
147.18018018018	0.0268119922093824\\
147.72972972973	0.0267135937515324\\
148.279279279279	0.0266159148529182\\
148.828828828829	0.0265189476497684\\
149.378378378378	0.0264226843924769\\
149.927927927928	0.0263271174435388\\
150.477477477477	0.0262322392755311\\
151.027027027027	0.0261380424691358\\
151.576576576577	0.0260445197112071\\
152.126126126126	0.0259516637928785\\
152.675675675676	0.0258594676077113\\
153.225225225225	0.0257679241498813\\
153.774774774775	0.0256770265124051\\
154.324324324324	0.025586767885403\\
154.873873873874	0.0254971415543983\\
155.423423423423	0.0254081408986531\\
155.972972972973	0.0253197593895378\\
156.522522522523	0.0252319905889349\\
157.072072072072	0.0251448281476763\\
157.621621621622	0.0250582658040124\\
158.171171171171	0.0249722973821122\\
158.720720720721	0.0248869167905958\\
159.27027027027	0.0248021180210949\\
159.81981981982	0.0247178951468442\\
160.369369369369	0.0246342423213004\\
160.918918918919	0.0245511537767899\\
161.468468468468	0.024468623823183\\
162.018018018018	0.0243866468465952\\
162.567567567568	0.0243052173081149\\
163.117117117117	0.0242243297425553\\
163.666666666667	0.0241439787572325\\
164.216216216216	0.0240641590307671\\
164.765765765766	0.0239848653119093\\
165.315315315315	0.0239060924183876\\
165.864864864865	0.0238278352357803\\
166.414414414414	0.0237500887164083\\
166.963963963964	0.02367284787825\\
167.513513513514	0.0235961078038777\\
168.063063063063	0.0235198636394134\\
168.612612612613	0.0234441105935063\\
169.162162162162	0.0233688439363287\\
169.711711711712	0.0232940589985923\\
170.261261261261	0.0232197511705819\\
170.810810810811	0.0231459159012091\\
171.36036036036	0.0230725486970829\\
171.90990990991	0.0229996451215985\\
172.459459459459	0.0229272007940429\\
173.009009009009	0.0228552113887175\\
173.558558558559	0.0227836726340775\\
174.108108108108	0.022712580311887\\
174.657657657658	0.0226419302563899\\
175.207207207207	0.0225717183534964\\
175.756756756757	0.0225019405399845\\
176.306306306306	0.0224325928027162\\
176.855855855856	0.0223636711778683\\
177.405405405405	0.0222951717501774\\
177.954954954955	0.0222270906521981\\
178.504504504505	0.0221594240635759\\
179.054054054054	0.022092168210332\\
179.603603603604	0.0220253193641623\\
180.153153153153	0.0219588738417479\\
180.702702702703	0.0218928280040795\\
181.252252252252	0.0218271782557924\\
181.801801801802	0.0217619210445147\\
182.351351351351	0.0216970528602261\\
182.900900900901	0.0216325702346289\\
183.45045045045	0.0215684697405303\\
184	0.0215047479912345\\
184.54954954955	0.0214414016399468\\
185.099099099099	0.0213784273791879\\
185.648648648649	0.0213158219402176\\
186.198198198198	0.0212535820924701\\
186.747747747748	0.0211917046429979\\
187.297297297297	0.0211301864359259\\
187.846846846847	0.0210690243519149\\
188.396396396396	0.0210082153076348\\
188.945945945946	0.0209477562552457\\
189.495495495496	0.0208876441818897\\
190.045045045045	0.0208278761091898\\
190.594594594595	0.0207684490927584\\
191.144144144144	0.0207093602217133\\
191.693693693694	0.0206506066182034\\
192.243243243243	0.0205921854369407\\
192.792792792793	0.0205340938647412\\
193.342342342342	0.0204763291200738\\
193.891891891892	0.0204188884526158\\
194.441441441441	0.020361769142817\\
194.990990990991	0.0203049685014701\\
195.540540540541	0.020248483869289\\
196.09009009009	0.0201923126164936\\
196.63963963964	0.0201364521424017\\
197.189189189189	0.0200808998750281\\
197.738738738739	0.0200256532706893\\
198.288288288288	0.0199707098136156\\
198.837837837838	0.0199160670155692\\
199.387387387387	0.0198617224154688\\
199.936936936937	0.0198076735790197\\
200.486486486486	0.0197539180983509\\
201.036036036036	0.019700453591657\\
201.585585585586	0.0196472777028469\\
202.135135135135	0.0195943881011975\\
202.684684684685	0.0195417824810133\\
203.234234234234	0.0194894585612913\\
203.783783783784	0.0194374140853913\\
204.333333333333	0.0193856468207118\\
204.882882882883	0.0193341545583702\\
205.432432432432	0.0192829351128894\\
205.981981981982	0.0192319863218879\\
206.531531531532	0.0191813060457761\\
207.081081081081	0.0191308921674563\\
207.630630630631	0.0190807425920285\\
208.18018018018	0.0190308552464996\\
208.72972972973	0.0189812280794984\\
209.279279279279	0.0189318590609939\\
209.828828828829	0.0188827461820191\\
210.378378378378	0.018833887454398\\
210.927927927928	0.0187852809104775\\
211.477477477477	0.0187369246028634\\
212.027027027027	0.0186888166041603\\
212.576576576577	0.0186409550067154\\
213.126126126126	0.0185933379223666\\
213.675675675676	0.0185459634821939\\
214.225225225225	0.0184988298362752\\
214.774774774775	0.0184519351534455\\
215.324324324324	0.0184052776210595\\
215.873873873874	0.0183588554447585\\
216.423423423423	0.0183126668482399\\
216.972972972973	0.0182667100730311\\
217.522522522523	0.018220983378266\\
218.072072072072	0.0181754850404654\\
218.621621621622	0.0181302133533204\\
219.171171171171	0.0180851666274794\\
219.720720720721	0.0180403431903374\\
220.27027027027	0.0179957413858293\\
220.81981981982	0.0179513595742264\\
221.369369369369	0.0179071961319345\\
221.918918918919	0.0178632494512969\\
222.468468468468	0.0178195179403987\\
223.018018018018	0.017776000022875\\
223.567567567568	0.017732694137721\\
224.117117117117	0.017689598739106\\
224.666666666667	0.017646712296189\\
225.216216216216	0.0176040332929376\\
225.765765765766	0.0175615602279495\\
226.315315315315	0.0175192916142761\\
226.864864864865	0.0174772259792494\\
227.414414414414	0.0174353618643109\\
227.963963963964	0.0173936978248426\\
228.513513513514	0.0173522324300014\\
229.063063063063	0.0173109642625551\\
229.612612612613	0.0172698919187209\\
230.162162162162	0.0172290140080061\\
230.711711711712	0.0171883291530515\\
231.261261261261	0.0171478359894767\\
231.810810810811	0.0171075331657273\\
232.36036036036	0.0170674193429247\\
232.90990990991	0.0170274931947179\\
233.459459459459	0.0169877534071373\\
234.009009009009	0.0169481986784505\\
234.558558558559	0.0169088277190202\\
235.108108108108	0.016869639251164\\
235.657657657658	0.0168306320090164\\
236.207207207207	0.0167918047383923\\
236.756756756757	0.0167531561966526\\
237.306306306306	0.0167146851525718\\
237.855855855856	0.0166763903862071\\
238.405405405405	0.0166382706887695\\
238.954954954955	0.0166003248624966\\
239.504504504505	0.0165625517205273\\
240.054054054054	0.0165249500867778\\
240.603603603604	0.0164875187958199\\
241.153153153153	0.0164502566927603\\
241.702702702703	0.0164131626331219\\
242.252252252252	0.0163762354827266\\
242.801801801802	0.01633947411758\\
243.351351351351	0.016302877423757\\
243.900900900901	0.0162664442972894\\
244.45045045045	0.0162301736440551\\
245	0.0161940643796682\\
245.54954954955	0.0161581154293713\\
246.099099099099	0.0161223257279286\\
246.648648648649	0.0160866942195211\\
247.198198198198	0.0160512198576421\\
247.747747747748	0.0160159016049955\\
248.297297297297	0.0159807384333941\\
248.846846846847	0.0159457293236602\\
249.396396396396	0.0159108732655271\\
249.945945945946	0.0158761692575416\\
250.495495495496	0.0158416163069685\\
251.045045045045	0.0158072134296958\\
251.594594594595	0.0157729596501413\\
252.144144144144	0.0157388540011601\\
252.693693693694	0.015704895523954\\
253.243243243243	0.0156710832679816\\
253.792792792793	0.0156374162908692\\
254.342342342342	0.0156038936583239\\
254.891891891892	0.0155705144440464\\
255.441441441441	0.0155372777296466\\
255.990990990991	0.0155041826045585\\
256.540540540541	0.0154712281659576\\
257.09009009009	0.0154384135186786\\
257.63963963964	0.0154057377751346\\
258.189189189189	0.0153732000552365\\
258.738738738739	0.0153407994863148\\
259.288288288288	0.0153085352030407\\
259.837837837838	0.0152764063473499\\
260.387387387387	0.0152444120683659\\
260.936936936937	0.0152125515223253\\
261.486486486487	0.0151808238725034\\
262.036036036036	0.015149228289141\\
262.585585585586	0.0151177639493723\\
263.135135135135	0.0150864300371529\\
263.684684684685	0.01505522574319\\
264.234234234234	0.0150241502648721\\
264.783783783784	0.0149932028062006\\
265.333333333333	0.0149623825777215\\
265.882882882883	0.0149316887964587\\
266.432432432432	0.0149011206858472\\
266.981981981982	0.014870677475668\\
267.531531531532	0.0148403584019832\\
268.081081081081	0.0148101627070722\\
268.630630630631	0.0147800896393687\\
269.18018018018	0.014750138453398\\
269.72972972973	0.0147203084097157\\
270.279279279279	0.0146905987748467\\
270.828828828829	0.0146610088212252\\
271.378378378378	0.0146315378271353\\
271.927927927928	0.0146021850766522\\
272.477477477477	0.0145729498595843\\
273.027027027027	0.014543831471416\\
273.576576576577	0.0145148292132511\\
274.126126126126	0.0144859423917567\\
274.675675675676	0.014457170319108\\
275.225225225225	0.0144285123129342\\
275.774774774775	0.0143999676962636\\
276.324324324324	0.0143715357974714\\
276.873873873874	0.014343215950226\\
277.423423423423	0.0143150074934378\\
277.972972972973	0.0142869097712071\\
278.522522522523	0.0142589221327737\\
279.072072072072	0.0142310439324664\\
279.621621621622	0.0142032745296537\\
280.171171171171	0.0141756132886942\\
280.720720720721	0.0141480595788886\\
281.27027027027	0.0141206127744315\\
281.81981981982	0.0140932722543639\\
282.369369369369	0.014066037402527\\
282.918918918919	0.0140389076075149\\
283.468468468468	0.01401188226263\\
284.018018018018	0.0139849607658367\\
284.567567567568	0.0139581425197175\\
285.117117117117	0.0139314269314285\\
285.666666666667	0.0139048134126555\\
286.216216216216	0.0138783013795713\\
286.765765765766	0.0138518902527925\\
287.315315315315	0.0138255794573379\\
287.864864864865	0.013799368422586\\
288.414414414414	0.0137732565822344\\
288.963963963964	0.0137472433742588\\
289.513513513513	0.0137213282408727\\
290.063063063063	0.0136955106284873\\
290.612612612613	0.0136697899876725\\
291.162162162162	0.0136441657731175\\
291.711711711712	0.0136186374435928\\
292.261261261261	0.0135932044619114\\
292.810810810811	0.0135678662948916\\
293.36036036036	0.0135426224133199\\
293.90990990991	0.0135174722919136\\
294.459459459459	0.0134924154092846\\
295.009009009009	0.0134674512479037\\
295.558558558559	0.0134425792940646\\
296.108108108108	0.0134177990378486\\
296.657657657658	0.0133931099730901\\
297.207207207207	0.0133685115973419\\
297.756756756757	0.0133440034118412\\
298.306306306306	0.0133195849214758\\
298.855855855856	0.0132952556347509\\
299.405405405405	0.013271015063756\\
299.954954954955	0.0132468627241324\\
300.504504504505	0.0132227981350407\\
301.054054054054	0.0131988208191292\\
301.603603603604	0.0131749303025023\\
302.153153153153	0.0131511261146891\\
302.702702702703	0.0131274077886124\\
303.252252252252	0.0131037748605586\\
303.801801801802	0.0130802268701472\\
304.351351351351	0.013056763360301\\
304.900900900901	0.0130333838772162\\
305.45045045045	0.0130100879703337\\
306	0.0129868751923097\\
306.54954954955	0.0129637450989874\\
307.099099099099	0.0129406972493683\\
307.648648648649	0.0129177312055846\\
308.198198198198	0.012894846532871\\
308.747747747748	0.0128720427995378\\
309.297297297297	0.0128493195769433\\
309.846846846847	0.0128266764394672\\
310.396396396396	0.0128041129644839\\
310.945945945946	0.0127816287323363\\
311.495495495496	0.0127592233263097\\
312.045045045045	0.0127368963326061\\
312.594594594595	0.0127146473403187\\
313.144144144144	0.0126924759414069\\
313.693693693694	0.0126703817306708\\
314.243243243243	0.0126483643057273\\
314.792792792793	0.0126264232669849\\
315.342342342342	0.0126045582176202\\
315.891891891892	0.0125827687635535\\
316.441441441441	0.0125610545134253\\
316.990990990991	0.0125394150785729\\
317.540540540541	0.0125178500730071\\
318.09009009009	0.0124963591133894\\
318.63963963964	0.0124749418190091\\
319.189189189189	0.0124535978117612\\
319.738738738739	0.0124323267161235\\
320.288288288288	0.0124111281591353\\
320.837837837838	0.012390001770375\\
321.387387387387	0.012368947181939\\
321.936936936937	0.0123479640284199\\
322.486486486487	0.0123270519468858\\
323.036036036036	0.0123062105768591\\
323.585585585586	0.0122854395602959\\
324.135135135135	0.0122647385415652\\
324.684684684685	0.0122441071674292\\
325.234234234234	0.0122235450870226\\
325.783783783784	0.0122030519518332\\
326.333333333333	0.0121826274156819\\
326.882882882883	0.0121622711347031\\
327.432432432432	0.012141982767326\\
327.981981981982	0.0121217619742546\\
328.531531531532	0.0121016084184498\\
329.081081081081	0.0120815217651097\\
329.630630630631	0.0120615016816518\\
330.18018018018	0.0120415478376942\\
330.72972972973	0.0120216599050378\\
331.279279279279	0.0120018375576478\\
331.828828828829	0.0119820804716363\\
332.378378378378	0.0119623883252447\\
332.927927927928	0.0119427607988258\\
333.477477477477	0.0119231975748269\\
334.027027027027	0.0119036983377727\\
334.576576576577	0.0118842627742479\\
335.126126126126	0.011864890572881\\
335.675675675676	0.011845581424327\\
336.225225225225	0.0118263350212517\\
336.774774774775	0.0118071510583148\\
337.324324324324	0.011788029232154\\
337.873873873874	0.0117689692413689\\
338.423423423423	0.0117499707865053\\
338.972972972973	0.0117310335700394\\
339.522522522523	0.0117121572963625\\
340.072072072072	0.0116933416717651\\
340.621621621622	0.0116745864044224\\
341.171171171171	0.0116558912043784\\
341.720720720721	0.0116372557835317\\
342.27027027027	0.0116186798556202\\
342.81981981982	0.0116001631362068\\
343.369369369369	0.0115817053426644\\
343.918918918919	0.0115633061941622\\
344.468468468468	0.0115449654116508\\
345.018018018018	0.0115266827178483\\
345.567567567568	0.0115084578372268\\
346.117117117117	0.0114902904959978\\
346.666666666667	0.011472180422099\\
347.216216216216	0.0114541273451805\\
347.765765765766	0.0114361309965915\\
348.315315315315	0.0114181911093667\\
348.864864864865	0.0114003074182136\\
349.414414414414	0.0113824796594989\\
349.963963963964	0.0113647075712358\\
350.513513513513	0.0113469908930711\\
351.063063063063	0.0113293293662725\\
351.612612612613	0.0113117227337161\\
352.162162162162	0.0112941707398739\\
352.711711711712	0.0112766731308012\\
353.261261261261	0.0112592296541245\\
353.810810810811	0.0112418400590297\\
354.36036036036	0.0112245040962495\\
354.90990990991	0.0112072215180519\\
355.459459459459	0.0111899920782282\\
356.009009009009	0.0111728155320813\\
356.558558558559	0.0111556916364145\\
357.108108108108	0.0111386201495194\\
357.657657657658	0.011121600831165\\
358.207207207207	0.0111046334425861\\
358.756756756757	0.0110877177464726\\
359.306306306306	0.0110708535069579\\
359.855855855856	0.0110540404896083\\
360.405405405405	0.011037278461412\\
360.954954954955	0.0110205671907683\\
361.504504504505	0.011003906447477\\
362.054054054054	0.010987296002728\\
362.603603603604	0.0109707356290904\\
363.153153153153	0.0109542251005025\\
363.702702702703	0.0109377641922614\\
364.252252252252	0.0109213526810126\\
364.801801801802	0.0109049903447403\\
365.351351351351	0.0108886769627572\\
365.900900900901	0.0108724123156942\\
366.45045045045	0.0108561961854914\\
367	0.0108400283553875\\
367.54954954955	0.0108239086099107\\
368.099099099099	0.010807836734869\\
368.648648648649	0.0107918125173403\\
369.198198198198	0.0107758357456639\\
369.747747747748	0.0107599062094301\\
370.297297297297	0.0107440236994719\\
370.846846846847	0.0107281880078553\\
371.396396396396	0.0107123989278703\\
371.945945945946	0.010696656254022\\
372.495495495496	0.0106809597820218\\
373.045045045045	0.0106653093087783\\
373.594594594595	0.0106497046323886\\
374.144144144144	0.0106341455521299\\
374.693693693694	0.0106186318684505\\
375.243243243243	0.0106031633829614\\
375.792792792793	0.0105877398984282\\
376.342342342342	0.010572361218762\\
376.891891891892	0.0105570271490118\\
377.441441441441	0.0105417374953558\\
377.990990990991	0.0105264920650935\\
378.540540540541	0.0105112906666374\\
379.09009009009	0.0104961331095051\\
379.63963963964	0.0104810192043115\\
380.189189189189	0.0104659487627603\\
380.738738738739	0.0104509215976371\\
381.288288288288	0.0104359375228009\\
381.837837837838	0.0104209963531766\\
382.387387387387	0.0104060979047476\\
382.936936936937	0.0103912419945481\\
383.486486486487	0.0103764284406556\\
384.036036036036	0.0103616570621834\\
384.585585585586	0.0103469276792733\\
385.135135135135	0.0103322401130885\\
385.684684684685	0.010317594185806\\
386.234234234234	0.0103029897206098\\
386.783783783784	0.0102884265416833\\
387.333333333333	0.0102739044742029\\
387.882882882883	0.0102594233443303\\
388.432432432432	0.0102449829792061\\
388.981981981982	0.0102305832069426\\
389.531531531532	0.0102162238566174\\
390.081081081081	0.0102019047582659\\
390.630630630631	0.0101876257428754\\
391.18018018018	0.010173386642378\\
391.72972972973	0.0101591872896439\\
392.279279279279	0.0101450275184753\\
392.828828828829	0.0101309071635995\\
393.378378378378	0.0101168260606629\\
393.927927927928	0.010102784046224\\
394.477477477477	0.0100887809577478\\
395.027027027027	0.0100748166335988\\
395.576576576577	0.0100608909130355\\
396.126126126126	0.0100470036362037\\
396.675675675676	0.0100331546441305\\
397.225225225225	0.0100193437787186\\
397.774774774775	0.0100055708827396\\
398.324324324324	0.00999183579982879\\
398.873873873874	0.00997813837447855\\
399.423423423423	0.00996447845203296\\
399.972972972973	0.00995085587868178\\
400.522522522523	0.00993727050145472\\
401.072072072072	0.00992372216821568\\
401.621621621622	0.00991021072765711\\
402.171171171171	0.00989673602929437\\
402.720720720721	0.00988329792346012\\
403.27027027027	0.00986989626129881\\
403.81981981982	0.00985653089476115\\
404.369369369369	0.00984320167659869\\
404.918918918919	0.00982990846035841\\
405.468468468468	0.00981665110037732\\
406.018018018018	0.00980342945177719\\
406.567567567568	0.00979024337045923\\
407.117117117117	0.00977709271309889\\
407.666666666667	0.00976397733714064\\
408.216216216216	0.00975089710079283\\
408.765765765766	0.00973785186302257\\
409.315315315315	0.00972484148355071\\
409.864864864865	0.00971186582284673\\
410.414414414414	0.00969892474212381\\
410.963963963964	0.00968601810333387\\
411.513513513513	0.00967314576916266\\
412.063063063063	0.00966030760302489\\
412.612612612613	0.00964750346905939\\
413.162162162162	0.00963473323212434\\
413.711711711712	0.00962199675779247\\
414.261261261261	0.00960929391234642\\
414.810810810811	0.00959662456277397\\
415.36036036036	0.00958398857676348\\
415.90990990991	0.00957138582269922\\
416.459459459459	0.00955881616965683\\
417.009009009009	0.00954627948739881\\
417.558558558559	0.00953377564636997\\
418.108108108108	0.00952130451769298\\
418.657657657658	0.00950886597316399\\
419.207207207207	0.00949645988524819\\
419.756756756757	0.00948408612707544\\
420.306306306306	0.00947174457243599\\
420.855855855856	0.00945943509577617\\
421.405405405405	0.00944715757219412\\
421.954954954955	0.00943491187743558\\
422.504504504505	0.00942269788788969\\
423.054054054054	0.00941051548058483\\
423.603603603604	0.00939836453318452\\
424.153153153153	0.00938624492398328\\
424.702702702703	0.0093741565319026\\
425.252252252252	0.00936209923648689\\
425.801801801802	0.00935007291789948\\
426.351351351351	0.00933807745691869\\
426.900900900901	0.00932611273493382\\
427.45045045045	0.0093141786339413\\
428	0.00930227503654077\\
428.54954954955	0.00929040182593126\\
429.099099099099	0.00927855888590737\\
429.648648648649	0.00926674610085543\\
430.198198198198	0.00925496335574983\\
430.747747747748	0.00924321053614917\\
431.297297297297	0.00923148752819267\\
431.846846846847	0.0092197942185964\\
432.396396396396	0.00920813049464971\\
432.945945945946	0.00919649624421153\\
433.495495495496	0.00918489135570684\\
434.045045045045	0.00917331571812307\\
434.594594594595	0.00916176922100658\\
435.144144144144	0.00915025175445913\\
435.693693693694	0.00913876320913438\\
436.243243243243	0.00912730347623448\\
436.792792792793	0.00911587244750659\\
437.342342342342	0.0091044700152395\\
437.891891891892	0.0090930960722602\\
438.441441441441	0.00908175051193058\\
438.990990990991	0.00907043322814409\\
439.540540540541	0.00905914411532237\\
440.09009009009	0.00904788306841205\\
440.63963963964	0.00903664998288143\\
441.189189189189	0.00902544475471729\\
441.738738738739	0.00901426728042164\\
442.288288288288	0.00900311745700857\\
442.837837837838	0.00899199518200104\\
443.387387387387	0.00898090035342781\\
443.936936936937	0.00896983286982024\\
444.486486486487	0.00895879263020928\\
445.036036036036	0.00894777953412234\\
445.585585585586	0.00893679348158026\\
446.135135135135	0.00892583437309429\\
446.684684684685	0.00891490210966309\\
447.234234234234	0.00890399659276973\\
447.783783783784	0.00889311772437876\\
448.333333333333	0.00888226540693323\\
448.882882882883	0.00887143954335183\\
449.432432432432	0.00886064003702594\\
449.981981981982	0.00884986679181679\\
450.531531531532	0.00883911971205261\\
451.081081081081	0.00882839870252576\\
451.630630630631	0.00881770366848996\\
452.18018018018	0.00880703451565749\\
452.72972972973	0.00879639115019641\\
453.279279279279	0.00878577347872779\\
453.828828828829	0.00877518140832303\\
454.378378378378	0.00876461484650111\\
454.927927927928	0.00875407370122591\\
455.477477477477	0.00874355788090353\\
456.027027027027	0.00873306729437964\\
456.576576576577	0.00872260185093688\\
457.126126126126	0.00871216146029217\\
457.675675675676	0.0087017460325942\\
458.225225225225	0.00869135547842079\\
458.774774774775	0.00868098970877635\\
459.324324324324	0.00867064863508934\\
459.873873873874	0.00866033216920978\\
460.423423423423	0.00865004022340667\\
460.972972972973	0.00863977271036558\\
461.522522522523	0.00862952954318612\\
462.072072072072	0.00861931063537956\\
462.621621621622	0.0086091159008663\\
463.171171171171	0.00859894525397353\\
463.720720720721	0.00858879860943282\\
464.27027027027	0.0085786758823777\\
464.81981981982	0.00856857698834133\\
465.369369369369	0.00855850184325413\\
465.918918918919	0.00854845036344145\\
466.468468468468	0.00853842246562128\\
467.018018018018	0.00852841806690189\\
467.567567567568	0.00851843708477963\\
468.117117117117	0.00850847943713658\\
468.666666666667	0.00849854504223835\\
469.216216216216	0.00848863381873183\\
469.765765765766	0.00847874568564295\\
470.315315315315	0.00846888056237451\\
470.864864864865	0.00845903836870398\\
471.414414414414	0.00844921902478128\\
471.963963963964	0.00843942245112669\\
472.513513513513	0.00842964856862868\\
473.063063063063	0.00841989729854174\\
473.612612612613	0.00841016856248434\\
474.162162162162	0.00840046228243676\\
474.711711711712	0.00839077838073904\\
475.261261261261	0.00838111678008891\\
475.810810810811	0.00837147740353973\\
476.36036036036	0.00836186017449843\\
476.90990990991	0.0083522650167235\\
477.459459459459	0.00834269185432297\\
478.009009009009	0.00833314061175242\\
478.558558558559	0.00832361121381296\\
479.108108108108	0.00831410358564929\\
479.657657657658	0.00830461765274774\\
480.207207207207	0.00829515334093429\\
480.756756756757	0.00828571057637267\\
481.306306306306	0.00827628928556244\\
481.855855855856	0.00826688939533705\\
482.405405405405	0.00825751083286198\\
482.954954954955	0.00824815352563284\\
483.504504504505	0.00823881740147352\\
484.054054054054	0.0082295023885343\\
484.603603603604	0.00822020841529006\\
485.153153153153	0.00821093541053839\\
485.702702702703	0.00820168330339781\\
486.252252252252	0.00819245202330596\\
486.801801801802	0.0081832415000178\\
487.351351351351	0.00817405166360382\\
487.900900900901	0.00816488244444829\\
488.45045045045	0.00815573377324747\\
489	0.00814660558100791\\
489.54954954955	0.00813749779904469\\
490.099099099099	0.00812841035897967\\
490.648648648649	0.00811934319273984\\
491.198198198198	0.00811029623255556\\
491.747747747748	0.00810126941095892\\
492.297297297297	0.00809226266078202\\
492.846846846847	0.00808327591515536\\
493.396396396396	0.00807430910750611\\
493.945945945946	0.00806536217155653\\
494.495495495496	0.00805643504132231\\
495.045045045045	0.00804752765111096\\
495.594594594595	0.00803863993552017\\
496.144144144144	0.00802977182943625\\
496.693693693694	0.0080209232680325\\
497.243243243243	0.00801209418676768\\
497.792792792793	0.00800328452138437\\
498.342342342342	0.00799449420790749\\
498.891891891892	0.00798572318264269\\
499.441441441441	0.00797697138217484\\
499.990990990991	0.00796823874336651\\
500.540540540541	0.00795952520335642\\
501.09009009009	0.00795083069955796\\
501.63963963964	0.00794215516965768\\
502.189189189189	0.0079334985516138\\
502.738738738739	0.00792486078365473\\
503.288288288288	0.00791624180427762\\
503.837837837838	0.00790764155224688\\
504.387387387387	0.00789905996659272\\
504.936936936937	0.00789049698660974\\
505.486486486487	0.00788195255185548\\
506.036036036036	0.00787342660214899\\
506.585585585586	0.00786491907756945\\
507.135135135135	0.00785642991845472\\
507.684684684685	0.00784795906539998\\
508.234234234234	0.00783950645925634\\
508.783783783784	0.00783107204112945\\
509.333333333333	0.00782265575237813\\
509.882882882883	0.00781425753461304\\
510.432432432432	0.00780587732969529\\
510.981981981982	0.00779751507973514\\
511.531531531532	0.00778917072709062\\
512.081081081081	0.00778084421436625\\
512.630630630631	0.0077725354844117\\
513.18018018018	0.00776424448032048\\
513.72972972973	0.00775597114542866\\
514.279279279279	0.00774771542331357\\
514.828828828829	0.00773947725779249\\
515.378378378378	0.00773125659292143\\
515.927927927928	0.00772305337299381\\
516.477477477477	0.00771486754253924\\
517.027027027027	0.00770669904632226\\
517.576576576577	0.00769854782934108\\
518.126126126126	0.00769041383682636\\
518.675675675676	0.00768229701423999\\
519.225225225225	0.00767419730727386\\
519.774774774775	0.00766611466184864\\
520.324324324324	0.00765804902411261\\
520.873873873874	0.00765000034044043\\
521.423423423423	0.00764196855743195\\
521.972972972973	0.00763395362191108\\
522.522522522523	0.00762595548092453\\
523.072072072072	0.00761797408174075\\
523.621621621622	0.00761000937184866\\
524.171171171171	0.00760206129895658\\
524.720720720721	0.00759412981099107\\
525.27027027027	0.00758621485609575\\
525.81981981982	0.00757831638263023\\
526.369369369369	0.00757043433916893\\
526.918918918919	0.00756256867450002\\
527.468468468468	0.00755471933762426\\
528.018018018018	0.00754688627775394\\
528.567567567568	0.00753906944431175\\
529.117117117117	0.00753126878692971\\
529.666666666667	0.0075234842554481\\
530.216216216216	0.00751571579991433\\
530.765765765766	0.00750796337058196\\
531.315315315315	0.00750022691790957\\
531.864864864865	0.0074925063925597\\
532.414414414414	0.00748480174539786\\
532.963963963964	0.00747711292749144\\
533.513513513514	0.00746943989010869\\
534.063063063063	0.00746178258471767\\
534.612612612613	0.00745414096298527\\
535.162162162162	0.00744651497677615\\
535.711711711712	0.00743890457815175\\
536.261261261261	0.0074313097193693\\
536.810810810811	0.00742373035288077\\
537.36036036036	0.00741616643133194\\
537.90990990991	0.00740861790756138\\
538.459459459459	0.00740108473459946\\
539.009009009009	0.00739356686566741\\
539.558558558559	0.00738606425417632\\
540.108108108108	0.00737857685372619\\
540.657657657658	0.007371104618105\\
541.207207207207	0.00736364750128769\\
541.756756756757	0.00735620545743529\\
542.306306306306	0.00734877844089394\\
542.855855855856	0.00734136640619397\\
543.405405405405	0.00733396930804895\\
543.954954954955	0.00732658710135481\\
544.504504504505	0.00731921974118887\\
545.054054054054	0.007311867182809\\
545.603603603604	0.00730452938165262\\
546.153153153153	0.00729720629333589\\
546.702702702703	0.00728989787365276\\
547.252252252252	0.0072826040785741\\
547.801801801802	0.00727532486424681\\
548.351351351351	0.00726806018699295\\
548.900900900901	0.00726081000330885\\
549.45045045045	0.00725357426986427\\
550	0.0072463529435015\\
};
\end{axis}

\pgfmathsetmacro{\X}{+5.2}
\pgfmathsetmacro{\Y}{-2.5}
\pgfmathsetmacro{\scale}{2}
\draw[ball color = black](-3.5*\scale+\X*\scale,4*\scale+\Y*\scale) circle (.0675*\scale);
\draw[ball color = black](-3.35*\scale+\X*\scale,4*\scale+\Y*\scale) circle (.0675*\scale);
\draw[ball color = black](-3.2*\scale+\X*\scale,4*\scale+\Y*\scale) circle (.0675*\scale);
\draw[ball color = black](-3.05*\scale+\X*\scale,4*\scale+\Y*\scale) circle (.0675*\scale);
\draw[ball color = black](-2.9*\scale+\X*\scale,4*\scale+\Y*\scale) circle (.0675*\scale);
\draw[ball color = black](-2.75*\scale+\X*\scale,4*\scale+\Y*\scale) circle (.0675*\scale);
\draw[ball color = black](-2.6*\scale+\X*\scale,4*\scale+\Y*\scale) circle (.0675*\scale);
\draw[ball color = black](-3.5*\scale+\X*\scale,3.85*\scale+\Y*\scale) circle (.0675*\scale);
\draw[ball color = black](-3.35*\scale+\X*\scale,3.85*\scale+\Y*\scale) circle (.0675*\scale);
\draw[ball color = black](-3.2*\scale+\X*\scale,3.85*\scale+\Y*\scale) circle (.0675*\scale);
\draw[ball color = black](-3.05*\scale+\X*\scale,3.85*\scale+\Y*\scale) circle (.0675*\scale);
\draw[ball color = black](-2.9*\scale+\X*\scale,3.85*\scale+\Y*\scale) circle (.0675*\scale);
\draw[ball color = black](-2.75*\scale+\X*\scale,3.85*\scale+\Y*\scale) circle (.0675*\scale);
\draw[ball color = black](-2.6*\scale+\X*\scale,3.85*\scale+\Y*\scale) circle (.0675*\scale);
\draw[ball color = black](-3.5*\scale+\X*\scale,3.7*\scale+\Y*\scale) circle (.0675*\scale);
\draw[ball color = black](-3.35*\scale+\X*\scale,3.7*\scale+\Y*\scale) circle (.0675*\scale);
\draw[ball color = black](-3.2*\scale+\X*\scale,3.7*\scale+\Y*\scale) circle (.0675*\scale);
\draw[ball color = black](-3.05*\scale+\X*\scale,3.7*\scale+\Y*\scale) circle (.0675*\scale);
\draw[ball color = black](-2.9*\scale+\X*\scale,3.7*\scale+\Y*\scale) circle (.0675*\scale);
\draw[ball color = black](-2.75*\scale+\X*\scale,3.7*\scale+\Y*\scale) circle (.0675*\scale);
\draw[ball color = black](-2.6*\scale+\X*\scale,3.7*\scale+\Y*\scale) circle (.0675*\scale);
\draw[ball color = black](-3.5*\scale+\X*\scale,3.55*\scale+\Y*\scale) circle (.0675*\scale);
\draw[ball color = black](-3.35*\scale+\X*\scale,3.55*\scale+\Y*\scale) circle (.0675*\scale);
\draw[ball color = black](-3.2*\scale+\X*\scale,3.55*\scale+\Y*\scale) circle (.0675*\scale);
\draw[ball color = black](-3.05*\scale+\X*\scale,3.55*\scale+\Y*\scale) circle (.0675*\scale);
\draw[ball color = black](-2.9*\scale+\X*\scale,3.55*\scale+\Y*\scale) circle (.0675*\scale);
\draw[ball color = black](-2.75*\scale+\X*\scale,3.55*\scale+\Y*\scale) circle (.0675*\scale);
\draw[ball color = black](-2.6*\scale+\X*\scale,3.55*\scale+\Y*\scale) circle (.0675*\scale);
\draw[ball color = black](-3.5*\scale+\X*\scale,3.4*\scale+\Y*\scale) circle (.0675*\scale);
\draw[ball color = black](-3.35*\scale+\X*\scale,3.4*\scale+\Y*\scale) circle (.0675*\scale);
\draw[ball color = black](-3.2*\scale+\X*\scale,3.4*\scale+\Y*\scale) circle (.0675*\scale);
\draw[ball color = black](-3.05*\scale+\X*\scale,3.4*\scale+\Y*\scale) circle (.0675*\scale);
\draw[ball color = black](-2.9*\scale+\X*\scale,3.4*\scale+\Y*\scale) circle (.0675*\scale);
\draw[ball color = black](-2.75*\scale+\X*\scale,3.4*\scale+\Y*\scale) circle (.0675*\scale);
\draw[ball color = black](-2.6*\scale+\X*\scale,3.4*\scale+\Y*\scale) circle (.0675*\scale);
\draw[ball color = black](-3.5*\scale+\X*\scale,3.25*\scale+\Y*\scale) circle (.0675*\scale);
\draw[ball color = black](-3.35*\scale+\X*\scale,3.25*\scale+\Y*\scale) circle (.0675*\scale);
\draw[ball color = black](-3.2*\scale+\X*\scale,3.25*\scale+\Y*\scale) circle (.0675*\scale);
\draw[ball color = black](-3.05*\scale+\X*\scale,3.25*\scale+\Y*\scale) circle (.0675*\scale);
\draw[ball color = black](-2.9*\scale+\X*\scale,3.25*\scale+\Y*\scale) circle (.0675*\scale);
\draw[ball color = black](-2.75*\scale+\X*\scale,3.25*\scale+\Y*\scale) circle (.0675*\scale);
\draw[ball color = black](-2.6*\scale+\X*\scale,3.25*\scale+\Y*\scale) circle (.0675*\scale);
\draw[ball color = black](-3.5*\scale+\X*\scale,3.1*\scale+\Y*\scale) circle (.0675*\scale);
\draw[ball color = black](-3.35*\scale+\X*\scale,3.1*\scale+\Y*\scale) circle (.0675*\scale);
\draw[ball color = black](-3.2*\scale+\X*\scale,3.1*\scale+\Y*\scale) circle (.0675*\scale);
\draw[ball color = black](-3.05*\scale+\X*\scale,3.1*\scale+\Y*\scale) circle (.0675*\scale);
\draw[ball color = black](-2.9*\scale+\X*\scale,3.1*\scale+\Y*\scale) circle (.0675*\scale);
\draw[ball color = black](-2.75*\scale+\X*\scale,3.1*\scale+\Y*\scale) circle (.0675*\scale);
\draw[ball color = black](-2.6*\scale+\X*\scale,3.1*\scale+\Y*\scale) circle (.0675*\scale);
\draw[dashed, color = red, line width = 1.5pt] (-3.5*\scale+\X*\scale,4*\scale+\Y*\scale) rectangle (-2.6*\scale+\X*\scale,3.1*\scale+\Y*\scale);
\draw[dashed, color = blue!75!white, line width = 1.5pt] (-3.35*\scale+\X*\scale,3.85*\scale+\Y*\scale) rectangle (-2.75*\scale+\X*\scale,3.25*\scale+\Y*\scale);
\draw[dashed, color = black!25!green, line width = 1.5pt] (-3.2*\scale+\X*\scale,3.7*\scale+\Y*\scale) rectangle (-2.9*\scale+\X*\scale,3.4*\scale+\Y*\scale);
\draw[|{stealth}-{stealth}|, color = red] (-3.5*\scale+\X*\scale,4.2*\scale+\Y*\scale) -- (-2.6*\scale+\X*\scale,4.2*\scale+\Y*\scale);
\draw[|{stealth}-{stealth}|, color = blue!75!white] (-3.7*\scale+\X*\scale,3.85*\scale+\Y*\scale) -- (-3.7*\scale+\X*\scale,3.25*\scale+\Y*\scale);
\draw[|{stealth}-{stealth}|, color = black!25!green] (-3.2*\scale+\X*\scale,2.9*\scale+\Y*\scale) -- (-2.9*\scale+\X*\scale,2.9*\scale+\Y*\scale);
\draw[|{stealth}-, color = black] (-2.4*\scale+\X*\scale,3.4825*\scale+\Y*\scale) -- (-2.4*\scale+\X*\scale,3.3225*\scale+\Y*\scale);
\draw[-{stealth}|, color = black] (-2.4*\scale+\X*\scale,3.7775*\scale+\Y*\scale) -- (-2.4*\scale+\X*\scale,3.6175*\scale+\Y*\scale);
\draw[color = black] (-2.4*\scale+\X*\scale,3.6175*\scale+\Y*\scale) -- (-2.4*\scale+\X*\scale,3.4825*\scale+\Y*\scale);
\node[color = red] at (-3.05*\scale+\X*\scale,4.3*\scale+\Y*\scale) {\(l_3\)};
\node[color = blue!75!white] at (-3.8*\scale+\X*\scale,3.55*\scale+\Y*\scale) {\(l_2\)};
\node[color = black!25!green] at (-3.05*\scale+\X*\scale,2.8*\scale+\Y*\scale) {\(l_1\)};
\node[color = black] at (-2.2*\scale+\X*\scale,3.55*\scale+\Y*\scale) {\(D_{PI}\)};

\end{tikzpicture}

%% file: Figures/Results_Exp/YZ-Projection.tikz
%
%
\begin{tikzpicture}

\begin{axis}[%
width=0.75\columnwidth,
height=0.75\columnwidth,
point meta min=0,
point meta max=0.015,
xmin=1,
xmax=33,
xtick={3.48,17,30.52},
xticklabels={{200},{0},{-200}},
xlabel style={font=\color{white!15!black}},
xlabel={y / \si{\micro\meter}},
ymin=1,
ymax=28,
ytick={1,13.5,17,26},
yticklabels={{0},{250},{320},{500}},
ylabel style={font=\color{white!15!black}},
ylabel={z / \si{\micro\meter}},
axis background/.style={fill=white},
axis x line*=bottom,
axis y line*=left,
xmajorgrids,
ymajorgrids,
legend style={legend cell align=left, align=left, draw=white!15!black},
colormap={mymap}{[1pt] rgb(0pt)=(0.2422,0.1504,0.6603); rgb(1pt)=(0.2444,0.1534,0.6728); rgb(2pt)=(0.2464,0.1569,0.6847); rgb(3pt)=(0.2484,0.1607,0.6961); rgb(4pt)=(0.2503,0.1648,0.7071); rgb(5pt)=(0.2522,0.1689,0.7179); rgb(6pt)=(0.254,0.1732,0.7286); rgb(7pt)=(0.2558,0.1773,0.7393); rgb(8pt)=(0.2576,0.1814,0.7501); rgb(9pt)=(0.2594,0.1854,0.761); rgb(11pt)=(0.2628,0.1932,0.7828); rgb(12pt)=(0.2645,0.1972,0.7937); rgb(13pt)=(0.2661,0.2011,0.8043); rgb(14pt)=(0.2676,0.2052,0.8148); rgb(15pt)=(0.2691,0.2094,0.8249); rgb(16pt)=(0.2704,0.2138,0.8346); rgb(17pt)=(0.2717,0.2184,0.8439); rgb(18pt)=(0.2729,0.2231,0.8528); rgb(19pt)=(0.274,0.228,0.8612); rgb(20pt)=(0.2749,0.233,0.8692); rgb(21pt)=(0.2758,0.2382,0.8767); rgb(22pt)=(0.2766,0.2435,0.884); rgb(23pt)=(0.2774,0.2489,0.8908); rgb(24pt)=(0.2781,0.2543,0.8973); rgb(25pt)=(0.2788,0.2598,0.9035); rgb(26pt)=(0.2794,0.2653,0.9094); rgb(27pt)=(0.2798,0.2708,0.915); rgb(28pt)=(0.2802,0.2764,0.9204); rgb(29pt)=(0.2806,0.2819,0.9255); rgb(30pt)=(0.2809,0.2875,0.9305); rgb(31pt)=(0.2811,0.293,0.9352); rgb(32pt)=(0.2813,0.2985,0.9397); rgb(33pt)=(0.2814,0.304,0.9441); rgb(34pt)=(0.2814,0.3095,0.9483); rgb(35pt)=(0.2813,0.315,0.9524); rgb(36pt)=(0.2811,0.3204,0.9563); rgb(37pt)=(0.2809,0.3259,0.96); rgb(38pt)=(0.2807,0.3313,0.9636); rgb(39pt)=(0.2803,0.3367,0.967); rgb(40pt)=(0.2798,0.3421,0.9702); rgb(41pt)=(0.2791,0.3475,0.9733); rgb(42pt)=(0.2784,0.3529,0.9763); rgb(43pt)=(0.2776,0.3583,0.9791); rgb(44pt)=(0.2766,0.3638,0.9817); rgb(45pt)=(0.2754,0.3693,0.984); rgb(46pt)=(0.2741,0.3748,0.9862); rgb(47pt)=(0.2726,0.3804,0.9881); rgb(48pt)=(0.271,0.386,0.9898); rgb(49pt)=(0.2691,0.3916,0.9912); rgb(50pt)=(0.267,0.3973,0.9924); rgb(51pt)=(0.2647,0.403,0.9935); rgb(52pt)=(0.2621,0.4088,0.9946); rgb(53pt)=(0.2591,0.4145,0.9955); rgb(54pt)=(0.2556,0.4203,0.9965); rgb(55pt)=(0.2517,0.4261,0.9974); rgb(56pt)=(0.2473,0.4319,0.9983); rgb(57pt)=(0.2424,0.4378,0.9991); rgb(58pt)=(0.2369,0.4437,0.9996); rgb(59pt)=(0.2311,0.4497,0.9995); rgb(60pt)=(0.225,0.4559,0.9985); rgb(61pt)=(0.2189,0.462,0.9968); rgb(62pt)=(0.2128,0.4682,0.9948); rgb(63pt)=(0.2066,0.4743,0.9926); rgb(64pt)=(0.2006,0.4803,0.9906); rgb(65pt)=(0.195,0.4861,0.9887); rgb(66pt)=(0.1903,0.4919,0.9867); rgb(67pt)=(0.1869,0.4975,0.9844); rgb(68pt)=(0.1847,0.503,0.9819); rgb(69pt)=(0.1831,0.5084,0.9793); rgb(70pt)=(0.1818,0.5138,0.9766); rgb(71pt)=(0.1806,0.5191,0.9738); rgb(72pt)=(0.1795,0.5244,0.9709); rgb(73pt)=(0.1785,0.5296,0.9677); rgb(74pt)=(0.1778,0.5349,0.9641); rgb(75pt)=(0.1773,0.5401,0.9602); rgb(76pt)=(0.1768,0.5452,0.956); rgb(77pt)=(0.1764,0.5504,0.9516); rgb(78pt)=(0.1755,0.5554,0.9473); rgb(79pt)=(0.174,0.5605,0.9432); rgb(80pt)=(0.1716,0.5655,0.9393); rgb(81pt)=(0.1686,0.5705,0.9357); rgb(82pt)=(0.1649,0.5755,0.9323); rgb(83pt)=(0.161,0.5805,0.9289); rgb(84pt)=(0.1573,0.5854,0.9254); rgb(85pt)=(0.154,0.5902,0.9218); rgb(86pt)=(0.1513,0.595,0.9182); rgb(87pt)=(0.1492,0.5997,0.9147); rgb(88pt)=(0.1475,0.6043,0.9113); rgb(89pt)=(0.1461,0.6089,0.908); rgb(90pt)=(0.1446,0.6135,0.905); rgb(91pt)=(0.1429,0.618,0.9022); rgb(92pt)=(0.1408,0.6226,0.8998); rgb(93pt)=(0.1383,0.6272,0.8975); rgb(94pt)=(0.1354,0.6317,0.8953); rgb(95pt)=(0.1321,0.6363,0.8932); rgb(96pt)=(0.1288,0.6408,0.891); rgb(97pt)=(0.1253,0.6453,0.8887); rgb(98pt)=(0.1219,0.6497,0.8862); rgb(99pt)=(0.1185,0.6541,0.8834); rgb(100pt)=(0.1152,0.6584,0.8804); rgb(101pt)=(0.1119,0.6627,0.877); rgb(102pt)=(0.1085,0.6669,0.8734); rgb(103pt)=(0.1048,0.671,0.8695); rgb(104pt)=(0.1009,0.675,0.8653); rgb(105pt)=(0.0964,0.6789,0.8609); rgb(106pt)=(0.0914,0.6828,0.8562); rgb(107pt)=(0.0855,0.6865,0.8513); rgb(108pt)=(0.0789,0.6902,0.8462); rgb(109pt)=(0.0713,0.6938,0.8409); rgb(110pt)=(0.0628,0.6972,0.8355); rgb(111pt)=(0.0535,0.7006,0.8299); rgb(112pt)=(0.0433,0.7039,0.8242); rgb(113pt)=(0.0328,0.7071,0.8183); rgb(114pt)=(0.0234,0.7103,0.8124); rgb(115pt)=(0.0155,0.7133,0.8064); rgb(116pt)=(0.0091,0.7163,0.8003); rgb(117pt)=(0.0046,0.7192,0.7941); rgb(118pt)=(0.0019,0.722,0.7878); rgb(119pt)=(0.0009,0.7248,0.7815); rgb(120pt)=(0.0018,0.7275,0.7752); rgb(121pt)=(0.0046,0.7301,0.7688); rgb(122pt)=(0.0094,0.7327,0.7623); rgb(123pt)=(0.0162,0.7352,0.7558); rgb(124pt)=(0.0253,0.7376,0.7492); rgb(125pt)=(0.0369,0.74,0.7426); rgb(126pt)=(0.0504,0.7423,0.7359); rgb(127pt)=(0.0638,0.7446,0.7292); rgb(128pt)=(0.077,0.7468,0.7224); rgb(129pt)=(0.0899,0.7489,0.7156); rgb(130pt)=(0.1023,0.751,0.7088); rgb(131pt)=(0.1141,0.7531,0.7019); rgb(132pt)=(0.1252,0.7552,0.695); rgb(133pt)=(0.1354,0.7572,0.6881); rgb(134pt)=(0.1448,0.7593,0.6812); rgb(135pt)=(0.1532,0.7614,0.6741); rgb(136pt)=(0.1609,0.7635,0.6671); rgb(137pt)=(0.1678,0.7656,0.6599); rgb(138pt)=(0.1741,0.7678,0.6527); rgb(139pt)=(0.1799,0.7699,0.6454); rgb(140pt)=(0.1853,0.7721,0.6379); rgb(141pt)=(0.1905,0.7743,0.6303); rgb(142pt)=(0.1954,0.7765,0.6225); rgb(143pt)=(0.2003,0.7787,0.6146); rgb(144pt)=(0.2061,0.7808,0.6065); rgb(145pt)=(0.2118,0.7828,0.5983); rgb(146pt)=(0.2178,0.7849,0.5899); rgb(147pt)=(0.2244,0.7869,0.5813); rgb(148pt)=(0.2318,0.7887,0.5725); rgb(149pt)=(0.2401,0.7905,0.5636); rgb(150pt)=(0.2491,0.7922,0.5546); rgb(151pt)=(0.2589,0.7937,0.5454); rgb(152pt)=(0.2695,0.7951,0.536); rgb(153pt)=(0.2809,0.7964,0.5266); rgb(154pt)=(0.2929,0.7975,0.517); rgb(155pt)=(0.3052,0.7985,0.5074); rgb(156pt)=(0.3176,0.7994,0.4975); rgb(157pt)=(0.3301,0.8002,0.4876); rgb(158pt)=(0.3424,0.8009,0.4774); rgb(159pt)=(0.3548,0.8016,0.4669); rgb(160pt)=(0.3671,0.8021,0.4563); rgb(161pt)=(0.3795,0.8026,0.4454); rgb(162pt)=(0.3921,0.8029,0.4344); rgb(163pt)=(0.405,0.8031,0.4233); rgb(164pt)=(0.4184,0.803,0.4122); rgb(165pt)=(0.4322,0.8028,0.4013); rgb(166pt)=(0.4463,0.8024,0.3904); rgb(167pt)=(0.4608,0.8018,0.3797); rgb(168pt)=(0.4753,0.8011,0.3691); rgb(169pt)=(0.4899,0.8002,0.3586); rgb(170pt)=(0.5044,0.7993,0.348); rgb(171pt)=(0.5187,0.7982,0.3374); rgb(172pt)=(0.5329,0.797,0.3267); rgb(173pt)=(0.547,0.7957,0.3159); rgb(175pt)=(0.5748,0.7929,0.2941); rgb(176pt)=(0.5886,0.7913,0.2833); rgb(177pt)=(0.6024,0.7896,0.2726); rgb(178pt)=(0.6161,0.7878,0.2622); rgb(179pt)=(0.6297,0.7859,0.2521); rgb(180pt)=(0.6433,0.7839,0.2423); rgb(181pt)=(0.6567,0.7818,0.2329); rgb(182pt)=(0.6701,0.7796,0.2239); rgb(183pt)=(0.6833,0.7773,0.2155); rgb(184pt)=(0.6963,0.775,0.2075); rgb(185pt)=(0.7091,0.7727,0.1998); rgb(186pt)=(0.7218,0.7703,0.1924); rgb(187pt)=(0.7344,0.7679,0.1852); rgb(188pt)=(0.7468,0.7654,0.1782); rgb(189pt)=(0.759,0.7629,0.1717); rgb(190pt)=(0.771,0.7604,0.1658); rgb(191pt)=(0.7829,0.7579,0.1608); rgb(192pt)=(0.7945,0.7554,0.157); rgb(193pt)=(0.806,0.7529,0.1546); rgb(194pt)=(0.8172,0.7505,0.1535); rgb(195pt)=(0.8281,0.7481,0.1536); rgb(196pt)=(0.8389,0.7457,0.1546); rgb(197pt)=(0.8495,0.7435,0.1564); rgb(198pt)=(0.86,0.7413,0.1587); rgb(199pt)=(0.8703,0.7392,0.1615); rgb(200pt)=(0.8804,0.7372,0.165); rgb(201pt)=(0.8903,0.7353,0.1695); rgb(202pt)=(0.9,0.7336,0.1749); rgb(203pt)=(0.9093,0.7321,0.1815); rgb(204pt)=(0.9184,0.7308,0.189); rgb(205pt)=(0.9272,0.7298,0.1973); rgb(206pt)=(0.9357,0.729,0.2061); rgb(207pt)=(0.944,0.7285,0.2151); rgb(208pt)=(0.9523,0.7284,0.2237); rgb(209pt)=(0.9606,0.7285,0.2312); rgb(210pt)=(0.9689,0.7292,0.2373); rgb(211pt)=(0.977,0.7304,0.2418); rgb(212pt)=(0.9842,0.733,0.2446); rgb(213pt)=(0.99,0.7365,0.2429); rgb(214pt)=(0.9946,0.7407,0.2394); rgb(215pt)=(0.9966,0.7458,0.2351); rgb(216pt)=(0.9971,0.7513,0.2309); rgb(217pt)=(0.9972,0.7569,0.2267); rgb(218pt)=(0.9971,0.7626,0.2224); rgb(219pt)=(0.9969,0.7683,0.2181); rgb(220pt)=(0.9966,0.774,0.2138); rgb(221pt)=(0.9962,0.7798,0.2095); rgb(222pt)=(0.9957,0.7856,0.2053); rgb(223pt)=(0.9949,0.7915,0.2012); rgb(224pt)=(0.9938,0.7974,0.1974); rgb(225pt)=(0.9923,0.8034,0.1939); rgb(226pt)=(0.9906,0.8095,0.1906); rgb(227pt)=(0.9885,0.8156,0.1875); rgb(228pt)=(0.9861,0.8218,0.1846); rgb(229pt)=(0.9835,0.828,0.1817); rgb(230pt)=(0.9807,0.8342,0.1787); rgb(231pt)=(0.9778,0.8404,0.1757); rgb(232pt)=(0.9748,0.8467,0.1726); rgb(233pt)=(0.972,0.8529,0.1695); rgb(234pt)=(0.9694,0.8591,0.1665); rgb(235pt)=(0.9671,0.8654,0.1636); rgb(236pt)=(0.9651,0.8716,0.1608); rgb(237pt)=(0.9634,0.8778,0.1582); rgb(238pt)=(0.9619,0.884,0.1557); rgb(239pt)=(0.9608,0.8902,0.1532); rgb(240pt)=(0.9601,0.8963,0.1507); rgb(241pt)=(0.9596,0.9023,0.148); rgb(242pt)=(0.9595,0.9084,0.145); rgb(243pt)=(0.9597,0.9143,0.1418); rgb(244pt)=(0.9601,0.9203,0.1382); rgb(245pt)=(0.9608,0.9262,0.1344); rgb(246pt)=(0.9618,0.932,0.1304); rgb(247pt)=(0.9629,0.9379,0.1261); rgb(248pt)=(0.9642,0.9437,0.1216); rgb(249pt)=(0.9657,0.9494,0.1168); rgb(250pt)=(0.9674,0.9552,0.1116); rgb(251pt)=(0.9692,0.9609,0.1061); rgb(252pt)=(0.9711,0.9667,0.1001); rgb(253pt)=(0.973,0.9724,0.0938); rgb(254pt)=(0.9749,0.9782,0.0872); rgb(255pt)=(0.9769,0.9839,0.0805)},
colorbar,
colorbar style={ylabel={$\sqrt{u^2 + v^2}$ / \si{\meter\per\second}}, ytick={0,0.005,0.01,0.015},},
]

\addplot[%
surf,
shader=interp, 
opacity=1,
colormap={mymap}{[1pt] rgb(0pt)=(0.2422,0.1504,0.6603); rgb(1pt)=(0.2444,0.1534,0.6728); rgb(2pt)=(0.2464,0.1569,0.6847); rgb(3pt)=(0.2484,0.1607,0.6961); rgb(4pt)=(0.2503,0.1648,0.7071); rgb(5pt)=(0.2522,0.1689,0.7179); rgb(6pt)=(0.254,0.1732,0.7286); rgb(7pt)=(0.2558,0.1773,0.7393); rgb(8pt)=(0.2576,0.1814,0.7501); rgb(9pt)=(0.2594,0.1854,0.761); rgb(11pt)=(0.2628,0.1932,0.7828); rgb(12pt)=(0.2645,0.1972,0.7937); rgb(13pt)=(0.2661,0.2011,0.8043); rgb(14pt)=(0.2676,0.2052,0.8148); rgb(15pt)=(0.2691,0.2094,0.8249); rgb(16pt)=(0.2704,0.2138,0.8346); rgb(17pt)=(0.2717,0.2184,0.8439); rgb(18pt)=(0.2729,0.2231,0.8528); rgb(19pt)=(0.274,0.228,0.8612); rgb(20pt)=(0.2749,0.233,0.8692); rgb(21pt)=(0.2758,0.2382,0.8767); rgb(22pt)=(0.2766,0.2435,0.884); rgb(23pt)=(0.2774,0.2489,0.8908); rgb(24pt)=(0.2781,0.2543,0.8973); rgb(25pt)=(0.2788,0.2598,0.9035); rgb(26pt)=(0.2794,0.2653,0.9094); rgb(27pt)=(0.2798,0.2708,0.915); rgb(28pt)=(0.2802,0.2764,0.9204); rgb(29pt)=(0.2806,0.2819,0.9255); rgb(30pt)=(0.2809,0.2875,0.9305); rgb(31pt)=(0.2811,0.293,0.9352); rgb(32pt)=(0.2813,0.2985,0.9397); rgb(33pt)=(0.2814,0.304,0.9441); rgb(34pt)=(0.2814,0.3095,0.9483); rgb(35pt)=(0.2813,0.315,0.9524); rgb(36pt)=(0.2811,0.3204,0.9563); rgb(37pt)=(0.2809,0.3259,0.96); rgb(38pt)=(0.2807,0.3313,0.9636); rgb(39pt)=(0.2803,0.3367,0.967); rgb(40pt)=(0.2798,0.3421,0.9702); rgb(41pt)=(0.2791,0.3475,0.9733); rgb(42pt)=(0.2784,0.3529,0.9763); rgb(43pt)=(0.2776,0.3583,0.9791); rgb(44pt)=(0.2766,0.3638,0.9817); rgb(45pt)=(0.2754,0.3693,0.984); rgb(46pt)=(0.2741,0.3748,0.9862); rgb(47pt)=(0.2726,0.3804,0.9881); rgb(48pt)=(0.271,0.386,0.9898); rgb(49pt)=(0.2691,0.3916,0.9912); rgb(50pt)=(0.267,0.3973,0.9924); rgb(51pt)=(0.2647,0.403,0.9935); rgb(52pt)=(0.2621,0.4088,0.9946); rgb(53pt)=(0.2591,0.4145,0.9955); rgb(54pt)=(0.2556,0.4203,0.9965); rgb(55pt)=(0.2517,0.4261,0.9974); rgb(56pt)=(0.2473,0.4319,0.9983); rgb(57pt)=(0.2424,0.4378,0.9991); rgb(58pt)=(0.2369,0.4437,0.9996); rgb(59pt)=(0.2311,0.4497,0.9995); rgb(60pt)=(0.225,0.4559,0.9985); rgb(61pt)=(0.2189,0.462,0.9968); rgb(62pt)=(0.2128,0.4682,0.9948); rgb(63pt)=(0.2066,0.4743,0.9926); rgb(64pt)=(0.2006,0.4803,0.9906); rgb(65pt)=(0.195,0.4861,0.9887); rgb(66pt)=(0.1903,0.4919,0.9867); rgb(67pt)=(0.1869,0.4975,0.9844); rgb(68pt)=(0.1847,0.503,0.9819); rgb(69pt)=(0.1831,0.5084,0.9793); rgb(70pt)=(0.1818,0.5138,0.9766); rgb(71pt)=(0.1806,0.5191,0.9738); rgb(72pt)=(0.1795,0.5244,0.9709); rgb(73pt)=(0.1785,0.5296,0.9677); rgb(74pt)=(0.1778,0.5349,0.9641); rgb(75pt)=(0.1773,0.5401,0.9602); rgb(76pt)=(0.1768,0.5452,0.956); rgb(77pt)=(0.1764,0.5504,0.9516); rgb(78pt)=(0.1755,0.5554,0.9473); rgb(79pt)=(0.174,0.5605,0.9432); rgb(80pt)=(0.1716,0.5655,0.9393); rgb(81pt)=(0.1686,0.5705,0.9357); rgb(82pt)=(0.1649,0.5755,0.9323); rgb(83pt)=(0.161,0.5805,0.9289); rgb(84pt)=(0.1573,0.5854,0.9254); rgb(85pt)=(0.154,0.5902,0.9218); rgb(86pt)=(0.1513,0.595,0.9182); rgb(87pt)=(0.1492,0.5997,0.9147); rgb(88pt)=(0.1475,0.6043,0.9113); rgb(89pt)=(0.1461,0.6089,0.908); rgb(90pt)=(0.1446,0.6135,0.905); rgb(91pt)=(0.1429,0.618,0.9022); rgb(92pt)=(0.1408,0.6226,0.8998); rgb(93pt)=(0.1383,0.6272,0.8975); rgb(94pt)=(0.1354,0.6317,0.8953); rgb(95pt)=(0.1321,0.6363,0.8932); rgb(96pt)=(0.1288,0.6408,0.891); rgb(97pt)=(0.1253,0.6453,0.8887); rgb(98pt)=(0.1219,0.6497,0.8862); rgb(99pt)=(0.1185,0.6541,0.8834); rgb(100pt)=(0.1152,0.6584,0.8804); rgb(101pt)=(0.1119,0.6627,0.877); rgb(102pt)=(0.1085,0.6669,0.8734); rgb(103pt)=(0.1048,0.671,0.8695); rgb(104pt)=(0.1009,0.675,0.8653); rgb(105pt)=(0.0964,0.6789,0.8609); rgb(106pt)=(0.0914,0.6828,0.8562); rgb(107pt)=(0.0855,0.6865,0.8513); rgb(108pt)=(0.0789,0.6902,0.8462); rgb(109pt)=(0.0713,0.6938,0.8409); rgb(110pt)=(0.0628,0.6972,0.8355); rgb(111pt)=(0.0535,0.7006,0.8299); rgb(112pt)=(0.0433,0.7039,0.8242); rgb(113pt)=(0.0328,0.7071,0.8183); rgb(114pt)=(0.0234,0.7103,0.8124); rgb(115pt)=(0.0155,0.7133,0.8064); rgb(116pt)=(0.0091,0.7163,0.8003); rgb(117pt)=(0.0046,0.7192,0.7941); rgb(118pt)=(0.0019,0.722,0.7878); rgb(119pt)=(0.0009,0.7248,0.7815); rgb(120pt)=(0.0018,0.7275,0.7752); rgb(121pt)=(0.0046,0.7301,0.7688); rgb(122pt)=(0.0094,0.7327,0.7623); rgb(123pt)=(0.0162,0.7352,0.7558); rgb(124pt)=(0.0253,0.7376,0.7492); rgb(125pt)=(0.0369,0.74,0.7426); rgb(126pt)=(0.0504,0.7423,0.7359); rgb(127pt)=(0.0638,0.7446,0.7292); rgb(128pt)=(0.077,0.7468,0.7224); rgb(129pt)=(0.0899,0.7489,0.7156); rgb(130pt)=(0.1023,0.751,0.7088); rgb(131pt)=(0.1141,0.7531,0.7019); rgb(132pt)=(0.1252,0.7552,0.695); rgb(133pt)=(0.1354,0.7572,0.6881); rgb(134pt)=(0.1448,0.7593,0.6812); rgb(135pt)=(0.1532,0.7614,0.6741); rgb(136pt)=(0.1609,0.7635,0.6671); rgb(137pt)=(0.1678,0.7656,0.6599); rgb(138pt)=(0.1741,0.7678,0.6527); rgb(139pt)=(0.1799,0.7699,0.6454); rgb(140pt)=(0.1853,0.7721,0.6379); rgb(141pt)=(0.1905,0.7743,0.6303); rgb(142pt)=(0.1954,0.7765,0.6225); rgb(143pt)=(0.2003,0.7787,0.6146); rgb(144pt)=(0.2061,0.7808,0.6065); rgb(145pt)=(0.2118,0.7828,0.5983); rgb(146pt)=(0.2178,0.7849,0.5899); rgb(147pt)=(0.2244,0.7869,0.5813); rgb(148pt)=(0.2318,0.7887,0.5725); rgb(149pt)=(0.2401,0.7905,0.5636); rgb(150pt)=(0.2491,0.7922,0.5546); rgb(151pt)=(0.2589,0.7937,0.5454); rgb(152pt)=(0.2695,0.7951,0.536); rgb(153pt)=(0.2809,0.7964,0.5266); rgb(154pt)=(0.2929,0.7975,0.517); rgb(155pt)=(0.3052,0.7985,0.5074); rgb(156pt)=(0.3176,0.7994,0.4975); rgb(157pt)=(0.3301,0.8002,0.4876); rgb(158pt)=(0.3424,0.8009,0.4774); rgb(159pt)=(0.3548,0.8016,0.4669); rgb(160pt)=(0.3671,0.8021,0.4563); rgb(161pt)=(0.3795,0.8026,0.4454); rgb(162pt)=(0.3921,0.8029,0.4344); rgb(163pt)=(0.405,0.8031,0.4233); rgb(164pt)=(0.4184,0.803,0.4122); rgb(165pt)=(0.4322,0.8028,0.4013); rgb(166pt)=(0.4463,0.8024,0.3904); rgb(167pt)=(0.4608,0.8018,0.3797); rgb(168pt)=(0.4753,0.8011,0.3691); rgb(169pt)=(0.4899,0.8002,0.3586); rgb(170pt)=(0.5044,0.7993,0.348); rgb(171pt)=(0.5187,0.7982,0.3374); rgb(172pt)=(0.5329,0.797,0.3267); rgb(173pt)=(0.547,0.7957,0.3159); rgb(175pt)=(0.5748,0.7929,0.2941); rgb(176pt)=(0.5886,0.7913,0.2833); rgb(177pt)=(0.6024,0.7896,0.2726); rgb(178pt)=(0.6161,0.7878,0.2622); rgb(179pt)=(0.6297,0.7859,0.2521); rgb(180pt)=(0.6433,0.7839,0.2423); rgb(181pt)=(0.6567,0.7818,0.2329); rgb(182pt)=(0.6701,0.7796,0.2239); rgb(183pt)=(0.6833,0.7773,0.2155); rgb(184pt)=(0.6963,0.775,0.2075); rgb(185pt)=(0.7091,0.7727,0.1998); rgb(186pt)=(0.7218,0.7703,0.1924); rgb(187pt)=(0.7344,0.7679,0.1852); rgb(188pt)=(0.7468,0.7654,0.1782); rgb(189pt)=(0.759,0.7629,0.1717); rgb(190pt)=(0.771,0.7604,0.1658); rgb(191pt)=(0.7829,0.7579,0.1608); rgb(192pt)=(0.7945,0.7554,0.157); rgb(193pt)=(0.806,0.7529,0.1546); rgb(194pt)=(0.8172,0.7505,0.1535); rgb(195pt)=(0.8281,0.7481,0.1536); rgb(196pt)=(0.8389,0.7457,0.1546); rgb(197pt)=(0.8495,0.7435,0.1564); rgb(198pt)=(0.86,0.7413,0.1587); rgb(199pt)=(0.8703,0.7392,0.1615); rgb(200pt)=(0.8804,0.7372,0.165); rgb(201pt)=(0.8903,0.7353,0.1695); rgb(202pt)=(0.9,0.7336,0.1749); rgb(203pt)=(0.9093,0.7321,0.1815); rgb(204pt)=(0.9184,0.7308,0.189); rgb(205pt)=(0.9272,0.7298,0.1973); rgb(206pt)=(0.9357,0.729,0.2061); rgb(207pt)=(0.944,0.7285,0.2151); rgb(208pt)=(0.9523,0.7284,0.2237); rgb(209pt)=(0.9606,0.7285,0.2312); rgb(210pt)=(0.9689,0.7292,0.2373); rgb(211pt)=(0.977,0.7304,0.2418); rgb(212pt)=(0.9842,0.733,0.2446); rgb(213pt)=(0.99,0.7365,0.2429); rgb(214pt)=(0.9946,0.7407,0.2394); rgb(215pt)=(0.9966,0.7458,0.2351); rgb(216pt)=(0.9971,0.7513,0.2309); rgb(217pt)=(0.9972,0.7569,0.2267); rgb(218pt)=(0.9971,0.7626,0.2224); rgb(219pt)=(0.9969,0.7683,0.2181); rgb(220pt)=(0.9966,0.774,0.2138); rgb(221pt)=(0.9962,0.7798,0.2095); rgb(222pt)=(0.9957,0.7856,0.2053); rgb(223pt)=(0.9949,0.7915,0.2012); rgb(224pt)=(0.9938,0.7974,0.1974); rgb(225pt)=(0.9923,0.8034,0.1939); rgb(226pt)=(0.9906,0.8095,0.1906); rgb(227pt)=(0.9885,0.8156,0.1875); rgb(228pt)=(0.9861,0.8218,0.1846); rgb(229pt)=(0.9835,0.828,0.1817); rgb(230pt)=(0.9807,0.8342,0.1787); rgb(231pt)=(0.9778,0.8404,0.1757); rgb(232pt)=(0.9748,0.8467,0.1726); rgb(233pt)=(0.972,0.8529,0.1695); rgb(234pt)=(0.9694,0.8591,0.1665); rgb(235pt)=(0.9671,0.8654,0.1636); rgb(236pt)=(0.9651,0.8716,0.1608); rgb(237pt)=(0.9634,0.8778,0.1582); rgb(238pt)=(0.9619,0.884,0.1557); rgb(239pt)=(0.9608,0.8902,0.1532); rgb(240pt)=(0.9601,0.8963,0.1507); rgb(241pt)=(0.9596,0.9023,0.148); rgb(242pt)=(0.9595,0.9084,0.145); rgb(243pt)=(0.9597,0.9143,0.1418); rgb(244pt)=(0.9601,0.9203,0.1382); rgb(245pt)=(0.9608,0.9262,0.1344); rgb(246pt)=(0.9618,0.932,0.1304); rgb(247pt)=(0.9629,0.9379,0.1261); rgb(248pt)=(0.9642,0.9437,0.1216); rgb(249pt)=(0.9657,0.9494,0.1168); rgb(250pt)=(0.9674,0.9552,0.1116); rgb(251pt)=(0.9692,0.9609,0.1061); rgb(252pt)=(0.9711,0.9667,0.1001); rgb(253pt)=(0.973,0.9724,0.0938); rgb(254pt)=(0.9749,0.9782,0.0872); rgb(255pt)=(0.9769,0.9839,0.0805)}, mesh/rows=33]
table[row sep=crcr,point meta=1*\thisrow{c}] {%
x	y	c\\
1	1	0\\
1	2	0\\
1	3	0\\
1	4	0\\
1	5	0\\
1	6	0\\
1	7	0\\
1	8	0\\
1	9	0\\
1	10	0.00354471720042361\\
1	11	0.00194235728545722\\
1	12	0.00113404342754968\\
1	13	0.00143915581854446\\
1	14	0.00254493922259051\\
1	15	0.00236445395419115\\
1	16	0.00137564682556679\\
1	17	0.000455144708664395\\
1	18	0.000344239441506452\\
1	19	0.00121096985134492\\
1	20	0.00183348221373226\\
1	21	0.00247120930595296\\
1	22	0.00274880384848138\\
1	23	0.00279686284675916\\
1	24	0.0027374822895602\\
1	25	0.00224822423212587\\
1	26	0.00158258647420986\\
1	27	0.00109093723713143\\
1	28	0.00133075714185303\\
1	29	0.000947551767473488\\
1	30	0.00128705158680267\\
2	1	0.000771425062128138\\
2	2	0.00086248639328536\\
2	3	0.000861941377413308\\
2	4	0.00171767044747251\\
2	5	0.00252682249445503\\
2	6	0.00132841718018\\
2	7	0.00281536443514169\\
2	8	0.00252545917631743\\
2	9	0.00236942089600593\\
2	10	0.00191906298663464\\
2	11	0.00101409748585937\\
2	12	0.000804034322007199\\
2	13	0.000580878087120271\\
2	14	0.000998563577006884\\
2	15	0.00051165984930326\\
2	16	0.000346064105997406\\
2	17	0.00111493768948452\\
2	18	0.00187458481691627\\
2	19	0.00262725874152791\\
2	20	0.00317708820741019\\
2	21	0.00359151448588442\\
2	22	0.00373886838241498\\
2	23	0.00362868767587486\\
2	24	0.00336349441978081\\
2	25	0.00276164723479326\\
2	26	0.00192866303448806\\
2	27	0.00112786750713487\\
2	28	0.00128735838186306\\
2	29	0.00106327015351881\\
2	30	0.00128804221253463\\
3	1	0.000725262637091825\\
3	2	0.000536699669295176\\
3	3	0.000635337888121322\\
3	4	0.00103543418349341\\
3	5	0.0017337252486484\\
3	6	0.000853058484277036\\
3	7	0.00140332267299763\\
3	8	0.00130809587511853\\
3	9	0.00129693083366094\\
3	10	0.00103017368203647\\
3	11	0.000754904715639451\\
3	12	0.000774640850976231\\
3	13	0.000489741133131652\\
3	14	0.000511432701926538\\
3	15	0.00113894721782666\\
3	16	0.00194026034535518\\
3	17	0.00260821150016841\\
3	18	0.00336188073379616\\
3	19	0.00394436564355587\\
3	20	0.00445127028044711\\
3	21	0.00458893439407576\\
3	22	0.00467298429708238\\
3	23	0.00444381646808964\\
3	24	0.00395207177855732\\
3	25	0.00323570746841537\\
3	26	0.0022794075370638\\
3	27	0.00118814431550058\\
3	28	0.00126995587169474\\
3	29	0.00120778288009136\\
3	30	0.00132960116565967\\
4	1	0.00061368587578508\\
4	2	0.000368091845461333\\
4	3	0.00043173578471488\\
4	4	0.000648282400810416\\
4	5	0.00082532324425223\\
4	6	0.000439516930952429\\
4	7	0.000672959979091396\\
4	8	0.000835108807341231\\
4	9	0.000832305262838328\\
4	10	0.000722242737207845\\
4	11	0.000678966079870252\\
4	12	0.00100677707973005\\
4	13	0.00139010859057088\\
4	14	0.00198951881508484\\
4	15	0.00276117004206411\\
4	16	0.00340407826298966\\
4	17	0.00407366030425587\\
4	18	0.00468032713186583\\
4	19	0.00518046640649794\\
4	20	0.00556560577476711\\
4	21	0.00555947994877616\\
4	22	0.00552267147817124\\
4	23	0.00515295329232771\\
4	24	0.00453332266025624\\
4	25	0.00366261423693048\\
4	26	0.00253344443530591\\
4	27	0.00127779598519998\\
4	28	0.00137586844143374\\
4	29	0.00136135642141961\\
4	30	0.00135257353882406\\
5	1	0.000520780030055774\\
5	2	0.000250190694153804\\
5	3	0.000350896336269541\\
5	4	0.000407929054844798\\
5	5	0.00032324385158306\\
5	6	0.000328014612476682\\
5	7	0.000498497371925009\\
5	8	0.000595923583843595\\
5	9	0.0007526744208485\\
5	10	0.00107885019588675\\
5	11	0.00152305275299837\\
5	12	0.00213566254484508\\
5	13	0.00279650484432818\\
5	14	0.00357727586670347\\
5	15	0.00429723694601916\\
5	16	0.00491076027113063\\
5	17	0.00541235241631454\\
5	18	0.00596722016773527\\
5	19	0.00633482704824553\\
5	20	0.00659699937804804\\
5	21	0.00645712239044677\\
5	22	0.00629303481444784\\
5	23	0.00578424335167902\\
5	24	0.00503550977157113\\
5	25	0.00407992716439585\\
5	26	0.0028038082606966\\
5	27	0.00136192481927707\\
5	28	0.00146250571595659\\
5	29	0.00150443547113433\\
5	30	0.00144258004947419\\
6	1	0.000497694267994156\\
6	2	0.000362549459250333\\
6	3	0.000496935626454423\\
6	4	0.000598216208295612\\
6	5	0.000673528035042968\\
6	6	0.000772110020971364\\
6	7	0.000962177600555148\\
6	8	0.00136043865604094\\
6	9	0.00174258273399138\\
6	10	0.00231550968127602\\
6	11	0.0030437484395867\\
6	12	0.003633626819086\\
6	13	0.00430248208274989\\
6	14	0.00507217577282691\\
6	15	0.00569795733219275\\
6	16	0.00630203107813174\\
6	17	0.00668293858669673\\
6	18	0.00714966931096201\\
6	19	0.00741085731928385\\
6	20	0.00755326103127585\\
6	21	0.00730724801535859\\
6	22	0.00700814653600306\\
6	23	0.00636036492593443\\
6	24	0.00551807847148778\\
6	25	0.00446104858092396\\
6	26	0.00303254079906767\\
6	27	0.00147620191671671\\
6	28	0.00158516062836523\\
6	29	0.00162627270711181\\
6	30	0.00158756270851674\\
7	1	0.000535383060263023\\
7	2	0.000601748320741386\\
7	3	0.000855449022382731\\
7	4	0.00113016034242632\\
7	5	0.00147315819294124\\
7	6	0.00178904068203202\\
7	7	0.00222393698854543\\
7	8	0.00277304445439024\\
7	9	0.00325023633226293\\
7	10	0.0039053979632114\\
7	11	0.00458818315520715\\
7	12	0.0051345926342804\\
7	13	0.00568316137276299\\
7	14	0.0064244451981573\\
7	15	0.00701639848591606\\
7	16	0.00756293335413777\\
7	17	0.00791170900628537\\
7	18	0.00818731874637465\\
7	19	0.00838231757863535\\
7	20	0.00842324505812435\\
7	21	0.00805375639576951\\
7	22	0.00764669731758898\\
7	23	0.00688463108075173\\
7	24	0.00598859326496837\\
7	25	0.00480124874375675\\
7	26	0.00321421267066387\\
7	27	0.00157759053039047\\
7	28	0.00179537061319748\\
7	29	0.00181606570875613\\
7	30	0.00173046536699689\\
8	1	0.000674342992066747\\
8	2	0.000845012122884624\\
8	3	0.00125739736088859\\
8	4	0.00183837962463432\\
8	5	0.00238112617496284\\
8	6	0.00290566197223788\\
8	7	0.00356885499155333\\
8	8	0.00405931001049083\\
8	9	0.00477121421350206\\
8	10	0.00522529822797961\\
8	11	0.00595673379471388\\
8	12	0.00638846057260817\\
8	13	0.00701172213394113\\
8	14	0.00758852669971284\\
8	15	0.00813615134666374\\
8	16	0.00869824048421869\\
8	17	0.0091097585150827\\
8	18	0.00915341966214028\\
8	19	0.00930314711563797\\
8	20	0.00917391198190574\\
8	21	0.00877433851858209\\
8	22	0.00823207331078629\\
8	23	0.00740715581217948\\
8	24	0.00640325537739686\\
8	25	0.00514887993375776\\
8	26	0.00337545624067741\\
8	27	0.00162661607856327\\
8	28	0.00197772179841363\\
8	29	0.00195813807280929\\
8	30	0.00165085002171473\\
9	1	0.000797368365235357\\
9	2	0.00108695170453039\\
9	3	0.00162027600245815\\
9	4	0.00250167822106199\\
9	5	0.00317632748982506\\
9	6	0.003811582113035\\
9	7	0.00471512325546952\\
9	8	0.00523139009371061\\
9	9	0.00604176023025386\\
9	10	0.00648115785293053\\
9	11	0.00725739307169326\\
9	12	0.00768224383250525\\
9	13	0.0083047411730467\\
9	14	0.00881433408535029\\
9	15	0.00931865647941877\\
9	16	0.00977142675122304\\
9	17	0.0100730729416417\\
9	18	0.0100745232848944\\
9	19	0.0101029071518129\\
9	20	0.00989608561913847\\
9	21	0.0094038914738179\\
9	22	0.00880426462105464\\
9	23	0.00785348807177232\\
9	24	0.00674318250400424\\
9	25	0.00543919192138776\\
9	26	0.00355374064414512\\
9	27	0.00163095578448915\\
9	28	0.00200091891150771\\
9	29	0.00202474791866571\\
9	30	0.00163227709661364\\
10	1	0.000927845890787937\\
10	2	0.0012935919649316\\
10	3	0.00193228059530575\\
10	4	0.00301644360850734\\
10	5	0.00391208111447614\\
10	6	0.00473890418725594\\
10	7	0.00574182342985559\\
10	8	0.00623809491808023\\
10	9	0.00704212026733164\\
10	10	0.00763148662820923\\
10	11	0.00835270749696292\\
10	12	0.00886902183679095\\
10	13	0.0093596018805103\\
10	14	0.00998440014562071\\
10	15	0.0103941098963142\\
10	16	0.0107480115926627\\
10	17	0.0109715711663364\\
10	18	0.0108512916652839\\
10	19	0.0108318749586401\\
10	20	0.0105415327390321\\
10	21	0.00994542345298195\\
10	22	0.00930390978321113\\
10	23	0.00824121312148609\\
10	24	0.00708986849846738\\
10	25	0.00570645299019263\\
10	26	0.00376922803087106\\
10	27	0.00165430230399305\\
10	28	0.00208891583434865\\
10	29	0.00202346479007656\\
10	30	0.00161208651910133\\
11	1	0.00101609572323577\\
11	2	0.00144095617406944\\
11	3	0.00217099038538921\\
11	4	0.00338233871935669\\
11	5	0.00454728059671984\\
11	6	0.00552414724038134\\
11	7	0.00645403419145719\\
11	8	0.00718360196406468\\
11	9	0.00790758345973045\\
11	10	0.00858254054921484\\
11	11	0.00927637093331002\\
11	12	0.00978971851013317\\
11	13	0.0101710550582506\\
11	14	0.0108375046482765\\
11	15	0.0112971591446907\\
11	16	0.0115553425810761\\
11	17	0.0117436932456747\\
11	18	0.0114647696950491\\
11	19	0.0114547967980774\\
11	20	0.0110889976613455\\
11	21	0.0104516363278587\\
11	22	0.00971772560178299\\
11	23	0.00860520146337536\\
11	24	0.00734305586401959\\
11	25	0.00591398846249449\\
11	26	0.003890374642151\\
11	27	0.00167653435280511\\
11	28	0.00217888763481284\\
11	29	0.00202599582669892\\
11	30	0.00161969863664205\\
12	1	0.00107502644267468\\
12	2	0.00155572861774102\\
12	3	0.00233921616286859\\
12	4	0.00372897031062122\\
12	5	0.0050277517868152\\
12	6	0.00614575492708122\\
12	7	0.00705876374006586\\
12	8	0.00804695066390913\\
12	9	0.00872061740974651\\
12	10	0.00932643083246585\\
12	11	0.0100085884548138\\
12	12	0.0104688892480154\\
12	13	0.0110352535443511\\
12	14	0.0115507464633662\\
12	15	0.0120193837535795\\
12	16	0.0122102767773623\\
12	17	0.0123633025250752\\
12	18	0.0120114052366166\\
12	19	0.011919044067978\\
12	20	0.0115733926673118\\
12	21	0.0108491877277063\\
12	22	0.0100290326120236\\
12	23	0.00888621509038799\\
12	24	0.00757208072092897\\
12	25	0.00609469371574166\\
12	26	0.00394451383635273\\
12	27	0.00171045240017355\\
12	28	0.00221228509348708\\
12	29	0.00197021696479829\\
12	30	0.00190958625105574\\
13	1	0.00112300158095547\\
13	2	0.0016390760449136\\
13	3	0.00241355980028672\\
13	4	0.00404834035351134\\
13	5	0.00537936283099895\\
13	6	0.00655408387132131\\
13	7	0.00764947478543326\\
13	8	0.00861563619781556\\
13	9	0.00941851497915236\\
13	10	0.00993010463133929\\
13	11	0.0104931845597611\\
13	12	0.0110478215583011\\
13	13	0.0118686025853682\\
13	14	0.0121190880424206\\
13	15	0.0125374481008318\\
13	16	0.0127528799232844\\
13	17	0.0128584892747099\\
13	18	0.0124978470712595\\
13	19	0.0122733677742248\\
13	20	0.0119153203961037\\
13	21	0.0111186373458684\\
13	22	0.0102941416738087\\
13	23	0.00916844476773045\\
13	24	0.00773830570687785\\
13	25	0.0062290833877337\\
13	26	0.00386068523949831\\
13	27	0.00179500863274586\\
13	28	0.00224108047029502\\
13	29	0.00186998801412217\\
13	30	0.00193546083099833\\
14	1	0.00116482422693969\\
14	2	0.00172181805141371\\
14	3	0.00245989899000209\\
14	4	0.00417257578812818\\
14	5	0.00564065208478825\\
14	6	0.00696054682550475\\
14	7	0.00805871186860004\\
14	8	0.0090495494828191\\
14	9	0.00998631809368783\\
14	10	0.0104856760046346\\
14	11	0.0108963277244344\\
14	12	0.0115058969167173\\
14	13	0.0123286347679293\\
14	14	0.0126116615901948\\
14	15	0.0129513821788402\\
14	16	0.0132072796135537\\
14	17	0.0131926238754147\\
14	18	0.0128843332048978\\
14	19	0.0125121047554714\\
14	20	0.0121571360849833\\
14	21	0.0113793761050491\\
14	22	0.0105455195163492\\
14	23	0.00934399019034785\\
14	24	0.00788476827170797\\
14	25	0.00631504315136568\\
14	26	0.00390175300461606\\
14	27	0.00186144116749079\\
14	28	0.00232146462226884\\
14	29	0.00185495188964143\\
14	30	0.00189361713105235\\
15	1	0.00121239477401371\\
15	2	0.00175494875309429\\
15	3	0.00251452452909889\\
15	4	0.00433323105283928\\
15	5	0.00581102612394303\\
15	6	0.00723178949889101\\
15	7	0.00837886821308564\\
15	8	0.00941078921236115\\
15	9	0.01039163025729\\
15	10	0.0109139233519188\\
15	11	0.0112650372600429\\
15	12	0.0117784891952856\\
15	13	0.0126668797899985\\
15	14	0.012991503929084\\
15	15	0.0132163343437295\\
15	16	0.0135227254661835\\
15	17	0.0134626601157327\\
15	18	0.013055305538289\\
15	19	0.0127952008990985\\
15	20	0.0123707307737374\\
15	21	0.0115934668095941\\
15	22	0.0106769899317085\\
15	23	0.00945549744788585\\
15	24	0.00799248143397255\\
15	25	0.00631299678518576\\
15	26	0.00400080196730433\\
15	27	0.00189175413277969\\
15	28	0.00226132155245885\\
15	29	0.00184548142683794\\
15	30	0.00170985718860085\\
16	1	0.00121948150903848\\
16	2	0.00178928397833445\\
16	3	0.00262911221997295\\
16	4	0.00447489996311243\\
16	5	0.00591739869089902\\
16	6	0.00731338890697361\\
16	7	0.00854906255521458\\
16	8	0.00964818010957522\\
16	9	0.0106106769766841\\
16	10	0.0111132379123246\\
16	11	0.0116251415376922\\
16	12	0.0120486710880049\\
16	13	0.0128693938008574\\
16	14	0.0132167305235773\\
16	15	0.0132885235613509\\
16	16	0.0136658118381384\\
16	17	0.0135914518209812\\
16	18	0.0132265245720609\\
16	19	0.012957004194664\\
16	20	0.0124385174822104\\
16	21	0.0116976495525116\\
16	22	0.0107188961483992\\
16	23	0.00953484579726904\\
16	24	0.00808871199785057\\
16	25	0.00629760916540455\\
16	26	0.00405910061197489\\
16	27	0.00188634835999293\\
16	28	0.00227347046751169\\
16	29	0.00180750350418772\\
16	30	0.00153157735637827\\
17	1	0.00121125379257401\\
17	2	0.00179797770739448\\
17	3	0.00276329773512974\\
17	4	0.00454173820253703\\
17	5	0.00598187649874548\\
17	6	0.00725765896989642\\
17	7	0.00857598561435524\\
17	8	0.00972113431984789\\
17	9	0.0105757085465531\\
17	10	0.0111918440982306\\
17	11	0.0118765316170238\\
17	12	0.012112579052329\\
17	13	0.0129253946473575\\
17	14	0.0133443601771002\\
17	15	0.0133527490713513\\
17	16	0.0136909395031563\\
17	17	0.0136303723047956\\
17	18	0.013272917114414\\
17	19	0.0129644218193042\\
17	20	0.0123549921226371\\
17	21	0.0117142298066358\\
17	22	0.0107388194047056\\
17	23	0.00960589982469241\\
17	24	0.00812810685599307\\
17	25	0.00630730328992112\\
17	26	0.00409644691552553\\
17	27	0.00197467358084234\\
17	28	0.00230308399740638\\
17	29	0.00193653900884352\\
17	30	0.00145298331877656\\
18	1	0.00118348143849018\\
18	2	0.00177501216194882\\
18	3	0.00279568848654233\\
18	4	0.00455655209698964\\
18	5	0.00597854531950468\\
18	6	0.00718943999217965\\
18	7	0.00848702523538637\\
18	8	0.00962058608094541\\
18	9	0.0104443965980476\\
18	10	0.0111217879783164\\
18	11	0.0119509231218642\\
18	12	0.0120099392640913\\
18	13	0.0129794549325259\\
18	14	0.0133029968818161\\
18	15	0.0134991782812373\\
18	16	0.0136167012687491\\
18	17	0.013606467021235\\
18	18	0.01329574440953\\
18	19	0.0129528884683027\\
18	20	0.0122906811351321\\
18	21	0.0117231761529174\\
18	22	0.010712723045515\\
18	23	0.00957712648880131\\
18	24	0.00810434423121132\\
18	25	0.00627568242602288\\
18	26	0.0041420236978196\\
18	27	0.00193199395437564\\
18	28	0.00242811086824088\\
18	29	0.00203290356807497\\
18	30	0.001603427508811\\
19	1	0.00119657277052913\\
19	2	0.00176114575645561\\
19	3	0.00280301227510441\\
19	4	0.00449181863895598\\
19	5	0.00597014195224436\\
19	6	0.00705353634472283\\
19	7	0.00827818531534004\\
19	8	0.00934390589539066\\
19	9	0.0100535368091686\\
19	10	0.0110983876182711\\
19	11	0.0116962535468175\\
19	12	0.0119872536816014\\
19	13	0.0127905012202138\\
19	14	0.0131162833247032\\
19	15	0.0133676260811605\\
19	16	0.01348396267307\\
19	17	0.0134412817313924\\
19	18	0.0132068532616966\\
19	19	0.0126917707257579\\
19	20	0.0122201949906507\\
19	21	0.0116574455735164\\
19	22	0.0106163295448996\\
19	23	0.00939777857509212\\
19	24	0.00800087023341483\\
19	25	0.0063145169115569\\
19	26	0.00418445340774923\\
19	27	0.00185629545288104\\
19	28	0.0024360815445601\\
19	29	0.00205088101495131\\
19	30	0.00161200436345995\\
20	1	0.00119823596847709\\
20	2	0.00172426058983142\\
20	3	0.00272638970641405\\
20	4	0.00433622875192518\\
20	5	0.00580892934037359\\
20	6	0.00688924884145851\\
20	7	0.00802895366783772\\
20	8	0.00890251891873957\\
20	9	0.00974093897954428\\
20	10	0.0108709586765055\\
20	11	0.0113246292863536\\
20	12	0.011846515407158\\
20	13	0.0124670817731813\\
20	14	0.0127334582041667\\
20	15	0.0130260254998754\\
20	16	0.0131828481720887\\
20	17	0.0131080020340151\\
20	18	0.0129943693442248\\
20	19	0.0123298585207608\\
20	20	0.0120914557904845\\
20	21	0.0114724404969922\\
20	22	0.0104762530442305\\
20	23	0.00921920770415891\\
20	24	0.00792368993793161\\
20	25	0.00629048272678834\\
20	26	0.00414418421654601\\
20	27	0.00179552054277872\\
20	28	0.00238197942486832\\
20	29	0.00196988173162921\\
20	30	0.00168489060911798\\
21	1	0.00118681392790131\\
21	2	0.00162181744865249\\
21	3	0.00258730377173647\\
21	4	0.00421951961889228\\
21	5	0.00552822074689115\\
21	6	0.00666715172495822\\
21	7	0.00769826539576254\\
21	8	0.00847861976804148\\
21	9	0.00929568312740965\\
21	10	0.0104558798822789\\
21	11	0.0109049874945126\\
21	12	0.0115261012659517\\
21	13	0.011982725157141\\
21	14	0.0122207346947083\\
21	15	0.0125741411812354\\
21	16	0.0128037110568774\\
21	17	0.0126841011623206\\
21	18	0.0126058984919298\\
21	19	0.0120876802352307\\
21	20	0.0118241041061022\\
21	21	0.0111734997296603\\
21	22	0.0102167503194053\\
21	23	0.00910027408522611\\
21	24	0.00775563062645128\\
21	25	0.00620082333650633\\
21	26	0.00406092491553855\\
21	27	0.0017969462827854\\
21	28	0.002388886599104\\
21	29	0.00189148768260665\\
21	30	0.00147176206572502\\
22	1	0.00110387759885515\\
22	2	0.00156323301285231\\
22	3	0.00234445186063368\\
22	4	0.00397026926724497\\
22	5	0.00509973193289304\\
22	6	0.00635970307798306\\
22	7	0.00721393250513994\\
22	8	0.00807008290708821\\
22	9	0.00887299135398374\\
22	10	0.00984876160650674\\
22	11	0.0103491385132059\\
22	12	0.0110323218511923\\
22	13	0.0113092358841425\\
22	14	0.0117373161842395\\
22	15	0.012039852386965\\
22	16	0.0122157202224576\\
22	17	0.0121910908841189\\
22	18	0.012107058152498\\
22	19	0.0118088950278253\\
22	20	0.0114495616941239\\
22	21	0.0108266208507451\\
22	22	0.00992179177082263\\
22	23	0.00892423198627831\\
22	24	0.00747427724155171\\
22	25	0.00604369971876456\\
22	26	0.00399106126448746\\
22	27	0.00176452829301688\\
22	28	0.00231153302453009\\
22	29	0.00189010446398717\\
22	30	0.0016157273568681\\
23	1	0.00102601454381109\\
23	2	0.0014949673126845\\
23	3	0.00218334553934187\\
23	4	0.00364942330892628\\
23	5	0.00467984870554529\\
23	6	0.00583321252920784\\
23	7	0.00671931182379382\\
23	8	0.00752950837829632\\
23	9	0.00821626234824607\\
23	10	0.0091219516391989\\
23	11	0.00963318485148196\\
23	12	0.0102604414761013\\
23	13	0.0105030321074475\\
23	14	0.0110123213565077\\
23	15	0.0111972008251134\\
23	16	0.0115227546080583\\
23	17	0.0115503569216818\\
23	18	0.0114930837688404\\
23	19	0.0112681002936727\\
23	20	0.0110490578856302\\
23	21	0.0103955873805205\\
23	22	0.00958148668104857\\
23	23	0.00860554473526456\\
23	24	0.00722544630277164\\
23	25	0.00579417050288906\\
23	26	0.003915823361881\\
23	27	0.00180663858062089\\
23	28	0.00233616279584485\\
23	29	0.00190810873210917\\
23	30	0.00180835866381182\\
24	1	0.000941640011013605\\
24	2	0.00135390027344646\\
24	3	0.00199635046868677\\
24	4	0.00331369146305293\\
24	5	0.00421353482062805\\
24	6	0.00517419508939649\\
24	7	0.00601263334275185\\
24	8	0.00673301014860069\\
24	9	0.00738212782509283\\
24	10	0.00811890847655132\\
24	11	0.00870342582287565\\
24	12	0.00924015212693559\\
24	13	0.00957414101713833\\
24	14	0.0101645025321957\\
24	15	0.01046937745653\\
24	16	0.0106497131215167\\
24	17	0.0108022094155555\\
24	18	0.0106941887035361\\
24	19	0.010689268252615\\
24	20	0.0104887374883365\\
24	21	0.00985293226990887\\
24	22	0.00915104724775446\\
24	23	0.00827934364764695\\
24	24	0.00694137338411218\\
24	25	0.00557037809007978\\
24	26	0.00371939436175444\\
24	27	0.00185003755239913\\
24	28	0.00237919958324989\\
24	29	0.001978302267171\\
24	30	0.00205263973402236\\
25	1	0.000836851224121191\\
25	2	0.00119882619382714\\
25	3	0.00175841177440133\\
25	4	0.00288721465731515\\
25	5	0.00365878367843165\\
25	6	0.00441936092055986\\
25	7	0.00517688908764931\\
25	8	0.00581250142503613\\
25	9	0.00633082738393419\\
25	10	0.00697480498816063\\
25	11	0.0076383223901065\\
25	12	0.00805815121381302\\
25	13	0.00837316057931946\\
25	14	0.00915664378971046\\
25	15	0.00954955612380708\\
25	16	0.00971943376031056\\
25	17	0.00992778774871304\\
25	18	0.00987363414756592\\
25	19	0.00999677634371045\\
25	20	0.0098161703360947\\
25	21	0.00924172559436162\\
25	22	0.00866484423311221\\
25	23	0.00784600315781332\\
25	24	0.00659879613918002\\
25	25	0.005308257563809\\
25	26	0.00362785578359412\\
25	27	0.00187893085111404\\
25	28	0.00224908296921843\\
25	29	0.00195353029351001\\
25	30	0.00212527214144526\\
26	1	0.000742465295287327\\
26	2	0.000998736139271986\\
26	3	0.0014608076280603\\
26	4	0.00232506252544381\\
26	5	0.00294868139106367\\
26	6	0.00354596455384615\\
26	7	0.00415953223129547\\
26	8	0.00475079955389378\\
26	9	0.00518471296460576\\
26	10	0.00580290189356813\\
26	11	0.00644252291850427\\
26	12	0.0067896711037405\\
26	13	0.00732582706582039\\
26	14	0.00799398932751067\\
26	15	0.00841606513552559\\
26	16	0.00876049871622357\\
26	17	0.00907737887716014\\
26	18	0.00909879842921318\\
26	19	0.00915395122219803\\
26	20	0.0091184336150521\\
26	21	0.00861838074239045\\
26	22	0.00819688053704631\\
26	23	0.00735346775162727\\
26	24	0.00629268554607703\\
26	25	0.00503691444129731\\
26	26	0.00358826887360778\\
26	27	0.00180209488975634\\
26	28	0.00216534169740233\\
26	29	0.00187897965354327\\
26	30	0.00195238799485917\\
27	1	0.000611104563081719\\
27	2	0.000779081948162123\\
27	3	0.00116175451344799\\
27	4	0.00173129526583562\\
27	5	0.00212932842053475\\
27	6	0.00250879576524807\\
27	7	0.00295727597102802\\
27	8	0.00346483954695327\\
27	9	0.00397486142143082\\
27	10	0.00452593260089635\\
27	11	0.00504098655924255\\
27	12	0.00547311390478223\\
27	13	0.00604017270674297\\
27	14	0.00663098721314248\\
27	15	0.00715110081445374\\
27	16	0.00751679962913741\\
27	17	0.00786068878707564\\
27	18	0.00821246271423461\\
27	19	0.00823007201046102\\
27	20	0.00821413686349253\\
27	21	0.00794244797603396\\
27	22	0.00760270516497538\\
27	23	0.00685544828841987\\
27	24	0.00589852384805927\\
27	25	0.00471103662287574\\
27	26	0.00342730198333573\\
27	27	0.00170014414197805\\
27	28	0.00207317516683662\\
27	29	0.00181873938741918\\
27	30	0.0017996534531893\\
28	1	0.000502981823721832\\
28	2	0.000596241683308679\\
28	3	0.000880087521141209\\
28	4	0.00110080334781536\\
28	5	0.00122219924443646\\
28	6	0.00143891801812615\\
28	7	0.00160028277056942\\
28	8	0.00199222627367681\\
28	9	0.00247002250400347\\
28	10	0.00303925894314141\\
28	11	0.00348778766210092\\
28	12	0.00399070593018702\\
28	13	0.00457493247497878\\
28	14	0.00516752402775943\\
28	15	0.00578143539659093\\
28	16	0.00615984605001771\\
28	17	0.00663620861397905\\
28	18	0.00710968699862044\\
28	19	0.00723247162757043\\
28	20	0.00734101473313845\\
28	21	0.00715019537033688\\
28	22	0.00687132140997161\\
28	23	0.00632016415080685\\
28	24	0.00543814692883494\\
28	25	0.00440242311186463\\
28	26	0.00319531540370918\\
28	27	0.00163329272041863\\
28	28	0.00193291478636715\\
28	29	0.00170690186822326\\
28	30	0.00175854130853828\\
29	1	0.00047139415201438\\
29	2	0.000475315660457483\\
29	3	0.000723068353028371\\
29	4	0.000596185183152634\\
29	5	0.000717221462835979\\
29	6	0.000724861974028134\\
29	7	0.000715938463475589\\
29	8	0.000937561759361281\\
29	9	0.00114204560509466\\
29	10	0.00153977082232299\\
29	11	0.00187723305552045\\
29	12	0.00244515726596\\
29	13	0.00300020845050053\\
29	14	0.00359113845048524\\
29	15	0.00431355249410776\\
29	16	0.00478942866055971\\
29	17	0.00535236007450914\\
29	18	0.00586288410937078\\
29	19	0.00617217579870488\\
29	20	0.00638833123951122\\
29	21	0.00631686751607748\\
29	22	0.00609277090925651\\
29	23	0.00569117082472364\\
29	24	0.00493737854213389\\
29	25	0.00408043076141641\\
29	26	0.00289566696127766\\
29	27	0.00152045667290283\\
29	28	0.00182330017741225\\
29	29	0.00158527138997995\\
29	30	0.00167183436735198\\
30	1	0.000595638247852938\\
30	2	0.000503830466839874\\
30	3	0.00084339927954173\\
30	4	0.000564676019287323\\
30	5	0.000697363476478666\\
30	6	0.000622713760601055\\
30	7	0.000553013439612441\\
30	8	0.000654578418173645\\
30	9	0.000454588345883997\\
30	10	0.000589507818192739\\
30	11	0.000727313626434165\\
30	12	0.0012531561419256\\
30	13	0.00147952189406224\\
30	14	0.00194973608595525\\
30	15	0.00268735765842731\\
30	16	0.00328929124702054\\
30	17	0.00396725359714562\\
30	18	0.0045043971766439\\
30	19	0.00494776795616505\\
30	20	0.00534581444371469\\
30	21	0.00543710247487531\\
30	22	0.00531648937873803\\
30	23	0.00504977157205985\\
30	24	0.00441906047778852\\
30	25	0.00369215507446418\\
30	26	0.00264692844587487\\
30	27	0.00135386930095897\\
30	28	0.00167193723535172\\
30	29	0.00142537257728177\\
30	30	0.00157837180973416\\
31	1	0.000761937167014465\\
31	2	0.000602345586580963\\
31	3	0.000914349578448845\\
31	4	0.000476501398866137\\
31	5	0.000731247551555927\\
31	6	0.000794168864548815\\
31	7	0.000534931468981089\\
31	8	0.000690467539247111\\
31	9	0.000629015648877612\\
31	10	0.000570343062020683\\
31	11	0.000715277905225684\\
31	12	0.000634836774155444\\
31	13	0.000521129989031998\\
31	14	0.000806257803183489\\
31	15	0.001175759559185\\
31	16	0.00186031525616288\\
31	17	0.00245277891651362\\
31	18	0.00311071542114879\\
31	19	0.00366960072720931\\
31	20	0.00416949578537723\\
31	21	0.00445606555195871\\
31	22	0.00446687883581664\\
31	23	0.0043574618655955\\
31	24	0.00390326653938525\\
31	25	0.00325165250423983\\
31	26	0.00237104315004298\\
31	27	0.00119955078820242\\
31	28	0.00145804917154817\\
31	29	0.00128124257945136\\
31	30	0.00142650385841692\\
32	1	0.000523667767038012\\
32	2	0.000704407294917996\\
32	3	0.000914415194013708\\
32	4	0.000549668306039208\\
32	5	0.000878589818061658\\
32	6	0.0010664573740908\\
32	7	0.000678239386473965\\
32	8	0.000716242999469544\\
32	9	0.000979989751731864\\
32	10	0.000715492473632225\\
32	11	0.000730281838815846\\
32	12	0.000564521213010602\\
32	13	0.000772412249805136\\
32	14	0.000530777735852109\\
32	15	0.000123206144470932\\
32	16	0.000581022850007231\\
32	17	0.000948823904541972\\
32	18	0.0017245884687784\\
32	19	0.00232826852551446\\
32	20	0.00293904385648644\\
32	21	0.00346010330506475\\
32	22	0.00356688314468479\\
32	23	0.00359719267517543\\
32	24	0.00329370745833071\\
32	25	0.0028103063976517\\
32	26	0.00207915707447384\\
32	27	0.00102560329686614\\
32	28	0.00119103315689278\\
32	29	0.00106776549598812\\
32	30	0.00128429016098512\\
33	1	0.000514016374549036\\
33	2	0\\
33	3	0\\
33	4	0.000467999700198263\\
33	5	0.00132665195547744\\
33	6	0.00159207914838406\\
33	7	0.00133191945534876\\
33	8	0.00114422768552924\\
33	9	0.00206164360768721\\
33	10	0.00130116548981672\\
33	11	0.00121058109430962\\
33	12	0.000588721543222727\\
33	13	0.000740952375049148\\
33	14	0.000207669531707571\\
33	15	0.00118573575987049\\
33	16	0.000912696676213403\\
33	17	0.000672962544766139\\
33	18	0.000210089001714789\\
33	19	0.000838781615477642\\
33	20	0.0016003681947561\\
33	21	0.00234956557454194\\
33	22	0.00268920529910749\\
33	23	0.00280983429371471\\
33	24	0.00273286287113538\\
33	25	0.00232490171713325\\
33	26	0.00171774671469957\\
33	27	0.000830041966987387\\
33	28	0.000938094462448708\\
33	29	0.000839169823382523\\
33	30	0.00106983918429253\\
};

\addplot [color=black,dashed,line width=1pt]
  table[row sep=crcr]{%
17	0\\
17	30\\
};

\addplot [color=black,dashed,line width=1pt]
  table[row sep=crcr]{%
2.13	17\\
31.87	17\\
};

\draw[fill=gray,color=gray,pattern=north west lines,pattern color=gray,rounded corners=2mm] (17-10.86,0) -- (17-10.86,1) -- (17-11.13,4.1) -- (17-11.43,6.65) --  (17-11.8,9.25) -- (17-12.78,11.8) -- (17-13.7,14.4) -- (17-14.88,16.95) -- (17-16.47,19.55) -- (17-17.86,22.1) -- (17-18.84,24.7) -- (17-19.55,27.25) -- (17-19.55,1) --  (17-19.55,0) -- cycle;
\draw[fill=gray,color=gray,pattern=north west lines,pattern color=gray,rounded corners=2mm] (17+10.86,0) -- (17+10.86,1) -- (17+11.13,4.1) -- (17+11.43,6.65) --  (17+11.8,9.25) -- (17+12.78,11.8) -- (17+13.7,14.4) -- (17+14.88,16.95) -- (17+16.47,19.55) -- (17+17.86,22.1) -- (17+18.84,24.7) -- (17+19.55,27.25) -- (17+19.55,1) --  (17+19.55,0) -- cycle;
\end{axis}
\end{tikzpicture}%

%% file: Figures/Results_Exp/2019-10-04_SuspensionPIV2/COMPARISON_V2DxxyMean_z=17.tikz
%
%
\definecolor{mycolor1}{rgb}{1.00000,0.00000,1.00000}%
\begin{tikzpicture}

\begin{axis}[%
width=1\linewidth,
height=.8\linewidth,
xmin=-220,
xmax=220,
xtick={-200,0,200},
xticklabels={200,0,-200},
xlabel={y / \si{\micro\meter}},
ymin=0,
ymax=0.017,
ylabel={u / \si{\meter\per\second}},
axis background/.style={fill=white},
legend style={at={(axis cs:-219,0.01695)},anchor=north west,legend cell align=left, align=left, draw=none}
]
\addplot [color=black, draw=none, mark=x, mark options={solid, black},mark size=2,line width=.5pt]
 plot [error bars/.cd, y dir = both, y explicit,error bar style={line width=1pt,solid}]
 table[row sep=crcr, y error plus index=2, y error minus index=3, x expr=\thisrowno{0} - 237.42]{%
29.6780802309513	0.00260716477068379	0.000102835152837599	0.000102835152837599\\
44.517120346427	0.00407362353270324	8.86198439808972e-05	8.86198439808972e-05\\
59.3561604619026	0.0054096833112971	9.02366673127819e-05	9.02366673127819e-05\\
74.1952005773783	0.00668293817131849	8.98628285117553e-05	8.98628285117553e-05\\
89.0342406928539	0.00791168262797348	0.00010672449153093	0.00010672449153093\\
103.87328080833	0.00910975815131776	9.41554885613982e-05	9.41554885613982e-05\\
118.712320923805	0.0100730676248262	7.59022489327243e-05	7.59022489327243e-05\\
133.551361039281	0.010971480097639	6.00170481446227e-05	6.00170481446227e-05\\
148.390401154757	0.0117436812891791	5.91573368324827e-05	5.91573368324827e-05\\
163.229441270232	0.0123628585843671	7.45189274387048e-05	7.45189274387048e-05\\
178.068481385708	0.0128574980281204	9.23558733867402e-05	9.23558733867402e-05\\
192.907521501184	0.0131926226961498	9.36023958242327e-05	9.36023958242327e-05\\
207.746561616659	0.0134616333236015	0.000106670243629203	0.000106670243629203\\
222.585601732135	0.013591423787086	9.35516858291408e-05	9.35516858291408e-05\\
237.42464184761	0.0136300907479947	0.000141664857451484	0.000141664857451484\\
252.263681963086	0.013606226388444	0.000107882566302564	0.000107882566302564\\
267.102722078562	0.0134410922376826	9.11411921089567e-05	9.11411921089567e-05\\
281.941762194037	0.0131079405422529	0.000148332632154968	0.000148332632154968\\
296.780802309513	0.0126841005067635	0.000192431459049569	0.000192431459049569\\
311.619842424989	0.0121910908586664	0.000195783035934483	0.000195783035934483\\
326.458882540464	0.0115503535604489	0.00015968223663787	0.00015968223663787\\
341.29792265594	0.0108021028175209	0.000160720022583938	0.000160720022583938\\
356.136962771416	0.00992770565494183	0.000168447989975506	0.000168447989975506\\
370.976002886891	0.00907732969910757	8.16761125599363e-05	8.16761125599363e-05\\
385.815043002367	0.00786056341501418	0.000164064523655583	0.000164064523655583\\
400.654083117843	0.00663603614306729	0.000169519975685706	0.000169519975685706\\
415.493123233318	0.00535232323103485	0.000106604202705363	0.000106604202705363\\
430.332163348794	0.00396723985951142	5.57975048320891e-05	5.57975048320891e-05\\
445.17120346427	0.0024521808740563	6.11444610587597e-05	6.11444610587597e-05\\
};
\addlegendentry{OPF}

\addplot [color=blue, draw=none, mark=triangle, mark options={solid, blue},mark size=2,line width=.5pt]
 plot [error bars/.cd, y dir = both, y explicit,error bar style={line width=1pt,solid}]
 table[row sep=crcr, y error plus index=2, y error minus index=3, x expr=\thisrowno{0} - 237.42]{%
29.6780802309513	0.0028556873808161	6.51540885776569e-05	6.51540885776569e-05\\
44.517120346427	0.00437653556082718	5.5304556507707e-05	5.5304556507707e-05\\
59.3561604619026	0.00576900490210086	6.11621505919313e-05	6.11621505919313e-05\\
74.1952005773783	0.00702771660725077	6.09380831717576e-05	6.09380831717576e-05\\
89.0342406928539	0.00818483496479172	5.59791173726509e-05	5.59791173726509e-05\\
103.87328080833	0.0092205146771096	4.28794284079714e-05	4.28794284079714e-05\\
118.712320923805	0.0101874788081713	3.99759863634053e-05	3.99759863634053e-05\\
133.551361039281	0.0110244201532128	4.95071068362662e-05	4.95071068362662e-05\\
148.390401154757	0.0117510719761382	5.88884559282743e-05	5.88884559282743e-05\\
163.229441270232	0.0123770809690371	5.8153750650547e-05	5.8153750650547e-05\\
178.068481385708	0.0129005236900026	4.45552168504282e-05	4.45552168504282e-05\\
192.907521501184	0.0132883891115321	3.37268639449828e-05	3.37268639449828e-05\\
207.746561616659	0.013541488593547	3.92990668940385e-05	3.92990668940385e-05\\
222.585601732135	0.013680774798438	3.56196439943446e-05	3.56196439943446e-05\\
237.42464184761	0.0137553668426138	2.69906897132355e-05	2.69906897132355e-05\\
252.263681963086	0.0137324789452942	3.01712677774901e-05	3.01712677774901e-05\\
267.102722078562	0.0135746776956835	2.458727180632e-05	2.458727180632e-05\\
281.941762194037	0.013309181388822	2.19173315996191e-05	2.19173315996191e-05\\
296.780802309513	0.0129228985908603	3.74345900977513e-05	3.74345900977513e-05\\
311.619842424989	0.0123974262907889	3.25110033649878e-05	3.25110033649878e-05\\
326.458882540464	0.0117828801153852	1.2238798350328e-05	1.2238798350328e-05\\
341.29792265594	0.0110242184925347	3.6402700662741e-05	3.6402700662741e-05\\
356.136962771416	0.010153677648187	2.71274887697626e-05	2.71274887697626e-05\\
370.976002886891	0.00915573796523961	2.13778008378851e-05	2.13778008378851e-05\\
385.815043002367	0.00799639190582124	2.79730484553653e-05	2.79730484553653e-05\\
400.654083117843	0.00674350949150249	2.72147571334092e-05	2.72147571334092e-05\\
415.493123233318	0.0053795733672346	2.16177888379132e-05	2.16177888379132e-05\\
430.332163348794	0.00390307166758252	3.89405590217606e-05	3.89405590217606e-05\\
445.17120346427	0.00232123115649796	3.2032206667143e-05	3.2032206667143e-05\\
};
\addlegendentry{SCL}

\addplot [color=red, draw=none, mark=square, mark options={solid,red},mark size=2,line width=.5pt]
 plot [error bars/.cd, y dir = both, y explicit,error bar style={line width=1pt,solid}]
 table[row sep=crcr, y error plus index=2, y error minus index=3, x expr=\thisrowno{0} - 237.42]{%
29.6780802309513	0.00268358163761903	0.000594687758052465	0.000594687758052465\\
59.3561604619026	0.00526757621551632	0.000369256769696819	0.000369256769696819\\
89.0342406928539	0.00748349096732381	0.000374315239120016	0.000374315239120016\\
118.712320923805	0.00931161341433895	0.000235955965767198	0.000235955965767198\\
148.390401154757	0.010758180885958	0.000226963786404966	0.000226963786404966\\
178.068481385708	0.0118136823098458	0.00017335506647731	0.00017335506647731\\
207.746561616659	0.0124779514043905	0.00012865951266372	0.00012865951266372\\
237.42464184761	0.0127355487450398	0.000122542472092979	0.000122542472092979\\
267.102722078562	0.0125393364431003	0.000276715008799	0.000276715008799\\
296.780802309513	0.0120053212778093	0.000493826904529593	0.000493826904529593\\
326.458882540464	0.0112126084829056	0.00033228372900207	0.00033228372900207\\
356.136962771416	0.0101336619414033	0.000227841776901383	0.000227841776901383\\
385.815043002367	0.00839335092670312	0.000207602002697538	0.000207602002697538\\
415.493123233318	0.00612435932026534	0.000197523391173016	0.000197523391173016\\
445.17120346427	0.00343267042218098	0.000314307625393084	0.000314307625393084\\
};
\addlegendentry{SP}

\end{axis}
\end{tikzpicture}%

%% file: Figures/Results_Exp/2019-10-04_SuspensionPIV2/COMPARISON_V2DxxzMean_y=17.tikz
%
%
\definecolor{mycolor1}{rgb}{1.00000,0.00000,1.00000}%
\def\S{100}
\begin{tikzpicture}
\begin{axis}[%
width=1\linewidth,
height=.8\linewidth,
xmin=0,
xmax=0.015,
xtick={0,0.005,0.01,0.015},
xticklabels={0,0.5,1,1.5},
xlabel={u / \si{\meter\per\second}},
ymin=0,
ymax=540,
ytick={0,125,250,375,500},
yticklabels = {0,125,250,375,500},
ylabel={z / \si{\micro\meter}},
axis background/.style={fill=white},
legend style={at={(axis cs:0.0005,490)},anchor=north west,legend cell align=left, align=left, draw=white!15!black,draw=none}
]
\addplot [color=black, draw=none, mark=x, mark options={solid, black},mark size=2,line width=.5]
plot [error bars/.cd, x dir=both, x explicit, error bar style={line width=1pt,solid}]
  table[row sep=crcr,x error plus index=2,x error minus index=2]{%
x	y	xerrpos	xerrneg\\
0.00120960909897216	0	0.00009242\\
0.00179753016834816	20	0.00007019\\
0.0027630779574539	40	0.0001446\\
0.00454147630602586	60	0.0001063\\
0.00598061889092172	80	0.0001588\\
0.00725764987662414	100	0.0001477\\
0.0085759800257712	120	0.0001063\\
0.00972046451872099	140	0.0001252\\
0.0105756260829293	160	0.000213\\
0.0111917430888786	180	0.0002623\\
0.0118764494907475	200	0.0001817\\
0.0121122427128095	220	0.0003232\\
0.0129252996501628	240	0.0001154\\
0.0133443470014649	260	0.0002551\\
0.013352250684886	280	0.0002615\\
0.0136909050044343	300	0.0001228\\
0.0136300907479947	320	0.0001417\\
0.0132720711068324	340	0.0001509\\
0.0129636859371451	360	0.0001007\\
0.0123549407493195	380	0.0001681\\
0.0117137435232395	400	0.00009687\\
0.0107387872610909	420	0.00007142\\
0.00960576440633409	440	0.00007799\\
0.00812801028773347	460	0.00006253\\
0.00630715705954965	480	0.00006965\\
0.00409644656840125	500	0.00007407\\
0.00197036000310305	520	0.0001131\\
};
\addlegendentry{OPF}

\addplot [color=blue, draw=none, mark=triangle, mark options={solid, blue},mark size=2,line width=0.5pt]
plot [error bars/.cd, x dir=both, x explicit, error bar style={line width=1pt,solid}]
  table[row sep=crcr,x error plus index=2,x error minus index=2]{%
x	y	xerrpos	xerrneg\\
0.00011362694737449	0	0.00006609\\
0.00130297327574969	20	0.0001075\\
0.00271595422038695	40	0.0001012\\
0.0040889323495355	60	0.0001734\\
0.00556291712344281	80	0.0001467\\
0.0068490570394264	100	0.0001037\\
0.0079295797183343	120	0.00008675\\
0.00879063543499536	140	0.0001315\\
0.00928586217124994	160	0.00008446\\
0.0100663915950072	180	0.0001331\\
0.0107477546751067	200	0.00007618\\
0.0114164968629393	220	0.00005746\\
0.0118246462825056	240	0.00005929\\
0.0122650295694171	260	0.000058\\
0.012978867094513	280	0.0000795\\
0.0135925193588404	300	0.0000488\\
0.0137553668426138	320	0.00002699\\
0.0136282781604957	340	0.00003138\\
0.0133494380491117	360	0.00004527\\
0.0126719206735237	380	0.00008582\\
0.0118000849553479	400	0.00004176\\
0.0107427119788506	420	0.00006665\\
0.00956885342641822	440	0.00008261\\
0.00812808930098163	460	0.0001183\\
0.00649208933583667	480	0.00009694\\
0.00420254720906232	500	0.00006815\\
0.0021312337772128	520	0.00005348\\
};
\addlegendentry{SCL}

\addplot [color=red, draw=none, mark=square, mark options={solid,red},mark size=2,line width=.5pt]
plot [error bars/.cd, x dir=both, x explicit, error bar style={line width=1pt,solid}]
  table[row sep=crcr,x error plus index=2,x error minus index=2]{%
x	y	xerrpos	xerrneg\\
0.000929294859823848	20	0.00009393\\
0.00251443387919381	40	0.0003119\\
0.00306880380065433	60	0.0001361\\
0.00494397627834819	80	0.0001083\\
0.00572217481128456	100	0.0001641\\
0.00700900933627069	120	0.0001607\\
0.00719132709955494	140	0.0003324\\
0.00805852698073761	160	0.0001945\\
0.00870638372640564	180	0.0004291\\
0.00969565200019229	200	0.0002887\\
0.0104803573330803	220	0.0001791\\
0.0112124492771071	240	0.0004011\\
0.0116530530935322	260	0.0003664\\
0.0121628996393019	280	0.0001796\\
0.011591477007864	300	0.0002456\\
0.0127355487450398	320	0.0001225\\
0.0124800411279091	340	0.0005598\\
0.0124049225252959	360	0.0003322\\
0.0114790989415314	380	0.0005411\\
0.0100385571145617	400	0.0004775\\
0.00964445524811258	420	0.0005542\\
0.00769174866205091	440	0.0008695\\
0.0075332575206466	460	0.0007028\\
0.00634456334639433	480	0.0002078\\
0.00479622253616627	500	0.0006162\\
};
\addlegendentry{SP}

\end{axis}
\end{tikzpicture}%

%% file: Figures/Results_Exp/2019-10-04_SuspensionPIV2/NoP-z_y=9.tikz
%
%
\definecolor{mycolor1}{rgb}{0.00000,0.44700,0.74100}%
\begin{tikzpicture}

\begin{axis}[%
width=1\columnwidth,
height=.8\columnwidth,
bar shift auto,
xmin=0,
xmax=45,
xlabel style={font=\color{white!15!black}},
xlabel={$N_I/N_{IW}$},
ymin=0,
ymax=540,
ylabel style={font=\color{white!15!black}},
ytick={0,125,250,375,500},
ylabel={z / \si{\micro\meter}},
axis background/.style={fill=white},
axis x line*=bottom,
axis y line*=left,
legend style={legend cell align=left, align=left, draw=white!15!black}
]
\addplot[xbar, bar width=16, fill=red, draw=black, area legend] table[row sep=crcr] {%
11.25	20\\
17.15	40\\
21.2	60\\
25.2	80\\
28.1	100\\
28.95	120\\
32.85	140\\
34.6	160\\
35.2	180\\
38.1	200\\
37.3	220\\
40.55	240\\
39.45	260\\
30.75	280\\
7.75	300\\
3.4	320\\
2.1	340\\
1.3	360\\
0.75	380\\
1.15	400\\
1.05	420\\
1	440\\
1.1	460\\
0.9	480\\
1.5	500\\
0.7	520\\
};
\end{axis}
\end{tikzpicture}%

%% file: SuspensionPIV.bbl
\begin{thebibliography}{31}
\providecommand{\natexlab}[1]{#1}
\providecommand{\url}[1]{\texttt{#1}}
\expandafter\ifx\csname urlstyle\endcsname\relax
  \providecommand{\doi}[1]{doi: #1}\else
  \providecommand{\doi}{doi: \begingroup \urlstyle{rm}\Url}\fi

\bibitem[Adrian(1991)]{Adrian1991}
R.~J. Adrian.
\newblock {Particle-Imaging Techniques for Experimental Fluid Mechanics}.
\newblock \emph{Annual Review of Fluid Mechanics}, 23\penalty0 (1):\penalty0
  261--304, 1991.
\newblock ISSN 0066-4189.
\newblock \doi{10.1146/annurev.fl.23.010191.001401}.

\bibitem[Adrian and Westerweel(2011)]{adrian2011particle}
R.~J. Adrian and J.~Westerweel.
\newblock \emph{{Particle Image Velocimetry}}.
\newblock Cambridge Aerospace Series. Cambridge University Press, 2011.
\newblock ISBN 9780521440080.

\bibitem[Anders et~al.(2019)Anders, Noto, Seilmayer, and Eckert]{Anders2019a}
S.~Anders, D.~Noto, M.~Seilmayer, and S.~Eckert.
\newblock {Spectral random masking: a novel dynamic masking technique for PIV
  in multiphase flows}.
\newblock \emph{Experiments in Fluids}, 60\penalty0 (4):\penalty0 1--6, 2019.
\newblock ISSN 07234864.
\newblock \doi{10.1007/s00348-019-2703-8}.
\newblock URL \url{http://dx.doi.org/10.1007/s00348-019-2703-8}.

\bibitem[Bailey and Yoda(2003)]{Bailey2003a}
B.~C. Bailey and M.~Yoda.
\newblock {An aqueous low-viscosity density- and refractive index-matched
  suspension system}.
\newblock \emph{Experiments in Fluids}, 35\penalty0 (1):\penalty0 1--3, 2003.
\newblock ISSN 07234864.
\newblock \doi{10.1007/s00348-003-0598-9}.

\bibitem[Blanc et~al.(2013)Blanc, Lemaire, Meunier, and Peters]{Blanc2013a}
F.~Blanc, E.~Lemaire, A.~Meunier, and F.~Peters.
\newblock {Microstructure in sheared non-Brownian concentrated suspensions}.
\newblock \emph{Journal of Rheology}, 57\penalty0 (1):\penalty0 273--292, 2013.
\newblock ISSN 0148-6055.
\newblock \doi{10.1122/1.4766597}.

\bibitem[Brenner(1966)]{Brenner1966}
H.~Brenner.
\newblock {Hydrodynamic Resistance of Particles at Small Reynolds Numbers}.
\newblock In T.~B. Drew, J.~W. Hoopes, and T.~Vermeulen, editors,
  \emph{Advances in Chemical Engineering}, volume~6 of \emph{Advances in
  Chemical Engineering}, pages 287--438. Academic Press, 1966.
\newblock \doi{https://doi.org/10.1016/S0065-2377(08)60277-X}.
\newblock URL
  \url{http://www.sciencedirect.com/science/article/pii/S006523770860277X}.

\bibitem[Byron and Variano(2013)]{Byron2013a}
M.~L. Byron and E.~A. Variano.
\newblock {Refractive-index-matched hydrogel materials for measuring
  flow-structure interactions}.
\newblock \emph{Experiments in Fluids}, 54\penalty0 (2), 2013.
\newblock ISSN 07234864.
\newblock \doi{10.1007/s00348-013-1456-z}.

\bibitem[Chen et~al.(2005)Chen, Mikami, and Nishikawa]{Chen2005a}
B.~Chen, F.~Mikami, and N.~Nishikawa.
\newblock {Experimental studies on transient features of natural convection in
  particles suspensions}.
\newblock \emph{International Journal of Heat and Mass Transfer}, 48\penalty0
  (14):\penalty0 2933--2942, 2005.
\newblock ISSN 00179310.
\newblock \doi{10.1016/j.ijheatmasstransfer.2004.11.016}.

\bibitem[Coupland and Pickering(1988)]{Coupland1988a}
J.~M. Coupland and C.~J. Pickering.
\newblock {Particle image velocimetry: Estimation of measurement confidence at
  low seeding densities}.
\newblock \emph{Optics and Lasers in Engineering}, 9\penalty0 (3-4):\penalty0
  201--210, 1 1988.
\newblock ISSN 01438166.
\newblock \doi{10.1016/S0143-8166(98)90003-3}.
\newblock URL
  \url{https://linkinghub.elsevier.com/retrieve/pii/S0143816698900033}.

\bibitem[Feng et~al.(1994)Feng, Joseph, and Hu]{Feng1994d}
J.~Feng, D.~D. Joseph, and H.~H. Hu.
\newblock {Direct simulation of initial value problems for the motion of solid
  bodies in a Newtonian fluid. part 2. Couette and Poiseuille flows}.
\newblock \emph{Journal of Fluid Mechanics}, 277:\penalty0 271--301, 1994.
\newblock ISSN 14697645.
\newblock \doi{10.1017/S0022112094002764}.

\bibitem[Funatani et~al.(2004)Funatani, Fujisawa, and Ikeda]{Funatani2004}
S.~Funatani, N.~Fujisawa, and H.~Ikeda.
\newblock {Simultaneous measurement of temperature and velocity using
  two-colour LIF combined with PIV with a colour CCD camera and its application
  to the turbulent buoyant plume}.
\newblock \emph{Measurement Science and Technology}, 15\penalty0 (5):\penalty0
  983--990, 2004.
\newblock ISSN 09570233.
\newblock \doi{10.1088/0957-0233/15/5/030}.

\bibitem[Guazzelli and Morris(2011)]{Guazzelli2011}
E.~Guazzelli and J.~F. Morris.
\newblock \emph{{A physical introduction to suspension dynamics}}, volume~45.
\newblock Cambridge University Press, 2011.
\newblock ISBN 9781139503938.

\bibitem[Haam et~al.(2000)Haam, Brodkey, Fort, Klaboch, Placnik, and
  Vanecek]{Haam2000b}
S.~J. Haam, R.~S. Brodkey, I.~Fort, L.~Klaboch, M.~Placnik, and V.~Vanecek.
\newblock {Laser Doppler anemometry measurements in an index of refraction
  matched column in the presence of dispersed beads: Part I}.
\newblock \emph{International Journal of Multiphase Flow}, 26\penalty0
  (9):\penalty0 1401--1418, 2000.
\newblock ISSN 03019322.
\newblock \doi{10.1016/S0301-9322(99)00094-4}.

\bibitem[Hassan and Dominguez-Ontiveros(2008)]{Hassan2008a}
Y.~A. Hassan and E.~E. Dominguez-Ontiveros.
\newblock {Flow visualization in a pebble bed reactor experiment using PIV and
  refractive index matching techniques}.
\newblock \emph{Nuclear Engineering and Design}, 238\penalty0 (11):\penalty0
  3080--3085, 2008.
\newblock ISSN 00295493.
\newblock \doi{10.1016/j.nucengdes.2008.01.027}.

\bibitem[Keane and Adrian(1990)]{Keane1990a}
R.~D. Keane and R.~J. Adrian.
\newblock {Optimization of particle image velocimeters. I. Double pulsed
  systems Optimization of particle image ve I oci m e te rs. Part I: Double
  pulsed systems}.
\newblock \emph{Meas. Sci. Technol. Meas. Sci. Technol}, 1\penalty0
  (1):\penalty0 1202--1215, 1990.
\newblock URL \url{http://iopscience.iop.org/0957-0233/1/11/013}.

\bibitem[Keane and Adrian(1992)]{Keane1992}
R.~D. Keane and R.~J. Adrian.
\newblock {Theory of cross-correlation analysis of PIV images}.
\newblock \emph{Applied Scientific Research}, 49\penalty0 (3):\penalty0
  191--215, 7 1992.
\newblock ISSN 0003-6994.
\newblock \doi{10.1007/BF00384623}.
\newblock URL \url{http://link.springer.com/10.1007/BF00384623}.

\bibitem[Kordel et~al.(2016)Kordel, Nowak, Skoda, and Hussong]{Kordel2016a}
S.~Kordel, T.~Nowak, R.~Skoda, and J.~Hussong.
\newblock {Combined density gradient and velocity field measurements in
  transient flows by means of Differential Interferometry and Long-range
  {$\mu$} PIV}.
\newblock \emph{Experiments in Fluids}, 57\penalty0 (9):\penalty0 1--11, 2016.
\newblock ISSN 07234864.
\newblock \doi{10.1007/s00348-016-2224-7}.

\bibitem[Koutsiaris et~al.(1999)Koutsiaris, Mathioulakis, and
  Tsangaris]{Koutsiaris1999b}
A.~G. Koutsiaris, D.~S. Mathioulakis, and S.~Tsangaris.
\newblock {Microscope PIV for velocity-field measurement of particle
  suspensions flowing inside glass capillaries}.
\newblock \emph{Measurement Science and Technology}, 10\penalty0 (11):\penalty0
  1037--1046, 1999.
\newblock ISSN 09570233.
\newblock \doi{10.1088/0957-0233/10/11/311}.

\bibitem[Lima et~al.(2006)Lima, Wada, Tsubota, and Yamaguchi]{Lima2006b}
R.~Lima, S.~Wada, K.~I. Tsubota, and T.~Yamaguchi.
\newblock {Confocal micro-PIV measurements of three-dimensional profiles of
  cell suspension flow in a square microchannel}.
\newblock \emph{Measurement Science and Technology}, 17\penalty0 (4):\penalty0
  797--808, 2006.
\newblock ISSN 13616501.
\newblock \doi{10.1088/0957-0233/17/4/026}.

\bibitem[Lindken and Merzkirch(2002)]{Lindken2002}
R.~Lindken and W.~Merzkirch.
\newblock {A novel PIV technique for measurements in multiphase flows and its
  application to two-phase bubbly flows}.
\newblock \emph{Experiments in Fluids}, 33\penalty0 (6):\penalty0 814--825,
  2002.
\newblock ISSN 07234864.
\newblock \doi{10.1007/s00348-002-0500-1}.

\bibitem[Loisel et~al.(2015)Loisel, Abbas, Masbernat, and Climent]{Loisel2015}
V.~Loisel, M.~Abbas, O.~Masbernat, and E.~Climent.
\newblock {Inertia-driven particle migration and mixing in a wall-bounded
  laminar suspension flow}.
\newblock \emph{Physics of Fluids}, 27\penalty0 (12), 2015.
\newblock ISSN 10897666.
\newblock \doi{10.1063/1.4936402}.
\newblock URL \url{http://dx.doi.org/10.1063/1.4936402}.

\bibitem[Olsen and Adrian(2000)]{Olsen2000}
M.~G. Olsen and R.~J. Adrian.
\newblock {Out-of-focus effects on particle image visibility and correlation in
  microscopic particle image velocimetry}.
\newblock \emph{Experiments in Fluids}, 29\penalty0 (7):\penalty0 S166--S174,
  12 2000.
\newblock ISSN 0723-4864.
\newblock \doi{10.1007/s003480070018}.
\newblock URL \url{http://link.springer.com/10.1007/s003480070018}.

\bibitem[Raffel et~al.(2007)Raffel, Willert, Wereley, and
  Kompenhans]{Raffel2007}
M.~Raffel, C.~E. Willert, S.~Wereley, and J.~Kompenhans.
\newblock \emph{{Particle Image Velocimetry: A Practical Guide}}.
\newblock Experimental Fluid Mechanics. Springer Berlin Heidelberg, 2007.
\newblock ISBN 9783540723073.

\bibitem[Scharnowski et~al.(2018)Scharnowski, Sciacchitano, and
  K{\"{a}}hler]{Scharnowski2018}
S.~Scharnowski, A.~Sciacchitano, and C.~J. K{\"{a}}hler.
\newblock {A new look on the “Valid Detection Probability” of PIV Vectors}.
\newblock In \emph{LXLASER2018: 19th International Symposium on the Application
  of Laser and Imaging Techniques to Fluid Mechanics}, 2018.

\bibitem[Skarman et~al.(1996)Skarman, Becker, and Wozniak]{Skarman1996a}
B.~Skarman, J.~Becker, and K.~Wozniak.
\newblock {Simultaneous 3D-PIV and temperature measurements using a new
  CCD-based holographic interferometer}, 1996.
\newblock ISSN 09555986.

\bibitem[Wang et~al.(2008)Wang, Song, Briscoe, and Makse]{Wang2008a}
P.~Wang, C.~Song, C.~Briscoe, and H.~A. Makse.
\newblock {Particle dynamics and effective temperature of jammed granular
  matter in a slowly sheared three-dimensional Couette cell}.
\newblock \emph{Physical Review E - Statistical, Nonlinear, and Soft Matter
  Physics}, 77\penalty0 (6):\penalty0 1--15, 2008.
\newblock ISSN 15393755.
\newblock \doi{10.1103/PhysRevE.77.061309}.

\bibitem[Westerweel(1993)]{Westerweel1993}
J.~Westerweel.
\newblock {Analysis of PIV interrogation with low-pixel resolution}.
\newblock In S.~S. Cha and J.~D. Trolinger, editors, \emph{Optical Diagnostics
  in Fluid and Thermal Flow}, volume 2005, pages 624--635. International
  Society for Optics and Photonics, SPIE, 12 1993.
\newblock \doi{10.1117/12.163745}.
\newblock URL \url{https://doi.org/10.1117/12.163745
  http://proceedings.spiedigitallibrary.org/proceeding.aspx?articleid=934839}.

\bibitem[Westerweel(1997)]{Westerweel1997b}
J.~Westerweel.
\newblock {Fundamentals of digital particle image velocimetry}.
\newblock \emph{Measurement Science and Technology}, 8\penalty0 (12):\penalty0
  1379--1392, 1997.
\newblock ISSN 09570233.
\newblock \doi{10.1088/0957-0233/8/12/002}.

\bibitem[Wiederseiner et~al.(2011)Wiederseiner, Andreini, Epely-Chauvin, and
  Ancey]{Wiederseiner2011a}
S.~Wiederseiner, N.~Andreini, G.~Epely-Chauvin, and C.~Ancey.
\newblock {Refractive-index and density matching in concentrated particle
  suspensions: A review}.
\newblock \emph{Experiments in Fluids}, 50\penalty0 (5):\penalty0 1183--1206,
  2011.
\newblock ISSN 07234864.
\newblock \doi{10.1007/s00348-010-0996-8}.

\bibitem[Willert(1996)]{Willert1996a}
C.~E. Willert.
\newblock {The fully digital evaluation of photographic PIV recordings}.
\newblock \emph{Flow, Turbulence and Combustion}, 56\penalty0 (2-3):\penalty0
  79--102, 1996.
\newblock ISSN 13866184.

\bibitem[Willert and Gharib(1991)]{Willert1991a}
C.~E. Willert and M.~Gharib.
\newblock {Digital particle image velocimetry}.
\newblock \emph{Experiments in Fluids}, 10\penalty0 (4):\penalty0 181--193,
  1991.
\newblock ISSN 07234864.
\newblock \doi{10.1007/BF00190388}.

\end{thebibliography}
